\documentclass[aps,prd,superscriptaddress,groupedaddress,preprintnumbers,floatfix,nofootinbib,12pt,tightenlines]{revtex4}  % for review and submission

\usepackage[utf8]{inputenc}
\usepackage{graphicx,comment}  % needed for figures
\graphicspath{{figures/}}
\usepackage{dcolumn}   % needed for some tables
\usepackage{bm}        % for math
\usepackage{amssymb}   % for math
\usepackage{multirow}

\usepackage{subfigure}
\usepackage{multirow}
\usepackage{slashed}
\usepackage{placeins}
\usepackage{epstopdf}
\usepackage{mathtools}
\usepackage{color}

\usepackage[T1]{fontenc}
\usepackage{amsfonts}    
\usepackage{amsmath}
\usepackage{dcolumn}

\usepackage{tikz}
\usetikzlibrary{shapes,arrows,chains}

\usepackage[hypertexnames=false]{hyperref}
\hypersetup{
    colorlinks=true,       % false: boxed links; true: colored links
    linkcolor=blue,          % color of internal links
    citecolor=blue,        % color of links to bibliography
    filecolor=blue,      % color of file links
    urlcolor=blue           % color of external links
}

\newcolumntype{d}[1]{D{.}{.}{1.8}}

\newcommand{\ket}[1]{\left| #1 \right>} % for Dirac bras
\newcommand{\braket}[2]{\left< #1 \vphantom{#2} \right|
 \left. #2 \vphantom{#1} \right>} % for Dirac brackets
\newcommand{\mbraket}[3]{\left< #1 \vphantom{#2#3} \right|
 #2 \left| #3 \vphantom{#1#2} \right>} % for Dirac matrix elements
\renewcommand{\v}[1]{\ensuremath{\mathbf{#1}}} % for vectors
\newcommand{\msbar}{\overline{\text{MS}}}

\begin{document}

\preprint{DESY 20-28}
\preprint{FERMILAB-PUB-20-123-T}
\preprint{ICCUB-20-007}
\preprint{MIT-CTP/5183}
\preprint{NT@UW-20-03}

\title{Charged multi-hadron systems in lattice QCD+QED}
\author{S.~R.~Beane}\affiliation{Department of Physics, University of Washington,Box 351560, Seattle, WA 98195, U.S.A.}
\author{W.~Detmold}\affiliation{Center for Theoretical Physics, Massachusetts Institute of Technology, Cambridge, MA 02139, U.S.A.}
\author{R.~Horsley} \affiliation{School of Physics and Astronomy, University of Edinburgh, Edinburgh EH9 3FD, UK}
\author{M.~Illa} \affiliation{Departament de Física Quàntica i Astrofísica, Institut de Ciències del Cosmos (ICCUB), Universitat de Barcelona, Martí Franquès 1, E08028 Barcelona, Spain}
\author{M.~Jafry}\affiliation{Department of Physics, University of Washington,Box 351560, Seattle, WA 98195, U.S.A.}
\author{D.~J.~Murphy}\affiliation{Center for Theoretical Physics, Massachusetts Institute of Technology, Cambridge, MA 02139, U.S.A.}
\author{Y.~Nakamura} \affiliation{RIKEN Center for Computational Science, Kobe, Hyogo 650-0047, Japan}
\author{H.~Perlt} \affiliation{Institut f\"ur Theoretische Physik, Universit\"at Leipzig, 04109 Leipzig, Germany}
\author{P.E.L.~Rakow} \affiliation{Theoretical Physics Division, Department of Mathematical Sciences,University of Liverpool, Liverpool L69 3BX, UK}
\author{G.~Schierholz} \affiliation{Deutsches Elektronen-Synchrotron DESY, 22603 Hamburg, Germany}
\author{P.~E.~Shanahan}\affiliation{Center for Theoretical Physics, Massachusetts Institute of Technology, Cambridge, MA 02139, U.S.A.}
\author{H.~St\"uben} \affiliation{Regionales Rechenzentrum, Universit\"at Hamburg, 20146 Hamburg, Germany}
\author{M.~L.~Wagman}\affiliation{Center for Theoretical Physics, Massachusetts Institute of Technology, Cambridge, MA 02139, U.S.A.}
\affiliation{Fermi National Accelerator Laboratory, Batavia, IL 60510, U.S.A.}
\author{F.~Winter} \affiliation{Jefferson Laboratory, 12000 Jefferson Avenue, Newport News, VA 23606, USA}
\author{R.~D.~Young}\affiliation{CSSM, Department of Physics, University of Adelaide, Adelaide SA 5005, Australia}
\author{J.~M.~Zanotti}\affiliation{CSSM, Department of Physics, University of Adelaide, Adelaide SA 5005, Australia}

\collaboration{NPLQCD and QCDSF collaborations}
\begin{abstract}
  Systems with the quantum numbers of up to twelve charged and neutral pseudoscalar mesons, as well as one-, two-, and three-nucleon systems, are studied using dynamical lattice quantum chromodynamics and quantum electrodynamics (QCD+QED) calculations and effective field theory.
QED effects on hadronic interactions are determined by comparing systems of charged and neutral hadrons after tuning the quark masses to remove strong isospin breaking effects.
A non-relativistic effective field theory, which perturbatively includes finite-volume Coulomb effects, is analyzed for systems of multiple charged hadrons and found to accurately reproduce the lattice QCD+QED results.
QED effects on charged multi-hadron systems beyond Coulomb photon exchange are determined by comparing the two- and three-body interaction parameters extracted from the lattice QCD+QED results for charged and neutral multi-hadron systems.

\end{abstract}

\maketitle

\section{Introduction}

The interplay between the strong and electromagnetic (EM) interactions of the Standard Model is subtle but central to the complexity of visible matter. While EM interactions are typically much weaker than the strong interactions, it is their competition with strong isospin-breaking effects that leads to the observed proton--neutron mass difference and determines the stability of atomic nuclei.
Moreover, the hierarchy between the length scales where strong and EM interactions are important leads to the existence of chemistry and all of the complexity that it entails.
Progress towards understanding the combined effects of the strong and EM interactions from first principles has been made in recent years, and the EM--strong interaction decomposition of the neutron-proton mass difference, $M_n - M_p$, has been calculated from the Standard Model for the first time using  the numerical  lattice formulation of quantum chromodynamics  (QCD) and quantum electrodynamics (QED)   \cite{Borsanyi:2013lga,Borsanyi:2014jba,Horsley:2015eaa}.
Coupled lattice QCD+QED calculations have also been performed for meson and baryon masses~\cite{Horsley:2015vla,Basak:2018yzz,Horsley:2019wha,Kordov:2019oer}, leptonic decay rates~\cite{Lubicz:2016xro,Giusti:2017dwk,DiCarlo:2019thl}, and the anomalous magnetic moment of the muon~\cite{Blum:2014oka,Blum:2017cer,Borsanyi:2017zdw,Boyle:2017gzv,Giusti:2017jof,Blum:2018mom,Bijnens:2019ejw,Blum:2019ugy,Giusti:2019xct,Borsanyi:2020mff}, as both QCD and QED must be accounted for in precision tests of the Standard Model. 

Systems of multiple charged hadrons exhibit rich phenomenology in which the interplay between QCD and QED effects is less well understood than for single-hadron systems.
The differences between $pp$, $np$ and $nn$ two-nucleon scattering channels are much more poorly constrained than their isospin average.
Lattice QCD+QED can potentially provide phenomenologically useful information on such isospin-breaking effects in nucleon-nucleon scattering;
furthermore, calculations with a range of quark masses and values of the QED fine structure constant can provide insight into the fine-tuning of QED effects and strong isospin breaking interactions, providing complementary information to experiment.
In large nuclei with charge $Z\sim1/\alpha$, where $\alpha$ is the EM fine structure constant, relativistic QED effects are expected to become nonperturbative and give rise to electron-positron pair-production and interesting consequences for the structure of elements~\cite{SCHWERDTFEGER2015551}.

QCD+QED studies of systems involving two or more protons are of great phenomenological interest, but calculations of large nuclei are difficult using current lattice QCD+QED techniques~\cite{Detmold:2019ghl}.
It is also interesting to consider systems of multiple charged mesons, which allow an investigation of similar QED effects without a number of the numerical complexities of multi-proton calculations.
In particular, charged multi-hadron systems include QED effects that become non-perturbative for hadron pairs with sufficiently small relative velocity $v \ll \alpha$.
All calculations of charged hadrons must also reconcile the presence of net charge in a finite volume (FV) with Gauss's law, which can be accomplished in several ways~\cite{Duncan:1996xy,Duncan:2004ys,Blum:2007cy,Hayakawa:2008an,Blum:2010ym,Aoki:2012st,Borsanyi:2013lga,Tantalo:2013maa,Davoudi:2014qua,Borsanyi:2014jba,Horsley:2015eaa,Endres:2015gda,Fodor:2015pna,Lucini:2015hfa,Fodor:2016bgu,Lubicz:2016xro,Giusti:2017dwk,Hansen:2018zre,Giusti:2019xct,Matzelle:2017qsw,Patella:2017fgk,Basak:2018yzz,Horsley:2019wha,Davoudi:2018qpl}.
The current study makes use of the QED$_L$ formulation~\cite{Hayakawa:2008an}, in which the photon spatial zero-mode is removed, as has been recently studied in detail for single-hadron systems in Refs.~\cite{Duncan:1996xy,Duncan:2004ys,Blum:2007cy,Hayakawa:2008an,Blum:2010ym,Aoki:2012st,Borsanyi:2013lga,Davoudi:2014qua,Borsanyi:2014jba,Horsley:2015eaa,Fodor:2015pna,Lucini:2015hfa,Fodor:2016bgu,Lubicz:2016xro,Giusti:2017dwk,Hansen:2018zre,Giusti:2019xct,Matzelle:2017qsw,Patella:2017fgk,Basak:2018yzz,Horsley:2019wha,Davoudi:2018qpl}.
Subtleties related to the nonlocality of zero-mode subtraction in QED$_L$ have been understood for single-hadron systems~\cite{Borsanyi:2014jba,Davoudi:2014qua,Fodor:2015pna,Lee:2015rua,Matzelle:2017qsw,Davoudi:2018qpl}, and similar methods are used to understand nonlocality in multi-hadron systems in this framework as described below.

In lattice QCD calculations, hadronic scattering phase shifts are determined by relating them to the FV energies of two-hadron systems using FV relations first derived by L\"uscher~\cite{Luscher:1986pf,Luscher:1990ux} and generalized in recent years (see Ref.~\cite{Briceno:2017max} for a review).
Understanding scattering in lattice QCD+QED is complicated by the lack of a FV formalism that includes nonperturbative Coulomb effects, and thereby relates FV energy shifts calculable in lattice QCD+QED to infinite-volume (Coulomb-corrected) scattering phase shifts.
In the QED$_L$ formulation, Ref.~\cite{Beane:2014qha} argues that Coulomb effects can be treated perturbatively for sufficiently small volumes where $L \ll 1/(\alpha M)$ and therefore the Bohr radius of the charged-particle system is larger than the infrared cutoff provided by the FV. 
However, the volume must also satisfy $L \gg 1/M$ in order for relativistic FV effects not accounted for in L\"uscher's analysis to be negligible.
For some systems there may be a window of intermediate-sized volumes where relativistic FV effects and nonperturbative Coulomb effects can both be neglected.
Non-relativistic effective field theory (EFT) can be used to investigate this issue and to relate FV energy shifts for systems with more than two particles to two-, three-, and higher-body EFT low-energy constants (LECs).

This work studies QCD+QED effects for systems of multiple hadrons including up to twelve charged or neutral mesons and up to three nucleons in multiple finite volumes. 
A larger than physical value of the fine-structure constant, $\alpha \simeq 0.1$, is used in order to increase the size of QED effects and permit the study of systems with total charge $Z$ satisfying $Z \alpha \gtrsim 1$.
As described in Ref.~\cite{Horsley:2015vla} and summarized below, the quark masses are tuned to remove strong isospin breaking effects so that splittings between particles that are degenerate in QCD can be identified as pure QED effects.
Calculations are performed in both lattice QCD+QED$_L$ (LQCD+QED$_L$) and in an EFT appropriate for non-relativistic charged hadrons, referred to as non-relativistic QED$_L$ (NRQED$_L$) below.
Two-body scattering lengths and three-body interactions of charged and neutral mesons are determined by tuning the hadronic interaction parameters appearing in the NRQED$_L$ Lagrangian to reproduce the LQCD+QED$_L$ FV energy levels.
QED effects on hadron interactions beyond Coulomb photon exchange are determined by comparing the two-body and three-body interaction parameters determined for charged and neutral mesons.
Studies of multi-baryon systems are used to probe volumes where $L > 1/(\alpha M)$ and Coulomb effects may not be perturbative according to the criteria of Ref.~\cite{Beane:2014qha}. 
It is noteworthy that no significant nonperturbative Coulomb effects or relativistic effects are seen at the level of precision of the results obtained here, even for systems with  $L > 1/(\alpha M)$ or $Z \alpha > 1$.

The practical window of $L$ where relativistic effects can be neglected and Coulomb effects can be treated perturbatively is explored by comparing LQCD+QED$_L$ and NRQED$_L$ results for a variety of FV systems.
From these studies, it is argued that this window requires the spatial extent $L$ of a cubic FV to satisfy
\begin{equation}
\begin{split}
   \frac{1}{ML} \ll 1, \hspace{20pt} \frac{\alpha M L}{4\pi} \ll 1.
\end{split}\label{eq:hierarchy4pi}
\end{equation}
The combination $\alpha M L / (4\pi)$ in Eq.~\eqref{eq:hierarchy4pi} that quantifies the size of FV Coulomb effects is equivalent to the infinite-volume Coulomb expansion parameter $\alpha / v$ for a pair of hadrons moving back-to-back with one unit of FV momentum, i.e. $\pm 2\pi / L$.
Eq.~\eqref{eq:hierarchy4pi} indicates that Coulomb effects can be treated perturbatively in current and future LQCD+QED$_L$ calculations over a wide range of volumes and in particular for systems with $L \sim 1/ (\alpha M)$, where the Bohr radius is commensurate with the volume.
This scaling differs from that suggested in Ref.~\cite{Beane:2014qha} by the factor of $4\pi$ in the denominator of the $\alpha M L / (4\pi)$.

This manuscript is structured as follows. Section \ref{sec:lattice} presents the details and methodology of the LQCD+QED$_{L}$ calculations that are performed.  Section \ref{sec:theory} discusses the construction of NRQED$_L$ for charged multi-hadron systems and provides the formalism needed to extract hadronic scattering lengths and other parameters from the results of these LQCD+QED$_L$ calculations. Section \ref{sec:results} describes the determination of hadronic scattering parameters from LQCD+QED$_L$ and presents results for multi-meson and multi-nucleon systems. Conclusions are presented in Section \ref{sec:conclusion}. Appendix~\ref{app:matching} contains further technical details on the matching between QED$_L$ and NRQED$_L$, and Appendix~\ref{app:fits} contains details on the fitting procedure used to extract energy levels from Euclidean correlation functions computed in LQCD+QED$_L$.

\section{Lattice QCD + QED}
\label{sec:lattice}

Lattice QCD+QED is a nonperturbative approach to QCD+QED that at intermediate stages implements an ultraviolet regulator defined by a lattice spacing $a$ (where $1/a$ is assumed to be much smaller than the QED Landau pole).
Calculations are performed in Euclidean spacetime with a cubic spatial volume of extent $L\times L \times L$ and a finite temporal extent $T$;
the quark, gluon, and photon fields satisfy periodic boundary conditions (PBCs) in all spatial directions.
Here, the QED$_L$ formalism~\cite{Hayakawa:2008an} is used to define charged-particle correlation functions as detailed below, which defines the LQCD+QED$_L$ formalism.
The LQCD+QED$_L$ gauge field configurations used were generated by the QCDSF collaboration. 
The full details of the generation of these ensembles are presented in Refs.~\cite{Horsley:2015eaa,Horsley:2015vla}. 
For completeness, relevant aspects of the ensemble generation are described below.

\subsection{Lattice action and parameters}

\begin{table}[t]
	\begin{ruledtabular}
	\begin{tabular}{cccccccc}
	$L^3\times T$ & $\beta$ & $\beta_E$ & $\kappa_u$ & $\kappa_d$ &  $\kappa_s$ &  $N_{\rm cfg}$& $N_{\rm src}$  \\ 
	\hline 
	$32^3\times64$ & 5.50 &  0.80 & 0.124362 & 0.121713 & 0.121713  &  1294 & 27 \\ 
	$48^3\times96$ & 5.50 &  0.80 & 0.124362 & 0.121713 & 0.121713  &  692 & 9 \\ 

\end{tabular} 
\end{ruledtabular}
   \caption{\label{tab:ensem} Parameters of the gauge field ensembles and calculations performed in this work. $L^3\times T$ is the Euclidean spacetime volume, $\beta$ is the $SU(3)$ gauge coupling as defined in Refs.~\cite{Symanzik:1983dc,Luscher:1984xn,Bietenholz:2011qq}, $\beta_E = 1/e^2$ is related to the QED gauge coupling appearing in Eq.~\eqref{eq:photon_action}, $\kappa_q$ are quark mass parameters for flavors $q\in\{u,d,s\}$ appearing in the quark action, Eq.~\eqref{eq:SFq}, $N_{\rm cfg}$ is the number of gauge field configurations analyzed in this work, and $N_{\rm src}$ is the average number of quark propagator sources randomly distributed on each gauge field configuration.}
\end{table}

The LQCD+QED$_L$ action used in this study is given by
\begin{align}
S=S_G+S_A+S_F,
\label{eq:action}
\end{align}
where $S_G$ is the tree-level Symanzik-improved $SU(3)$ gauge action as defined in Refs.~\cite{Symanzik:1983dc,Luscher:1984xn,Bietenholz:2011qq}.
The quark dynamics are encoded by an ${\cal O}(a)$-improved stout link
non-perturbative clover (SLiNC) action:
\begin{equation}
   \begin{split}
      S_F =& \sum_{q \in \{u,d,s\} } \sum_x \left\{ \frac{1}{2} \sum_\mu \left[
\bar{q}(x)(\gamma_\mu-1)e^{-ie_q A_\mu(x)} \tilde{U}_\mu(x)q(x+\hat
{\mu}) \right. \right. \\
      & \left. \hspace{100pt} \left.- \bar{q}(x)(\gamma_\mu+1) e^{ie_qA_\mu(x-\hat{\mu})}
\tilde{U}_\mu^\dagger(x- \hat{\mu})q(x- \hat{\mu})\right]
\right. \\
      &  \left. \hspace{70pt} + \frac{1}{2 \kappa_q} \bar{q}(x)q(x)-\frac{1}{4}c_{SW}
\sum_{\mu \nu} \bar{q}(x)\sigma_{\mu \nu} F_{\mu \nu} q(x) \right\},
\end{split}\label{eq:SFq}
\end{equation}
where $\tilde{U}_\mu$ is a single-iterated stout-smeared
$SU(3)$ link \cite{Morningstar:2003gk}, $A_\mu$ is a non-compact $U(1)$ gauge field, and the $U(1)$ quark charges are given by
$e_q$.
The field-strength $F_{\mu \nu}$ appearing in the Sheikholeslami-Wohlert or ``clover term''~\cite{Sheikholeslami:1985ij} involves the unsmeared $SU(3)$ gauge-field as in Ref.~\cite{Cundy:2009yy}.
The clover coefficient $c_{SW}$ was non-perturbatively determined for pure QCD in Ref.~\cite{Cundy:2009yy}.
Electromagnetic gauge fields are not included in the clover term,
however with these simulation parameters, the ${\cal O}(\alpha a)$ effects are no larger than the
residual ${\cal O}(a^2)$ effects of pure QCD.
The photon action is described by the non-compact form:
\begin{equation}\label{eq:photon_action}
S_A = \frac{1}{2e^2} \sum_{x,\mu<\nu}
   \left(A_\mu(x)+A_\nu(x+\mu)-A_\mu(x+\nu)-A_\nu(x) \vphantom{\frac{1}{2}} \right)^2\,
\end{equation}
where $e$ is the $U(1)$ gauge coupling corresponding to $\alpha = e^2 / (4\pi)$. Gauge fixing and the treatment of zero modes are discussed in Sec.~\ref{sec:zero}.

The parameters of the lattice action
were tuned by identifying a point of approximate $SU(3)$ flavor symmetry, where the average
light-quark mass takes its physical value---see
Ref.~\cite{Bietenholz:2011qq} for further discussion.
With dynamical QED, this is complicated by the fact that the quark
charges explicitly break the $SU(3)$ flavor symmetry.
An approximate $SU(3)$ flavor symmetry is then realized by tuning the quark mass
parameters such that the {\em connected} flavour-neutral pseudoscalar mesons\footnote{Since $\kappa_s = \kappa_d$ and $e_s = e_d$, this theory exhibits exact
  $U$-spin symmetry, and therefore the connected
  $d\bar{d}$ and $s\bar{s}$ correlation functions are identical to the $d\bar{s}$ correlation functions
The connected part of the $u\bar{u}$ correction function can be interpreted in a partially quenched theory, see Ref.~\cite{Horsley:2015vla} for further discussion.}
are 
degenerate.
As it is inspired by Dashen's theorem \cite{Dashen:1969eg},
this prescription for separating strong and electromagnetic effects is known as
the ``Dashen scheme'' \cite{Horsley:2015vla}.
This scheme preserves Dashen's theorem, whereby the neutral pseudoscalar mesons 
are
protected from receiving an electromagnetic self-energy correction in the chiral limit.
To reach the physical quark masses and charges, the $SU(3)$ flavor symmetric point can then be lowered to the physical value before the symmetry is broken to fix the individual quark masses.
In this exploratory work a single set of approximately $SU(3)$ flavor symmetric parameters is used.

The Dashen scheme provides a natural framework for separating QED
and QCD effects; at the $SU(3)$ symmetric point, any splittings among pure-QCD multiplets are
identified as pure QED effects.
The explicit breaking of the quark masses from this point then allows
one to independently isolate the effects of strong (or quark mass)
flavor symmetry breaking and QED effects.
The action parameters used in this work correspond to the $SU(3)$ symmetric point and are presented in Table~\ref{tab:ensem}.

\subsection{Gauge fixing and the $U(1)$ zero mode}\label{sec:zero}

The correlation functions of charged particles are not gauge invariant and hence ensemble-averaged quantities are only meaningful within some gauge fixing prescription.
In this work, Landau gauge is adopted by enforcing the condition
\begin{equation}
\sum_\mu \left(A_\mu(x)-A_\mu(x-\hat{\mu})\right)=0.
\end{equation}
This condition can be imposed after generation of the gauge fields.
However, the Landau gauge condition leaves the 4-dimensional zero mode unconstrained.
These uniform background fields do not contribute to the gauge action, however they do couple to the fermionic action, Eq.~(\ref{eq:SFq}).
The action remains invariant under discrete shifts of the zero mode in units of $2\pi/L_\mu$.
This redundant gauge degree of freedom can be eliminated by mapping the 4-dimensional zero modes onto the finite interval $-\pi/L_\mu<\widetilde{A}_\mu(k=0)\le \pi/L_\mu$ \cite{Gockeler:1991bu}, where $\widetilde{A}_\mu(k)$ is the Fourier transform of $A_\mu(x)$ defined explicitly in Eq.~\eqref{eq:AFV}.
The leading effect of these zero modes is to induce a charge-dependent twist of the single-hadron momenta.
This small energy shift can be corrected in single-hadron states \cite{Horsley:2015eaa}, however this would introduce an unnecessary complication in the analysis of many-body interactions.
Instead, this work adopts the QED$_L$ prescription \cite{Hayakawa:2008an,Borsanyi:2014jba} of setting the spatial (3-dimensional) zero modes of the gauge potential to zero on every timeslice.

With respect to the action used to generate the gauge configurations, the elimination of the 3-dimensional zero modes before computing quark propagators is not a gauge symmetry.
As a consequence, there is a partial-quenching effect whereby the valence quarks experience a different zero mode to the quarks in the sea.
The zero modes cannot affect closed fermion loops, and their only contribution to the determinant will be associated with fermion lines wrapping around the boundary of the lattice.
Consequently, this partial-quenching effect is exponentially suppressed in $m_\pi L$ and is negligible in comparison to the power-law FV effects studied in this work.

\subsection{Correlation functions}

The particular  gauge field ensembles used in this work are described in Table \ref{tab:ensem}; along with the parameters used in their generation, the number of configurations and the average number of correlation function source locations that are used per configuration are also reported. 
Up and down/strange (equivalent for the masses used here) quark propagators are computed from each of the randomly chosen source locations  using three-dimensionally Jacobi smeared sources~\cite{Allton:1993wc} (100 iterations with $\rho=0.21$) using a solver tolerance of $10^{-12}$. Local and smeared fields are used in the sink interpolating operator to construct smeared-point (SP) and smeared-smeared (SS) correlation functions with the former providing cleaner signals in all cases.

FV energy levels are determined by analyzing two-point correlation functions
\begin{eqnarray}
G_h(t; {\bf x}_0) =\left\langle \sum_{\bf x}  \tilde\chi_h({\bf x},t) \chi_h^\dagger({\bf x}_0,0)\right \rangle ,
\end{eqnarray}
where $\chi_h$ ($\tilde\chi_h$) is a source (sink) interpolating operator with the quantum numbers of the state being considered and ${\bf x}_0$ is the source location. Two-point correlation functions are constructed for $n\in\{1,\ldots,12\}$ charged pions ($u\bar{d}$) and neutral kaons ($s\bar{d}$) 
using the techniques developed in Refs.~\cite{Beane:2007es, Detmold:2008fn, Detmold:2010au, Detmold:2011kw, Detmold:2012wc}. At the sink, each color-singlet meson bilinear is separately projected to zero momentum; for example, for a system of $n$ pions ($h = n\pi^+$),
\begin{eqnarray}
G_{n\pi^+}(t; {\bf x}_0) =\left\langle  \left(\sum_{\bf x} \bar{u}({\bf x},t) \gamma_5 d({\bf x},t) \right)^n \left(u({\bf x}_0,0)\gamma_5 \bar d({\bf x}_0,0)\right)^n \right\rangle .
\end{eqnarray}

The correlation functions for the single proton and neutron make use of standard local  interpolating operators $\chi_{p;\alpha} =\epsilon^{ijk} (u^i C \gamma_5 d^j) u^k_\alpha$  and $\chi_{n;\alpha} =\epsilon^{ijk} (d^i C \gamma_5 u^j) d^k_\alpha$ where the parentheses indicate  contraction of spin indices. For two-baryon and three-baryon systems, the contraction techniques of Refs.~\cite{Detmold:2012eu,Beane:2012vq} are used, again with each baryon separately projected to zero momentum at the sink. 

\subsection{Finite-volume energy level determinations}

Finite-volume energy levels are extracted from the correlation functions for each system using correlated fits to their time dependence. For the multi-meson systems, which factorize easily into color-singlet components, thermal contributions where one component propagates forward in time and another propagates backwards in time are particularly relevant. The corresponding functional forms that are used to fit these correlation functions are 
\begin{equation}
   f_{n M}(t,\v{E},\v{Z}) = \sum_{m=0}^n Z_{n M; m} e^{-E_{m M} t} e^{-E_{(n-m)M}(T-t)}  + \sum_{e=1}^{N_{\rm states}-1} Z_{n M}^{(e)} e^{-E_{n M}^{(e)} t}\,, \label{eq:spectral}
\end{equation}
where $M \in \{ \pi^+,\ \overline{K^0} \}$ labels the type of meson, $N_{\rm states}$ is the total number of non-thermal states included in the fit, $E_{n M}$ is the ground-state energy of the system with the quantum numbers of $n$ mesons of type $M$, $Z_{n M; m}$ is the overlap factor describing the amplitude of thermal contributions with $m$ forwards propagating mesons and $n-m$ backwards propagating mesons, and non-thermal excited states are also included with energies $E_{n M}^{(e)}$ and overlap factors $Z_{n M}^{(e)}$.
In practice, fits with $N_{\rm states} \in \{ 1,2,3 \}$ are used in this work.
The vectors $\v{E}$ and $\v{Z}$ indicate the energy and overlap factor parameters to be constrained in the fit. 
In order to determine the many parameters of these fitting functions, fits are performed sequentially for increasing $n$, with the energies $E_{m M}$ for $m<n$ used as input for $f_{n M}$ as in Ref.~\cite{Detmold:2011kw}. 
Thermal effects arising from excited states are not found to be significant.

For baryon systems, statistical noise grows rapidly with the temporal separation between the source and the sink, and the contributions of thermal states are negligible with respect to the statistical uncertainties. For these systems a simpler fit function is used:
\begin{equation}
  f_b(t,\v{E},\v{Z})= Z_b e^{-E_b t} + \sum_{e=1}^{N_{\rm states} - 1} Z_b^{(e)}e^{-E_b^{(e)} t}\,,\label{eq:baryon}
\end{equation}
where $b$ labels the type of baryon system and, as for meson systems, the second term corresponds to  a sum over the excited states included in the fitting model.
This work investigates single-nucleon systems with $b \in \{ p, n \}$ as well as two-nucleon systems with $b \in \{ nn, np({}^1S_0), np({}^3S_1), pp \}$ and three-nucleon systems with $b \in \{ {}^3\text{H}, {}^3\text{He} \}$.

Best-fit parameters are determined from minimization of the correlated $\chi^2$ function
\begin{eqnarray}\label{eq:chi2}
   \chi^2 (\v{E},\v{Z}) = \sum_{t,t' = t_{\rm min}}^{t_{\rm max}} (G(t) - f(t,\v{E},\v{Z}))(\mathcal{S}(\lambda^*)^{-1})_{t,t'}(G(t') - f(t',\v{E},\v{Z}))
\end{eqnarray}
for the appropriate correlation functions, $G(t)$, and fit function, $f \in \{ f_{nM}, f_b \}$, $[t_{\rm min}, t_{\rm max}]$ is the range of times included in the fit, and  $\mathcal{S}(\lambda^*)$ is an estimate of the covariance matrix described below.
Finite sample-size fluctuations may make the sample covariance matrix ill-conditioned, and shrinkage techniques~\cite{stein1956,LEDOIT2004365} are used to obtain a numerical stable inverse covariance matrix.
Following the application of shrinkage to LQCD in Ref.~\cite{Rinaldi:2019thf}, the covariance matrix including shrinkage is defined as $\mathcal{S}(\lambda) = (1-\lambda)\mathcal{C} + \lambda \mathcal{T}$, where $\mathcal{C}$ is the bootstrap covariance matrix and $\mathcal{T} = \text{diagonal}(\mathcal{C})$, and therefore $\chi^2$-minimization with $\mathcal{S}(\lambda)$ interpolates between correlated $\chi^2$-minimization for $\lambda = 0$ and uncorrelated $\chi^2$-minimization for $\lambda = 1$.
The optimal shrinkage parameter $\lambda^*$ appearing in Eq.~\eqref{eq:chi2} is chosen to maximize the expected closeness to the true covariance matrix and defined in Eq.~\eqref{eq:optshrinkage}, see Appendix~\ref{app:fits} and Ref.~\cite{LEDOIT2004365} for further discussion.

Systematic uncertainties arise from the dependence of the fits on the functional forms and time ranges that are used. To address these, the time ranges are systematically sampled, and fits with and without excited states are attempted, with an information criterion used to select the appropriate number of excited states to include for each fit range choice.
A weighted average of the results from all successful fits passing various reliability checks is used to determine the final results.
Further details are presented in Appendix \ref{app:fits}.

\begin{figure}
  \centering
     \begin{minipage}[c][150pt][t]{70pt} $1\ \pi^+\newline L/a=48$ \end{minipage}
     \includegraphics[width=.40\textwidth]{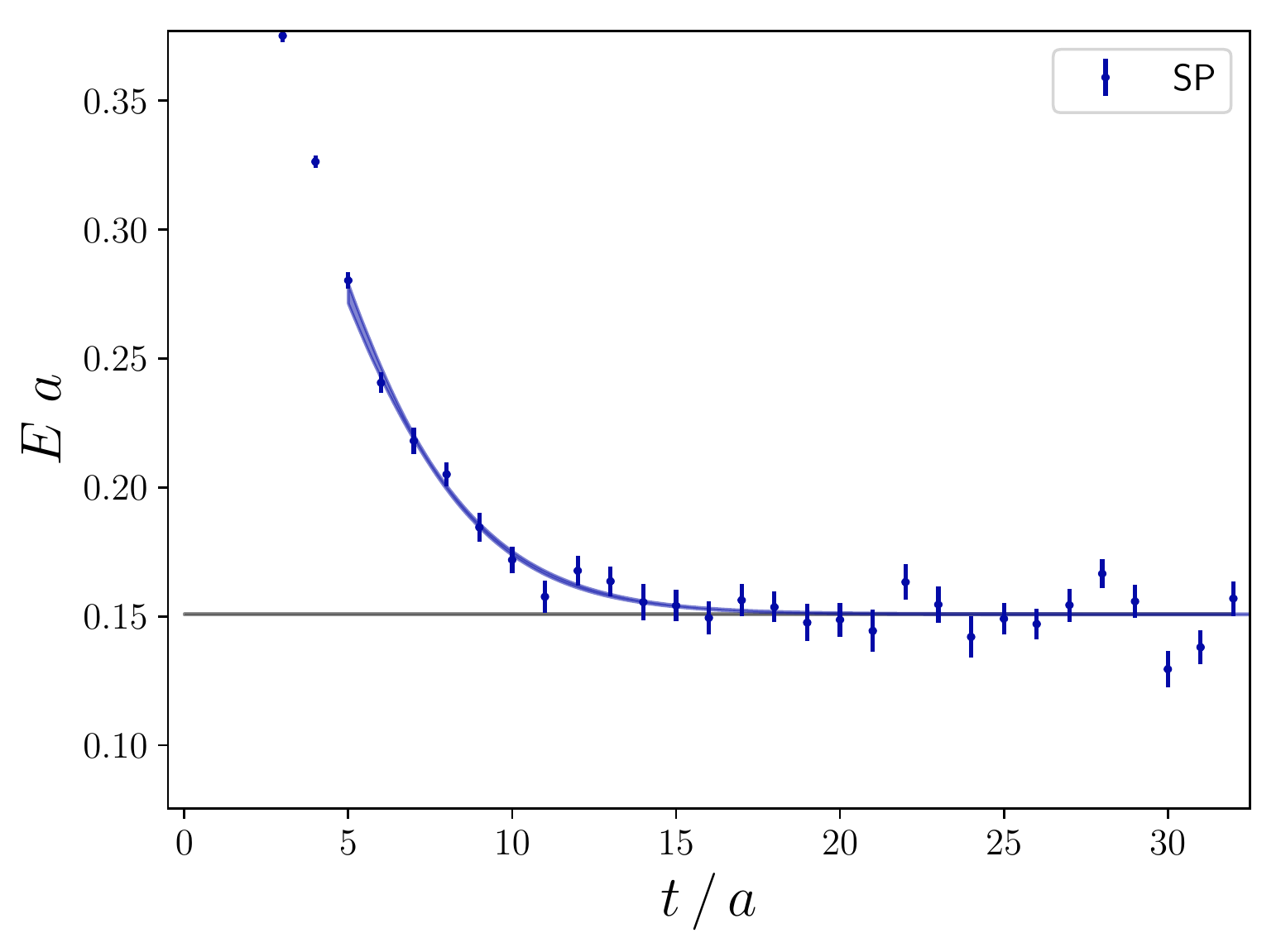}\hspace{10pt}
     \includegraphics[width=.40\textwidth]{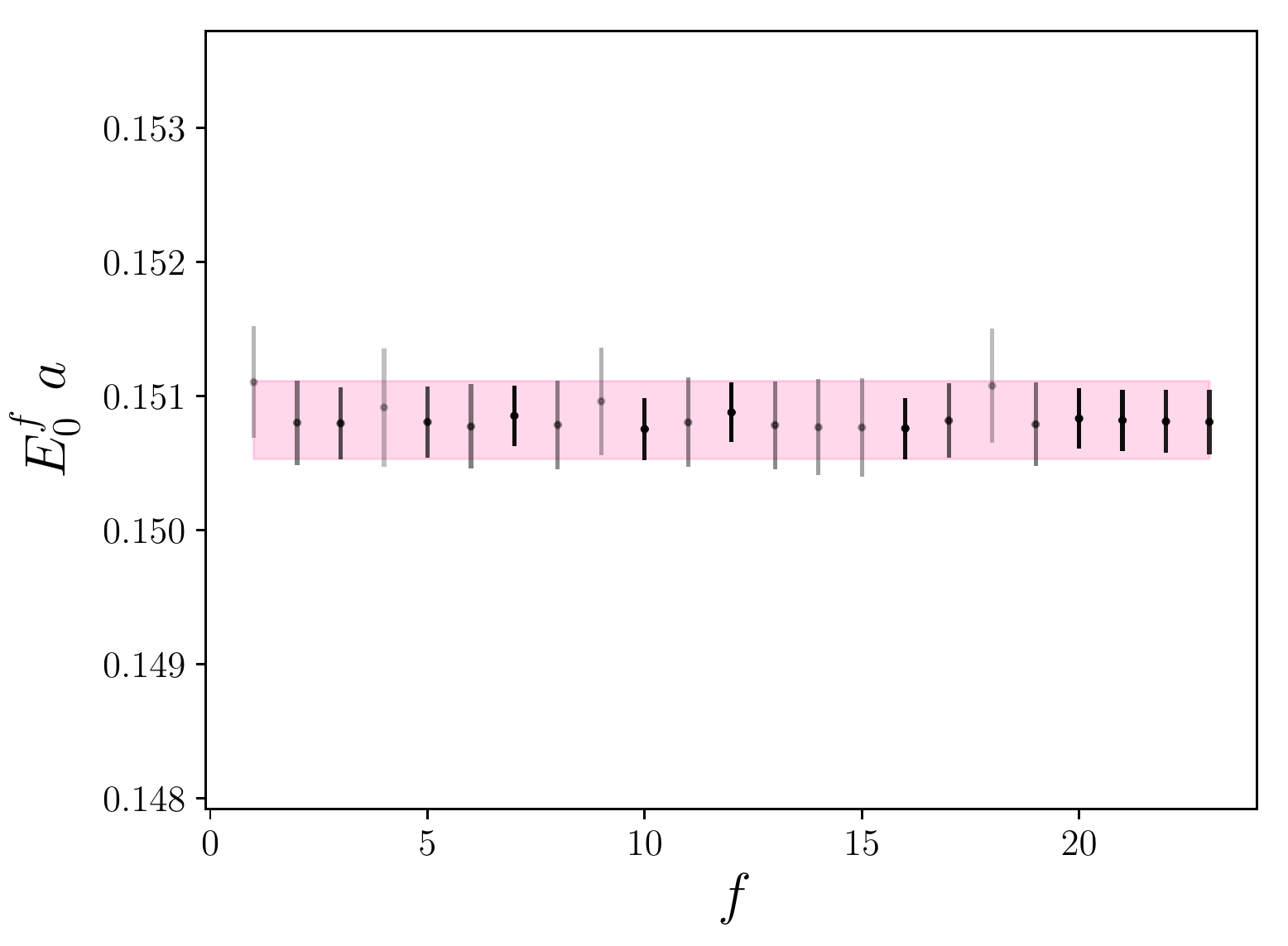} \\\vspace{-65pt}
     \begin{minipage}[c][150pt][t]{70pt} $2\ \pi^+\newline L/a=48$ \end{minipage}
     \includegraphics[width=.40\textwidth]{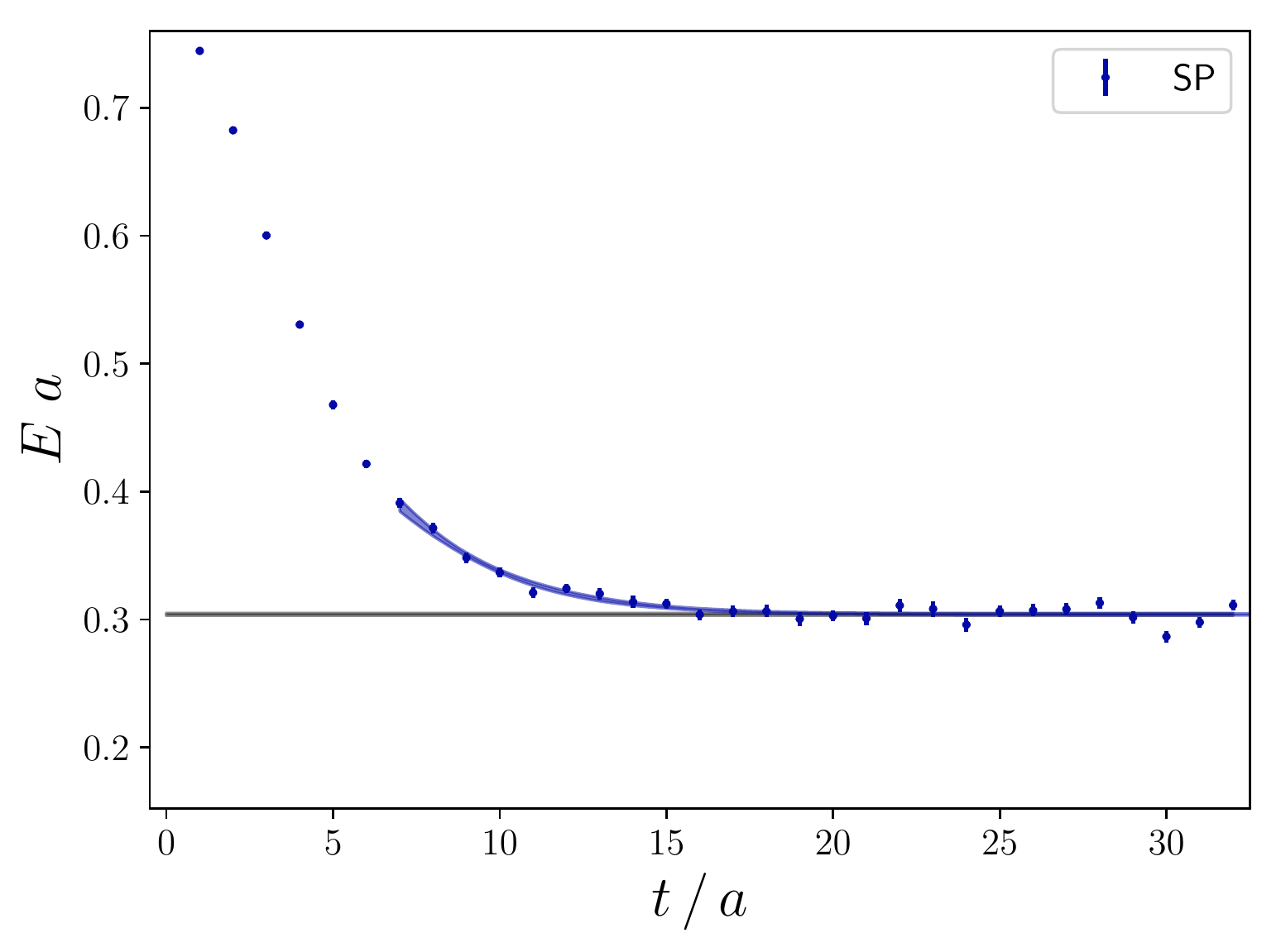}\hspace{10pt}
     \includegraphics[width=.40\textwidth]{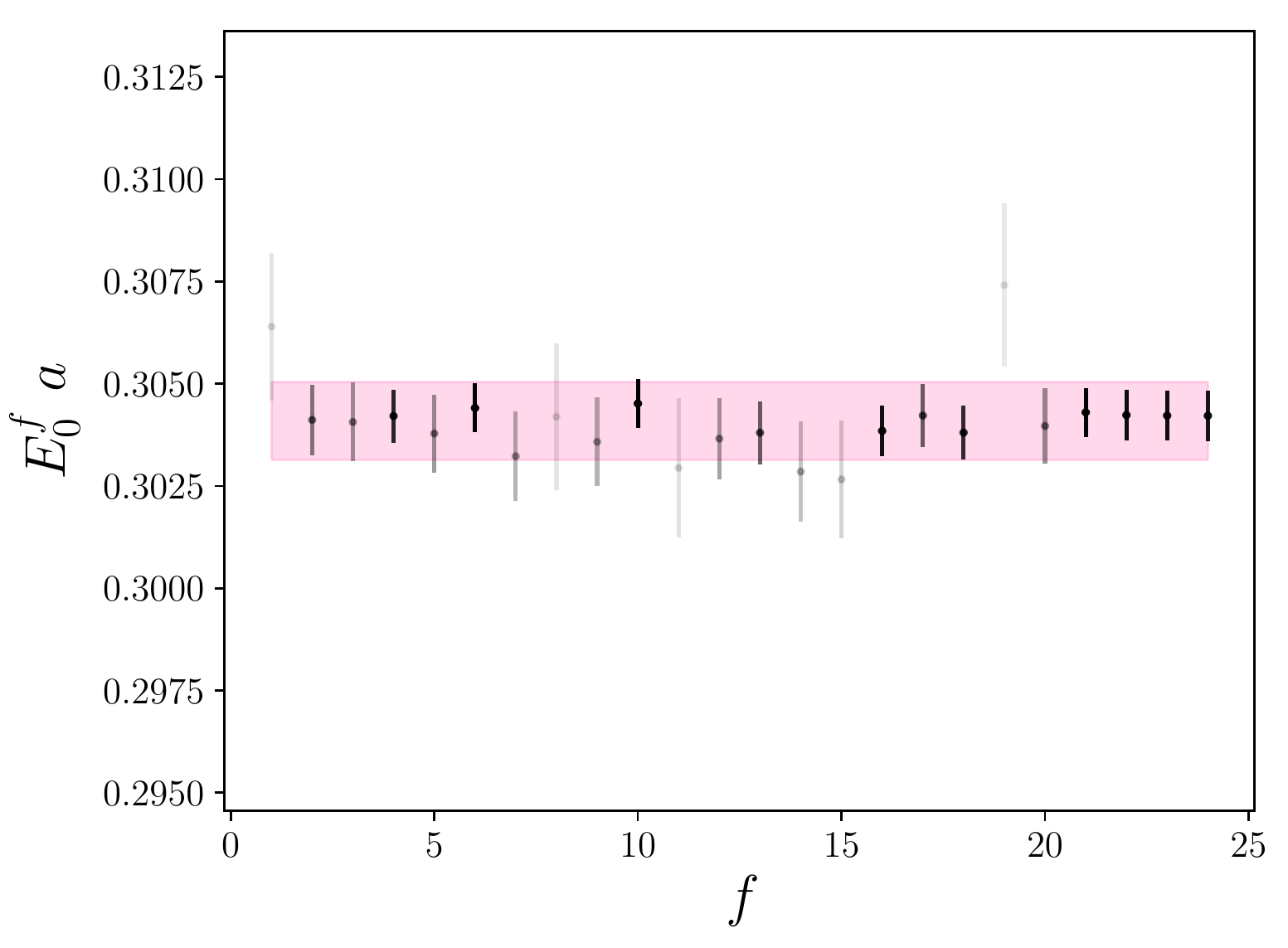} \\\vspace{-65pt}
     \begin{minipage}[c][150pt][t]{70pt} $3\ \pi^+\newline L/a=48$ \end{minipage}
     \includegraphics[width=.40\textwidth]{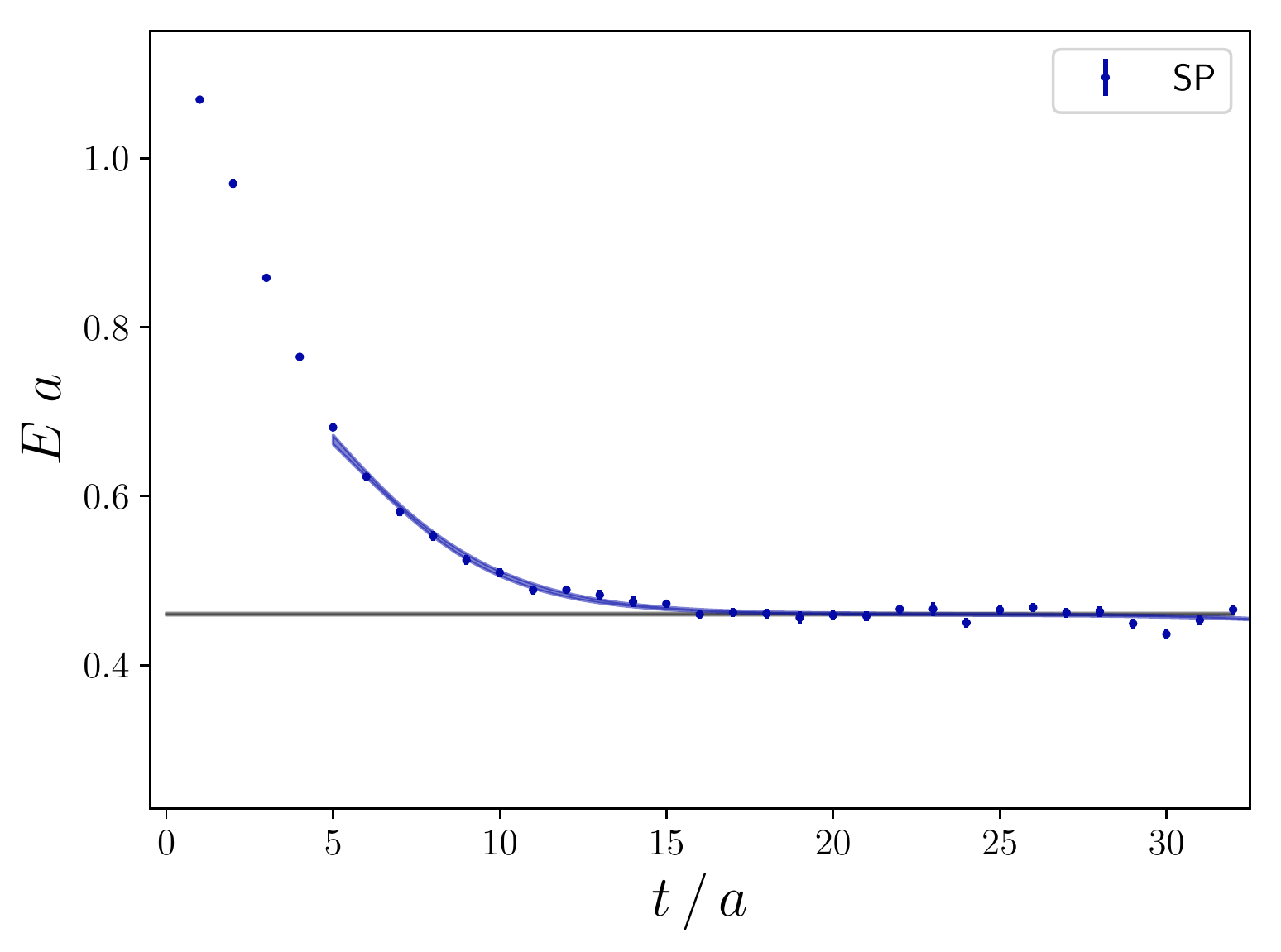}\hspace{10pt}
     \includegraphics[width=.40\textwidth]{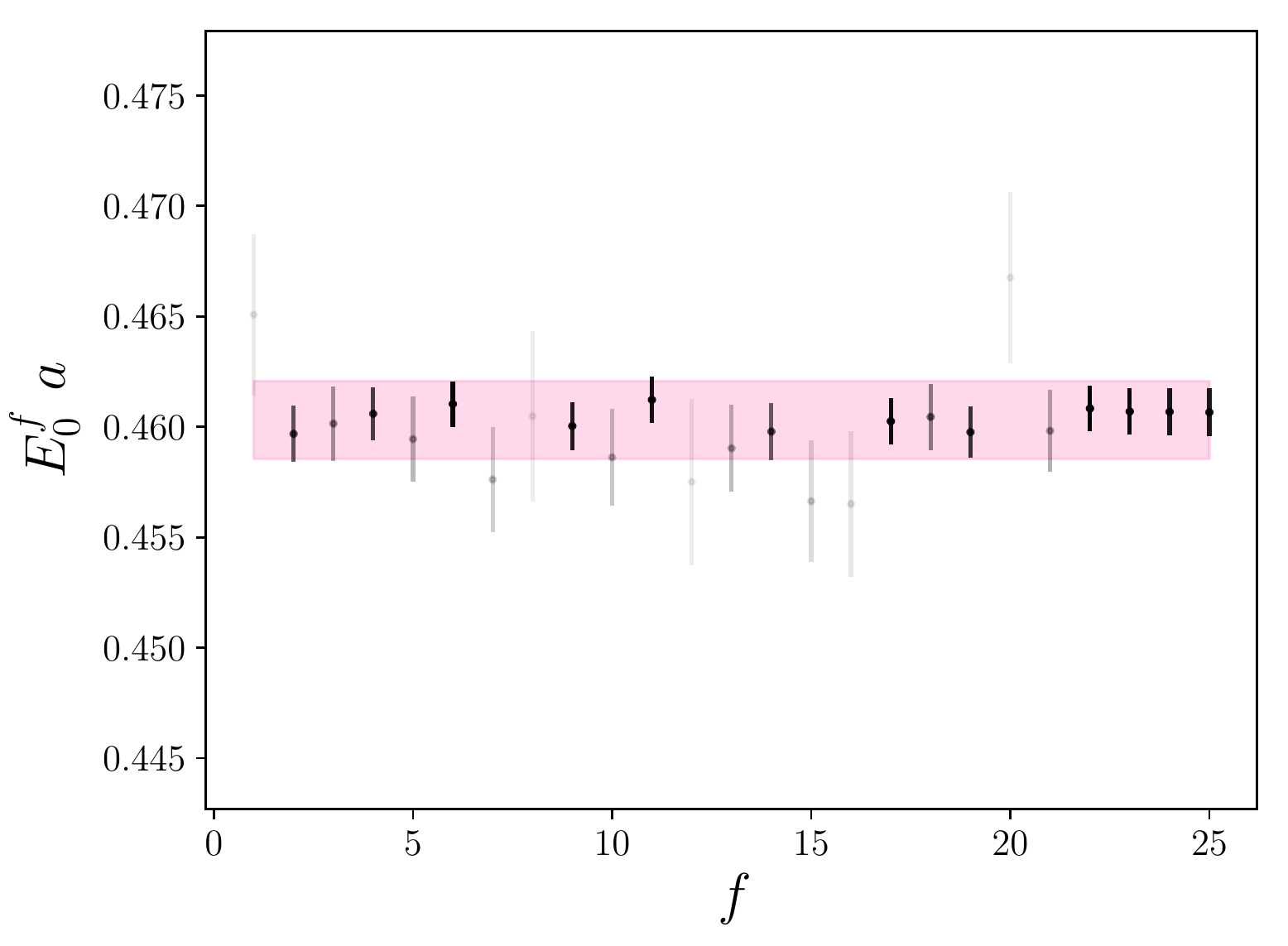} \\\vspace{-65pt}
     \begin{minipage}[c][150pt][t]{70pt} $4\ \pi^+\newline L/a=48$ \end{minipage}
     \includegraphics[width=.40\textwidth]{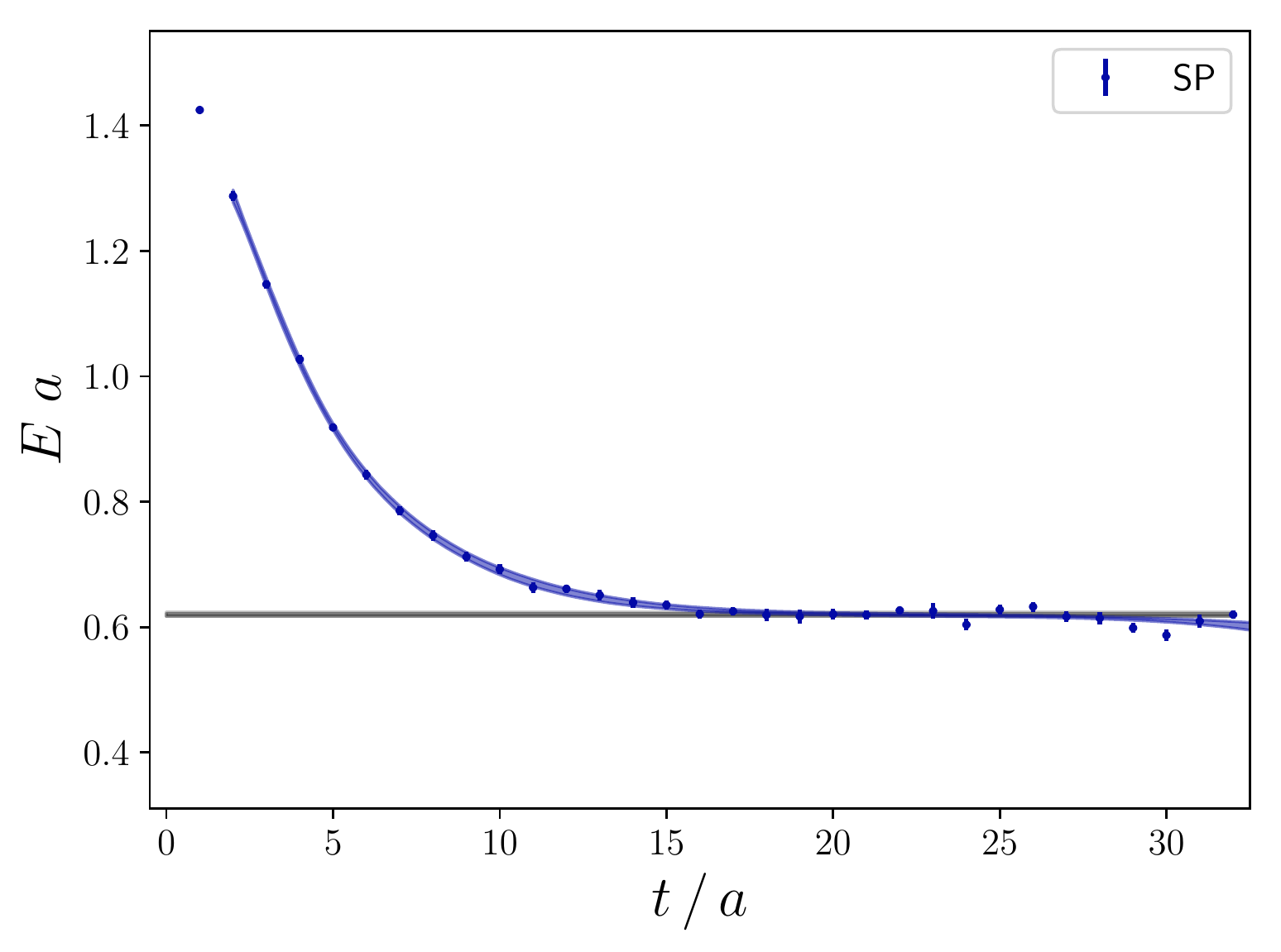}\hspace{10pt}
     \includegraphics[width=.40\textwidth]{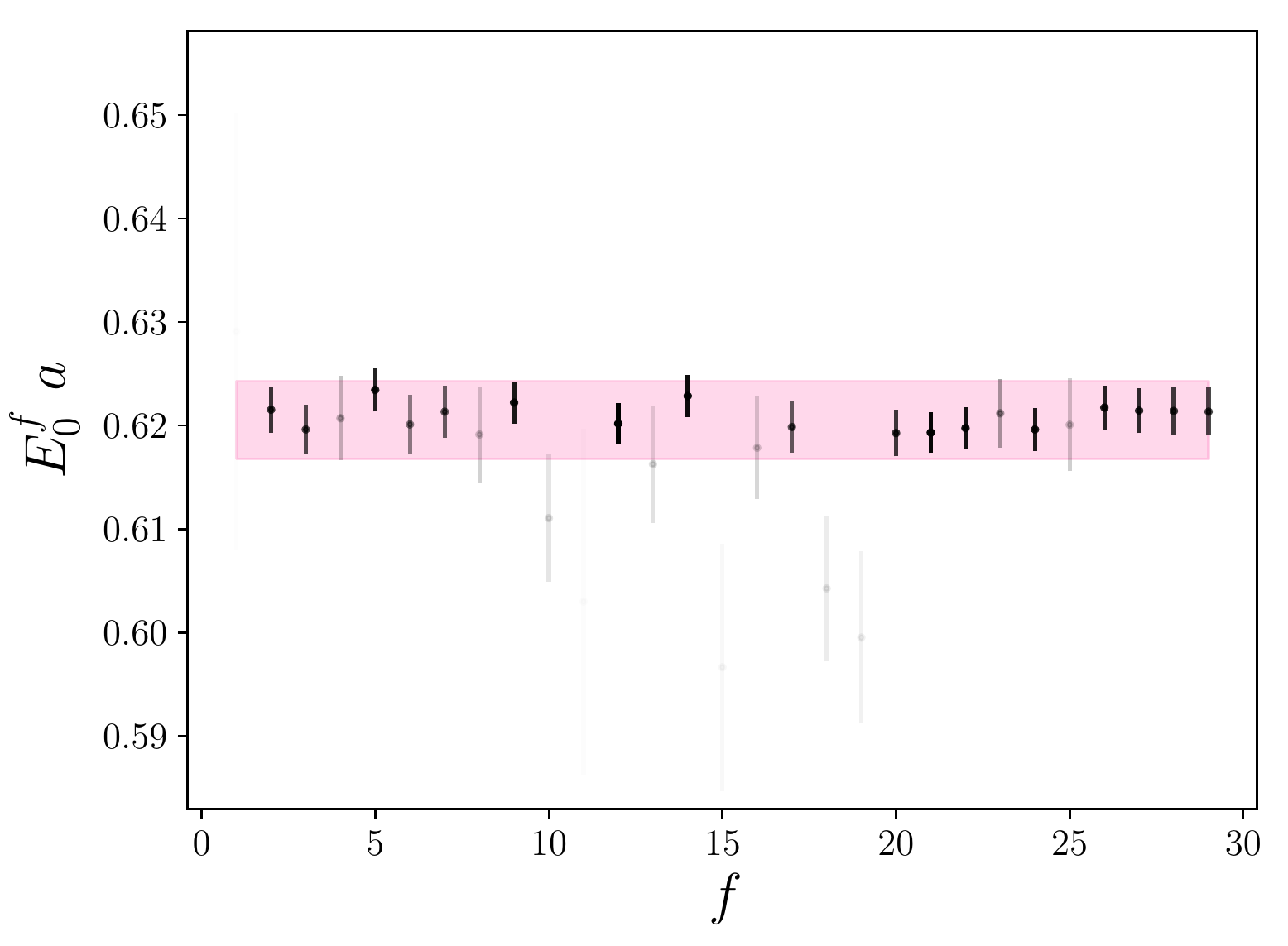} \\\vspace{-65pt}
     \caption{Fit results for $n\in\{1,\dots,4\}$ $\pi^+$ systems with $L/a = 48$. Blue points in the left plots show LQCD+QED$_L$  results for $E(t)$ defined in Eq.~\eqref{eq:EMdef}. Blue bands show $67\%$ confidence intervals from fits to Eq.~\eqref{eq:spectral} described in Appendix~\ref{app:fits}. Horizontal light (dark) gray bands show the statistical (total) uncertainty of the fitted ground-state energy. The right plot shows ground-state energy results from each successful fit range with opacity equal to their relative weight in the average determining the total statistical plus systematic uncertainty shown as a pink band.
    \label{fig:pion48a}}
\end{figure}

\begin{figure}
  \centering
     \begin{minipage}[c][150pt][t]{70pt} $5\ \pi^+\newline L/a=48$ \end{minipage}
     \includegraphics[width=.40\textwidth]{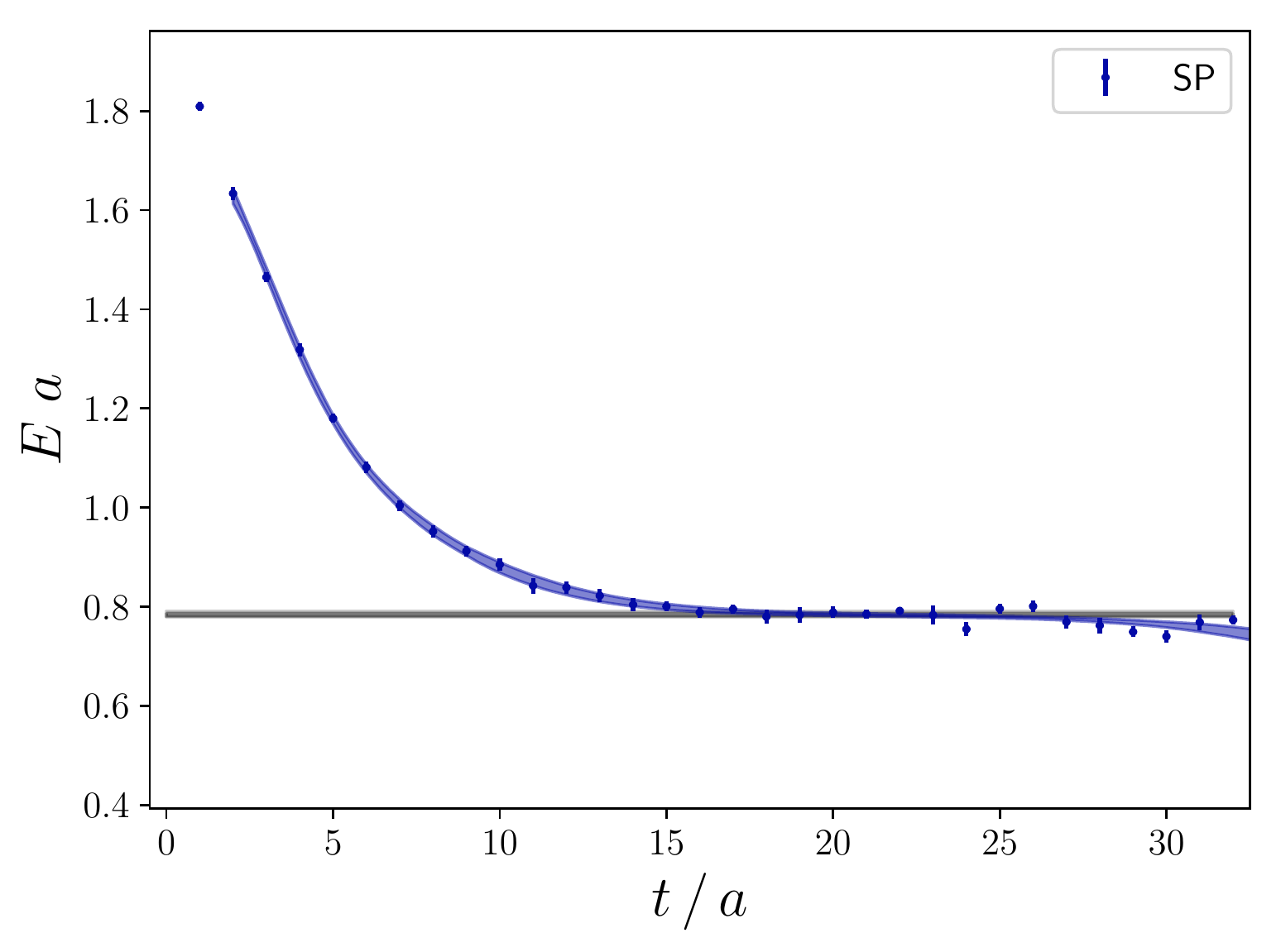}\hspace{10pt}
     \includegraphics[width=.40\textwidth]{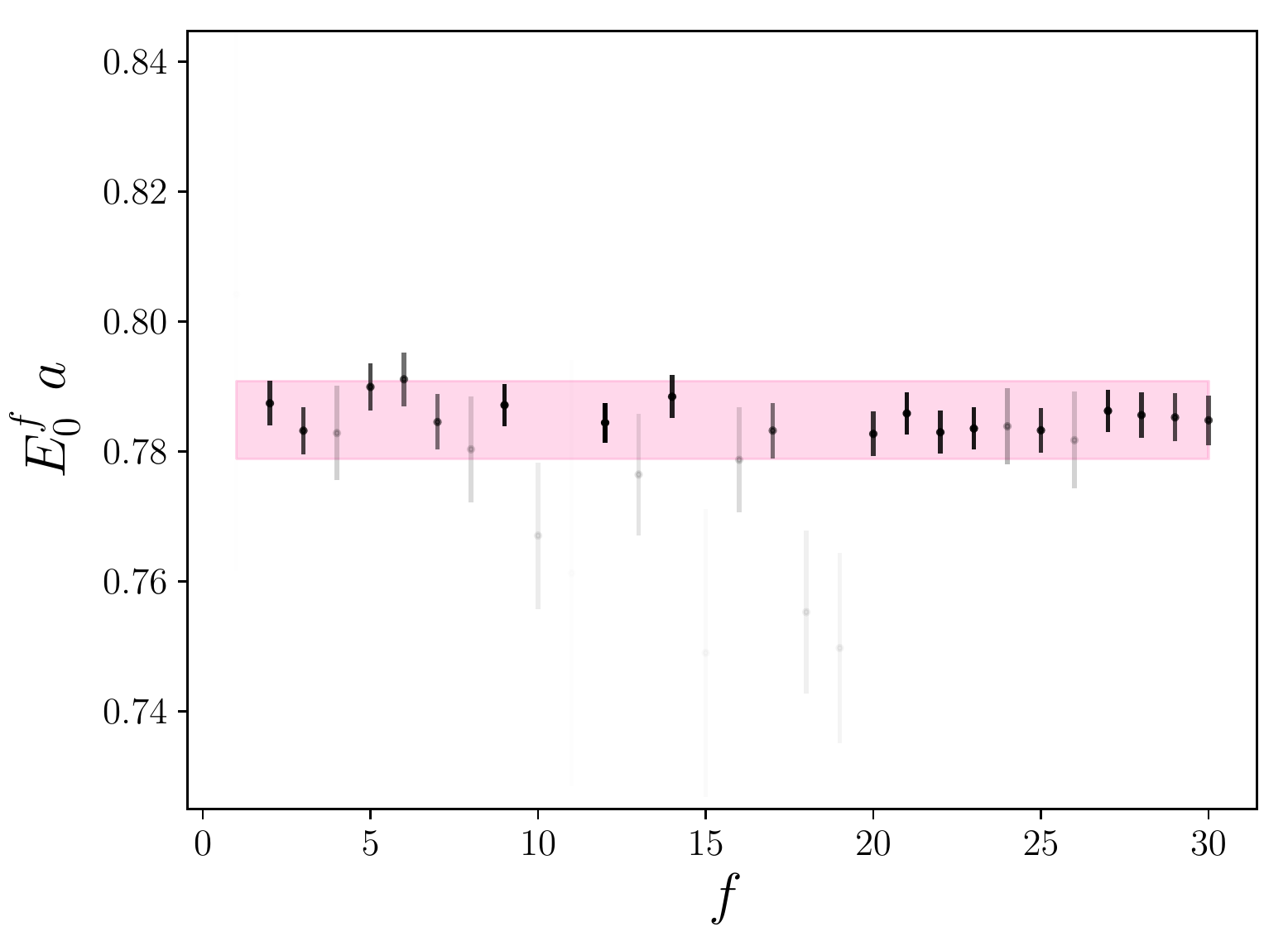} \\\vspace{-65pt}
     \begin{minipage}[c][150pt][t]{70pt} $6\ \pi^+\newline L/a=48$ \end{minipage}
     \includegraphics[width=.40\textwidth]{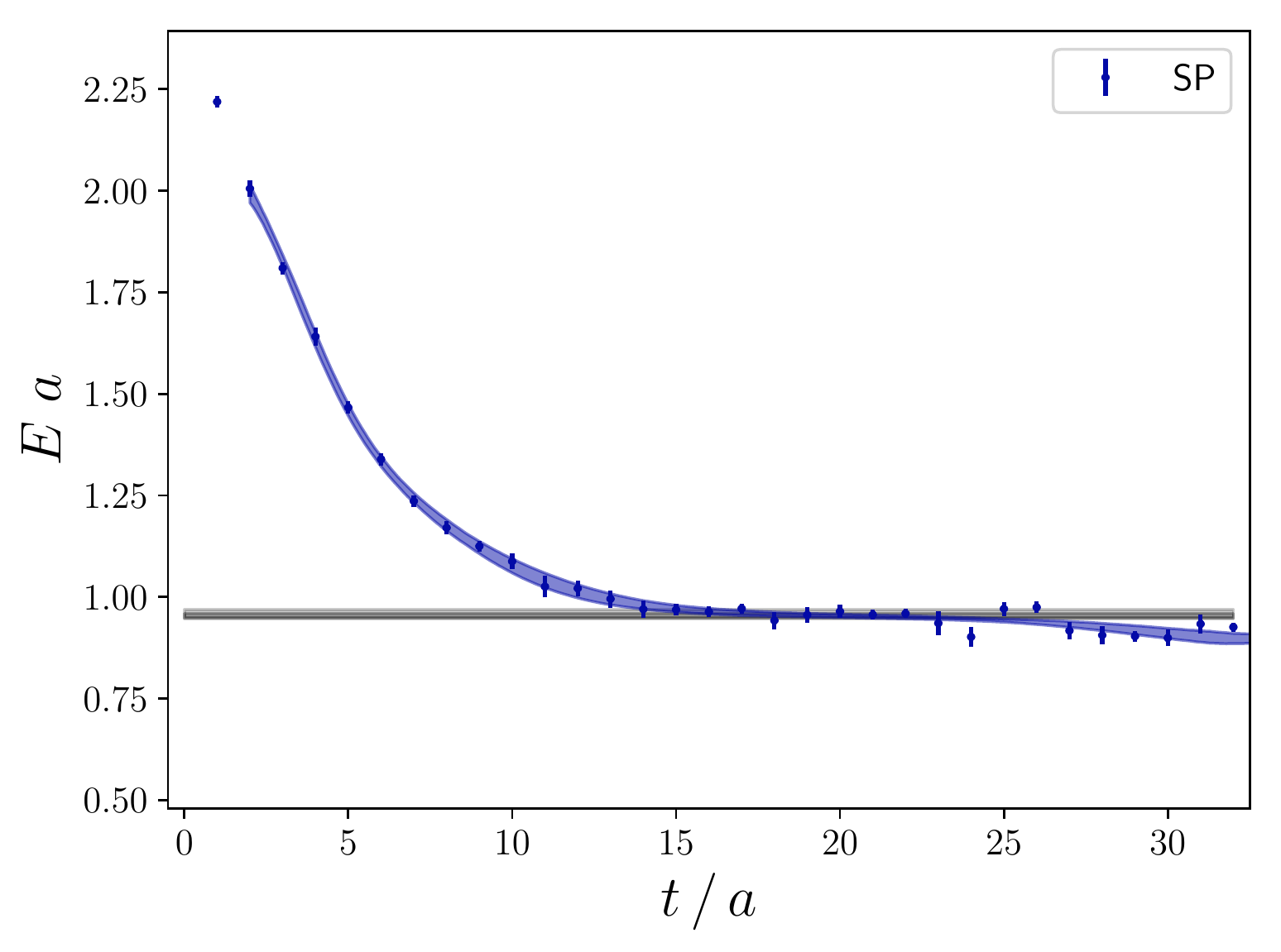}\hspace{10pt}
     \includegraphics[width=.40\textwidth]{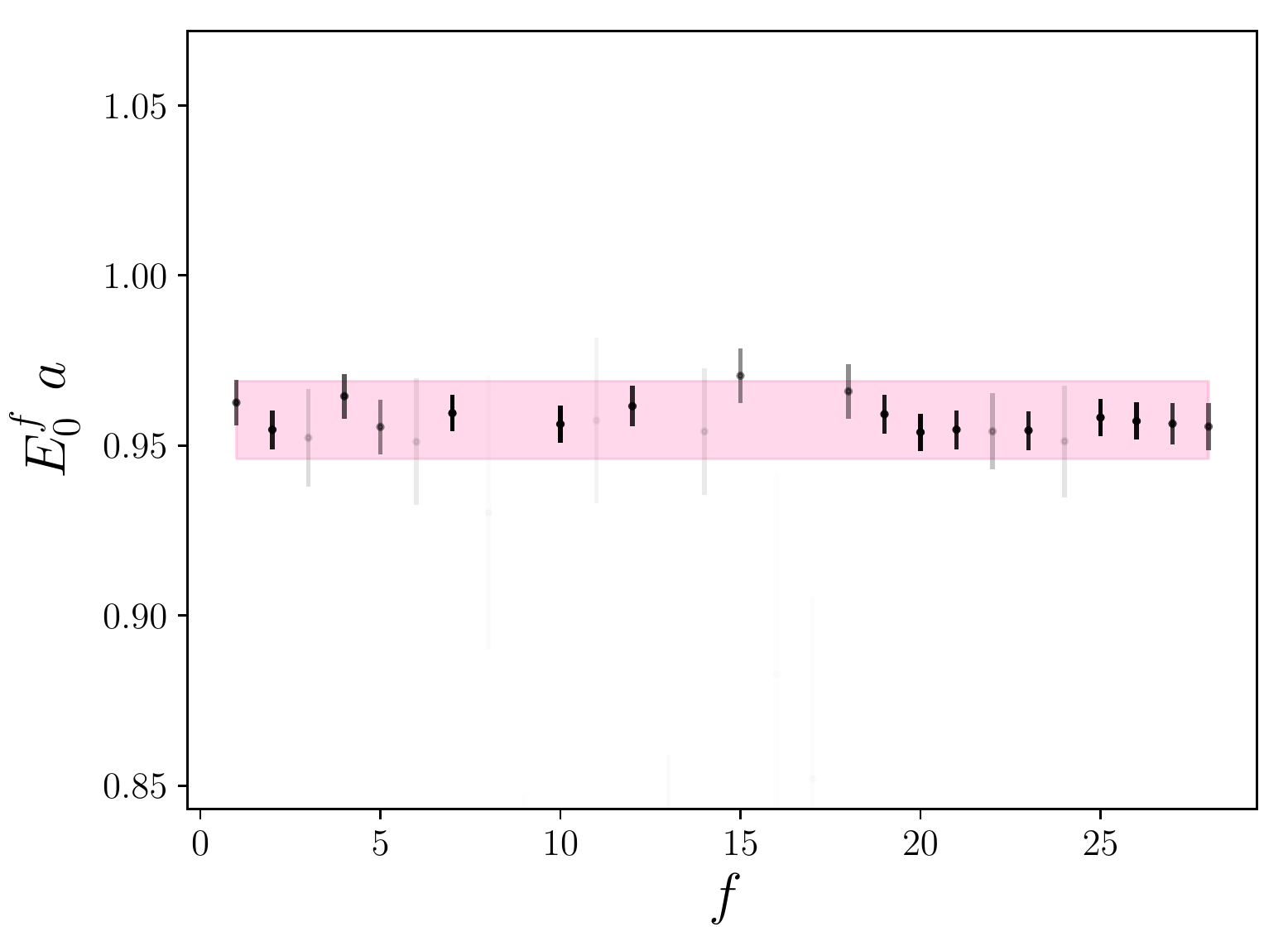} \\\vspace{-65pt}
     \begin{minipage}[c][150pt][t]{70pt} $7\ \pi^+\newline L/a=48$ \end{minipage}
     \includegraphics[width=.40\textwidth]{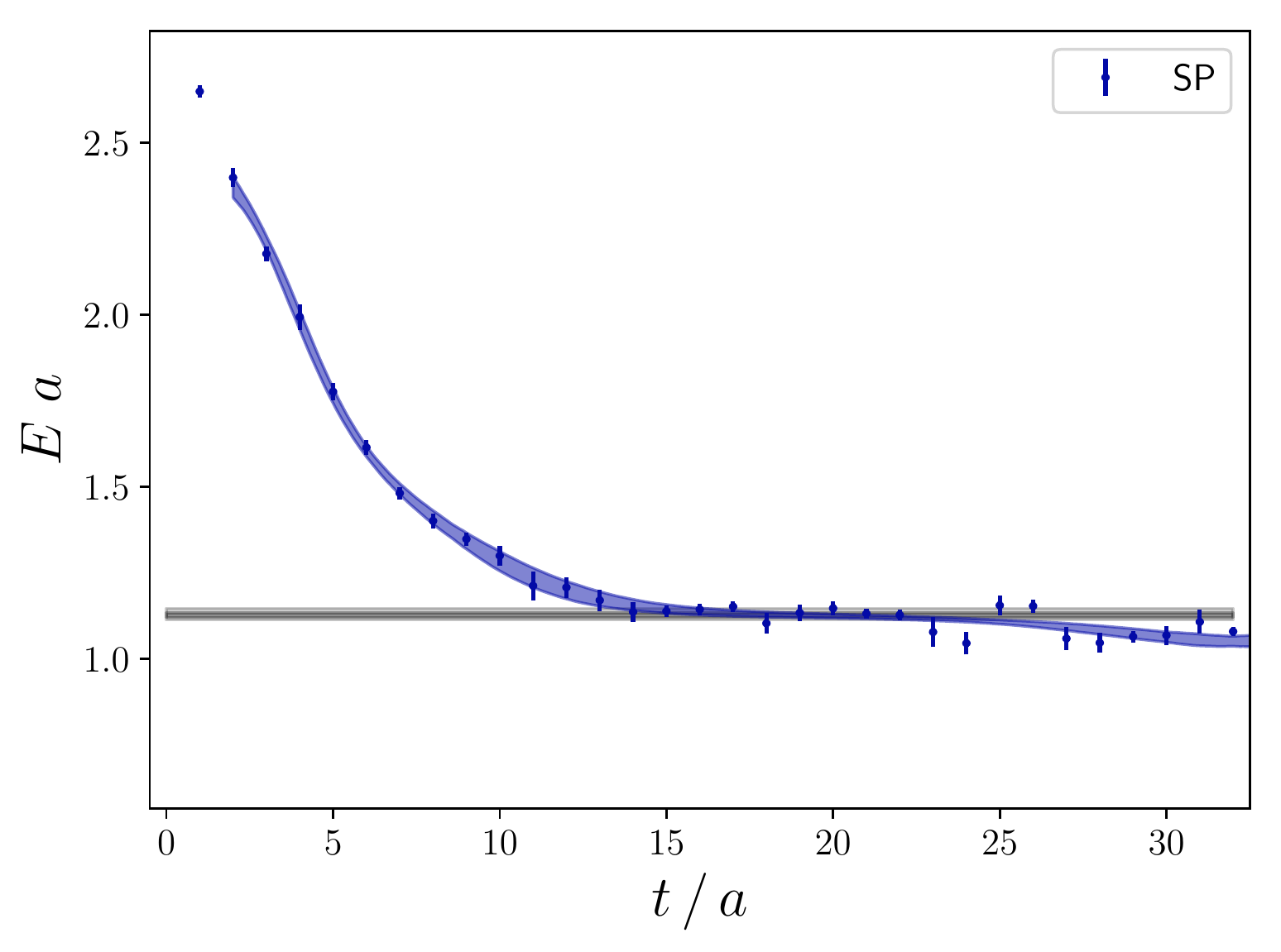}\hspace{10pt}
     \includegraphics[width=.40\textwidth]{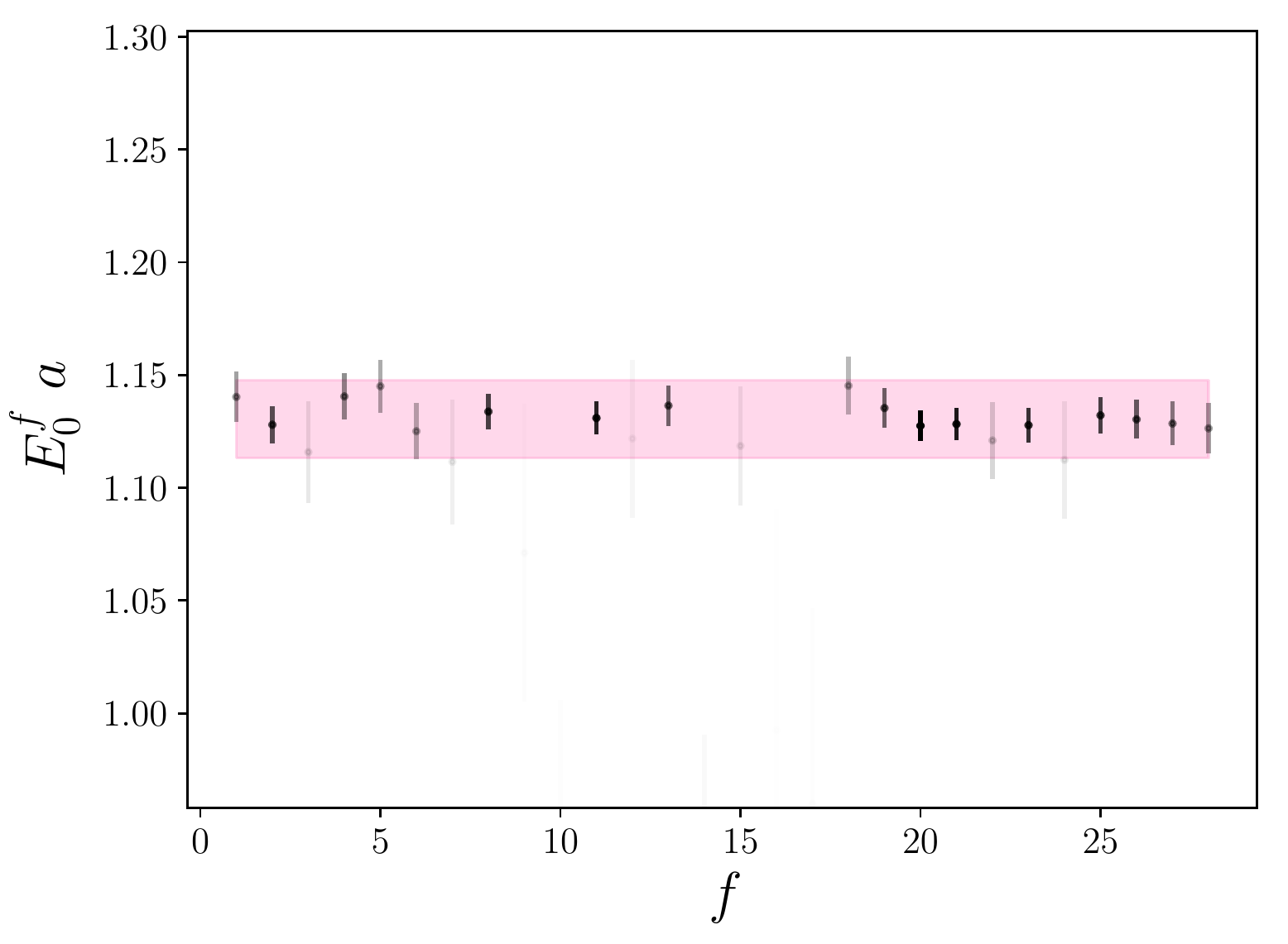} \\\vspace{-65pt}
     \begin{minipage}[c][150pt][t]{70pt} $8\ \pi^+\newline L/a=48$ \end{minipage}
     \includegraphics[width=.40\textwidth]{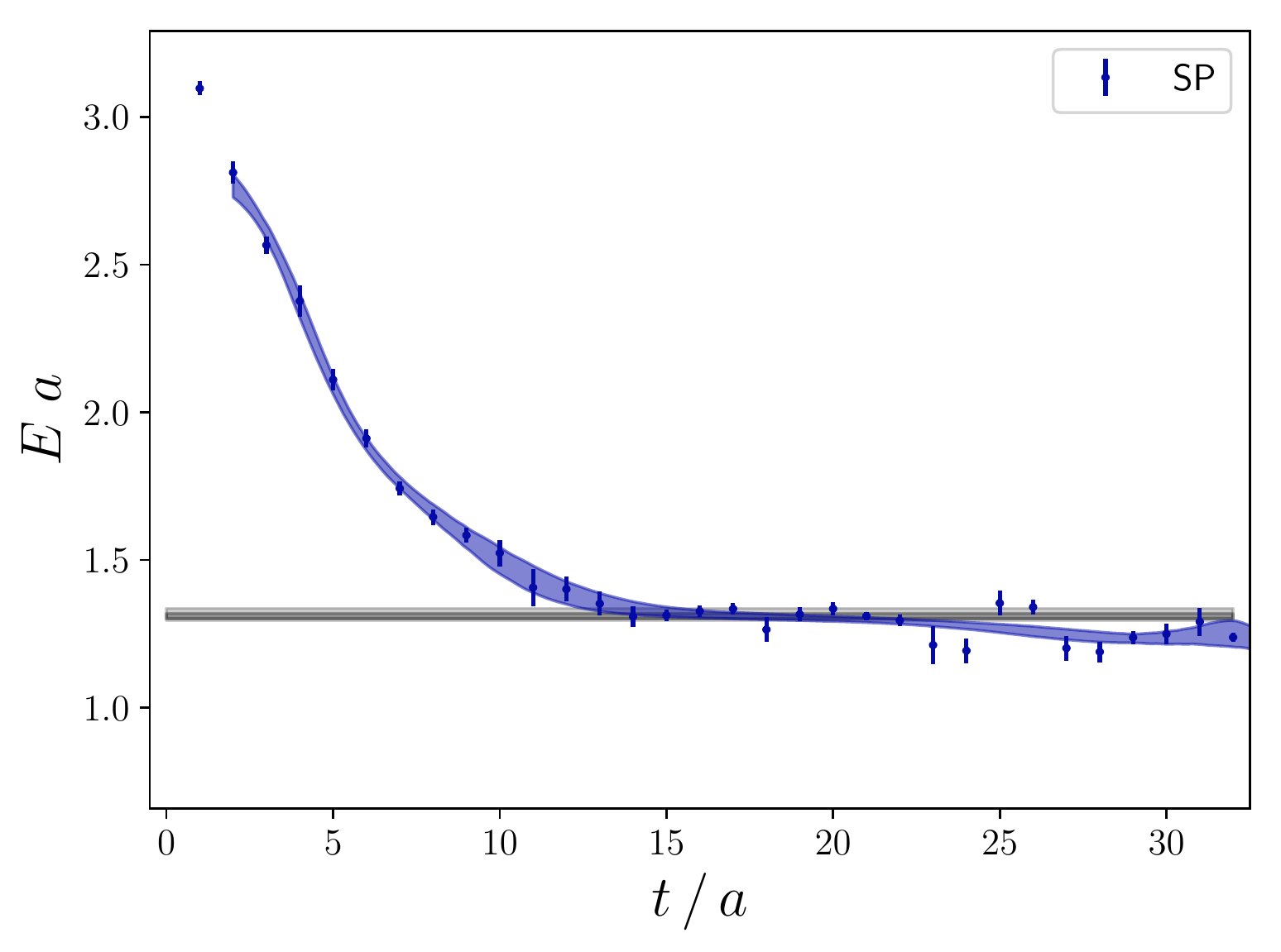}\hspace{10pt}
     \includegraphics[width=.40\textwidth]{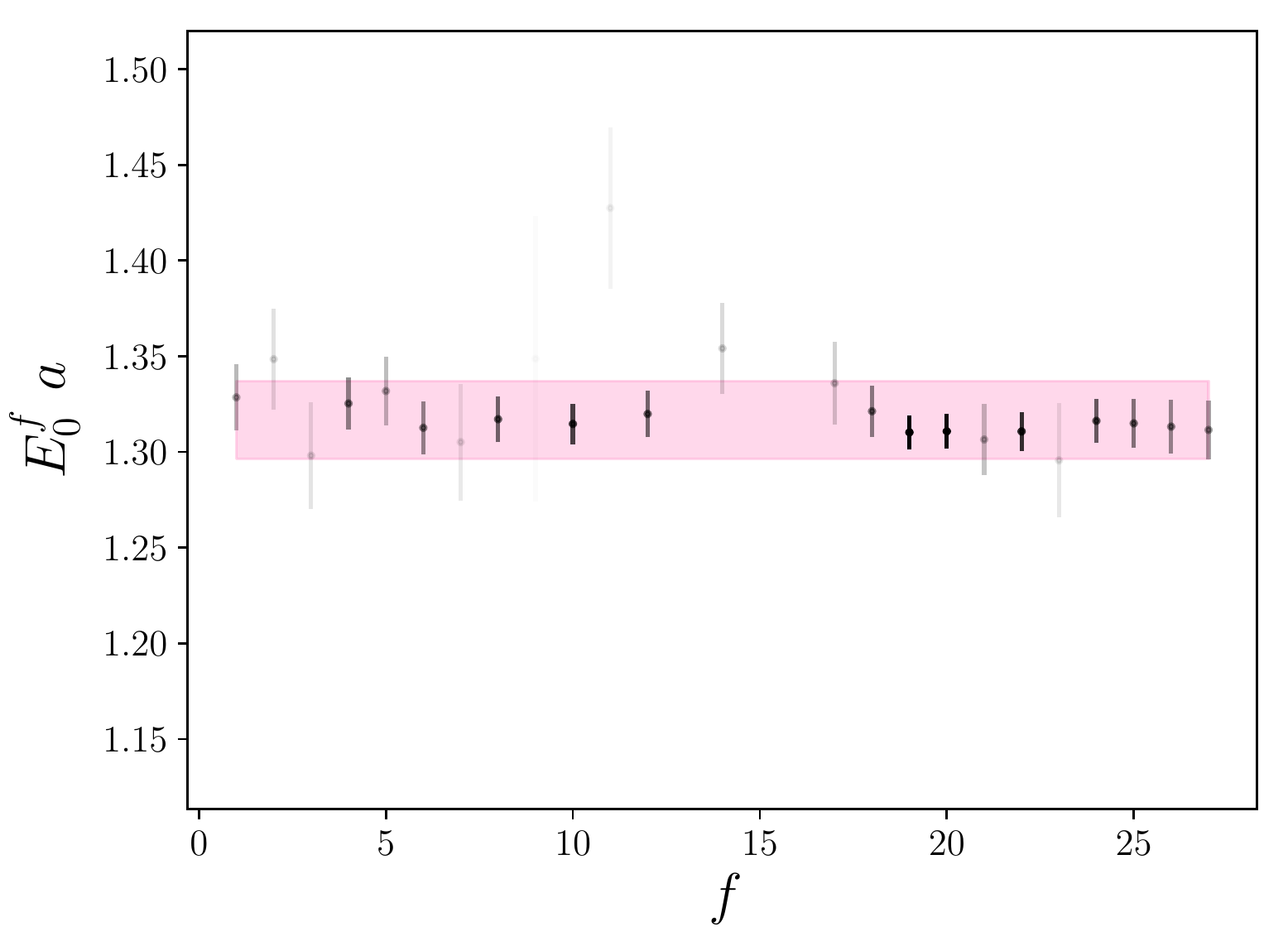} \\\vspace{-65pt}
   \caption{Fit results for systems of $n\in\{5,\dots,8\}$ $\pi^+$ mesons for the $L/a = 48$ lattice volume. The figures are analogous to Fig.~\ref{fig:pion48a}, see Appendix~\ref{app:fits} for a definition of the fitting procedure employed.
    \label{fig:pion48b}}
\end{figure}

\begin{figure}
  \centering
     \begin{minipage}[c][150pt][t]{70pt} $9\ \pi^+\newline L/a=48$ \end{minipage}
     \includegraphics[width=.40\textwidth]{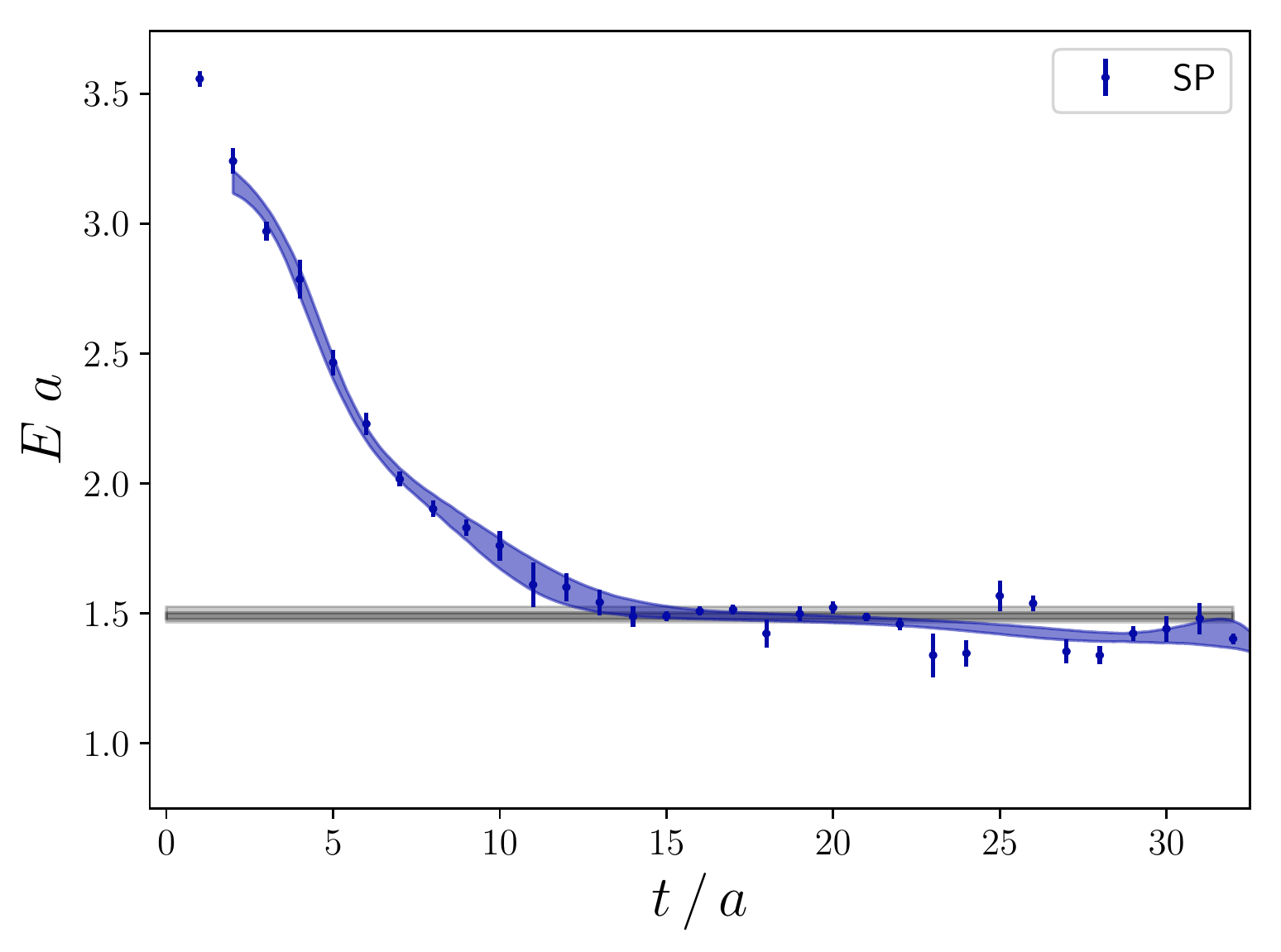}\hspace{10pt}
     \includegraphics[width=.40\textwidth]{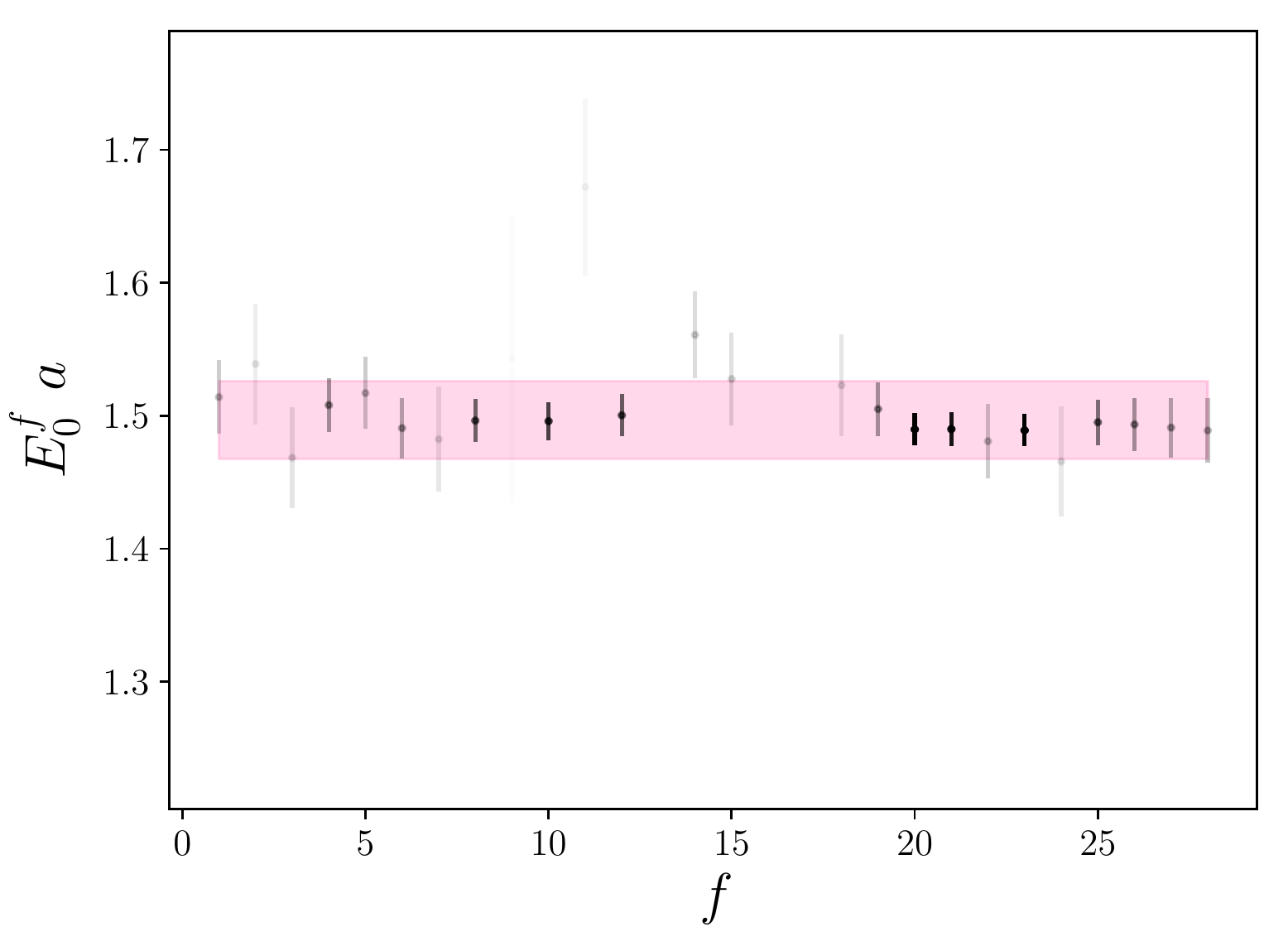} \\\vspace{-65pt}
     \begin{minipage}[c][150pt][t]{70pt} $10\ \pi^+\newline L/a=48$ \end{minipage}
     \includegraphics[width=.40\textwidth]{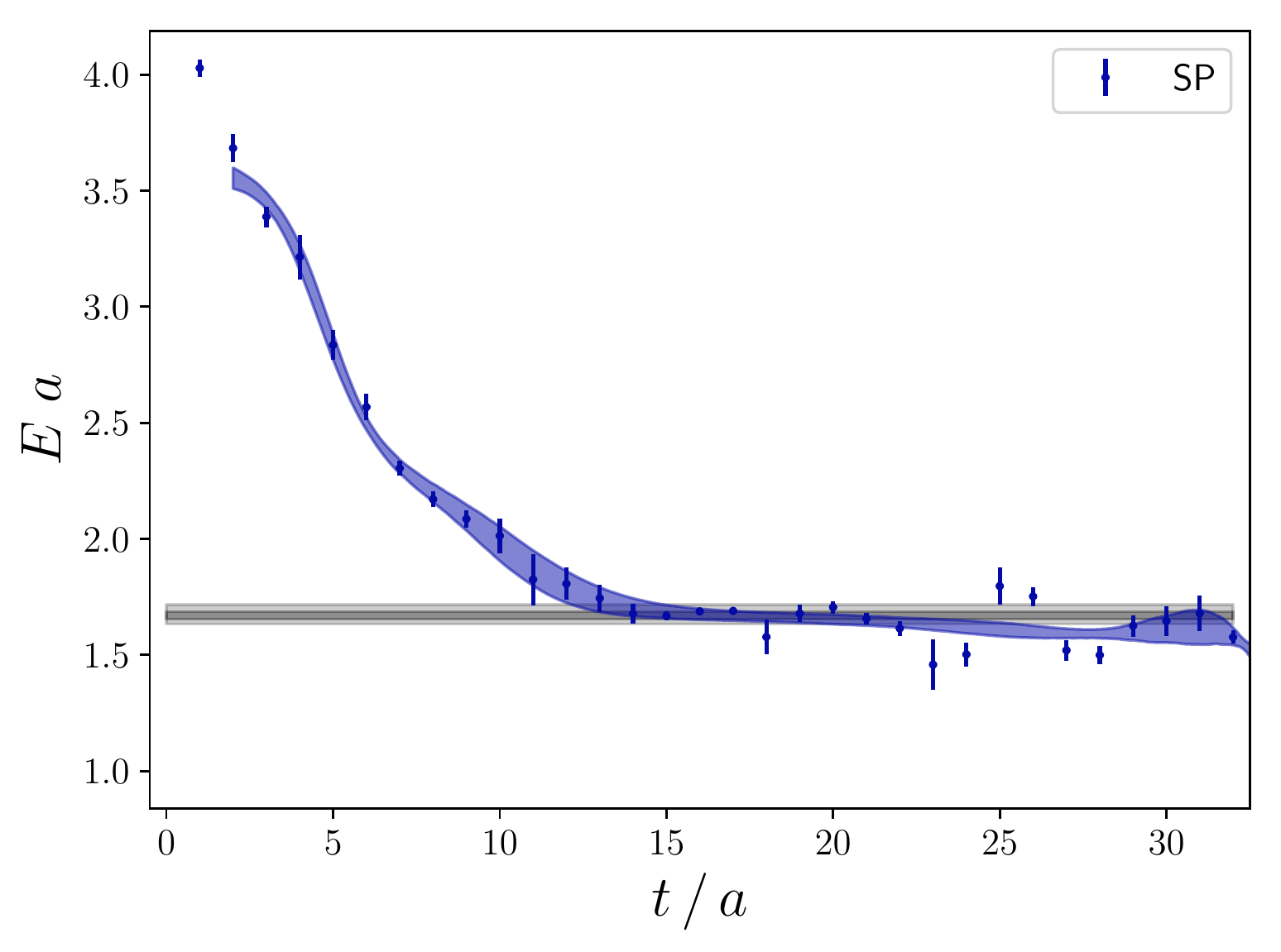}\hspace{10pt}
     \includegraphics[width=.40\textwidth]{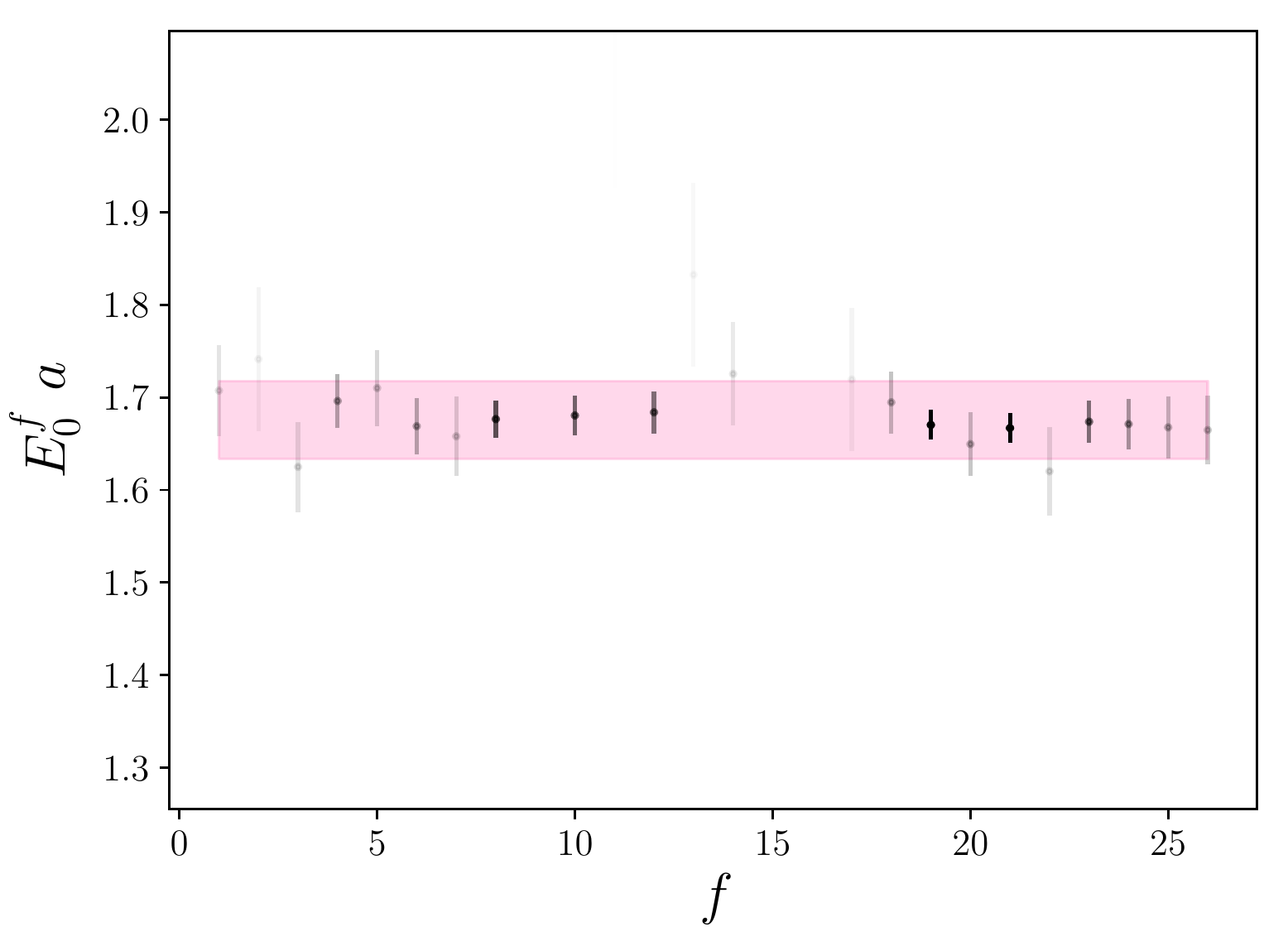} \\\vspace{-65pt}
     \begin{minipage}[c][150pt][t]{70pt} $11\ \pi^+\newline L/a=48$ \end{minipage}
     \includegraphics[width=.40\textwidth]{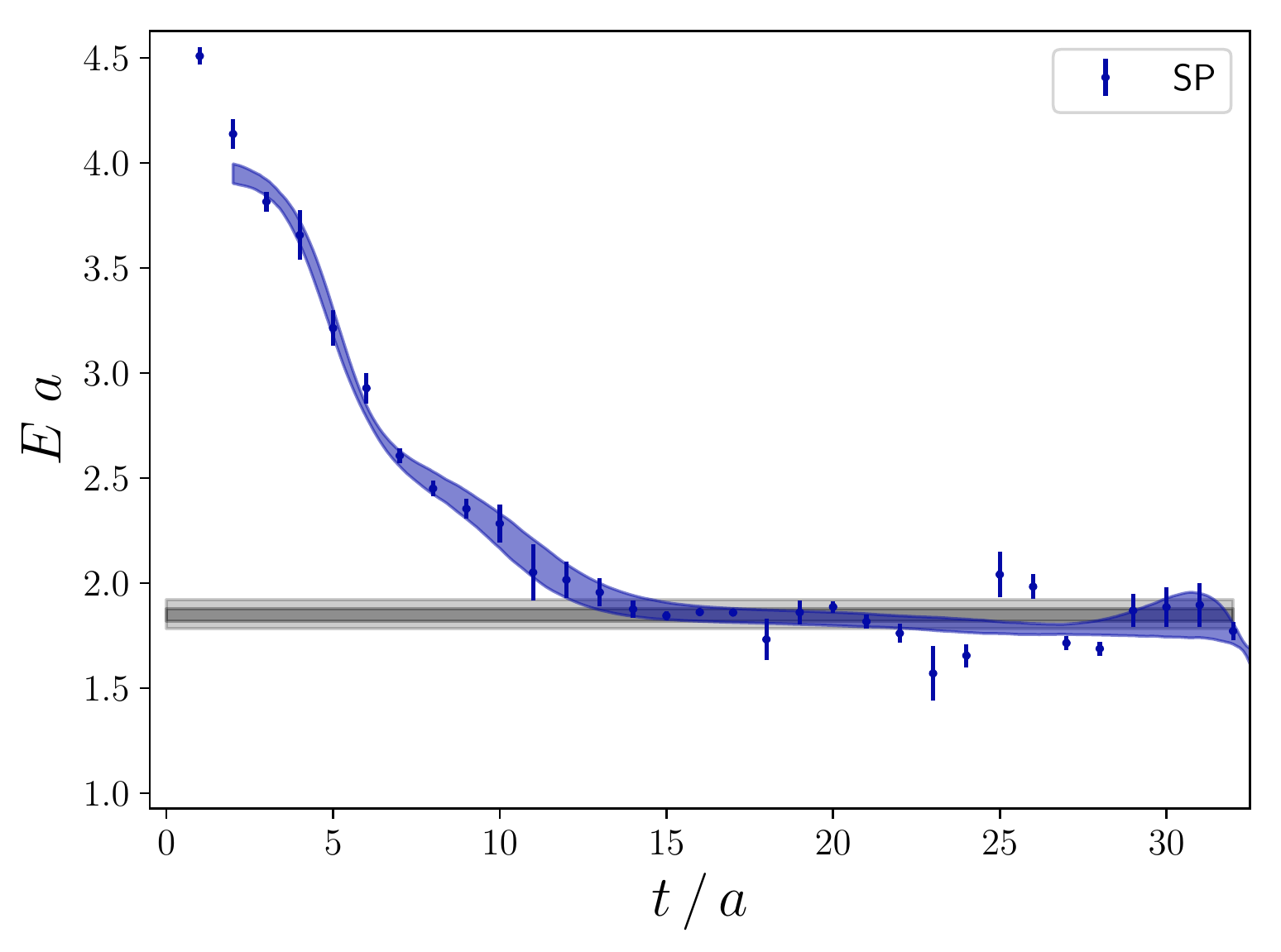}\hspace{10pt}
     \includegraphics[width=.40\textwidth]{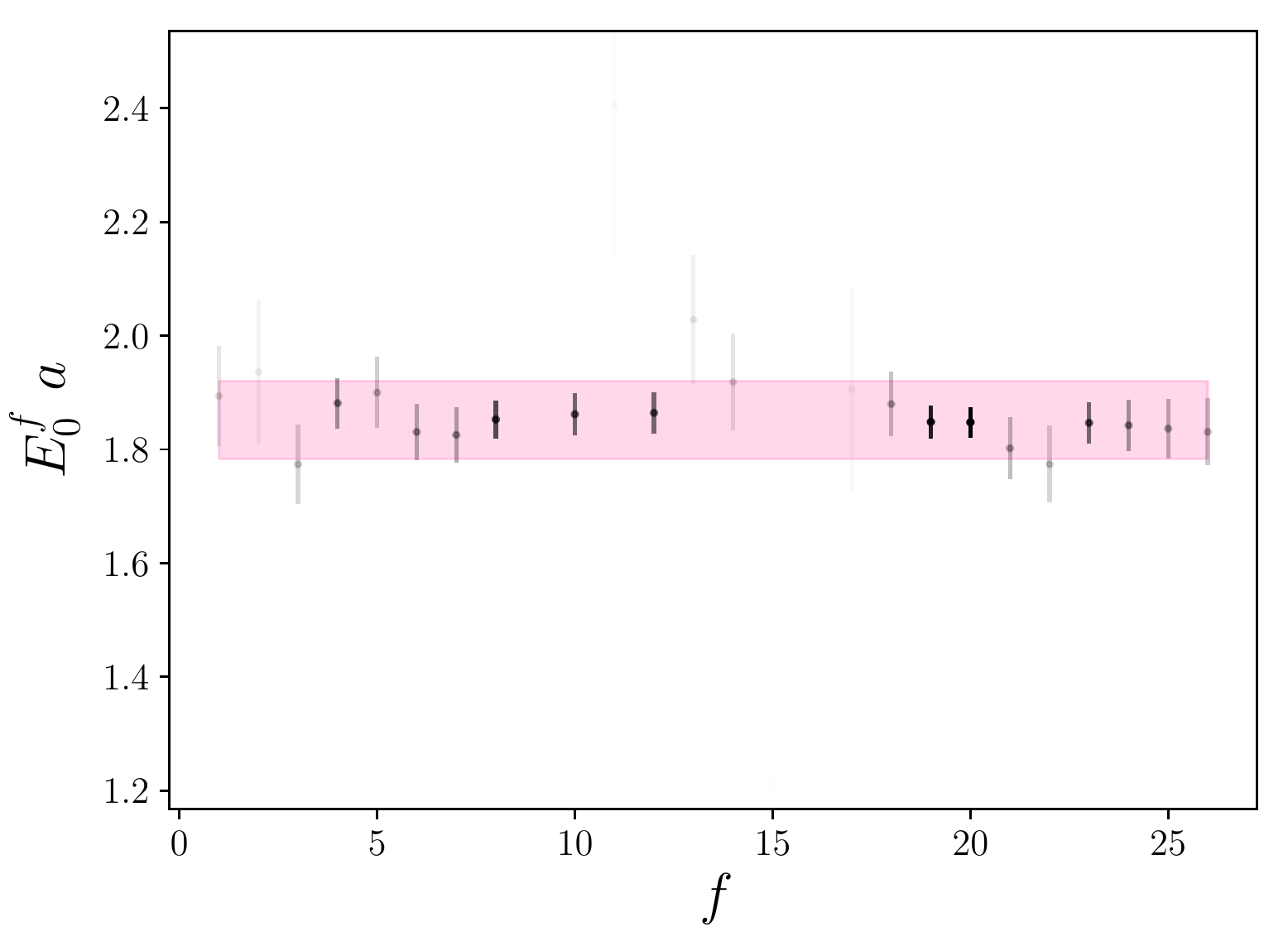} \\\vspace{-65pt}
     \begin{minipage}[c][150pt][t]{70pt} $12\ \pi^+\newline L/a=48$ \end{minipage}
     \includegraphics[width=.40\textwidth]{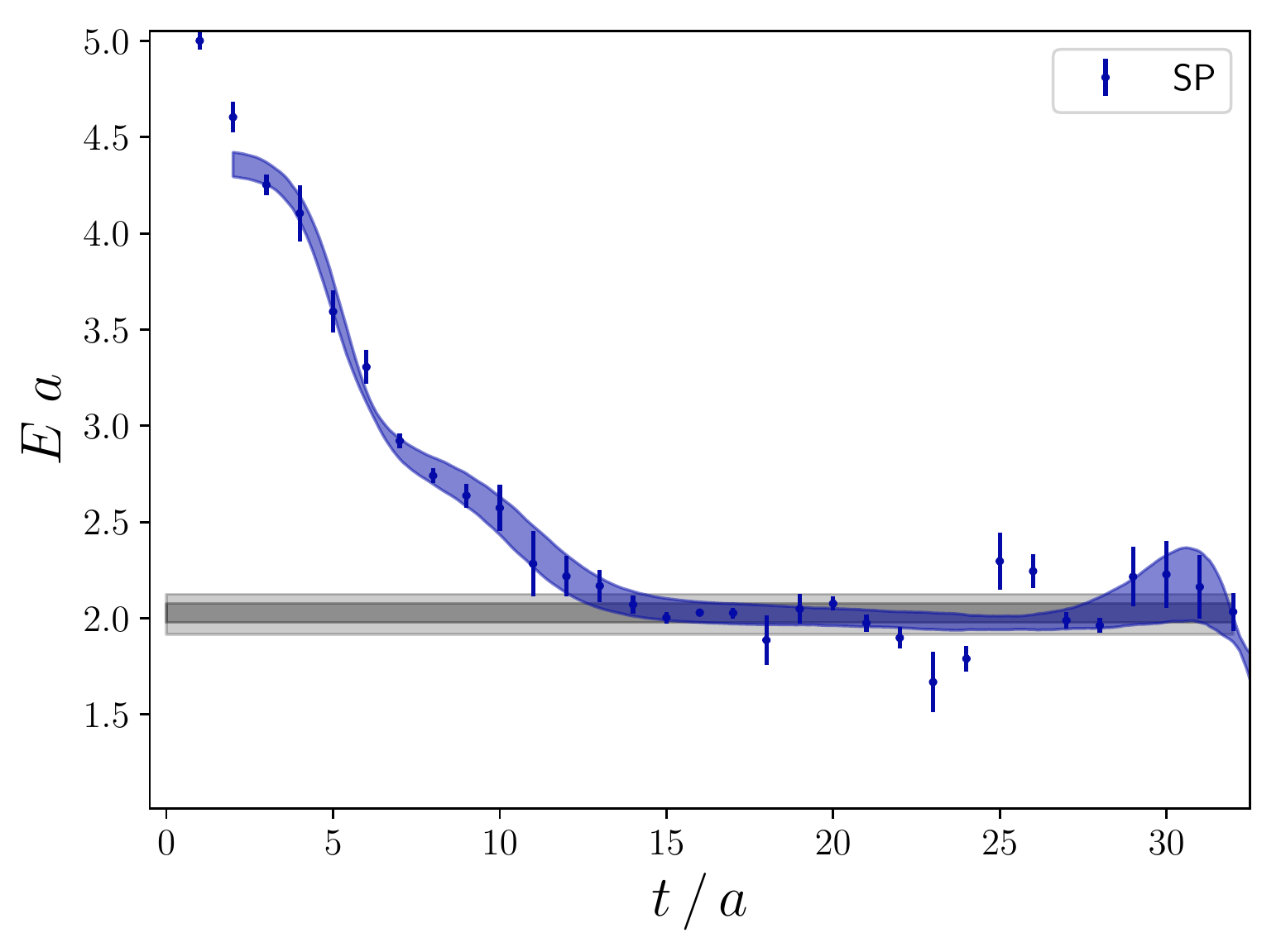}\hspace{10pt}
     \includegraphics[width=.40\textwidth]{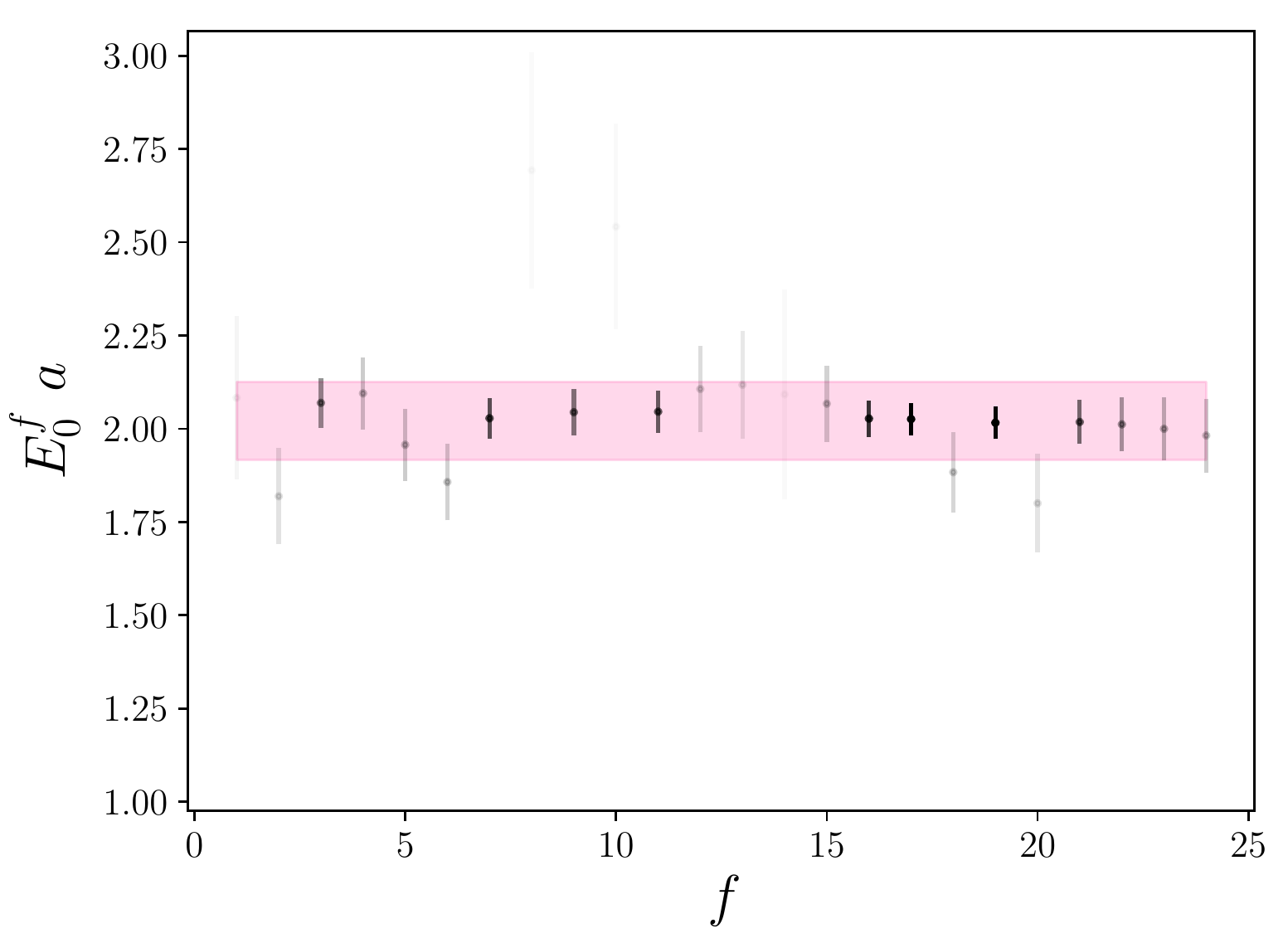} \\\vspace{-65pt}
   \caption{Fit results for systems of $n\in\{9,\dots,12\}$ $\pi^+$ mesons for the $L/a = 48$ lattice volume. The figures are analogous to Fig.~\ref{fig:pion48a}, see Appendix~\ref{app:fits} for a definition of the fitting procedure employed.
    \label{fig:pion48c}}
\end{figure}

\begin{table}[t!]
   \begin{tabular}{|c||c|c||c|c|} \hline
      & \multicolumn{2}{c||}{$a   E_{n\overline{K}^0} \vphantom{\frac{1}{2\frac{1}{2}}}$ }  & \multicolumn{2}{c|}{$a   E_{n\pi^+}$} \\\hline
      $n$ & $L/a = 32$  & $L/a = 48$  & $L/a= 32$  & $L/a = 48$ \\\hline\hline
      1   & 0.13910(45) & 0.13921(30) & 0.14921(49) & 0.15082(29) \\\hline
      2   & 0.2864(19)  & 0.2808(10)  & 0.3058(22)  & 0.30409(95) \\\hline
      3   & 0.4431(32)  & 0.4250(17)  & 0.4713(37)  & 0.4603(18)  \\\hline
      4   & 0.6164(86)  & 0.5730(39)  & 0.6535(81)  & 0.6205(37)  \\\hline
      5   & 0.803(14)   & 0.7245(63)  & 0.8479(13)  & 0.7848(60)  \\\hline
      6   & 1.024(23)   & 0.886(11)   & 1.076(21)   & 0.958(11)   \\\hline
      7   & 1.253(32)   & 1.049(14)   & 1.316(27)   & 1.130(17)   \\\hline
      8   & 1.527(63)   & 1.227(19)   & 1.591(49)   & 1.317(20)   \\\hline
      9   & 1.82(11)    & 1.404(26)   & 1.865(81)   & 1.497(29)   \\\hline
      10  & 2.28(17)    & 1.597(41)   & 2.17(12)    & 1.676(42)   \\\hline
      11  & 1.8(1.1)    & 1.796(58)   & 2.54(19)    & 1.852(68)   \\\hline
      12  & 2.7(1.1)    & 2.002(79)   & 3.20(20)    & 2.02(10)    \\\hline
   \end{tabular}
   \caption{Ground-state energy results for systems of $n\in\{1,\ldots,12\}$ neutral $\overline{K}^0$ mesons and charged $\pi^+$ mesons with lattice volumes $L/a$ shown. Results are determined by fitting Eq.~\eqref{eq:spectral} to the LQCD+QED$_L$ Euclidean correlation functions using the methods described in Appendix~\ref{app:fits}. Dimensionful energy results can be obtained using the lattice spacing $a = 0.068(2) \text{ fm}$ obtained for this gauge field ensemble in Ref.~\cite{Horsley:2015eaa}.  \label{tab:mesonM} }
\end{table}

%%%%%%%%%%%%%%%%%%%%%%%%%%%%%%%%%%%%%%%%%%%%%%%%%%%%%%%%%%%%%%%%%%%%%%%%%
\begin{figure}[t]
  \centering
  \includegraphics[width=.7\textwidth]{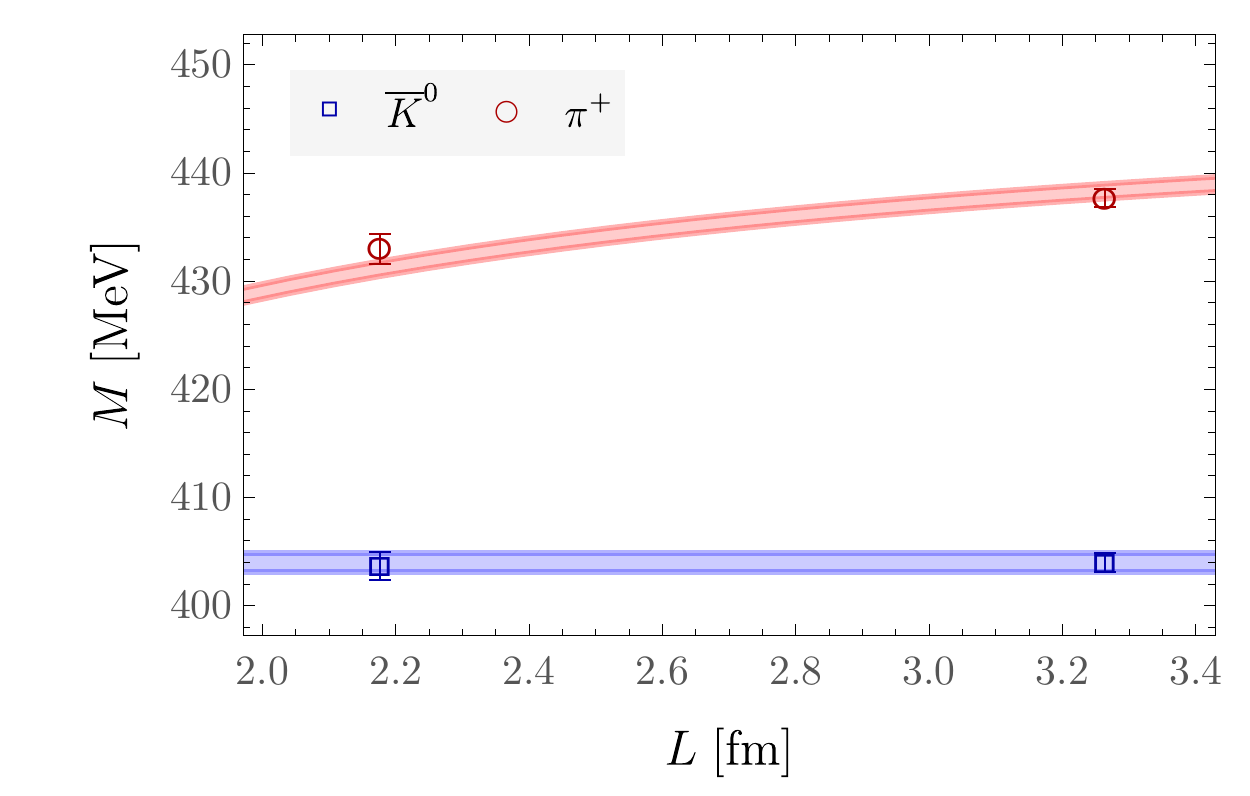}
  \caption{The blue and red points show the LQCD+QED$_L$ results from Table~\ref{tab:mesonM} for the $\overline{K}^0$ and $\pi^+$ single-particle energies for the $L/a\in\{32,48\}$ volumes. The blue band shows a constant fit to $E_{\overline{K}^0}$ results, and the red band shows a fit to QED$_L$ power-law FV effects at $\mathcal{O}(\alpha/(ML)^2)$ derived in Refs.~\cite{Borsanyi:2014jba,Davoudi:2014qua} and presented in Eq.~\eqref{eq:EpiNLO}. The width of the bands corresponds to 67\% confidence intervals estimated using bootstrap resampling. %Significant deviations between the extrapolated fit results and LQCD+QED$_L$ results for both $\overline{K}^0$ and $\pi^+$ on the $L/a = 24$ volume suggest that exponentially suppressed relativistic FV effects are significant for the $L/a = 24$ volume as discussed in Refs.~\cite{Horsley:2015eaa}.
    \label{fig:1_meson}}
\end{figure}
%%%%%%%%%%%%%%%%%%%%%%%%%%%%%%%%%%%%%%%%%%%%%%%%%%%%%%%%%%%%%%%%%%%%%%%%%

For one- and two-nucleon systems, combined fits to the SS and SP correlation functions are performed using common energies but different overlap factors for local and smeared sources.
For mesons and three-nucleon systems, combined fits performed in this way are only marginally more precise than fits using SP correlation functions alone, and fits using only SP correlation functions are therefore used in what follows for simplicity.

An effective energy plot that removes thermal effects from backwards propagating states for $n=1$ systems, as well as constant contributions from thermal effects on $n \geq 2$ correlation functions, is defined as
\begin{equation}
   \begin{split}
      E(t) = \text{ArcCosh} \left( \frac{G(t) - G(t+4)}{2 \left[G(t+1) - G(t+3) \right] } \right).
   \end{split}\label{eq:EMdef}
\end{equation}
Note that this effective ground-state energy function differs from the form $\text{ArcCosh}\left[ (G(t+1) + G(t-1))/(2G(t)) \right]$ commonly used to remove thermal effects from single-pion correlation functions. The advantage of Eq.~\eqref{eq:EMdef} is that constant terms present in the spectral representation Eq.~\eqref{eq:spectral} for $n \geq 2$ mesons exactly cancel in the correlation function differences appearing in Eq.~\eqref{eq:EMdef}, and in particular Eq.~\eqref{eq:EMdef} is able to exactly isolate the ground-state energy from a two-meson correlation function including thermal effects of the form $e^{-E t} + e^{-E(T-t)} + \text{const}$.
Additional terms with more complicated $t$ dependence arise for $n \geq 3$ meson correlation functions as shown in Eq.~\eqref{eq:spectral}. Effective energies for $n \geq 3$ systems will therefore include contamination from thermal effects and will only plateau to the true ground-state energy in the region $t \ll T$ where thermal effects can be neglected.

Figs.~\ref{fig:pion48a}-\ref{fig:pion48c} show effective energy plots, including correlation function results and fit results for $E(t)$ for each case of $n\in \{1,\ldots,12\}\pi^+$ on the $L/a=48$ lattice volume as well as the ground-state energy results from all successful fit range choices.
Analogous results for the $L/a=32$ volume and for $n = \{1,\ldots,12\}\overline{K}^0$ systems on the $L/a \in \{32,48\}$ volumes are shown in Appendix~\ref{app:fits}.
The resulting ground-state energies for $n\in \{1,\ldots,12\}$ meson systems on both lattice volumes are shown in Table~\ref{tab:mesonM}.

\subsection{Hadron mass results}

Single-meson masses computed with the gauge field ensembles used here have already been presented in Ref.~\cite{Horsley:2015eaa}; the meson mass results obtained using the fitting produce employed here are shown for completeness in Fig.~\ref{fig:1_meson}.
The $\overline{K}^0$ energy is equal within statistical uncertainties for the $L/a = 48$ and $L/a=32$ lattice volumes. 
Volume dependence in $E_{\pi^+}(L)$ and differences in the FV single-particle energies $E_{\pi^+}(L) - E_{\overline{K}^0}(L)$, however, are clearly visible.
Because of the quark mass tuning described in Sec.~\ref{sec:lattice}, which is designed to remove strong isospin breaking effects, these energy differences are ascribed to QED effects.
In order to interpret these QED effects, results for $E_{\pi^+}(L)$ can be compared to the $\mathcal{O}(\alpha,\ L^{-2})$ prediction of QED$_L$~\cite{Borsanyi:2014jba} or equivalently NRQED$_L$~\cite{Davoudi:2014qua},
\begin{equation}
   \begin{split}
      E_{\pi^+}(L) = m_\pi + \frac{\alpha}{2 L} c_1 \left( 1 + \frac{2}{m_\pi L} \right) + \mathcal{O}\left( \alpha^2,\ L^{-3} \right),
   \end{split}\label{eq:EpiNLO}
\end{equation}
where $c_1 = -0.266596$ is a geometric constant that does not depend on the system under consideration.
Nonlocal effects from zero mode subtraction and charge radius effects introducing a new parameter both arise at next-to-next-to-next-to-leading-order (N$^3$LO) and are neglected here.
Fitting the $L/a\in\{32,48\}$ results to Eq.~\eqref{eq:EpiNLO} gives 
\begin{equation}
   \begin{split}
      a\ m_{\pi^+} = 0.15419(29), \hspace{20pt} m_{\pi^+} = 449(1)(13)\text{ MeV},
   \end{split}\label{eq:mpi}
\end{equation}
where the lattice spacing $a=0.068(2) \text{ fm}$ is determined in Ref.~\cite{Horsley:2015eaa}. The first uncertainty in each term of Eq.~\eqref{eq:mpi} includes statistical and systematic fitting uncertainties as detailed in Appendix~\ref{app:fits}, and the second uncertainty in the expression for $m_{\pi^+}$ arises from the uncertainty in $a$.
A constant fit to the $L/a\in\{32,48\}$ results for $E_{\overline{K}^0}(L)$ gives 
\begin{equation}
   \begin{split}
      a\ m_{\overline{K}^0} = 0.13918(25), \hspace{20pt} m_{\overline{K}^0} =  404(1)(12)\text{ MeV}.
   \end{split}\label{eq:mK}
\end{equation}
For the values of the quark masses and $\alpha \simeq 0.1$ used in this work, strong isospin breaking effects approximately vanish and the difference between the charged and neutral pseudoscalar meson masses $m_{\pi^+} - m_{\overline{K}^0} = 45(2)$ MeV is attributed entirely to QED effects.
It is noteworthy that QED effects account for approximately $10\%$ of the $\pi^+$ mass with these parameters.

\begin{table}[t!]
	\begin{tabular}{|c||c|c|} \hline
      & \multicolumn{2}{c|}{$a   E_{b}$ }   \\\hline
      $b$ & $L/a = 32$  & $L/a = 48$   \\\hline\hline
      $p$             & 0.3997(43) & 0.3998(34)  \\\hline
      $n$            & 0.3988(43) & 0.3962(30) \\\hline
      $pp$            & 0.816(19)  & 0.816(16) \\\hline 
      $np({}^1S_0)$  & 0.807(20)  & 0.820(13)  \\\hline
      $nn$           & 0.806(18)  & 0.823(12)  \\\hline
      $np({}^3S_1)$   & 0.825(16)  & 0.811(12)  \\\hline
      ${}^3\text{He}$ & 1.294(77)  & 1.268(42)  \\\hline
      ${}^3\text{H}$ & 1.291(53)  & 1.294(31)  \\\hline
	\end{tabular}
   \caption{Ground-state energies of systems of $n\in\{1,2,3\}$ protons and neutrons determined by fitting Eq.~\eqref{eq:baryon} to the LQCD+QED$_L$ Euclidean correlation functions as described in Appendix~\ref{app:fits}. \label{tab:baryon} }
\end{table}

%%%%%%%%%%%%%%%%%%%%%%%%%%%%%%%%%%%%%%%%%%%%%%%%%%%%%%%%%%%%%%%%%%%%%%%%%
\begin{figure}[t]
  \centering
  \includegraphics[width=.7\textwidth]{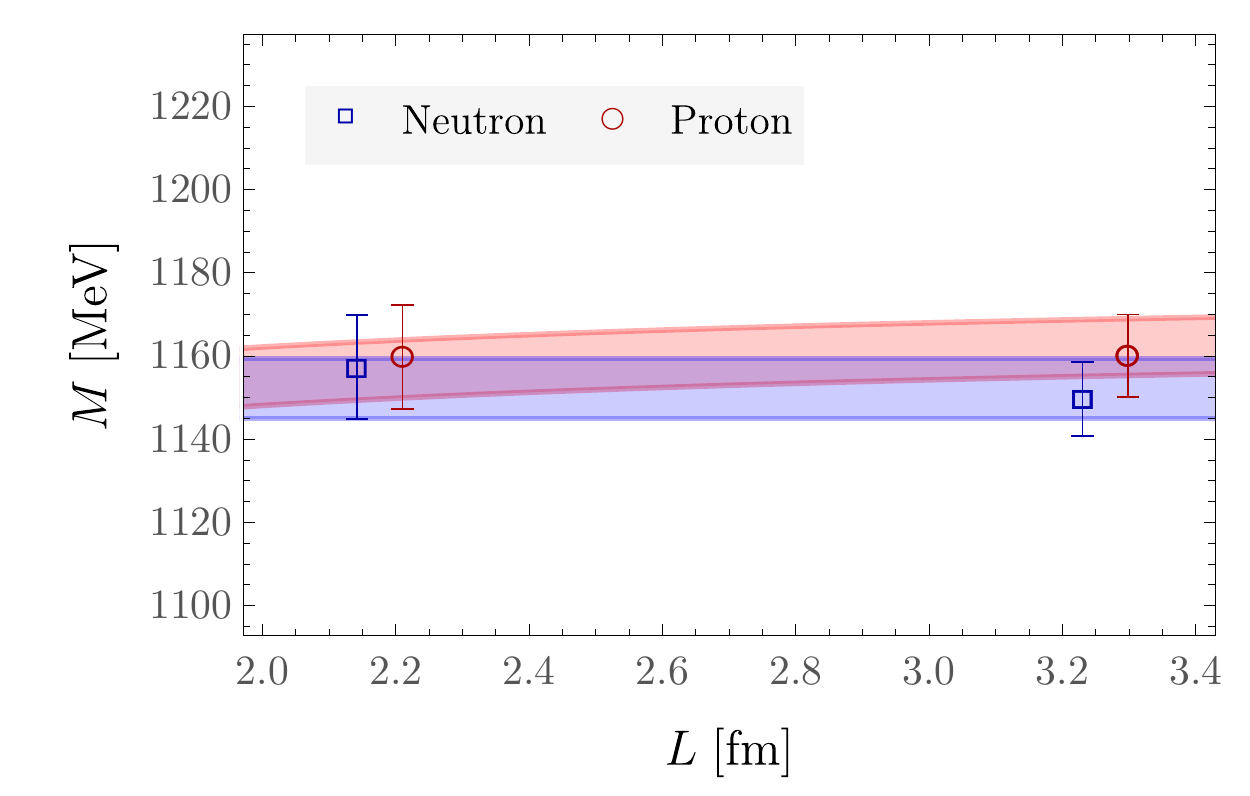} \;
  \caption{     
  The blue and red points show the LQCD+QED$_L$ results from Table~\ref{tab:baryon} for the neutron and proton single-particle energies for the $L/a\in\{32,48\}$ volumes. The blue band shows a fit to a constant for the $L/a\in\{32,48\}$ results for the neutron, and the red band shows a fit to Eq.~\eqref{eq:EpiNLO} (which is valid for arbitrary charged hadrons as well as the pion) for the proton. The band widths correspond to 67\% bootstrap confidence intervals.  A small horizontal offset is applied symmetrically to proton and neutron results.
  \label{fig:1_baryon}}
\end{figure}
%%%%%%%%%%%%%%%%%%%%%%%%%%%%%%%%%%%%%%%%%%%%%%%%%%%%%%%%%%%%%%%%%%%%%%%%%

Results for the proton and neutron ground-state energies are shown in Table~\ref{tab:baryon} and Fig.~\ref{fig:1_baryon}.
Volume dependence of the neutron mass is expected to be exponentially suppressed and is found to be consistent with zero within statistical uncertainties.
The proton mass includes power-law FV corrections from QED effects identical to those shown for the $\pi^+$ in Eq.~\eqref{eq:EpiNLO}.
These relativistic FV effects are suppressed by $\mathcal{O}\left((M_p L)^{-1}\right)$ and are therefore suppressed in $E_p(L)$ compared to $E_{\pi^+}(L)$ by $m_{\pi^+} / M_p \ll 1$.
The proton energy is found to have mild FV effects consistent with zero and with Eq.~\eqref{eq:EpiNLO}.
Fits to the NLO QED$_L$ prediction for $E_p(L)$ and to a constant for $E_n(L)$ give the results
   \begin{align}\label{eq:baryonM}
      a M_p &= 0.4037(23), \hspace{20pt} M_p =  1171(7)(34) \text{ MeV}, \\
      a M_n &= 0.3971(25), \hspace{20pt} M_n =  1152(7)(34) \text{ MeV},
   \end{align}
where the uncertainties are as defined for the $\pi^+$ after Eq.~\eqref{eq:mpi}.
Combining these results gives $M_p - M_n = 20(10)$ MeV.
This result is approximately ten times larger than the QED contribution to the proton-neutron mass difference at the physical values of the quark masses and $\alpha$~\cite{Borsanyi:2013lga,Borsanyi:2014jba,Horsley:2015eaa},
which is consistent with the expected linear dependence of the proton-neutron mass difference on $\alpha$, given the value $\alpha \simeq 0.1$ that is used here.

\section{Finite-volume non-relativistic QED}
\label{sec:theory}

Hadronic EFT results relating FV energy levels to the LECs appearing in EFT Lagrangians are useful for interpreting LQCD+QED$_L$ results, as evidenced by the use of fits to Eq.~\eqref{eq:EpiNLO} to describe the volume dependence of single-hadron energies and to extract $m_{\pi^+}$ and $M_p$ at $L=\infty$.
Analogous EFT results for charged multi-hadron systems are needed to extract hadronic scattering information from LQCD+QED$_L$ results for multi-hadron FV energy levels, but EFT for charged multi-hadron systems is complicated by the presence of Coulomb interactions that are nonperturbative for hadron pairs with sufficiently small relative momentum as discussed below.
Further complications arise for non-relativistic EFTs for QCD+QED$_L$ because of the nonlocality inherent in zero-mode subtraction.
However, non-relativistic EFTs have the advantage that FV energy shifts for systems of arbitrary particle number can be computed more simply than in relativistic EFTs, where the relation between scattering parameters and FV energy levels is not yet known for systems of four or more hadrons. 
This work therefore pursues the application of non-relativistic EFT to charged multi-hadron FV systems, and this section develops the formalism necessary for studying multi-hadron systems in NRQED$_L$ including nonlocal effects arising from zero-mode subtraction.

\subsection{Finite-volume formalism for two charged hadrons}

\label{sec:scattering}

Interactions of two electrically neutral particles with mass $M$ and relative momentum $2 \v{p}$ are described at low energies by a scattering phase shift $\delta(p)$, where $p = |\v{p}|$.
The phase shift is an analytic function of the center-of-mass energy $E^* = 2 \sqrt{p^2 + M^2}$ for energies below both the $t$-channel cut and the $s$-channel inelastic particle production threshold.
For $p \ll M$, the energy can be described by the non-relativistic expansion $E^*= 2M + p^2/M + \ldots $, and below the $t$-channel cut and inelastic threshold the phase shift admits a convergent effective range expansion $p \cot \delta(p) = -\frac{1}{a} + \frac{r}{2}p^2 + \ldots$, where $a$ is the scattering length (not to be confused with the lattice spacing $a$ appearing elsewhere) and $r$ is the effective range.
For neutral particles, this expansion is straightforwardly reproduced in terms of pionless EFT~\cite{vanKolck:1998bw,Chen:1999tn}, a theory of hadrons interacting via contact interactions organized in powers of derivatives.

Interactions of electrically charged particles are complicated by the fact that the $t$-channel cut associated with one-photon-exchange and the inelastic photon production threshold start at $p=0$.
This leads to $p=0$ singularities in contributions to the scattering amplitude from Coulomb ladder diagrams describing iterated one-photon-exchange, shown in Fig.~\ref{fig:coulomb}.
Increasingly higher-loop Coulomb ladder diagrams are suppressed by powers of $\alpha$, but the $p=0$ singularity becomes more severe.
The loop expansion for Coulomb ladder diagrams corresponds to a perturbative expansion in powers of
\begin{equation}
   \begin{split}
     \eta(p) \equiv  \frac{\alpha}{v(p)} = \frac{\alpha M}{2 p},
   \end{split}\label{eq:etadef}
\end{equation}
where $v$ is the relative velocity of the two charged particles and $p\ll M$ is assumed throughout this section.
For $\eta(p) \gtrsim 1$, QED becomes nonperturbative and Coulomb ladder diagrams must be resummed to all orders in $\alpha$.
As shown in nonrelativistic quantum mechanics by Bethe~\cite{Bethe:1949yr}, and in EFT by Kong and Ravndal~\cite{Kong:1998sx,Kong:1999sf}, the resummed scattering amplitude is nonanalytic in $\eta$ and the effective range expansion is modified as
\begin{equation}
   \begin{split}
      \left(\frac{2\pi \eta}{e^{2\pi \eta} - 1}\right) p \cot \delta(p) + 2 p \eta \left[ \text{Re}\psi(i \eta) - \ln(\eta) \right] = -\frac{1}{a_C} + \frac{1}{2}r_C p^2 + \ldots,
   \end{split}\label{eq:ERE}
\end{equation}
where $\psi(x) = \Gamma^\prime(x)/\Gamma(x)$ is the digamma function, $a_C$ is the Coulomb-corrected scattering length, and $r_C$ is the effective range of the charged particle system.
To leading order in $\alpha$, and at all orders in $\eta$ and in the four-particle contact interactions as in pionless EFT, the charged particle effective range is unaffected by QED interactions and $r_C = r$~\cite{Kong:1999sf}.
Conversely, the contact interaction associated with the scattering length must be renormalized to absorb divergences from Coulomb ladder diagrams.
In the $\msbar$ scheme, the running coupling that would be identified with the scattering length in the absence of QED is related to the Coulomb-corrected scattering length by~\cite{Kong:1999sf}
\begin{equation}
   \begin{split}
      \frac{1}{a^{\msbar}(\mu)} = \frac{1}{a_C} + \alpha M \left[ \ln\left(\frac{\mu \sqrt{\pi}}{\alpha M}\right) + 1 - \frac{3}{2}\gamma_E - \frac{1}{2}\mu r \right],
   \end{split}\label{eq:runninga}
\end{equation}
and can be understood as the physical scattering length with Coulomb effects from infrared length scales $> 1/\mu$ removed.

The FV two-particle energy spectrum can be related to the scattering phase shift in non-relativistic quantum mechanics~\cite{Huang:1957im} and in quantum field theory~\cite{Luscher:1986pf,Luscher:1990ux}.
In a finite spatial volume of size $L^3$ with PBCs, the system exhibits reduced spatial symmetries characterized by covariance under the cubic group, and the momentum carried by a free particle is quantized as $\v{p} = 2 \pi \v{n} / L$ with $\v{n} \in \mathbb{Z}_3$.
In QED$_L$, zero-mode subtraction implies that the momentum carried by a photon propagator is restricted to $p \geq 2\pi / L$.
The expansion parameter $\eta(p)$ in Eq.~\eqref{eq:etadef} is therefore restricted to
\begin{equation}
   \begin{split}
      \eta(p) \leq \eta_L \equiv \frac{\alpha M L}{4 \pi}
   \end{split}\label{eq:etaFV}
\end{equation}
in QED$_L$ Coulomb ladder diagrams.
For sufficiently small volumes, $\eta \leq \eta_L \ll 1$ and Coulomb photon exchange can be treated perturbatively.

The $\mathcal{O}(\alpha)$ quantization condition relating the $s$-wave scattering amplitude to the two-particle spectrum in the $A_1^+$ representation of the cubic group in non-relativistic EFT in the approximation of negligible partial wave mixing was derived in Ref.~\cite{Beane:2014qha},
\begin{equation}
   \begin{split}
      \left(\frac{2\pi \eta}{e^{2\pi \eta} - 1}\right) p \cot \delta(p) + 2 p \eta \left[ \text{Re}\psi(i \eta) - \ln(\eta) \right] = \frac{1}{\pi L}\mathcal{S}^C(p) + \alpha M \left[ \ln\left( \frac{4\pi}{\alpha M L} \right) - \gamma_E \right],
   \end{split}\label{eq:QC}
\end{equation}
where $\eta(p)$ is defined in Eq.~\eqref{eq:etadef} and
\begin{subequations}
   \begin{align}
      \mathcal{S}^C(x) &= \mathcal{S}(x) - \frac{\alpha M L}{4\pi^3} \mathcal{S}_2(x) + \frac{\alpha M a_C^2 r_C}{\pi^2 L^2} \mathcal{I} \  \mathcal{S}(x)^2 + O(\alpha^2), \\
      \mathcal{S}(x) &= \lim_{\Lambda_n \rightarrow \infty} \sum_{\v{n}\in\mathbb{Z}^3}^{|\v{n}|<\Lambda_n} \frac{1}{|\v{n}|^2 - x^2} - 4\pi \Lambda_n, \\
      \mathcal{S}_2(x) &= \lim_{\Lambda_n \rightarrow \infty} \sum_{\v{n}\in\mathbb{Z}^3}^{|\v{n}|<\Lambda_n} \sum_{\v{m}\in\mathbb{Z}^3 \setminus \{\v{n}\}} \frac{1}{|\v{n}|^2 - x^2} \frac{1}{|\v{m}|^2 - x^2} \frac{1}{|\v{n} - \v{m}|^2} - 4\pi^4 \ln \Lambda_n ,
   \end{align}\label{eq:sums}
\end{subequations}
where $\mathcal{I}\approx -8.9136$ is a geometric constant whose evaluation is detailed in  Refs.~\cite{Luscher:1986pf} and sums over integer triplets $\v{n}$ are restricted to $|\v{n}| \leq \Lambda_n$ where $\Lambda_n$ is a cutoff and the $\Lambda_n \rightarrow \infty$ limit should be taken as indicated.
LQCD+QED$_L$ results for the $p$ associated with FV energy levels below three-particle thresholds can be related to charged particle scattering phase shifts by Eq.~\eqref{eq:QC} and used to constrain parametrizations of the phase shift such as Eq.~\eqref{eq:ERE}.
Since Eq.~\eqref{eq:QC} neglects exponentially small FV effects present in small volumes, as well as $(\alpha M L)^n$ effects from Coulomb ladder diagrams with $n$ photon propagators that must be resummed for sufficiently large volumes, Eq.~\eqref{eq:QC} is necessarily valid only for an intermediate range of $L$.
An important goal of this work is to test the range of volumes over which Eq.~\eqref{eq:QC} can be reliably used to extract Coulomb-corrected scattering parameters from LQCD+QED$_L$ results.

\subsection{Charged two-hadron systems in NRQED$_L$}\label{sec:2body}

Eq.~\eqref{eq:QC} can be perturbatively expanded in powers of $a_C / L$ and other higher-order effective range expansion coefficients.
Ref.~\cite{Beane:2014qha} determined this expansion to $\mathcal{O}(\alpha  M L)$ and $\mathcal{O}(a_C / L)^3$ under the assumption that $\alpha M L \ll a_C / L \ll 1$.
Following  Refs.~\cite{Beane:2007qr,Detmold:2008gh}, the same result can be derived in non-relativistic quantum mechanics using a Hamiltonian that perturbatively includes the effects of relativity and allows straightforward generalization to many-particle systems.
The NRQED$_L$ Lagrangian is given by
\begin{equation}
   \begin{split}
     \mathcal{L} &= \psi^\dagger \left( i D_0 - \frac{\v{D}^2}{2M} \right) \psi -  \frac{1}{2}\left( \frac{4\pi a}{M} \right) (\psi^\dagger \psi)^2  - \frac{\eta_3(\mu)}{3!} (\psi^\dagger \psi)^3 + \mathcal{L}_{\gamma}^\xi + \mathcal{L}_{r}.
   \end{split}\label{eq:NRQEDL}
\end{equation}
In this expression $\psi$ is a non-relativistic hadron field, $D_\mu = \partial_\mu + i e Q A_\mu$ where $Q$ is the electric charge operator, the gauge-fixed photon Lagrangian $\mathcal{L}_{\gamma}^{\xi}$ is given in Eq.~\eqref{eq:Lgamma}, $\eta_3(\mu)$ is a renormalization-scale-dependent coupling associated with short-range three-body interactions, and four- and higher-body interactions are neglected.
$\mathcal{L}_{r}$ includes effective range contributions and relativistic corrections involving two derivatives, and is given by
\begin{equation}
   \begin{split}
      \mathcal{L}_r &= \psi^\dagger \left(  \frac{\v{D}^4}{8M^3} \right) \psi -  \frac{1}{2}\left( \frac{\pi a}{M} \right) \left( a r - \frac{1}{M^2} \right) (\psi^\dagger \psi)(\psi^\dagger \v{D}^2 \psi + \psi \v{D}^2 \psi^\dagger).
   \end{split}\label{eq:NRQEDLr}
\end{equation}
The coefficient of the $(\psi^\dagger \psi)(\psi^\dagger \mathbf{D}^2 \psi + \psi \mathbf{D}^2 \psi^\dagger)$ operator in Eq.~\eqref{eq:NRQEDLr} can be fixed by demanding that the strong interaction EFT given by replacing $D_\mu$ with $\partial_\mu$ reproduces Eq.~\eqref{eq:QC} with $\alpha = 0$ when both are expanded perturbatively in powers of $L^{-1}$ to\footnote{The Lagrangian in Eq.~\eqref{eq:NRQEDLr} can also be obtained by studying the nonrelativistic limit of relativistic scalar field theory as in Refs.~\cite{Namjoo:2017nia,Braaten:2018lmj} .} $\mathcal{O}(L^{-6})$.
This $\mathcal{O}(L^{-6})$ threshold expansion of L{\"u}scher's quantization condition is given in Ref.~\cite{Hansen:2015zta} and is verified below to be reproduced by the non-relativistic EFT defined by Eqs.~\eqref{eq:NRQEDL}-\eqref{eq:NRQEDLr}\footnote{Refs.~\cite{Beane:2007qr,Detmold:2008gh} include the operators in Eq.~\eqref{eq:NRQEDLr}, but a factor of 2 discrepancy in the coefficient of the last term leads to a difference in the $\mathcal{O}(L^{-6})$ threshold expansion result.}.
Operators in $\mathcal{L}_r$ lead to contributions to the threshold expansion suppressed by $ar/L^2$ as well as relativistic effects suppressed by $\mathcal{O}\left( (ML)^{-2} \right)$ that will be neglected in the power counting schemes discussed below. 
Additional relativistic corrections to Eq.~\eqref{eq:NRQEDL} arise from photon loops and operators with four and more derivatives, but these give rise to FV effects suppressed by powers of $\mathcal{O}\left( (ML)^{-1} \right)$ that will be neglected  as discussed below and detailed in Appendix~\ref{app:matching}. 

Introducing Fourier transformed fields
\begin{equation}
   \begin{split}
      \widetilde{\psi}_{\v{k}} = \frac{1}{L^{3/2}} \int d^3x \ e^{i \v{k}\cdot\v{x} } \psi(\v{x}), \hspace{20pt} \psi(\v{x}) = \frac{1}{L^{3/2}} \sum_{\v{k}} e^{-i \v{k}\cdot \v{x}} \widetilde{\psi}_{\v{k}},
   \end{split}\label{eq:psiFT}
\end{equation}
the FV Hamiltonian for NRQED$_L$ is given by
\begin{equation}
   \begin{split}
       H  &= \sum_{\v{k}} \widetilde{\psi}^\dagger_{\v{k}} \left( \frac{\v{k}^2}{2M}  \right) \widetilde{\psi}_{\v{k}} + H_{\gamma}^\xi + H_{\rm int},
   \end{split}\label{eq:Hdef}
\end{equation}
where $H_\gamma^\xi$ is the gauge-fixed photon Hamiltonian, whose explicit form will not be used below, and
\begin{equation}
   \begin{split}
      H_{\rm int} &=  - \sum_{\v{k}} \widetilde{\psi}^\dagger_{\v{k}}   \left( \frac{\v{k}^4}{8M^3} \right) \widetilde{\psi}_{\v{k}} + \frac{1}{2 L^3} \sum_{\v{p}^\prime, \v{p}, \v{Q}} V(\v{p},\v{p}^\prime) \widetilde{\psi}^\dagger_{\v{Q}-\v{p}^\prime} \widetilde{\psi}^\dagger_{\v{Q}+\v{p}^\prime} \widetilde{\psi}_{\v{Q}-\v{p}} \widetilde{\psi}_{\v{Q}+\v{p}} \\
        &\hspace{20pt} + \frac{\eta_3(\mu)}{(3!) L^6}  \sum_{\v{Q},\v{p},\v{p}^\prime,\v{q},\v{q}^\prime} \widetilde{\psi}_{\v{Q}+\v{p}^\prime}^\dagger \widetilde{\psi}_{\v{Q}+\v{q}^\prime}^\dagger \widetilde{\psi}_{\v{Q}-\v{p}^\prime-\v{q}^\prime}^\dagger \widetilde{\psi}_{\v{Q}+\v{p}} \widetilde{\psi}_{\v{Q}+\v{q}} \widetilde{\psi}_{\v{Q}-\v{p}-\v{q}},
   \end{split}\label{eq:Hint}
\end{equation}
where the two-body potential includes strong interaction and Coulomb terms,
\begin{equation}
   \begin{split}
     V(\v{p}^\prime,\v{p})  &= \frac{4\pi}{M}\left(a + \frac{a}{4}\left(a r - \frac{1}{M^2}\right) (\v{p}^2 + \v{p}^{\prime 2}) + \ldots \right) + \frac{4\pi \alpha}{|\v{p}^{\prime} - \v{p}|^2} \left( 1 - \delta_{\v{p},\v{p}^\prime} \right).
   \end{split}\label{eq:potential}
\end{equation}
Relativistic corrections to the potential, including photon loop effects as well as nonlocal effects of zero-mode removal, can in principle be calculated by carrying out the FV analog of higher-order matching between QED, NRQED, and potential NRQED  (pNRQED)~\cite{Caswell:1985ui,Pineda:1997bj,Pineda:1998kn}, but these terms give rise to $\mathcal{O}(ML)^{-1}$ effects that are neglected below.

Specializing to the case of identical bosons, operators are associated with the fields in Eq.~\eqref{eq:Hdef} and are defined to satisfy commutation relations $[\widetilde{\psi}^\dagger_{\v{p}^\prime} ,\widetilde{\psi}_{\v{p}} ] = \delta_{\v{p},\v{p}^\prime}$.
FV two-particle states defined by
\begin{equation}
   \begin{split}
      \ket{\v{p}_1, \v{p}_2} = \frac{1}{\sqrt{2}} \widetilde{\psi}_{\v{p}_1}^\dagger \widetilde{\psi}_{\v{p}_2}^\dagger \ket{0}
   \end{split}\label{eq:states2}
\end{equation}
satisfy $\braket{\v{p}_1,\v{p}_2}{\v{p}_1,\v{p}_2} = 1$.
The ground state of the two-particle system in its center-of-mass frame is $\ket{\v{p}_1 = \v{0}, \v{p}_2 = \v{0}}$.
Splitting the Hamiltonian into kinetic energy terms and an interaction term $H_{\rm int}$ that will be treated perturbatively, the ground-state FV energy shift at leading order in Rayleigh-Schr{\"o}dinger perturbation theory is then given by
\begin{equation}
   \begin{split}
     \Delta E^{\text{LO}} = \mbraket{\v{0},\v{0}}{H_{\rm int}}{\v{0},\v{0}} = \frac{1}{L^3} V(\v{0},\v{0})  = \frac{4\pi a}{ML^3} .
   \end{split}\label{eq:DeltaELO}
\end{equation}
Standard perturbation theory techniques can be used to extended this result to higher orders in $a/L$ and $\alpha$.
At next-to-next-to-leading order (NNLO) in $H_{\rm int}$, the result is given by
\begin{equation}
  \begin{split}
     \Delta E^{\text{NNLO,PC1}} &= \frac{4\pi a_C}{ML^3} \left\lbrace 1  -  \left( \frac{a_C}{\pi L} \right)   \mathcal{I} + \left( \frac{a_C}{\pi L} \right)^2  \left[ \mathcal{I}^2    -   \mathcal{J} \right]  \right\rbrace\\
                         &\hspace{10pt}  + \frac{\alpha}{\pi L} \left\lbrace - \left( \frac{a_C}{\pi L} \right) 2\mathcal{J} + \left( \frac{a_C}{\pi L} \right)^2 \left\lbrace 2 \mathcal{I}\mathcal{J} + \mathcal{R}_{22} - 2 \mathcal{K} + 4\pi^4\left[ \ln\left( \frac{4\pi}{\alpha M L} \right) - \gamma_E \right] \right\rbrace \right. \\
    &\hspace{50pt}  \left. - \left( \frac{\alpha M L}{4\pi^3} \right) \mathcal{K}  + \left( \frac{a_C}{\pi L} \right)\left( \frac{\alpha M L}{4\pi^3} \right)\left[ 2 \mathcal{R}_{24} + \mathcal{J}^2 - \mathcal{L} \right] + \left( \frac{\alpha M L}{4\pi^3} \right)^2 \mathcal{R}_{44} \right\rbrace ,
  \end{split}\label{eq:NNLO2body1}
\end{equation}
where PC1 is a label for the power counting scheme discussed below and higher-order corrections in $a_C/L \sim r/L$, as well as relativistic effects suppressed by $\mathcal{O}\left( (ML)^{-1}\right)$, have been neglected.
All sum-integral differences $\mathcal{I}$, $\mathcal{J}$, $\mathcal{K}$, $\mathcal{L}$, and $\mathcal{R}_{nm}$ which appear in this expression are evaluated\footnote{See also Refs.~\cite{Luscher:1986pf,Luscher:1990ux,Gockeler:2012yj,Leskovec:2012gb,Beane:2014qha} for more details on evaluating FV sums.} in Ref.~\cite{Beane:2014qha}, which includes an evaluation of Eq.~\eqref{eq:NNLO2body1} in EFT to leading order in $\alpha M L$ and to higher order in $a/L$, except for $\mathcal{R}_{44}$ which is a convergent sum given by $\mathcal{R}_{44} \approx 55.47$.

Eq.~\eqref{eq:NNLO2body1} can be expressed as an expansion in $a_C / L$ and the FV Coulomb parameter $\eta_L$ defined in Eq.~\eqref{eq:etaFV},
\begin{equation}
  \begin{split}
     \Delta E^{\text{NNLO,PC1}} &= \frac{4\pi a_C}{ML^3} \left\lbrace 1  -  \left( \frac{a_C}{\pi L} \right)   \mathcal{I} + \left( \frac{a_C}{\pi L} \right)^2  \left[ \mathcal{I}^2    -   \mathcal{J} \right]  \right\rbrace\\
     &\hspace{10pt}  + \frac{4 \eta_L}{ML^2} \left\lbrace - \left( \frac{a_C}{\pi L} \right) 2\mathcal{J} + \left( \frac{a_C}{\pi L} \right)^2 \left[ 2 \mathcal{I}\mathcal{J} + \mathcal{R}_{22} - 2 \mathcal{K} - 4\pi^4\left[ \ln\left( \eta_L \right) + \gamma_E \right] \right] \right. \\
                         &\hspace{60pt}  \left. - \left( \frac{\eta_L}{\pi^2} \right) \mathcal{K}  + \left( \frac{a_C}{\pi L} \right)\left( \frac{\eta_L}{\pi^2} \right)\left[ 2 \mathcal{R}_{24} + \mathcal{J}^2 - \mathcal{L} \right] + \left( \frac{\eta_L}{\pi^2} \right)^2 \mathcal{R}_{44} \right\rbrace .
  \end{split}\label{eq:NNLO2body}
\end{equation}
This resembles a double power series expansion in the parameters $\eta_L$ and $a_C/L$, suggesting that the NRQED$_L$ threshold expansion should provide a good approximation of FV energy shifts for
\begin{equation}
   \begin{split}
      \text{PC1: } \eta_L \sim \frac{a_C}{L} \ll 1 \ .
   \end{split}\label{eq:pc1}
\end{equation}
It is noteworthy that $\eta_L \ll 1$ only requires $\alpha M L \ll 4\pi$ and is less restrictive than the condition $\alpha M L \ll 1$ discussed in Ref.~\cite{Beane:2014qha}.
In the matching to the LQCD+QED$_L$ simulations discussed below, the power counting $\eta_L \sim a_C / L$ numerically overestimates the size of QED effects, particularly on the smaller volumes studied, and the alternative power counting\footnote{The description of FV effects on charged hadron masses in QED$_L$ as a dual expansion in $\alpha$ and $(ML)^{-1}$ and the possibility of using alternative power countings in LQCD+QED$_L$ calculations depending on the values of these parameters is explored in Ref.~\cite{Matzelle:2017qsw}.}
\begin{equation}
   \begin{split}
      \text{PC2: } \eta_L^{1/2} \sim \frac{a_C}{L} \ll 1,
   \end{split}\label{eq:pc2}
\end{equation}
will also be used in fits to LQCD+QED$_L$ results.
Higher-order results for the threshold expansion for short-range contact interactions without QED~\cite{Beane:2007qr,Detmold:2008gh,Hansen:2015zta,Hansen:2016fzj} can be used to extend Eq.~\eqref{eq:NNLO2body} from NNLO in Eq.~\eqref{eq:pc1} to N$^3$LO in Eq.~\eqref{eq:pc2}, 
\begin{equation}
  \begin{split}
     \Delta E^{\text{N$^3$LO,PC2}} &= \frac{4\pi \overline{a}_C}{ML^3} \left\lbrace 1  -  \left( \frac{\overline{a}_C}{\pi L} \right)   \mathcal{I} + \left( \frac{\overline{a}_C}{\pi L} \right)^2  \left[ \mathcal{I}^2    -   \mathcal{J} \right]  - \left( \frac{\overline{a}_C}{\pi L} \right)^3 \left[ \mathcal{I}^3 - 3 \mathcal{I}\mathcal{J} + \mathcal{K} \right] \right\rbrace\\
     &\hspace{10pt}  + \frac{4\eta_L}{ML^2} \left\lbrace - \left( \frac{\overline{a}_C}{\pi L} \right) 2\mathcal{J} - \left( \frac{\eta_L}{\pi^2} \right) \mathcal{K} \right. \\
     &\hspace{60pt} \left. + \left( \frac{\overline{a}_C}{\pi L} \right)^2 \left[ 2 \mathcal{I}\mathcal{J} + \mathcal{R}_{22} - 2 \mathcal{K} - 4\pi^4\left[ \ln\left( \eta_L \right) + \gamma_E \right] \right] \right\rbrace.
  \end{split}\label{eq:N3LO2body}
\end{equation}
The parameter $\overline{a}_C(L)$ is equal to $a_C$ plus a $1/L^3$ suppressed correction arising from the interaction terms in Eq.~\eqref{eq:NRQEDLr}, 
\begin{equation}
   \begin{split}
      a_C = \overline{a}_C(L) - \frac{2\pi \overline{a}_C(L)^2}{L^3}\left( \overline{a}_C(L) r - \frac{1}{2M^2} \right),
   \end{split}\label{eq:aCbardef}
\end{equation}
where $(ML)^{-2}$ suppressed effects are shown for completeness and lead to agreement with the $\mathcal{O}(L^{-6})$ strong interaction relativistic threshold expansion of Ref.~\cite{Hansen:2015zta} after taking\footnote{The operators in Eq.~\eqref{eq:NRQEDLr} lead to additional effective range and relativistic corrections to the right-hand-side of Eq.~\eqref{eq:aCbardef}, namely an additional term of the form $-2a_C^2 r \eta_L \mathcal{I} / (ML^5) - \eta_L^2 \mathcal{J} / (4\pi^2 M^3 L^4)$, but these are higher order than the effective range term in Eq.~\eqref{eq:aCbardef} according to the power counting of Eq.~\eqref{eq:pc2}.}  $\alpha = 0$.

In principle, LQCD+QED$_L$ results for $\pi^+\pi^+$ FV energy shifts on multiple lattice volumes could be used to extract both $a_C$ and $r$ by constraining the $\mathcal{O}(L^{-6})$ difference between $\overline{a}_C$ and $a_C$.
In the LQCD+QED$_L$ calculations discussed below, $(\overline{a}_C(L) - a_C)/a_C$ can be estimated at LO in chiral perturbation theory ($\chi$PT) to be $2\%$ and $8\%$ for the $L/a=48$ and $L/a=32$ lattice volumes respectively.
To see whether $\overline{a}_C - a_C$ can be reliably determined, this estimate must be compared with an estimate of relativistic effects neglected in Eq.~\eqref{eq:N3LO2body}, which as discussed in Sec.~\ref{sec:zeromode} below modify the dominant QED FV effects by $\mathcal{O}\left( (ML)^{-1} \right)$.
For $\pi^+\pi^+$ systems on the $L/a =48$ lattice volume, the dominant (NLO) QED effect in Eq.~\eqref{eq:N3LO2body} amounts to a shift $\overline{a}_C \rightarrow \overline{a}_C - 2\mathcal{J} (\eta_L / \pi^2) \overline{a}_C$, and radiation photon effects can be estimated to lead to a $\mathcal{O}\left( (ML)^{-1} \right)$ suppressed shift in $(\overline{a}_C(L) - a_C)/a_C$ of order   $\sim 2\mathcal{J} (\eta_L / \pi^2)/(ML) = \mathcal{J} \alpha/(2\pi^3) \sim 3\%$ for $\alpha \simeq 0.1$.
Relativistic effects are therefore comparable to $\overline{a}_C - a_C$ for $\pi^+\pi^+$ systems and prevent the effective range contribution to $\overline{a}_C(L)$ from being disentangled from other FV effects neglected in Eq.~\eqref{eq:N3LO2body}.
Therefore, $\overline{a}_C(L) - a_C$ will be neglected when fitting $\pi^+\pi^+$ LQCD+QED$_L$ results to Eq.~\eqref{eq:N3LO2body}.

\subsection{Zero-mode effects}\label{sec:zeromode}

The derivation of Eq.~\eqref{eq:QC} in Ref.~\cite{Beane:2014qha} and the form of the NRQED$_L$ Lagrangian in Eq.~\eqref{eq:NRQEDL} assume that charged particle NRQED$_L$ contact interactions in FV are equal to their infinite-volume counterparts up to exponentially suppressed corrections.
More recently, however, it has been shown in Refs.~\cite{Borsanyi:2014jba,Davoudi:2014qua,Fodor:2015pna,Lee:2015rua,Matzelle:2017qsw,Davoudi:2018qpl} that this assumption is violated in the single-particle sector of NRQED$_L$ because of the inherent nonlocality of zero-mode subtraction.
NRQED$_L$ parameters can be obtained by calculating masses, scattering amplitudes, or other observables in both NRQED$_L$ and QED$_L$ and tuning the parameters of the NRQED$_L$ Lagrangian to reproduce QED$_L$ results.
In QED$_L$, on-shell photon exchange leads to power-law FV effects on charged particle masses suppressed by powers of $\alpha$ and $1/(ML)$ that have been studied in Refs~\cite{Borsanyi:2014jba,Davoudi:2014qua,Fodor:2015pna,Lee:2015rua,Matzelle:2017qsw,Davoudi:2018qpl}.
The $\mathcal{O}\left( \alpha / (ML) \right)$ and $\mathcal{O} \left( \alpha / (ML)^{2} \right)$ corrections are independent of the structure of the charged particle and are described by one-loop diagrams in both QED$_L$ and NRQED$_L$.
At order $\mathcal{O}\left(\alpha / (ML)^{3}\right)$, structure-dependent effects involving magnetic moments and charge radii arise. 
Nonlocal effects from zero-mode subtraction also enter at order $\mathcal{O}\left( \alpha / (ML)^{3}\right)$ because zero-mode subtraction leads to power-law FV effects from off-shell antiparticle modes in QED$_L$ that are not reproduced by NRQED$_L$ loop diagrams.
These effects can be included in NRQED$_L$ by adjusting the Lagrangian to include additional particle-antiparticle interactions~\cite{Fodor:2015pna,Lee:2015rua}, or more simply by adjusting the coefficients of mass operators in the NRQED$_L$ Lagrangian by factors proportional to $\alpha/(ML)^{3}$~\cite{Davoudi:2018qpl}.
For charged scalars, although not for charged fermions, these effects vanish in the charged particle rest frame~\cite{Davoudi:2018qpl}.
This non-decoupling of antiparticle modes is a consequence of the nonlocality of NRQED$_L$, and for EFTs with breakdown scale $\Lambda$ generically produces $\mathcal{O}\left( \alpha / (\Lambda L)^3 \right)$ corrections to LECs~\cite{Davoudi:2018qpl}.

Nonlocal effects from zero-mode subtraction could lead to power-law FV effects that modify four-hadron contact interaction couplings proportional to $a_C$ in NRQED$_L$.
Considering the $\mathcal{O}(ML)$ enhancement of FV effects associated with Coulomb ladder diagrams, it is necessary to analyze nonlocal effects of zero-mode subtraction on $a_C$ in order to determine whether Eq.~\eqref{eq:QC} is modified within the order of approximation considered.
The effects of zero-mode subtraction on $a_C$ can be determined by matching any QED$_L$ and NRQED$_L$ correlation functions sensitive to four-hadron contact interactions.
One-particle FV self-energies only receive contributions from four-hadron contact interactions in $\mathcal{O}(\alpha^2)$ diagrams containing closed loops of particle-antiparticle pairs.
Two-particle FV Green's functions receive contributions from four-hadron contact interactions at $\mathcal{O}(\alpha^0)$, and it is convenient to calculate nonlocal FV effects on four-hadron contact interactions in NRQED$_L$ by directly matching two-particle FV Green's functions in QED$_L$ and NRQED$_L$.
This matching is detailed in Appendix~\ref{app:matching} and summarized below.

\begin{figure}
  \centering
  \includegraphics[width=.8\textwidth]{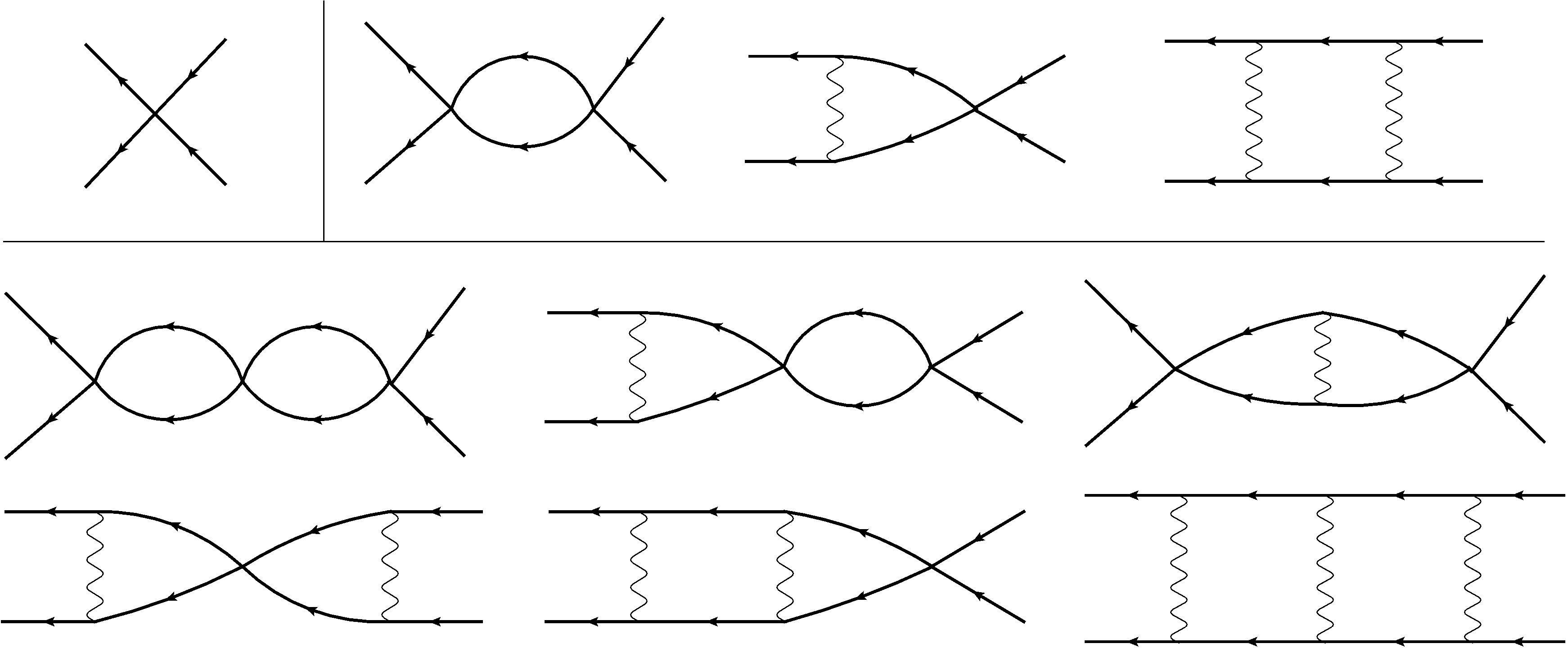}
   \caption{The strong-interaction and Coulomb scattering diagrams contributing to the two-body FV energy shift in NRQED$_L$. The top-left section shows the LO diagram . The top-right section shows the NLO diagrams. The bottom section shows the NNLO diagrams in the power counting of Eq.~\eqref{eq:pc1}. Diagrams that vanish because of zero-mode subtraction, including the tree-level one-photon-exchange diagram, are not shown.
    \label{fig:coulomb}}
\end{figure}

\begin{figure}
  \centering
  \includegraphics[width=.8\textwidth]{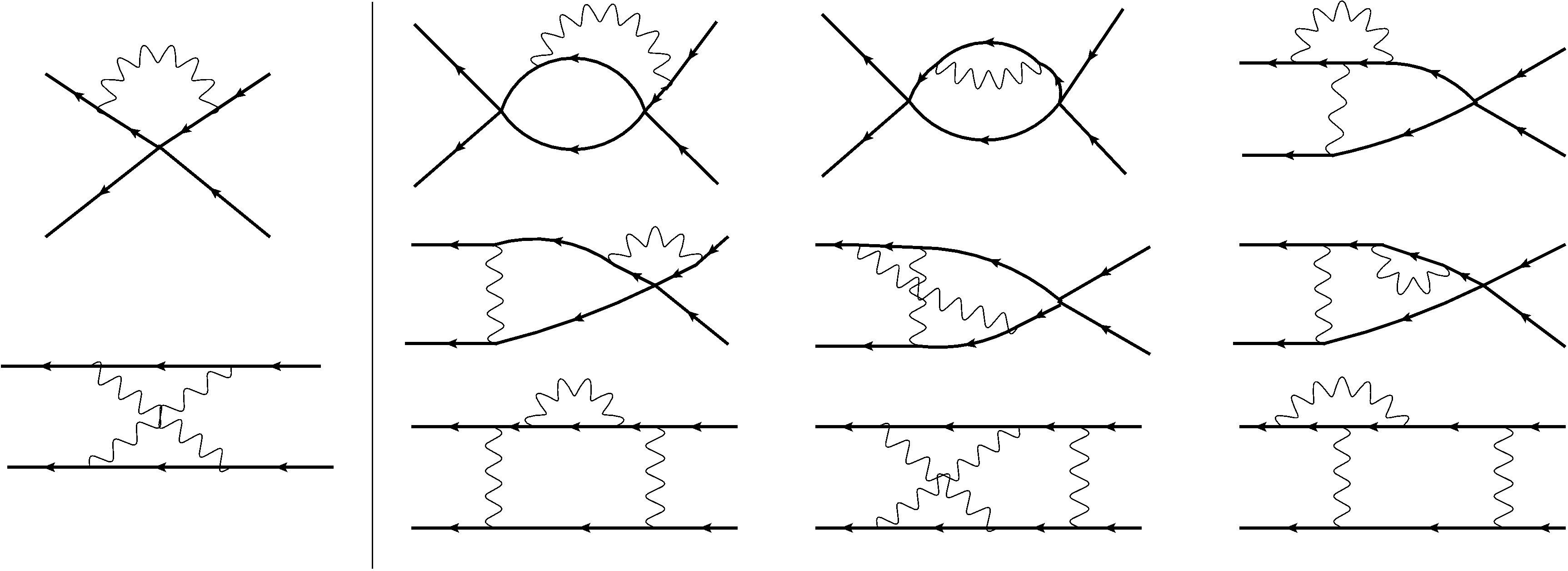}
   \caption{Radiation photon diagrams making power-suppressed contributions to the two-body FV energy shift in NRQED$_L$. The left and right sections show NLO and NNLO diagrams in the power counting of Eq.~\eqref{eq:pc1}, respectively, which lead to FV effects suppressed by $\mathcal{O}\left( (ML)^{-1} \right)$.
    \label{fig:radiation}}
\end{figure}

The $\mathcal{O}(ML)$ enhancement of Coulomb ladder diagrams arises when one intermediate particle propagator is placed on-shell and another intermediate particle propagator is nearly on-shell with virtuality $k^2 / M$, as compared with virtuality $k$ when a photon propagator is placed on-shell.
This can be explicitly seen by comparing the integrands of the ``box'' diagram (rightmost top row in Fig.~\ref{fig:coulomb}) with the ``crossed-box'' diagram (leftmost bottom row in Fig.~\ref{fig:radiation}).
In NRQED$_L$, the box diagram involves the integral
\begin{equation}
   \begin{split}
      & i\int \frac{dk^0}{2\pi} \left( \frac{1}{k^0 - k^2/2M + i \epsilon} \right) \left( \frac{1}{-k^0 - k^2/2M + i \epsilon} \right) \left(\frac{1}{(k^0)^2 - k^2 + i\epsilon} \right)^2 \\
      &= -\left( \frac{M}{k^2} \right) \left(\frac{1}{(k^2/2M)^2 - k^2}\right) + \frac{d}{dk^0}\left[ \left(\frac{1}{k^0 - |k|}\right)^2 \left( \frac{1}{k^0 - k^2/2M} \right) \left( \frac{1}{-k^0 - k^2/2M}\right) \right]_{k^0 \rightarrow -k}\\
      &= \frac{-M}{k^6}\left[ 1 + \mathcal{O}(k^2/M^3) \right] - \frac{3}{4k^5}\left[ 1 + \mathcal{O}(k^2/M^2) \right].
   \end{split}\label{eq:box}
\end{equation}
In this expression, the first contribution involving the particle pole places the second particle propagator nearly on-shell with kinetic energy $k^2/2M$. 
The second contribution from the photon double pole gives both particle propagators off-shell kinetic energies $k$.
When FV effects are computed, $k$ is replaced by the quantized values $2\pi n /L$ with $n\in \mathbb{Z}_3$, and (after adding all necessary UV counterterms) amplitude suppression by powers of $k/M$ implies suppression of FV effects by the corresponding power of $(ML)^{-1}$.
The NRQED$_L$ crossed-box diagram involves the integral
\begin{equation}
   \begin{split}
      &i\int \frac{dk^0}{2\pi} \left( \frac{1}{k^0 - k^2/2M + i \epsilon} \right)^2 \left( \frac{1}{(k^0)^2 - k^2 + i\epsilon} \right)^2 \\
      &= \frac{d}{dk^0}\left[ \left(\frac{1}{k^0 - |k|}\right)^2 \left(\frac{1}{k^0 - k^2/2M}\right)^2 \right]_{k^0\rightarrow -k}\\
      &= \frac{3}{4k^5}\left[ 1 + \mathcal{O}(k^2/M^2) \right]\,,
   \end{split}\label{eq:crossbox}
\end{equation}
where only the photon pole contributes and leads to particle propagators with off-shell kinetic energies $k$. FV effects associated with the crossed-box diagram are therefore suppressed by $\mathcal{O}\left( (ML)^{-1} \right)$ compared to the dominant contribution of the box diagram.
 
\begin{figure}
  \centering
  \includegraphics[width=.8\textwidth]{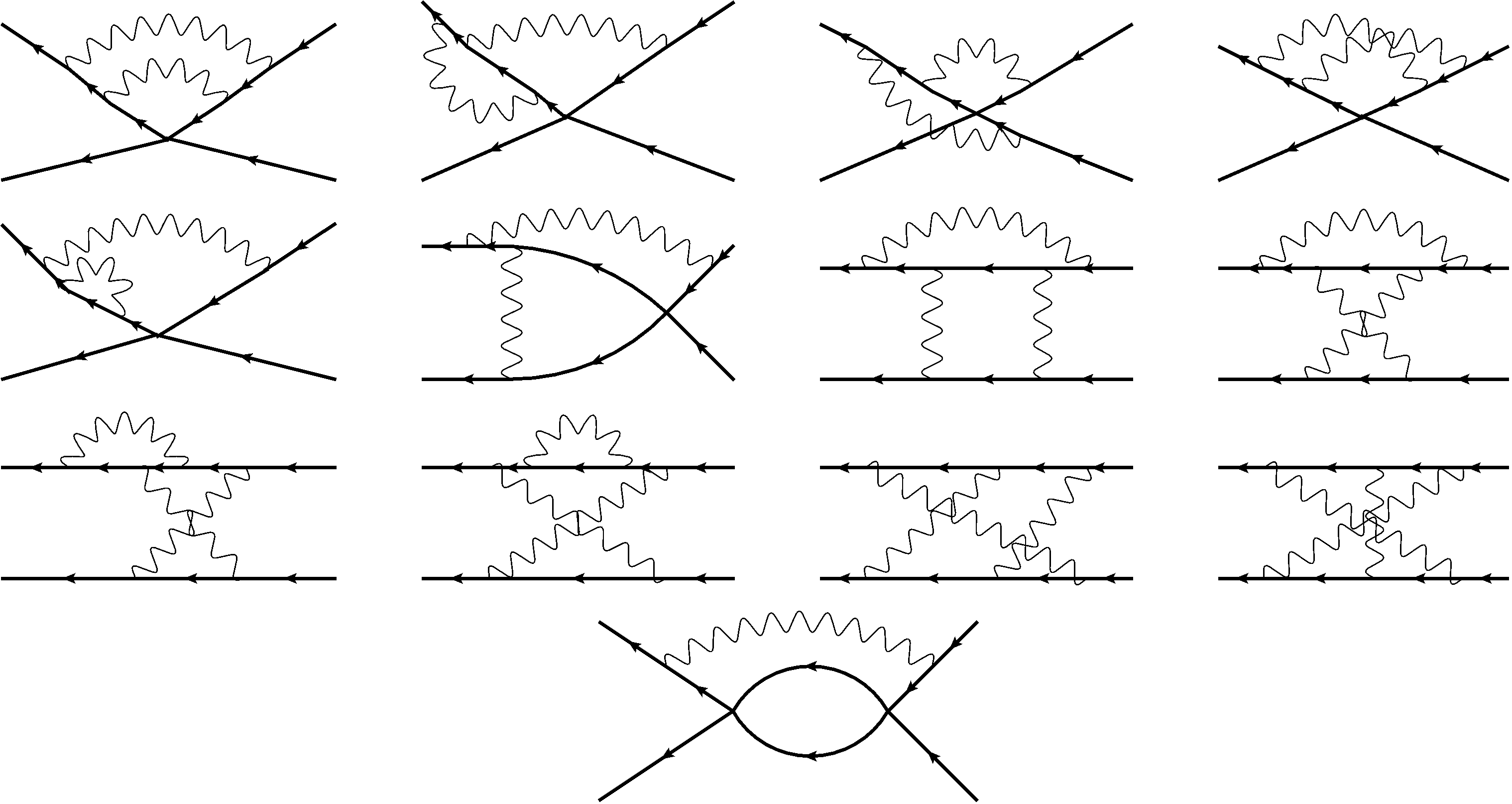}
   \caption{Radiation photon diagrams, including the jellyfish diagram on the forth row, that appear at NNLO in the power counting of Eq.~\eqref{eq:pc1} and lead to FV effects suppressed by $\mathcal{O}\left( (ML)^{-2} \right)$.
    \label{fig:radiation2}}
\end{figure}

The $\mathcal{O}(ML)$ power enhancement of the box diagram is only present in the diagrams involving repeated $s$-channel Coulomb interactions shown in Fig.~\ref{fig:coulomb} but not in the diagrams in Figs.~\ref{fig:radiation}-\ref{fig:radiation2} or other diagrams involving particle-antiparticle pair creation that vanish non-relativistically\footnote{More details on the power-counting of diagrams involving FV photon exchange are given in Ref.~\cite{Beane:2014qha} and discussions of analogous power counting arguments for radiation pions are given in Ref.~\cite{Mehen:1999hz}.}.
Furthermore, the $\mathcal{O}(ML)$ enhancement only occurs when the intermediate-state charged particles are both nearly on-shell.
As detailed in Appendix~\ref{app:matching}, power-law FV effects in QED$_L$ that are not reproduced by loop diagrams with the leading order NRQED$_L$ Lagrangian arise from antiparticle poles where intermediate states have a large virtuality of order $2M$.
These do not receive the $\mathcal{O}(ML)$ enhancement of particle pole contributions, and zero-mode effects are found to be suppressed by $\mathcal{O}\left(\alpha / (ML)^{3}\right)$.
Matching between QED$_L$ and NRQED$_L$ is explicitly performed for charged scalars in Appendix~\ref{app:matching}, and zero-mode effects are found to modify the four-scalar coupling in NRQED$_L$ at order $\mathcal{O}\left( \alpha/(ML)^{3} \right)$ for boosted systems but not to modify the coupling for scalars at rest at this order.
This shows that Eq.~\eqref{eq:QC} is valid for charged scalars in NRQED$_L$ up to $\mathcal{O}(\alpha^2)$ effects and $\mathcal{O}(ML)^{-1}$ relativistic effects.
The vanishing of these effects in the center-of-mass frame arises from cancellations due to the specific form of the scalar-photon vertex functions.
It is possible that for charged fermions nonlocal effects also arise at $\mathcal{O}\left( \alpha / (ML)^{3} \right)$ in the charged fermion rest frame.
The $\mathcal{O}\left( \alpha/(ML)^3 \right)$ suppression factor is consistent with the physical arguments of Ref.~\cite{Davoudi:2018qpl} that nonlocal effects arise from interactions between the subtracted zero-mode, which can be interpreted as a uniform background charge density that ensures Gauss's law is satisfied for the FV system~\cite{Hayakawa:2008an}, and high-energy modes that have been integrated out of the EFT, and therefore that FV effects from zero-mode subtraction are suppressed by $\alpha$ times the inverse volume.

\subsection{Charged many-hadron systems in NRQED$_L$}

\begin{figure}
  \centering
  \includegraphics[width=.8\textwidth]{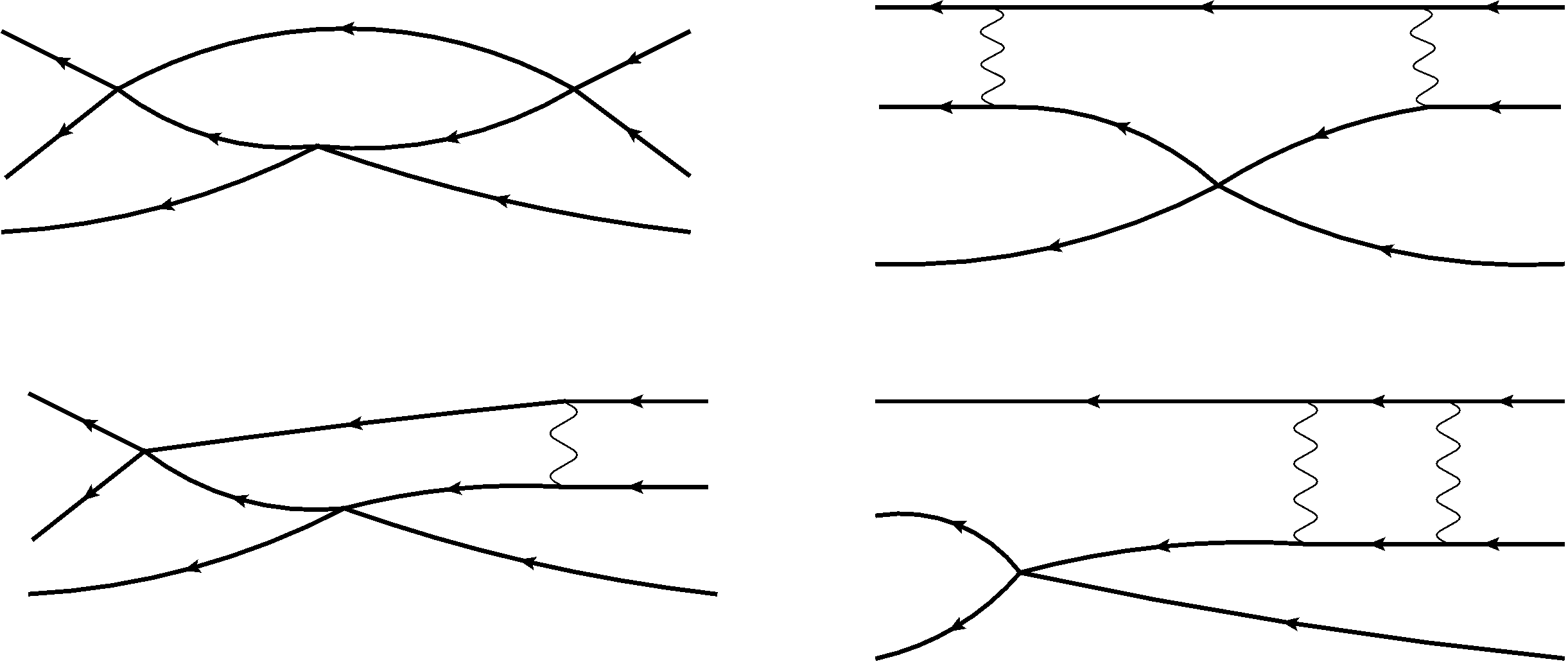}
   \caption{The strong-interaction and Coulomb diagrams contributing to three- and higher-body FV energy shifts in NRQED$_L$ at leading order in $(ML)^{-1}$ and NNLO in short-range and Coulomb interactions in the power counting of Eq.~\eqref{eq:pc1}. Diagrams that vanish because of zero-mode subtraction, partially disconnected diagrams involving pairs of two-body interactions among four and more particles, and diagrams that involve on-shell internal propagators and vanish in Rayleigh-Schr{\"o}dinger perturbation theory, are not shown.
    \label{fig:coulomb_many}}
\end{figure}

\begin{figure}
  \centering
  \includegraphics[width=.8\textwidth]{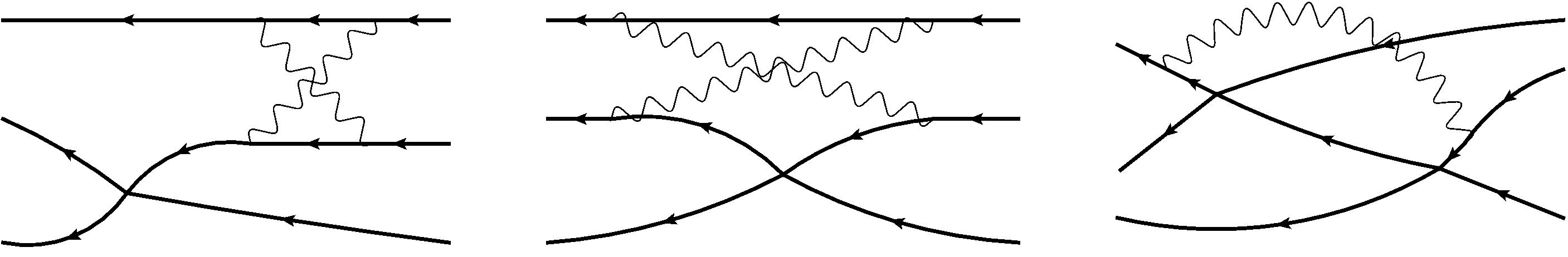}
   \caption{Radiation photon diagrams with negligible contributions to three- and higher-body FV energy shifts in NRQED$_L$ at NNLO. The left and center diagrams are suppressed by $\mathcal{O}\left( (ML)^{-1}\right)$, while the right diagram is suppressed by $\mathcal{O}\left( (ML)^{-2}\right)$. As in Fig.~\ref{fig:coulomb_many}, partially disconnected diagrams and diagrams that vanish because of zero-mode subtraction or kinematical constraints are not shown.
    \label{fig:radiation_many}}
\end{figure}

The two-particle energy shifts in Eqs.~\eqref{eq:NNLO2body}-\eqref{eq:N3LO2body} can be extended to a threshold expansion for FV effects on systems of $n$ non-relativistic particles using Rayleigh-Schr{\"o}dinger perturbation theory.
Unit-normalized many-particle states are given by
\begin{equation}
   \begin{split}
      \ket{\v{p}_1,\ldots,\v{p}_n} = \frac{1}{\sqrt{n}}\widetilde{\psi}^\dagger_{\v{p}_1}\times \ldots \times \widetilde{\psi}^\dagger_{\v{p}_2} \ket{0},
   \end{split}\label{eq:statenorm}
\end{equation}
and the leading order FV energy shift for the ground state of $n$ identical bosons in the center-of-mass frame is given by
\begin{equation}
   \begin{split}
     \mbraket{\v{0},\ldots \v{0}}{H_{\rm int}}{\v{0},\ldots \v{0}} = \frac{1}{L^3}{n \choose 2}V(\v{0},\v{0}) + \frac{1}{L^6}{n\choose 3}\eta_3(\mu) .
   \end{split}\label{eq:manyLO}
\end{equation}
Working to NNLO in the power counting of Eq.~\eqref{eq:pc1}, the energy shift of an $n$-hadron state is equal to ${n \choose 2}$ times the two-body energy shift of Eq.~\eqref{eq:NNLO2body}, plus additional contributions from induced three-body and four-body forces shown in Fig.~\ref{fig:coulomb_many}.
Additional diagrams associated with radiation photon exchange shown in Fig.~\ref{fig:radiation_many} are suppressed by $\mathcal{O}\left( (ML)^{-1}\right)$.
The resulting threshold expansion for the FV energy shift for a system of $n$ like-charged hadrons at rest is given by
\begin{equation}
  \begin{split}
     \Delta E_n^{\text{NNLO,PC1}} &= \frac{4\pi a_C}{ML^3} {n \choose 2} \left\lbrace 1  -  \left( \frac{a_C}{\pi L} \right)   \mathcal{I} + \left( \frac{a_C}{\pi L} \right)^2  \left[ \mathcal{I}^2    + (2n-5)   \mathcal{J} \right]  \right\rbrace\\
     &\hspace{15pt}  + \frac{4 \eta_L}{ML^2} {n \choose 2} \left\lbrace - \left( \frac{a_C}{\pi L} \right) 2\mathcal{J}    - \left( \frac{\eta_L}{\pi^2} \right) \mathcal{K} + \left( \frac{\eta_L}{\pi^2} \right)^2 \mathcal{R}_{44}\right. \\
     &\hspace{90pt}  \left.  + \left( \frac{a_C}{\pi L} \right)^2 \left[ 2 \mathcal{I}\mathcal{J} + \mathcal{R}_{22} + (4n - 10) \mathcal{K} - 4\pi^4\left[ \ln\left( \eta_L \right) + \gamma_E \right] \right] \right. \\
     &\hspace{90pt} \left.   + \left( \frac{a_C}{\pi L} \right)\left( \frac{\eta_L}{\pi^2} \right)\left[ 2 \mathcal{R}_{24} + \mathcal{J}^2 + (2n - 5) \mathcal{L} \right] \right\rbrace ,
  \end{split}\label{eq:NNLO}
\end{equation}
where omitted terms are: quartic or higher in $\eta_L \sim a_C / L$; relativistic effects suppressed by $\mathcal{O}\left( (ML)^{-1} \right)$; or three-body contact interactions where $\eta_3 \sim a_C^4/M$ is assumed so that $\mathcal{O}(L^{-6})$ terms of the strong interaction threshold expansion appear at the same order.

At N$^3$LO in the power counting of Eq.~\eqref{eq:pc2} there are contributions from short-distance three-body interactions well as the induced few-body interactions discussed above.
These introduce a new free parameter $\eta_3(\mu)$ that parametrizes the strength of six-particle operators in the NRQED$_L$ Hamiltonian, Eq.~\eqref{eq:Hdef}.
Combining Eq.~\eqref{eq:NNLO} with the N$^3$LO results for the QCD threshold expansion from Ref.~\cite{Beane:2007qr}, the N$^3$LO FV energy shift for a system of $n$ like-charged hadrons in the power counting of Eq.~\eqref{eq:pc2} is given by
\begin{equation}
  \begin{split}
     \Delta E_n^{\text{N$^3$LO,PC2}} &= \frac{4\pi \overline{a}_C}{ML^3} {n \choose 2} \left\lbrace 1  -  \left( \frac{\overline{a}_C}{\pi L} \right)   \mathcal{I} + \left( \frac{\overline{a}_C}{\pi L} \right)^2  \left[ \mathcal{I}^2    + (2n-5)   \mathcal{J} \right]  \right. \\
     &\hspace{80pt} \left. - \left( \frac{\overline{a}_C}{\pi L} \right)^3 \left[ \mathcal{I}^3 + (2n - 7) \mathcal{I}\mathcal{J} + (5n^2 - 41n + 63)\mathcal{K} \right] \right\rbrace\\
     &\hspace{20pt}  + \frac{4\eta_L}{ML^2} {n \choose 2} \left\lbrace - \left( \frac{\overline{a}_C}{\pi L} \right) 2\mathcal{J}  - \left( \frac{\eta_L}{\pi^2} \right) \mathcal{K} \right. \\
     &\hspace{100pt} \left. + \left( \frac{\overline{a}_C}{\pi L} \right)^2 \left[ 2 \mathcal{I}\mathcal{J} + \mathcal{R}_{22} + (4n - 10) \mathcal{K} - 4\pi^4\left[ \ln\left( \eta_L \right) + \gamma_E \right] \right] \right\rbrace  \\
     &\hspace{20pt} + {n \choose 3} \frac{1}{L^6} \left[ \eta_3(\mu) + \frac{64\pi \overline{a}_C^4}{M} (3\sqrt{3} - 4\pi) \ln(\mu L) - \frac{96 \overline{a}_C^4}{\pi^2 M} \mathcal{S}_{\rm MS} \right]. 
  \end{split}\label{eq:N3LO}
\end{equation}

The non-relativistic EFT three-body coupling $\eta_3(\mu)$ is renormalization scheme  and scale dependent.
The scale-dependence of $\eta_3(\mu)$ cancels the explicit $\ln(\mu L)$ scale-dependence shown in Eq.~\eqref{eq:N3LO}, and the scheme-dependence is compensated by scheme-dependence in the finite term $\mathcal{S}_{\rm MS}$.
This scale-dependence arises from the ambiguity in separating short-distance three-body interactions described by contact operators in NRQED$_L$ from long-distance two-body rescattering effects.
In relativistic theories this ambiguity does not arise, and the particle mass plays the role of the scale $\mu$ in relativistic descriptions of the $L^{-6} \ln L$ term in the three-body threshold expansion derived for generic relativistic field theories in Ref.~\cite{Hansen:2016fzj}.
Equating the $\mathcal{O}(L^{-6})$ term in the relativistic threshold expansion of the three-particle threshold amplitude $\mathcal{M}_{\rm 3,th}$ in Ref.~\cite{Hansen:2016fzj} with the corresponding $\mathcal{O}(L^{-6})$ nonrelativistic threshold expansion provides a relation between the non-relativistic coupling $\eta_3(\mu)$ and the scale-independent $\mathcal{M}_{\rm 3,th}$ in the absence of QED,
\begin{equation}
   \begin{split}
      \eta_3(\mu) &= -\frac{\mathcal{M}_{\rm 3,th}}{48 M^3} + \frac{64 \pi a^4}{M}(3\sqrt{3} - 4\pi)\ln\left(\frac{M}{2\pi \mu}\right)  + \frac{48 a^2 \pi^2}{M^3} + \frac{48 a^3 \pi^2 r}{M} \\
                  &\hspace{20pt}+ \frac{12 a^4}{\pi^2 M} \mathcal{S}_{3} + \frac{768 \pi^3 a^3}{M^2}\mathcal{C}_3,
   \end{split}\label{eq:etaRel}
\end{equation}
where $a$ is the scattering length for a neutral two-particle system, $\mathcal{C}_3 = -0.05806$ is a FV sum evaluated in Ref.~\cite{Hansen:2016fzj}, and $\mathcal{S}_{3} = 571.398$ is related to $\mathcal{S}_{\rm MS}$, evaluated in Ref.~\cite{Detmold:2008gh}, and to other FV sums from Ref.~\cite{Hansen:2016fzj} by $\mathcal{S}_{3} = \mathcal{C}_F + \mathcal{C}_4 + \mathcal{C}_5 + 8\mathcal{S}_{\rm MS}$.
QED effects will modify Eq.~\eqref{eq:etaRel}, but these modifications can be neglected at the EFT order considered here.
Below, QED effects on three-body forces will be studied by comparing the three-body interaction parameters extracted from LQCD+QED$_L$ results for systems of charged and neutral mesons.

\section{Results for charged multi-hadron systems} \label{sec:results}

This section combines the LQCD+QED$_L$ results from Section~\ref{sec:lattice} with the NRQED$_L$ results from Section~\ref{sec:theory} in order to obtain  QCD+QED predictions for scattering lengths and other hadronic interaction parameters at the values of the quark masses and $\alpha$ used here.

\subsection{Charged meson scattering}

\label{sec:2meson}

\begin{table}[t!]
   \begin{tabular}{|c||c|c||c|c|} \hline
      & \multicolumn{2}{c||}{$a  \Delta E_{n\overline{K}^0}\vphantom{\frac{1}{2\frac{1}{2}}}$}  & \multicolumn{2}{c|}{$a  \Delta E_{n\pi^+}$} \\\hline
      $n$ & $L/a = 32$ & $L/a = 48$ & $L/a= 32$  & $L/a = 48$ \\\hline\hline
      2   & 0.0087(13) & 0.00241(62) & 0.0080(16) & 0.00256(60) \\\hline
      3   & 0.0268(24) & 0.0074(15)  & 0.0249(28) & 0.0080(14)  \\\hline
      4   & 0.0622(66) & 0.0163(39)  & 0.0588(62) & 0.0179(38)  \\\hline
      5   & 0.107(11)  & 0.0286(62)  & 0.103(12)  & 0.0313(63)  \\\hline
      6   & 0.177(20)  & 0.050(11)   & 0.175(21)  & 0.053(11)   \\\hline
      7   & 0.267(30)  & 0.073(14)   & 0.263(28)  & 0.075(17)   \\\hline
      8   & 0.399(62)  & 0.113(20)   & 0.382(51)  & 0.111(22)   \\\hline
      9   & 0.53(11)   & 0.151(28)   & 0.495(86)  & 0.141(33)   \\\hline
      10  & 0.75(28)   & 0.206(42)   & 0.64(12)   & 0.167(47)   \\\hline
      11  & 0.5(1.2)   & 0.265(57)   & 0.80(25)   & 0.191(74)   \\\hline
      12  & 0.3(1.8)   & 0.331(77)   & 0.76(46)   & 0.21(12)    \\\hline
   \end{tabular}
   \caption{FV energy shift results for systems of $n\in\{2,\ldots,12\}$ neutral $\overline{K}^0$ mesons and charged $\pi^+$ mesons for lattice volumes with $L/a \in \{32,48\}$. Results are determined by taking correlated differences between LQCD+QED$_L$ ground-state energies during the fit range sampling procedure described in Appendix~\ref{app:fits}. \label{tab:mesonFV} }
\end{table}

The LQCD+QED$_L$ results for the FV spectrum results in Table~\ref{tab:mesonM} can be used to constrain the low-energy EFTs for charged and neutral meson interactions.
It is convenient to focus on results for the FV energy shifts
\begin{equation}
   \begin{split}
      \Delta E_{nM}(L) = E_{nM}(L) - n E_{M}(L),\hspace{20pt} M \in \{ \overline{K}^0, \ \pi^+\}.
   \end{split}\label{eq:DeltaE}
\end{equation}
Results for correlated differences between $n$-particle ground-state energies and $n$ times the one-particle ground-state energy, as defined in Eq.~\eqref{eq:DeltaE}, are more precise than $n$-particle energies alone. Furthermore, this subtraction nonperturbatively removes single-particle FV effects from $n$-meson FV energy shift results.
For multi-$\pi^+$ systems, LQCD+QED$_L$ results for these FV energy shifts can be identified with the interaction energy shifts $\Delta E_n$ computed perturbatively in NRQED$_L$ in Sec.~\ref{sec:theory}.
For multi-$\overline{K}^0$ systems, LQCD+QED$_L$ results can be identified with the same EFT results after setting $\alpha$ to zero. 
%Differences of the $n$-meson energies from $n$ free mesons therefore enable the extraction of scattering phase shift information. 
In the numerical LQCD+QED$_L$ calculation, FV energy shifts are computed in a correlated manner using bootstrap resampling as detailed in Appendix~\ref{app:fits}, and the results are shown in Table~\ref{tab:mesonFV}.
To access QED-specific effects, the differences of these differences between the $n$ charged pions and $n$ neutral kaon systems are also computed similarly.
The double subtraction suppresses any strong isospin breaking effects arising from mistuning of the quark masses for different charge quarks.
In the numerical LQCD+QED$_L$ calculation, correlated differences of FV energy shifts are computed using bootstrap resampling as detailed in Appendix~\ref{app:fits}, and the results are shown in Table~\ref{tab:mesonFVdiff}.

\begin{table}[t!]
   \begin{tabular}{|c||c|c|c|} \hline
      & \multicolumn{2}{c|}{$a  \Delta E_{n\pi^+}- a  \Delta E_{n\overline{K}^0}  \vphantom{\frac{1}{2\frac{1}{2}}}$} \\\hline
      $n$ & $L/a = 32$   & $L/a = 48$ \\\hline\hline
      2   & -0.0006(12)  & 0.00009(20)  \\\hline
      3   & -0.0054(29)  & -0.00009(86) \\\hline 
      4   & -0.0070(68)  & 0.0008(16)   \\\hline
      5   & -0.013(11)   & 0.0012(31)   \\\hline
      6   & -0.022(22)   & 0.0001(67)   \\\hline
      7   & -0.040(39)   & -0.003(11)   \\\hline
      8   & -0.065(96)   & -0.016(22)   \\\hline
      9   & -0.10(17)    & -0.028(25)   \\\hline
      10  & -0.14(30)    & -0.045(37)   \\\hline
      11  & 0.30(95)     & -0.070(59)   \\\hline
      12  & 0.3(1.1)     & -0.081(66)   \\\hline
   \end{tabular}
   \caption{FV energy shift differences between systems of $n\in\{2,\ldots,12\}$ charged $\pi^+$ and neutral $\overline{K}^0$ mesons  for lattice volumes with $L/a \in \{32,48\}$. Results are obtained by taking correlated differences between fitted energies as in Table~\ref{tab:mesonFV}. \label{tab:mesonFVdiff} }
\end{table}

\begin{figure}[t]
  \centering
  \includegraphics[width=.7\textwidth]{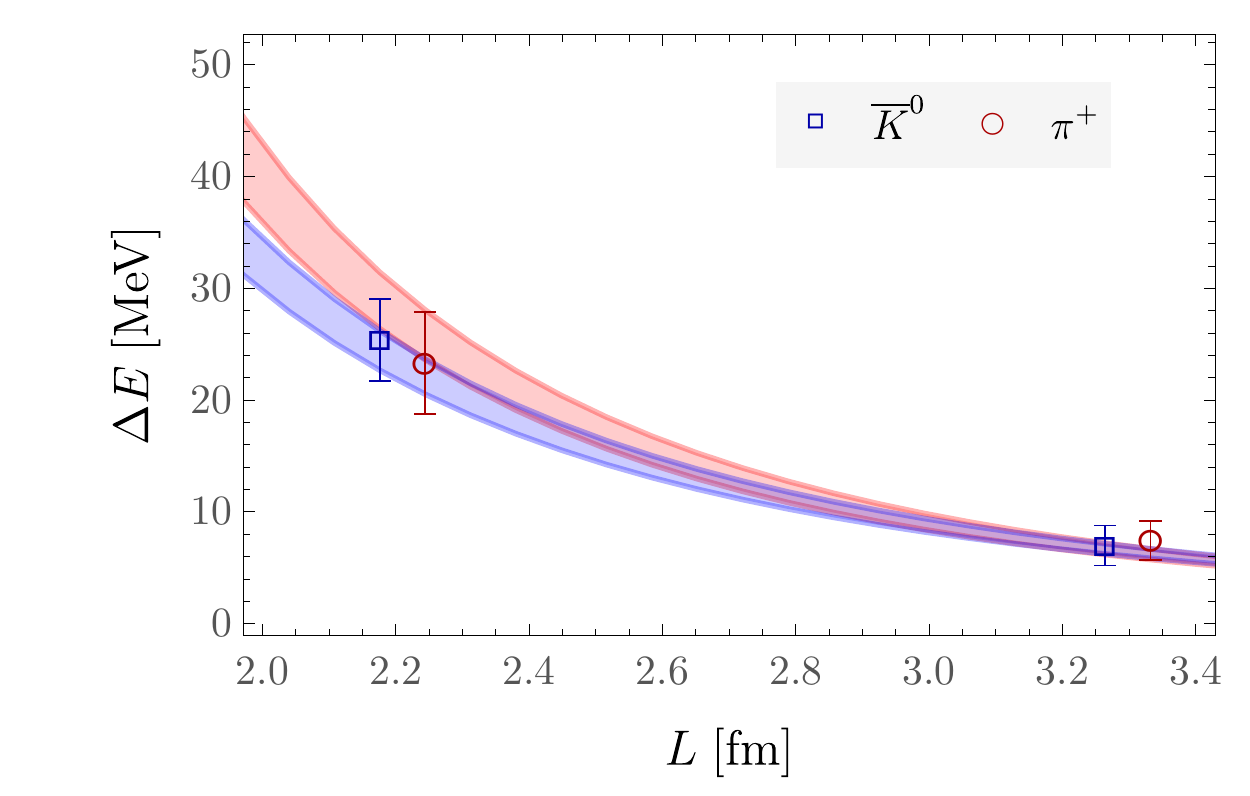}
   \caption{The blue and red points show the LQCD+QED$_L$ results from Table~\ref{tab:mesonFV} for the $\overline{K}^0\overline{K}^0$ and $\pi^+\pi^+$ FV energy shifts for the $L/a \in \{32,48\}$ volumes. The red band shows the NRQED$_L$ predictions of Eq.~\eqref{eq:NNLO2body} using the best result for $a_C^{\pi^+\pi^+}$ in Eq.~\eqref{eq:fit12b} obtained by fitting the $L/a \in \{32,\; 48\}$ results to Eq.~\eqref{eq:NNLO2body}.   The blue band shows the prediction of Eq.~\eqref{eq:NNLO2body} with $\alpha = 0$ using the best-fit result for $a^{\overline{K}^0\overline{K}^0}$ in Eq.~\eqref{eq:fit12b}. As in Fig.~\ref{fig:1_meson}, the widths of the bands correspond to 67\% confidence intervals estimated using bootstrap resampling. A small horizontal offset is applied symmetrically to $\pi^+$ and $\overline{K}^0$ results.
    \label{fig:2_meson}}
\end{figure}

Results for the two-particle FV energy shifts $\Delta E_{\pi^+\pi^+}$ and $\Delta E_{\overline{K}^0 \overline{K}^0}$ are shown in Fig.~\ref{fig:2_meson}.
Both energy shifts are clearly resolved from zero with relative uncertainties in the range of $15$ - $25\%$ for both volumes, although $\Delta E_{\pi^+\pi^+}$ and $\Delta E_{\overline{K}^0 \overline{K}^0}$ on a given volume are indistinguishable.
The small magnitude of QED effects on $\Delta E_{\pi^+\pi^+}$ might appear surprising because $\alpha m_\pi L \sim 0.74$ for this volume, but as discussed in Sec.~\ref{sec:theory} the appropriate FV analog of the Coulomb expansion parameter is $\eta_L = \alpha m_\pi L / (4\pi) \sim 0.06$ for the $L/a = 48$ volume.
Eqs.~\eqref{eq:NNLO2body}-\eqref{eq:N3LO2body} therefore predict that in addition to differences arising from $a_C^{\pi^+\pi^+} \neq a^{\overline{K}^0\overline{K}^0}$, NLO corrections from Coulomb photon exchange modify the LO FV energy shift by $\sim 20\%$ on the $L/a=48$ lattice volume, which is not expected to be distinguishable given the statistical uncertainties on $\Delta E_{\pi^+\pi^+}$ and $\Delta E_{\overline{K}^0 \overline{K}^0}$.
This expectation is consistent with LQCD+QED$_L$ results, as shown in Fig.~\ref{fig:2_meson}.

%%%%%%%%%%%%%%%%%%%%%%%%%%%%%%%%%%%%%%%%%%%%%%%%%%%%%%%%%%%%%%%%%%%%%%%%%
%%%%%%%%%%%%%%%%%%%%%%%%%%%%%%%%%%%%%%%%%%%%%%%%%%%%%%%%%%%%%%%%%%%%%%%%%

The scattering lengths $a_C^{\pi^+\pi^+}$ and $a^{\overline{K}^0\overline{K}^0}$ can be extracted from a combined fit to the results for $\Delta E_{\pi^+\pi^+}$ and $\Delta E_{\overline{K}^0 \overline{K}^0}$ shown in Table~\ref{tab:mesonFV} and the results for the precisely determined correlated differences $\Delta E_{\pi^+\pi^+} - \Delta E_{\overline{K}^0 \overline{K}^0}$ shown in Table~\ref{tab:mesonFVdiff}.
Fitting to Eq.~\eqref{eq:NNLO2body}, as is appropriate for the power counting PC1 of Eq.~\eqref{eq:pc1}, gives the results,
\begin{equation}
   \begin{split}
      \text{NNLO, PC1}:\hspace{20pt} a^{\overline{K}^0\overline{K}^0} m_{\overline{K}^0} = 0.335(26), \hspace{20pt}  a^{\pi^+ \pi^+}_C m_{\overline{K}^0} = 0.463(41),
   \end{split}\label{eq:fit12b} \end{equation}
where the common scale $m_{\overline{K}^0}$ has been included for both $\pi^+\pi^+$ and $\overline{K}^0\overline{K}^0$ to facilitate comparison of $a^{\overline{K}^0\overline{K}^0}$ and $a^{\pi^+\pi^+}_C$.
The lowest-order QED effect on $\Delta E_{\pi^+\pi^+}$ from Coulomb photon exchange in Eq.~\eqref{eq:NNLO2body} decreases $\Delta E_{\pi^+\pi^+}$ compared to the FV energy shift for neutral particles.\footnote{Coulomb photon exchange leads to a decrease in the energy of a system of $\pi^+$ mesons in QED$_L$ because of zero-mode subtraction. Physically, the energy decrease can be understood as arising from attraction between the charged particle system and the uniform background of opposite charge associated with zero-mode subtraction~\cite{Hayakawa:2008an}. Formally, zero-mode subtraction removes the LO one-photon-exchange diagrams associated with repulsion between charged particles. The dominant QED contribution therefore arises at NLO and necessarily lowers the ground-state energy since it appears at second order in perturbation theory.}
The scattering length results in Eq.~\eqref{eq:fit12b} show that this effect from Coulomb photon exchange competes with additional QED effects that lead to $a_C^{\pi^+\pi^+} > a^{\overline{K}^0\overline{K}^0}$.
Fitting to Eq.~\eqref{eq:N3LO2body}, as is appropriate for the power counting PC2 of Eq.~\eqref{eq:pc2}, gives consistent results,
\begin{equation}
   \begin{split}
      \text{N$^3$LO, PC2}:\hspace{20pt} a^{\overline{K}^0\overline{K}^0} m_{\overline{K}^0} = 0.332(27), \hspace{20pt}  a^{\pi^+ \pi^+}_C m_{\overline{K}^0} = 0.465(42),
   \end{split}\label{eq:fit22b}
\end{equation}
demonstrating that the fit is not overly sensitive to higher-order terms absent in one or the other power counting.

Both Eq.~\eqref{eq:NNLO2body} and Eq.~\eqref{eq:N3LO2body} neglect relativistic effects from radiation photon exchange leading to $\mathcal{O}\left( (ML)^{-1} \right)$ FV effects.
   These effects are estimated in Sec.~\ref{sec:2body} to lead to a shift in $a_C^{\pi^+\pi^+}$ of order $\sim 3\%$, which is smaller than the $6$ - $9\%$ statistical uncertainty on $a_C^{\pi^+\pi^+}$ in Eqs.~\eqref{eq:fit12b}-\eqref{eq:fit22b} and can be consistently neglected.

\subsection{Charged multi-nucleon systems}

\label{sec:2baryon}

Two-proton states receive QED$_L$ FV effects from Coulomb photon exchange proportional to $\alpha M_p L$ that are enhanced compared with those in the $\pi^+ \pi^+$ case discussed above.
For both lattice volumes $\alpha M_p L > 1$, and according to the scaling estimates of Ref.~\cite{Beane:2014qha} Coulomb effects should be nonperturbative.
However, $\eta_L^p = \alpha M_p L / (4\pi) \sim 0.15$ for the $L/a=48$ lattice volume and the NRQED$_L$ results in Sec.~\ref{sec:theory} expressed as power series in $\eta_L^p$ show signs of convergence.
   Examining the QED$_L$ contributions proportional to $a_C^{pp} / L$ in Eq.~\eqref{eq:NNLO2body}, the NLO mixed strong-Coulomb contribution is suppressed compared to the LO strong contribution by $(\eta_L^p / \pi^2)2\mathcal{J} \sim 0.51$, while the corresponding NNLO contribution is suppressed compared to the LO contribution by $(\eta_L^p/\pi^2)(2\mathcal{R}_{24} + \mathcal{J}^2 -\mathcal{L}) \sim 0.15$.
This suggests that Coulomb effects should be perturbative and subdominant compared to strong-interaction FV effects, which are enhanced by the large size of baryon-baryon scattering lengths~\cite{Beane:2009py,Beane:2011iw,Beane:2012vq,Yamazaki:2012hi,Beane:2013br,Orginos:2015aya,Berkowitz:2015eaa,Yamazaki:2015asa,Wagman:2017tmp}.
The quantization condition in Eq.~\eqref{eq:QC}, which neglects $\mathcal{O}( (\eta_L^p)^2 )$ perturbative Coulomb effects and $(ML)^{-1}$ relativistic effects but is nonperturbative in strong interaction effects, is therefore needed to relate the $pp$ FV energy shifts to the infinite volume $pp$ phase shift and determine $a_C^{pp}$.

\begin{table}[t!]
	\begin{tabular}{|c||c|c|} \hline
      & \multicolumn{2}{c|}{$a \Delta   E_{b}$ }   \\\hline
      $b$ & $L/a = 32$  & $L/a = 48$   \\\hline\hline
      $pp$            & 0.008(14)   & 0.011(11)   \\\hline 
      $np({}^1S_0)$   & 0.000(17)   & 0.017(10)   \\\hline
      $nn$            & 0.002(14)   & 0.021(10)   \\\hline
      $np({}^3S_1)$   & 0.017(13)   & 0.010(10)   \\\hline
      ${}^3\text{He}$ & -0.011(96)  & 0.038(56)   \\\hline
      ${}^3\text{H}$  & 0.015(75)   & 0.080(45)   \\\hline
	\end{tabular}
   \caption{FV energy shift results for systems of $n\in\{1,2,3\}$ protons and neutrons determined by fitting Eq.~\eqref{eq:baryon} to LQCD+QED$_L$ Euclidean correlation function results as described in Appendix~\ref{app:fits}. \label{tab:baryonFV} }
\end{table}

%%%%%%%%%%%%%%%%%%%%%%%%%%%%%%%%%%%%%%%%%%%%%%%%%%%%%%%%%%%%%%%%%%%%%%%%%
\begin{figure}[t]
  \centering
  \includegraphics[width=.7\textwidth]{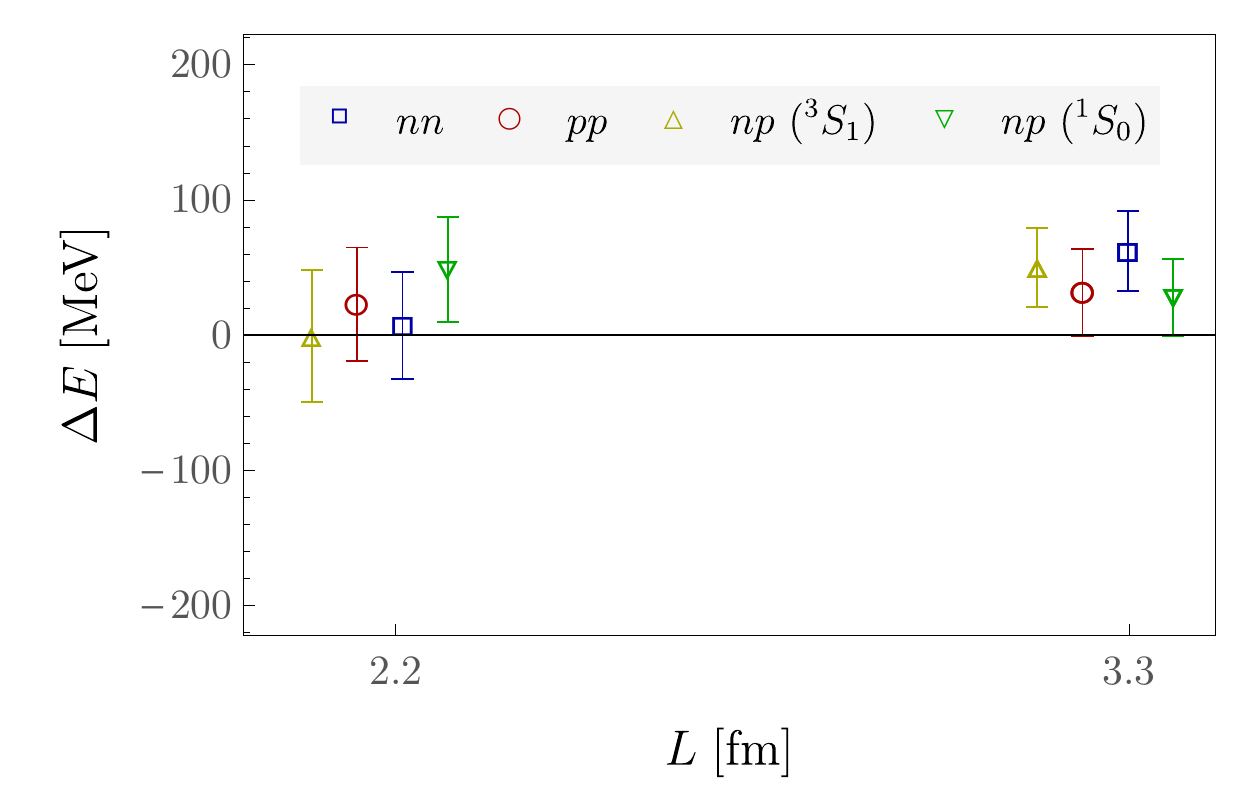}
  \caption{     
     Points show the LQCD+QED$_L$ results from Table~\ref{tab:baryonFV} for FV energy shifts determined from the correlated differences of two-nucleon and one-nucleon ground-state energies for the neutron-neutron and proton-proton FV energy shifts for the $L/a\in\{32,48\}$ volumes.   A smaller (larger) horizontal offset is applied symmetrically to $pp$ and $nn$ ($np({}^1S_0)$ and $np({}^3S_1)$) results.
  \label{fig:2_baryon}}
\end{figure}
%%%%%%%%%%%%%%%%%%%%%%%%%%%%%%%%%%%%%%%%%%%%%%%%%%%%%%%%%%%%%%%%%%%%%%%%%

   The two-nucleon isospin $I=1$ systems $pp$, $nn$, $np({}^1S_0)$, as well as the deuteron, are studied on both lattice volumes.
   Relatively clean signals are seen for each system, and their ground-state energies are determined with total (statistical plus fitting systematic) uncertainties at the $2\%$ level as shown in Table~\ref{tab:baryon}.
   FV energy-level shifts are determined from combined analyses of the two-nucleon and single-nucleon correlation functions as described in Appendix~\ref{app:fits}, and fit results for all systems are shown in Appendix~\ref{sec:fitresults}.
   As shown in Table.~\ref{tab:baryonFV} and Fig.~\ref{fig:2_baryon}, the statistical precision of this calculation is insufficient to resolve either the proton-proton or neutron-neutron FV energy shift from zero on either volume studied.
   Resolving non-zero FV shifts of the $\mathcal{O}(10 \text{ MeV})$ size expected for two-nucleon systems without QED at these quark masses at a 95\% confidence level for the $pp$, $nn$, and $np$ systems on the $L/a=32$ lattice volume would require statistical ensembles approximately $\sim 50$ - $100$ times larger  than the one used here, as estimated by extrapolating the uncertainties of the four different two-nucleon systems considered assuming $1/\sqrt{N}$ scaling of uncertainties.
Determination of $a_C^{pp}$ through the QED$_L$ quantization condition of Eq.~\eqref{eq:QC} is therefore left to future work.

The two $I=1/2$ three-nucleon systems ${}^3$He and ${}^3$H are also investigated, and their ground-state energies are given in Table~\ref{tab:baryon}.
Correlated differences between three-nucleon ground-state energies and the sums of their constituent nucleon masses are show in Table~\ref{tab:baryonFV}.
As with the two-nucleon systems, fit results are shown in Appendix~\ref{sec:fitresults}.
The results for ${}^3$He and ${}^3$H are not precise enough to allow FV effects to be reliably determined.
Precision in the three-nucleon sector is significantly worse than in the two-nucleon sector, as expected. 
   The absolute size of FV energy shifts is also expected to be larger for three-nucleon systems than two-nucleon systems, and for instance resolving an $\mathcal{O}(50\text{ MeV})$ FV energy shift at a 95\% confidence level for the ${}^3$He and ${}^3$H systems on the $L/a=32$ lattice volume would require a statistical ensemble approximately $\sim 100$ times larger than the one considered here, based on an extrapolation analogous to that described for the two-nucleon case.
   
Future high-precision LQCD+QED$_L$ calculations of these multi-nucleon systems will provide insight into QED effects on nucleon-nucleon and three-nucleon interactions through a determination of the ${}^3$He - ${}^3$H binding-energy difference and its decomposition into QED and strong isospin breaking effects from LQCD+QED$_L$.

\subsection{Systems of many charged mesons}

\label{sec:multi}

Multi-pion correlation functions do not suffer from significant exponential signal-to-noise degradation with increasing particle number and can be used to study QED in the regime where the charge $Z \sim 1/\alpha$.
In particular, the correlation functions for systems with $n \leq 12$ $\pi^+$ mesons described in Sec.~\ref{sec:lattice} can be used to study systems with $Z \alpha \leq 1.2$, reaching a charge density of $n / L^3 \sim 1.2 \text{ fm}^{-3}$.
The dominant strong interactions and QED effects on many-particle FV energy shifts in Eqs.~\eqref{eq:NNLO}-\eqref{eq:N3LO} both scale with $n^2$, and both $\Delta E_{n \pi^+}$ and $\Delta E_{n \overline{K}^0}$ can be extracted for larger $n$ with better relative precision than from the $n=2$ case discussed in Sec.~\ref{sec:2meson}.

\begin{figure}[t]
   \subfigure[~Multi-meson FV energy shifts for the $L/a = 32$ lattice volume.]{
  \centering
  \includegraphics[width=.7\textwidth]{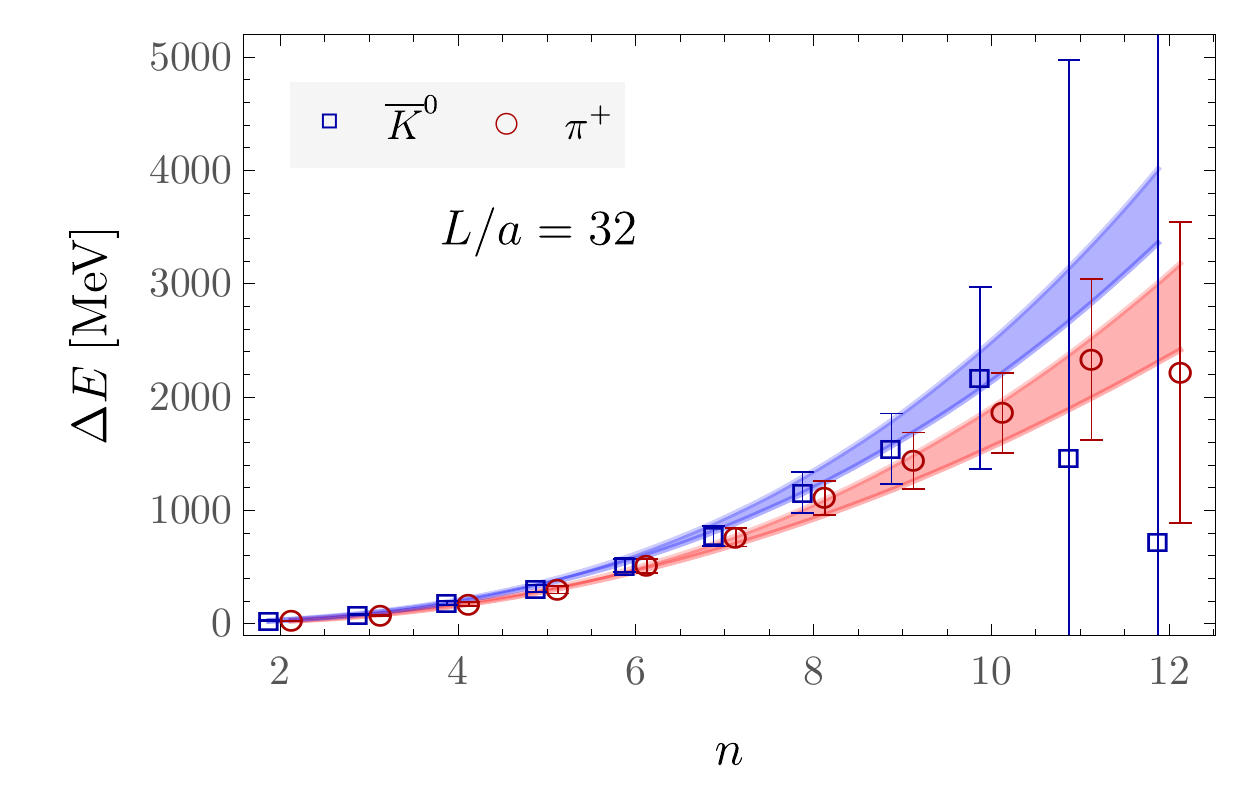}
   }\quad
   \subfigure[~Multi-meson FV energy shifts for the $L/a = 48$ lattice volume.]{
  \includegraphics[width=.7\textwidth]{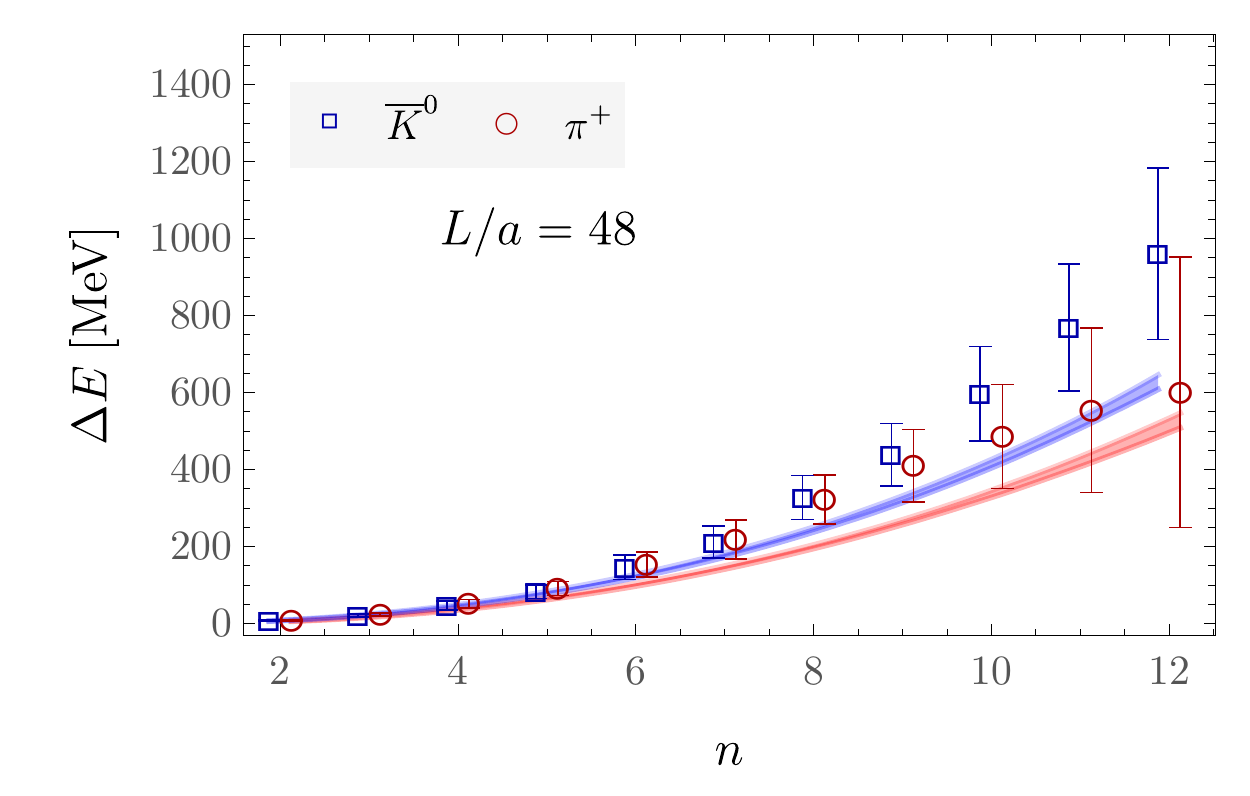}
   }\quad
   \caption{ Points show the LQCD+QED$_L$ results from Table~\ref{tab:mesonFV} for the FV energy shifts of multi-$\overline{K}^0$ and multi-$\pi^+$ meson systems as a function of meson number $n$ on both lattice volumes. Shaded bands show 67\% bootstrap confidence intervals for the predictions of Eq.~\eqref{eq:N3LO} for the best-fit parameters $\left\{ a_C^{\pi^+\pi^+},\ a^{\overline{K}^0\overline{K}^0},\ \eta_{3}^{\pi^+\pi^+\pi^+}(m_{\overline{K}^0}),\ \eta_3^{\overline{K}^0\overline{K}^0\overline{K}^0}(m_{\overline{K}^0}) \right\}$ obtained from a global fit to the $L/a\in\{32,48\}$ results for $n\in\{2,\ldots,12\}$ mesons in Tables~\ref{tab:mesonFV}-\ref{tab:mesonFVdiff} as described in the main text. A small horizontal offset is applied symmetrically to $\pi^+$ and $\overline{K}^0$ results.
  \label{fig:many_meson}}
\end{figure}

\begin{figure}[t]
  \centering
   \subfigure[~$\pi^+$ and $\overline{K}^0$ energy shift differences for the $L/a = 32$ lattice volume.]{
  \centering
  \includegraphics[width=.7\textwidth]{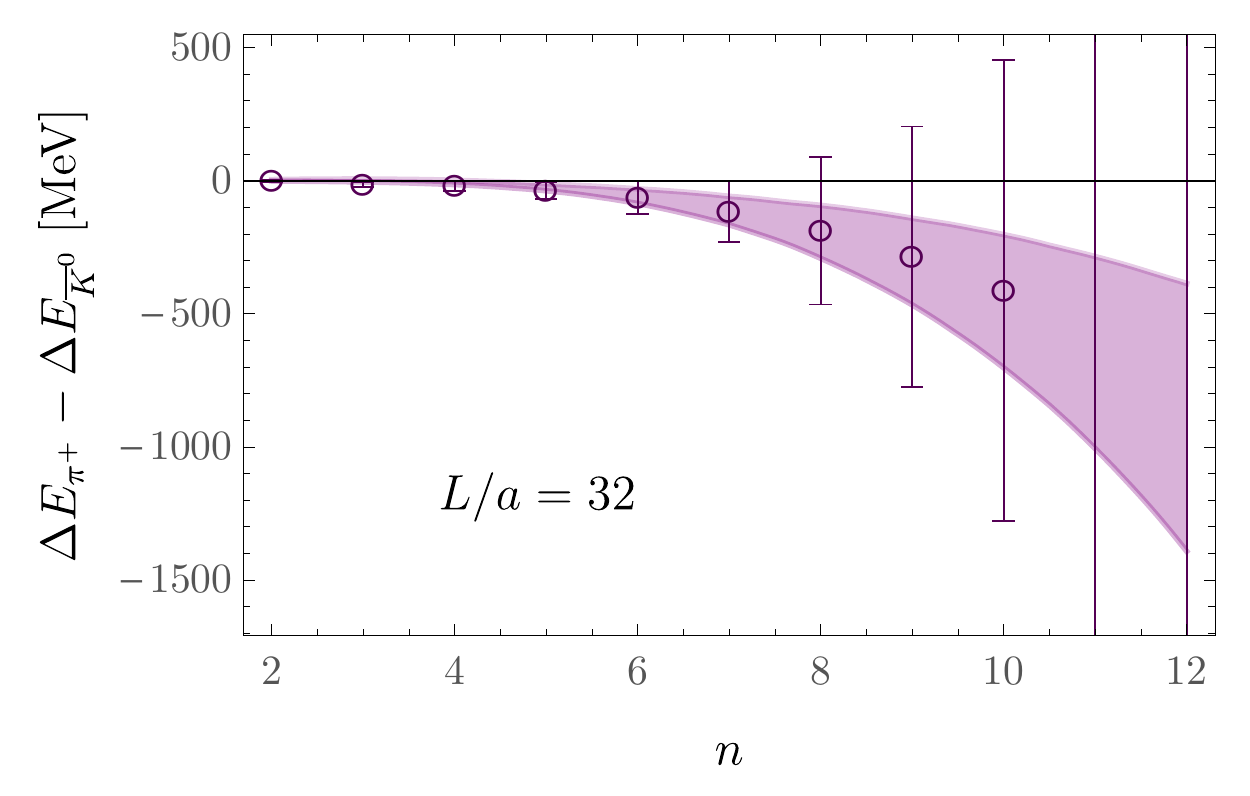}
   }\quad
   \subfigure[~$\pi^+$ and $\overline{K}^0$ energy shift differences for the $L/a = 48$ lattice volume.]{
  \includegraphics[width=.7\textwidth]{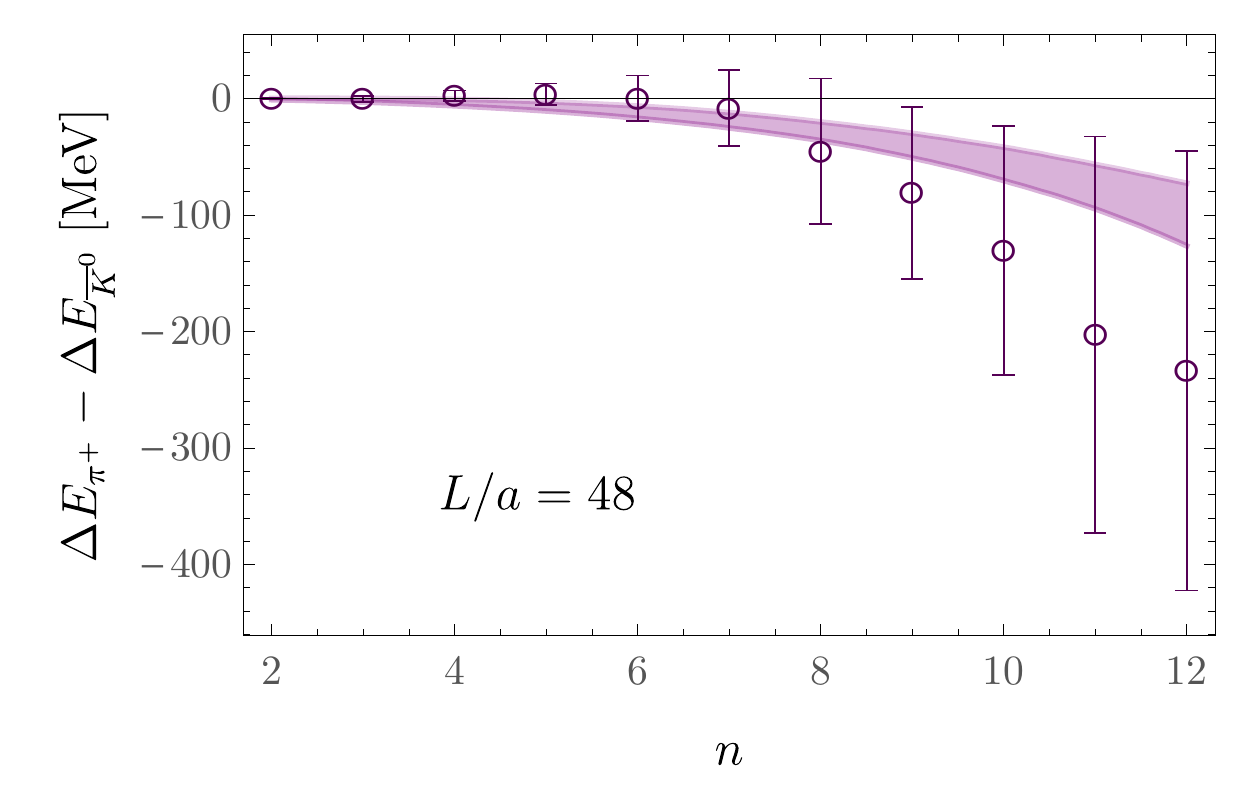}
   }\quad
  \caption{Points show the LQCD+QED$_L$ results from Table~\ref{tab:mesonFVdiff} for the FV energy-shift differences between multi-$\pi^+$ and multi-$\overline{K}^0$ systems as a function of meson number $n$ on both lattice volumes. The shaded band shows the 67\% bootstrap confidence interval for the corresponding prediction of Eq.~\eqref{eq:N3LO} for the best-fit parameters $\left\{ a_C^{\pi^+\pi^+},\ a^{\overline{K}^0\overline{K}^0},\ \eta_{3}^{\pi^+\pi^+\pi^+}(m_{\overline{K}^0}),\ \eta_3^{\overline{K}^0\overline{K}^0\overline{K}^0}(m_{\overline{K}^0}) \right\}$ obtained from a global fit to the $L/a\in\{32,48\}$ results for $n\in\{2,\ldots,12\}$ mesons, given in Tables~\ref{tab:mesonFV}-\ref{tab:mesonFVdiff}, as described in the main text.
  \label{fig:many_meson_diff}}
\end{figure}

Results for $\Delta E_{n \pi^+}$, $\Delta E_{n\overline{K}^0}$, and the correlated differences $\left(\Delta E_{n\pi^+} - \Delta E_{n\overline{K}^0}\right)$ for $n\in\{2,\ldots,12\}$ are shown in Tables~\ref{tab:mesonM}-\ref{tab:mesonFVdiff}.
QED effects leading to non-zero $\left(\Delta E_{n \pi^+} - \Delta E_{n \overline{K}^0}\right)$ can be resolved to better than $1 \sigma$ on the $L/a=32$ lattice volume for $3 \leq n \leq 7$, and on the $L/a = 48$ lattice volume for $n \geq 9$.
These 33 FV energy shifts and correlated differences $\Delta E_{n \pi^+}$, $\Delta E_{n\overline{K}^0}$, and $\left(\Delta E_{n\pi^+} - \Delta E_{n\overline{K}^0}\right)$ can be used to constrain the low-energy interaction parameters $\left\{ a_C^{\pi^+\pi^+},\ a^{\overline{K}^0\overline{K}^0},\ \eta_{3}^{\pi^+\pi^+\pi^+}(m_{\overline{K}^0}),\ \eta_3^{\overline{K}^0\overline{K}^0\overline{K}^0}(m_{\overline{K}^0}) \right\}$ appearing in Eqs.~\eqref{eq:NNLO}-\eqref{eq:N3LO}.\footnote{The renormalization scale used to evaluate $\eta_3(\mu)$ should be chosen close to the ``high'' energy scale where NRQED$_L$ is matched to QED$_L$ in order to avoid large logarithms that can worsen EFT convergence. For simplicity, $\mu = m_{\overline{K}^0}$ is used as the renormalization scale for $\eta_3$ throughout this work.}
Fits to the NNLO expression given in Eq.~\eqref{eq:NNLO} in PC1, which includes $\mathcal{O}(\eta_L^3)$ Coulomb effects but neglects three-body forces, underpredict LQCD+QED$_L$ energy-shift results for $n \gtrsim 8$ meson systems on both lattice volumes and obtain a minimum $\chi^2 / N_{\text{dof}} \sim 1.3$.
The N$^3$LO expression Eq.~\eqref{eq:N3LO} in PC2 includes additional free parameters related to three-body forces not present at NNLO.
Including three-body force parameters improves the quality of the fit, and Eq.~\eqref{eq:N3LO} provides a better description of the LQCD+QED$_L$ results with $\chi^2 / N_{\text{dof}} \sim 0.8$.
Fit results using the N$^3$LO expression Eq.~\eqref{eq:N3LO} and uncertainties computed using bootstrap resampling of the global fitting procedure are compared to LQCD+QED$_L$ results for the $\pi^+$ and $\overline{K}^0$ energy shifts in Figs.~\ref{fig:many_meson}-\ref{fig:many_meson_diff}.
The results for the meson scattering lengths are consistent with, but more precise than, the results obtained from the two-meson FV energy shifts alone,
\begin{equation}
   \begin{split}
      \text{N$^3$LO, PC2}:\hspace{20pt} a^{\overline{K}^0\overline{K}^0} m_{\overline{K}^0} = 0.337(19), \hspace{20pt} a^{\pi^+ \pi^+}_C  m_{\overline{K}^0} = 0.464(29).
   \end{split}\label{eq:fit1}
\end{equation}
It is noteworthy that results with $Z \alpha \geq 1$ can be fit by the NRQED$_L$ formula given in Eq.~\eqref{eq:N3LO} without additional modifications to account for relativistic QED effects or additional nonperturbative effects.
Some tensions between LQCD+QED$_L$ results and N$^3$LO NRQED$_L$ fits can be observed for $n \gtrsim 8$ meson systems on the $L/a = 48$ lattice volume in Fig.~\ref{fig:many_meson}; however, since these tensions are more significant for multi-$\overline{K}^0$ than multi-$\pi^+$ systems they are unlikely to be signals of nonperturbative QED effects and might result from correlations between LQCD+QED$_L$ results with different $n$ not accounted for in the fitting procedure employed here.

%%%%%%%%%%%%%%%%%%%%%%%%%%%%%%%%%%%%%%%%%%%%%%%%%%%%%%%%%%%%%%%%%%%%%%%%%
%%%%%%%%%%%%%%%%%%%%%%%%%%%%%%%%%%%%%%%%%%%%%%%%%%%%%%%%%%%%%%%%%%%%%%%%%

%%%%%%%%%%%%%%%%%%%%%%%%%%%%%%%%%%%%%%%%%%%%%%%%%%%%%%%%%%%%%%%%%%%%%%%%%
%%%%%%%%%%%%%%%%%%%%%%%%%%%%%%%%%%%%%%%%%%%%%%%%%%%%%%%%%%%%%%%%%%%%%%%%%

The three-pion scattering amplitude was calculated at LO in $\chi$PT in Ref.~\cite{Blanton:2019vdk} and is given by $\mathcal{M}_{\rm 3,th} = 108 m_{\overline{K}^0}^2/f_{\overline{K}^0}^4$ (with conventions for the LO Lagrangian such that $f_{\pi^+} \sim 130$ MeV).\footnote{The $SU(2)$ $\chi$PT result for $\mathcal{M}_{\rm 3,th}$ given in Ref.~\cite{Blanton:2019vdk} is valid for tree-level $\overline{K}^0\overline{K}^0$ scattering after reinterpreting the $SU(2)$ isospin $\chi$PT Lagrangian in terms of the $SU(2)$ $V$-spin doublet $(\pi^+, \overline{K}^0)$. More formally, the $V$-spin analog of $G$-parity acts as $(\pi^+,\overline{K}^0) \rightarrow (K^0, -\pi^0)$ and $(K^0, -\pi^0) \rightarrow (-\pi^+, \overline{K}^0)$ and therefore relates the three-body contact operators of the $SU(2)$ isospin and $SU(2)$ $V$-spin Lagrangians by $(\pi^+ \pi^-)^3 \rightarrow (\overline{K}^0 K^0)^3$. For the $SU(3)$ flavor symmetric quark mass scheme used here, $V$-spin is an exact symmetry of the leading order $\chi$PT Lagrangian broken only by QED corrections at higher orders.}
This can be combined with Eq.~\eqref{eq:etaRel} to provide a $\chi$PT prediction for the non-relativistic $3\overline{K}^0$ contact interaction,
\begin{equation}
   \begin{split}
      \eta_3^{\overline{K}^0\overline{K}^0\overline{K}^0}(\mu) &= \frac{3}{4 m_{\overline{K}^0} f_{\overline{K}^0}^4} + \frac{m_{\overline{K}^0}}{64 \pi^3 f_{\overline{K}^0}^8}(3\sqrt{3} - 4\pi)\ln\left(\frac{m_{\overline{K}^0}}{2\pi \mu}\right) \\
                                                                &\hspace{20pt} + \frac{3 m_{\overline{K}^0}^3}{1024 \pi^6 f_{\overline{K}^0}^8}\mathcal{S}_{\rm 3,th} + \frac{3m_{\overline{K}^0}}{2 f_{\overline{K}^0}^6}\mathcal{C}_3 ,
   \end{split}\label{eq:etaLO}
\end{equation}
where $\mathcal{S}_{\rm 3,th}$ and $\mathcal{C}_3$ are constants defined below Eq.~\eqref{eq:etaRel}, and to obtain a result entirely in terms of $m_{\overline{K}^0}$ and $f_{\overline{K}^0}$, the LO $\chi$PT relations~\cite{Bijnens:1997vq}
\begin{equation}
   \begin{split}
      a^{\overline{K}^0\overline{K}^0} m_{\overline{K}^0} = \frac{m_{\overline{K}^0}^2}{8 \pi f_{\overline{K}^0}^2},
      \hspace{20pt}   r^{\overline{K}^0\overline{K}^0} a^{\overline{K}^0\overline{K}^0} m^2_{\overline{K}^0} = 3 \ ,
   \end{split}\label{eq:LO2body}
\end{equation}
have been used.
The first of these relations also allows $f_{\overline{K}^0}$ and therefore $\eta_3^{\overline{K}^0\overline{K}^0\overline{K}^0}(\mu)$ to be predicted numerically at LO in $\chi$PT for the quark masses used in this work.
Inserting the LQCD+QED$_L$ results for $m_{\overline{K}^0} a^{\overline{K}^0\overline{K}^0}$ from Eq.~\eqref{eq:fit1} into the first relation in Eq.~\eqref{eq:LO2body} provides a prediction for $f_{\overline{K}^0}$ at the parameters of this LQCD+QED$_L$ calculation that is valid at LO in $\chi$PT:
\begin{equation}
   \begin{split}
      af_{\overline{K}^0} = 0.0476(35), \hspace{20pt} f_{\overline{K}^0} = 139(10)(4)\text{ MeV}.
   \end{split}\label{eq:fKLO}
\end{equation}
Here, the first uncertainty is statistical and the second uncertainty is from the uncertainty in the lattice spacing.
In this calculation $m_{\overline{K}^0} = 404(1)(12)$ MeV is between the physical pion and kaon masses; this can be compared with $f_{\pi^+} \sim 130$ MeV and $f_{\overline{K}^0} \sim 156$ MeV extracted from experiments~\cite{Tanabashi:2018oca}.
Inserting this result for $f_{\overline{K}^0}$, and the result for $a^{\overline{K}^0\overline{K}^0}$ in Eq.~\eqref{eq:fit1}, into Eq.~\eqref{eq:etaLO} then gives the numerical result $m_{\overline{K}^0} f_{\overline{K}^0}^4 \eta_3^{\overline{K}^0\overline{K}^0\overline{K}^0}(m_{\overline{K}^0}) = 0.62(27)$.
This result is valid at LO in $\chi$PT and can be compared to LQCD+QED$_L$ results for $\eta_3^{\overline{K}^0\overline{K}^0\overline{K}^0}(m_{\overline{K}^0})$ combined with the result for $f_{\overline{K}^0}$ given in Eq.~\eqref{eq:fKLO}.
%

%%%%%%%%%%%%%%%%%%%%%%%%%%%%%%%%%%%%%%%%%%%%%%%%%%%%%%%%%%%%%%%%%%%%%%%%%
\begin{figure}[t]
  \centering
  \includegraphics[width=.7\textwidth]{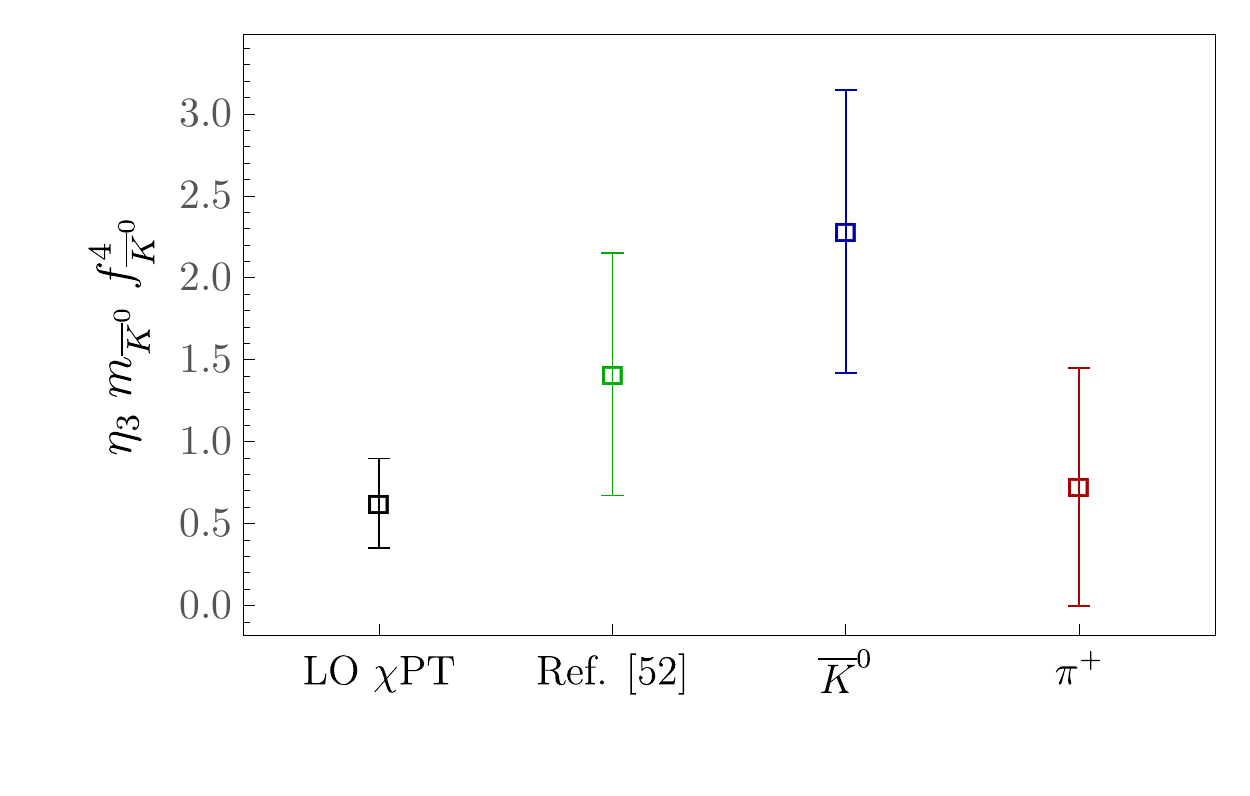}
   \caption{The blue and red points show the best-fit values and 67\% bootstrap confidence intervals for the three-body interaction parameters for neutral $\overline{K}^0\overline{K}^0\overline{K}^0$ and charged $\pi^+\pi^+\pi^+$ systems respectively, including a common normalization factor of $f_{\overline{K}^0}^4 m_{\overline{K}^0}$ to obtain a dimensionless quantity. The green point shows the LQCD result of Ref.~\cite{Beane:2007es}, which was obtained in a calculation using a pseudoscalar meson mass of 352 MeV similar to $m_{\overline{K}^0}$ here. The black point shows the LO $\chi$PT prediction of Eq.~\eqref{eq:etaLO} multiplied by the same normalization factor. 
  \label{fig:mesons_3body}}
\end{figure}
%%%%%%%%%%%%%%%%%%%%%%%%%%%%%%%%%%%%%%%%%%%%%%%%%%%%%%%%%%%%%%%%%%%%%%%%%

Dimensionless LQCD+QED$_L$ results for the three-body coupling that are expected to be $\mathcal{O}(1)$ in $\chi$PT are given by
\begin{equation}
   \begin{split}
       \eta_3^{\overline{K}^0\overline{K}^0\overline{K}^0}(m_{\overline{K}^0}) m_{\overline{K}^0} f_{\overline{K}^0}^4 = 2.28(86), \hspace{20pt}   \eta_{3}^{\pi^+ \pi^+ \pi^+}(m_{\overline{K}^0}) m_{\overline{K}^0} f_{\overline{K}^0}^4 = 0.72(73).
   \end{split}\label{eq:fit3eta}
\end{equation}
The result of this work for $\eta_3^{\overline{K}^0\overline{K}^0\overline{K}^0}(m_{\overline{K}^0})$  is consistent within $1\sigma$ uncertainties with the LQCD results of Ref.~\cite{Beane:2007es}, which were obtained in a calculation using quark masses similar to those used in this work, corresponding to a pion mass of 352 MeV, and extracted $\eta_3$ by fitting to the same $\mathcal{O}\left(L^{-6}\right)$ threshold expansion as used here for multi-$\overline{K}^0$ systems.
The corresponding result for $\eta_{3}^{\pi^+ \pi^+ \pi^+}(m_{\overline{K}^0})$ is about $2\sigma$ smaller than $\eta_3^{\overline{K}^0\overline{K}^0\overline{K}^0}(m_{\overline{K}^0})$, although it is also consistent within $1\sigma$ uncertainties with the LQCD results of Ref.~\cite{Beane:2007es}.
Results for $\eta_3^{\overline{K}^0\overline{K}^0\overline{K}^0}(m_{\overline{K}^0})$ and $\eta_{3}^{\pi^+ \pi^+ \pi^+}(m_{\overline{K}^0})$ as well as comparisons to LO $\chi$PT and to the LQCD results of Ref.~\cite{Beane:2007es} are shown in Fig.~\ref{fig:mesons_3body}.
Instead of using a global fit to extract the scattering lengths and three-body interaction parameters as above, one can instead fix the scattering lengths to the results obtained using fits to $\Delta E_{\pi^+\pi^+}$, $\Delta E_{\overline{K}^0\overline{K}^0}$ shown in Eq.~\eqref{eq:fit22b} and extract the three-body interaction parameters from fits to multi-meson results on each volume separately in order to provide an estimate of higher-order FV corrections. Using this alternative fitting procedure to extract the three-body interaction parameters from fits using only the $L/a = 32$ results gives   $\eta_3^{\overline{K}^0\overline{K}^0\overline{K}^0}(m_{\overline{K}^0}) m_{\overline{K}^0} f_{\overline{K}^0}^4 = 2.1(9)$ and  $\eta_{3}^{\pi^+ \pi^+ \pi^+}(m_{\overline{K}^0}) m_{\overline{K}^0} f_{\overline{K}^0}^4 = 0.5(9)$, while using only the $L/a = 48$ results gives $\eta_3^{\overline{K}^0\overline{K}^0\overline{K}^0}(m_{\overline{K}^0}) m_{\overline{K}^0} f_{\overline{K}^0}^4 = 7.0(2.6)$ and  $\eta_{3}^{\pi^+ \pi^+ \pi^+}(m_{\overline{K}^0}) m_{\overline{K}^0} f_{\overline{K}^0}^4 = 5.6(2.4)$, indicating that the $L/a=32$ results are primarily responsible for constraining the three-body interaction parameters and suggesting that higher-order FV corrections to the extracted three-body interaction parameters may be significant. Additional consistency checks on the three-body interaction determinations are discussed in Ref. [53], where combinations of $n$-meson FV energy shifts with fixed $n$ are presented that isolate the three-body interaction parameter in the threshold expansion. In all cases, differences between global fit results and results for three-body interactions parameters obtained with fixed $n$ and fixed $L/a$ in this way are smaller than the corresponding differences between global fits results and the combined fits to all $n$ and fixed $L/a$ discussed above.

Differences between  $\eta_{3}^{\pi^+ \pi^+ \pi^+}(m_{\overline{K}^0})$ and $\eta_3^{\overline{K}^0\overline{K}^0\overline{K}^0}(m_{\overline{K}^0})$ might arise from QED effects on $\mathcal{M}_{\rm 3,th}$ beyond LO in $\chi$PT, or from QED effects on the matching between $\mathcal{M}_{\rm 3,th}^{\pi^+\pi^+\pi+}$ and $\eta_{3}^{\pi^+ \pi^+ \pi^+}(m_{\overline{K}^0})$.
Differences between the extracted $\eta_{3}^{\pi^+ \pi^+ \pi^+}(m_{\overline{K}^0})$ and $\eta_3^{\overline{K}^0\overline{K}^0\overline{K}^0}(m_{\overline{K}^0})$ might also spuriously arise from mismodeling of $\mathcal{O}\left((ML)^{-1}\right)$ relativistic effects estimated in Sec.~\ref{sec:2body} to modify FV energy shifts by $~\sim 3\%$.
This estimated shift is larger or comparable to the statistical uncertainties on the three-body energy shift for all $n\pi^+$ systems on the $L/a = 48$ volume and $n\pi^+$ systems with $n\lesssim 6$ on the $L/a = 32$ volume.
Given these systematic uncertainties in conjunction with the statistical uncertainties on $\eta_{3}^{\pi^+ \pi^+ \pi^+}(m_{\overline{K}^0})$ and $\eta_3^{\overline{K}^0\overline{K}^0\overline{K}^0}(m_{\overline{K}^0})$, the results of this work do not provide significant evidence for differences in non-relativistic short-range three-meson interactions arising from QED effects.

\section{Conclusions}\label{sec:conclusion}

In this work, lattice QCD+QED$_L$ has been used to study systems of up to 12 charged or neutral mesons as well as systems of one, two, and three nucleons. Calculations were performed in two lattice volumes with charge-dependent quark masses tuned such that strong isospin breaking effects are negligible and energy differences between charged and neutral systems are primarily QED effects.
While the ground-state energies of two- and three-nucleon systems are determined with few-percent-level precision, QED effects leading to differences between two-nucleon and three-nucleon FV energies are not resolved.
Significantly higher precision will be needed in future calculations of QED effects in multi-nucleon systems.
Differences between charged and neutral FV energy shifts are resolved at the level of $1$-$2\sigma$ for systems of $3$-$12$ mesons, demonstrating the presence of QED effects on meson-meson interactions.
Analysis of the FV energy levels for multi-meson systems using non-relativistic EFT has allowed the extraction of the $\pi^+\pi^+$ and $\overline{K}^0\overline{K}^0$ scattering lengths as well as the $3 \pi^+$ and $3 \overline{K}^0$ interaction parameters.
Differences between $a_C^{\pi^+\pi^+}$ and $a^{\overline{K}^0 \overline{K}^0}$ are clearly resolved,
demonstrating that additional QED effects on meson-meson interactions can be resolved beyond the Coulomb photon exchange explicitly included in the EFT.
Differences between the three-body interactions for charged and neutral mesons are not well resolved.

The QED effects on multi-meson systems determined from LQCD+QED$_L$ in this work are well described by NRQED$_L$ results that incorporate short-range two- and three-body contact interactions as well as perturbative Coulomb photon exchange.
Although Coulomb photon exchange must be treated nonperturbatively in sufficiently large volumes, the expansion parameter describing the size of FV Coulomb effects is found to be $\alpha / v = \alpha M L/(4\pi)$ by examining the convergence pattern of the NRQED$_L$ expansion.
This includes a numerically significant factor of $1/(4\pi)$ compared to the parameter $\alpha M L$ discussed in Ref.~\cite{Beane:2014qha}.
For systems with unphysically large $\alpha$ and quark masses such as those studied here, $\alpha M L / (4\pi) \ll 1$ is satisfied for volumes satisfying $L \ll 20\text{ fm}$ for nucleons ($L \ll 50\text{ fm}$ for pions), and Coulomb corrections to the LO strong interaction FV energy shift appear perturbative for $L \lesssim 6\text{ fm}$ for nucleons ($L \lesssim 15\text{ fm}$ for pions).
For calculations with physical $\alpha$ and quark masses, FV Coulomb effects are reduced by a factor of 20 for nucleons (40 for pions) and are expected to be perturbative for all practically accessible lattice volumes.
The EFT analysis of this work also demonstrates that NRQED$_L$ results for FV energy shifts are unaffected by the complications of photon zero-mode subtraction up to effects suppressed by $\mathcal{O}\left( (ML)^{-3} \right)$ that are consistently neglected along with other relativistic effects.
Future LQCD+QED$_L$ calculations, especially those using lighter quark masses and/or smaller values of the QED fine structure constant, can therefore be interpreted using hadronic EFTs with perturbative Coulomb effects with a similar procedure to the one undertaken here.
Such calculations will give insight into the quark mass dependence of QED effects on meson-meson interactions, and, combined with higher-precision calculations of multi-nucleon FV energy levels, will permit first principles predictions of QED effects on nucleon-nucleon interactions and QED effects in light nuclei.

\section*{Acknowledgements}

We thank the other members of the NPLQCD collaboration, in particular Z.~Davoudi and M.J.~Savage, for discussions during the initial stages of this work. 
Calculations in this project were performed using the Hyak High Performance Computing and Data Ecosystem at the University of Washington (https://itconnect.uw.edu/research/hpc/), supported, in part, by the U.S. National Science Foundation Major Research Instrumentation Award, Grant Number 0922770, and on clusters at MIT with support from the NEC Corporation Fund.  
The numerical configuration generation (using the BQCD lattice  QCD program \cite{Haar:2017ubh}) was carried out on the IBM BlueGene/Q and HP Tesseract using DIRAC 2 resources  (EPCC, Edinburgh, UK), the IBM BlueGene/Q (NIC, J\"ulich, Germany)  and the Cray XC40 at HLRN (The North-German Supercomputer  Alliance), the NCI National Facility in Canberra, Australia  (supported by the Australian Commonwealth Government) and Phoenix (University of Adelaide).
The Chroma software library~\cite{Edwards:2004sx} was used in the data analysis.
SRB is supported in part by the U.S.~Department of Energy, Office of Science, Office of Nuclear Physics under grant Contract Number DE-FG02-97ER-41014.
WD, PES, and MLW are supported in part by the U.S.~Department of Energy, Office of Science, Office of Nuclear Physics under grant Contract Number DE-SC0011090.  
WD is also supported within the framework of the TMD Topical Collaboration of the U.S.~Department of Energy, Office of Science, Office of Nuclear Physics, and  by the SciDAC4 award DE-SC0018121.
RH is supported by STFC through grant ST/P000630/1.
MI is supported by the Universitat de Barcelona through the scholarship APIF, by the Spanish Ministerio de Economia y Competitividad (MINECO) under Project No. MDM-2014-0369 of ICCUB and with additional European FEDER funds, under the contract FIS2017-87534-P.
HP is supported by DFG Grant No. PE 2792/2-1.
PELR is supported in part by the STFC under contract ST/G00062X/1.
GS is supported by DFG Grant No. SCHI 179/8-1.
PES is supported in part by the National Science Foundation under CAREER Award 1841699.
MLW was supported in part by an MIT Pappalardo Fellowship.
RDY and JMZ are supported by the Australian Research Council Grants FT120100821, FT100100005, DP140103067 and DP190100297.  
This manuscript has been authored by Fermi Research Alliance, LLC under Contract No. DE-AC02-07CH11359 with the U.S. Department of Energy, Office of Science, Office of High Energy Physics.

\appendix

\section{Matching NRQED$_L$ and QED$_L$}\label{app:matching}

Neglecting relativistic effects suppressed by $M^{-1}$ (including nonlocal effects from zero-mode subtraction discussed in Refs.~\cite{Borsanyi:2014jba,Davoudi:2014qua,Fodor:2015pna,Lee:2015rua,Matzelle:2017qsw,Davoudi:2018qpl}), the NRQED$_L$ Lagrangian is identical for bosons and fermions. For a particle with charge $Q=1$, it is given by
\begin{equation}
   \begin{split}
      \mathcal{L}^{\text{NRQED}_L} &= \psi^\dagger \left( i D_0 - \frac{D_i D^i}{2M} \right) \psi -  \frac{2\pi a}{M} (\psi^\dagger \psi)^2  + \mathcal{L}_{\gamma}^\xi \\
      &= \psi^\dagger \left( i \partial_0 - \frac{\partial_i \partial^i}{2M} \right) \psi - e A_0 (\psi^\dagger \psi) -  \frac{2\pi a}{M} (\psi^\dagger \psi)^2  + \mathcal{L}_{\gamma}^\xi,
   \end{split}\label{eq:NRQEDLapp}
\end{equation}
where in this section we work in Minkowski spacetime with $(-+++)$ signature, $A_\mu = A_\mu^\dagger$ is the photon field, $F_{\mu\nu} = \partial_\mu A_\mu - \partial_\nu A_\mu$ is the field strength tensor, $D_\mu = \partial_\mu + i A_\mu$, and in generic $R_\xi$ gauge the photon Lagrangian is
\begin{equation}
  \begin{split}
    \mathcal{L}_\gamma^\xi &=  -\frac{1}{4}F_{\mu\nu}F^{\mu\nu} + \frac{1}{2\xi }(\partial_\mu A^\mu)^2.
  \end{split}\label{eq:Lgamma}
\end{equation}
Results for the Landau gauge QCD+QED$_L$ calculations performed in the main text are obtained by setting $\xi=0$.
In what follows, a finite spatial volume of extent $L^3$ with PBCs is considered.
Zero mode subtraction can be implemented in NRQED$_L$ by defining the FV photon field as a Fourier transform of the zero-mode subtracted momentum-space field,
\begin{equation}
   \begin{split}
      A_\mu(x) = \int \frac{dp^0}{2\pi}\frac{1}{L^3}\sum_{\v{n} \in \mathbb{Z}^3 \setminus \{ \v{0} \}} e^{-ip^0 x^0 + \frac{2\pi i}{L}\v{n}\cdot\v{x}} \tilde{A}_\mu(p^0,\v{n}), \hspace{20pt} \tilde{A}_\mu(p^0,\v{n}) = \int d^4x e^{ip^0 x^0 - \frac{2\pi i}{L}\v{n}\cdot\v{x}} A_\mu(x),
   \end{split}\label{eq:AFV}
\end{equation}
where $\v{n} \in \mathbb{Z}^3 \setminus \{ \v{0} \}$ excludes the photon zero mode.
Zero-mode subtraction can be defined at the path integral level as a constraint on the photon field~\cite{Davoudi:2018qpl}, but for perturbative matching with QED$_L$ Eq.~\eqref{eq:NRQEDLapp}-\eqref{eq:AFV} define NRQED$_L$.

The NRQED$_L$ massive particle propagator $G^{\text{NRQED}_L}$ is given by
\begin{equation}
  \begin{split}
    G^{\text{NRQED}_L}(p^0, \v{n}) &= \int d^4x e^{ip^0x^0 - \frac{2\pi i}{L}\v{n}\cdot\v{x}} \left< \psi(x) \psi^\dagger(0) \right>  = \frac{i}{p^0 - \frac{1}{2M}\left(\frac{2\pi \v{n}}{L}\right)^2 + i \epsilon}.
  \end{split}\label{eq:GNRQEDL}
\end{equation}
The photon propagator is given by
\begin{equation}
  \begin{split}
  G_{\mu\nu}^{\gamma_L}(p^0, \v{n}) &= \int dx^0 \sum_\v{x} e^{ip^0x^0 - \frac{2\pi i}{L}\v{p}\cdot\v{x}} \left< A_\mu(x) A_\nu(0) \right> = \frac{i \left[ g_{\mu\nu} - (1-\xi) \frac{p_\mu p_\nu}{p^2}\right]}{(p^0)^2 - \left(\frac{2\pi \v{n}}{L}\right)^2 + i \epsilon}.
  \end{split}\label{eq:GgammaL}
\end{equation}
Introducing Fourier transformed fields,
\begin{equation}
   \begin{split}
      \tilde{\psi}(p^0,\v{n}) = \int d^4x e^{ip^0x^0 - \frac{2\pi i}{L}\v{n}\cdot\v{x}} \psi(x), \hspace{20pt} \psi(x) = \int \frac{dp^0}{2\pi} \frac{1}{L^3}\sum_{\v{n} \in \mathbb{Z}^3} e^{-ip^0x^0 + \frac{2\pi i}{L}\v{n}\cdot\v{x}} \tilde{\psi}(p^0,\v{n}).
   \end{split}\label{eq:FTdef}
\end{equation}
Matching between QED$_L$ and NRQED$_L$ is performed for four-point correlation functions describing (off-shell) particles with energy $M$ and three-momentum $\v{p} \equiv \frac{2\pi}{L} \v{r}$ with $\v{r} \in \mathbb{Z}_3$,
\begin{equation}
  \begin{split}
     \mathcal{M}^{\text{NRQED}_L} &= G^{\text{NRQED}_L}(0,\v{r})^{-4} \left< \tilde{\psi}(0,\v{r})^2 \tilde{\psi}(0,\v{r})^{\dagger 2} \right> . \\
  \end{split}\label{eq:MNRLOdef}
\end{equation}
At tree level this correlation function is given by
\begin{equation}
  \begin{split}
    \mathcal{M}^{\text{NRQED}_L}_{\rm LO} &= -\frac{8\pi a}{M} .
  \end{split}\label{eq:MNRLO}
\end{equation}
The choice $\v{r}\neq 0$, which does not affect the tree-level correlation function, is made in order to regulate IR divergences in the one-loop correlation function.
Although one-photon-exchange contributions lead to an IR divergence (regulated for instance by considering non-zero momentum transfer) in the NRQED analog of Eq.~\eqref{eq:MNRLO}, one-photon-exchange contributions to the NRQED$_L$ amplitude in Eq.~\eqref{eq:MNRLOdef} vanish because of zero-mode subtraction.

Matching is performed by expanding $\mathcal{M}^{\text{NRQED}_L}$ and its QED$_L$ analog $\mathcal{M}^{\text{QED}_L}$ perturbatively in the small parameters $\alpha$, $1/(ML)$, and $a/L$ and defining higher-order terms in the NRQED$_L$ Lagrangian so that QED$_L$ and NRQED$_L$ agree order by order in this expansion.
Matching will be performed to leading order in $1/(ML)$ and third order in $\eta_L = \alpha M L / (4\pi)$ and $a/L$.
This corresponds to NNLO in the power counting of Eq.~\eqref{eq:pc1} and is equivalent to the non-relativistic limit with up to two-loop contact interactions and Coulomb photon exchange.
This matching is also sufficient for N$^3$LO in the power counting of Eq.~\eqref{eq:pc2}, which decreases the relative importance of Coulomb photon exchange and only requires one-loop Coulomb photon exchange effects.

This LO NRQED$_L$ amplitude in Eq.~\eqref{eq:MNRLO} can be straightforwardly matched to its QED$_L$ analog.
The scalar QED$_L$ Lagrangian is
\begin{equation}
   \begin{split}
      \mathcal{L}^{\text{QED}_L} &= -(D_\mu\varphi)^\dagger D^\mu\varphi - M^2 \varphi^\dagger \varphi  - 8\pi a M (\varphi^\dagger \varphi)^2  + \mathcal{L}_{\gamma}^\xi \\
     &= -(\partial_\mu\varphi)^\dagger \partial^\mu\varphi - M^2 \varphi^\dagger \varphi + i e A^\mu ( \varphi^\dagger \partial_\mu \varphi - (\partial_\mu \varphi)^\dagger \varphi) \\
     &\hspace{20pt} - e^2 A_\mu A^\mu \varphi^\dagger \varphi    -  8\pi a M (\varphi^\dagger \varphi)^2  + \mathcal{L}_{\gamma}^\xi,
   \end{split}\label{eq:QEDL}
\end{equation}
where $\mathcal{L}_\gamma^\xi$ is defined in Eq.~\eqref{eq:Lgamma}.
The fields appearing in the QED$_L$ and NRQED$_L$ Lagrangians are related by
\begin{equation}
  \begin{split}
    \psi(x) = \sqrt{2 M} e^{i M t} \varphi(x).
  \end{split}\label{eq:psidef}
\end{equation}
The scalar QED$_L$ particle propagator is
\begin{equation}
  \begin{split}
    G^{\text{QED}_L}(p^0, \v{n}) &= \int dx^0 \sum_{\v{x}} e^{i p^\mu x_\mu} \left< \varphi(x) \varphi^\dagger(0) \right> = \frac{i}{(p^0)^2 - \left( \frac{2\pi \v{n}}{L} \right)^2 - M^2 + i \epsilon}.
  \end{split}\label{eq:GQEDL}
\end{equation}
The QED$_L$ photon propagator is identical to the NRQED$_L$ propagator Eq.~\eqref{eq:GgammaL}.
Introducing Fourier transformed fields $\tilde{\varphi}$ as in Eq.~\eqref{eq:FTdef}, the FV two-particle amplitude in QED$_L$ normalized identically to the NRQED$_L$ amplitude is given by
\begin{equation}
  \begin{split}
     \mathcal{M}^{\text{QED}_L}_{\rm LO} &= \frac{1}{4M^2} G^{\text{QED}_L}(M,\v{r})^{-4} \left< \tilde{\varphi}(M,\v{r})^2 \tilde{\varphi}(M,\v{r})^{\dagger 2} \right> \\
    &= -\frac{8\pi a}{M} ,
  \end{split}\label{eq:MLO}
\end{equation}
in agreement with Eq.~\eqref{eq:MNRLO}. This demonstrates that the four-fermion contact interactions in Eq.~\eqref{eq:NRQEDLapp} and Eq.~\eqref{eq:QEDL} are normalized consistently at tree level.

Corrections to the LO amplitude arise from one-loop diagrams shown in Figs.~\ref{fig:coulomb}-\ref{fig:radiation2}.
The one-loop diagrams shown in Fig.~\ref{fig:coulomb} involve similar sums/integrals over loop momenta and differ only in the number of photon propagators $n_\gamma$ and contact interactions present.
The contribution to the amplitude from each diagram is denoted $\mathcal{M}^{\text{NRQED}_L}_{\rm NLO}(n_\gamma)$ where $n_\gamma \in \{ 0,\ 1,\ 2\}$ labels the numbers of photon propagators present,
\begin{equation}
   \begin{split}
      &\mathcal{M}^{\text{NRQED}_L}_{\rm NLO}(n_\gamma) =  \frac{i 16 \pi^2 a^2 }{M^2 L^3}\left( \frac{2 \alpha M}{a} \right)^{n_\gamma} \sum_{\v{n}}{}^{(\prime)} \int \frac{dk^0}{2\pi}  \left( \frac{1 }{ k^0 - \frac{2\pi^2(\v{n} - \v{r})^2}{ML^2} + i \epsilon} \right)  \\
      &\hspace{40pt} \times   \left( \frac{1}{ k^0 + \frac{2\pi^2(\v{n} + \v{r})^2}{ML^2} - i \epsilon  } \right) \left(\frac{1 - (1-\xi)\frac{(k^0)^2}{(k^0)^2 - \frac{4\pi^2\v{n}^2}{L^2} + i\epsilon}}{ (k^0)^2 - \frac{4\pi^2\v{n}^2}{L^2} + i\epsilon } \right)^{n_\gamma} \\
      &\hspace{20pt}= -\frac{16 \pi^2 a^2 }{M^2 L^3}\left( \frac{\alpha M}{a} \right)^{n_\gamma} \sum_{\v{n}}{}^{(\prime)}  \left[ \text{Res}\left(- \frac{2\pi^2(\v{n} - \v{r})^2}{ML^2} + i\epsilon \right) +  \text{Res}\left(-\frac{2\pi|\v{n}|}{L} + i\epsilon \right) \right],
   \end{split}\label{eq:MNRNLOdef}
\end{equation}
where  $\text{Res}(x)$ indicates the residue of the integrand in the first line at the pole $x$ and $\sum_{\v{n}}{}^{(\prime)}$ corresponds to $\sum_{\v{n}\in\mathbb{Z}^3}$ for $n_\gamma = 0$ and to $\sum_{\v{n}\in\mathbb{Z}^3\setminus \{\v{0}\}}$ for $n_\gamma \geq 1$.
After taking the $\epsilon \rightarrow 0$ limit, the residue at the particle pole $2\pi^2(\v{n}+\v{r})^2/(ML^2)$ is given by 
\begin{equation}
   \begin{split}
      \text{Res}\left(-\frac{2\pi^2(\v{n}-\v{r})^2}{ML^2}  \right) &= -\frac{ML^2}{4\pi^2(\v{n}^2 + \v{p}^2)} \left[  \frac{1}{ \frac{4\pi^4 (\v{n} - \v{r})^4}{M^2 L^4}  - \frac{4\pi^2\v{n}^2}{L^2} } - \frac{(1 - \xi)\frac{4\pi^4(\v{n}-\v{r})^4}{M^2L^4}}{\left( \frac{4\pi^4 (\v{n} - \v{r})^4}{M^2 L^4}  - \frac{4\pi^2 \v{n}^2}{L^2} \right)^2} \right]^{n_\gamma}  \\
      &= - \frac{ML^2}{4\pi^2} \left( - \frac{L^2}{4\pi^2} \right)^{n_\gamma} \frac{1}{(\v{n}^2 + \v{r}^2)\v{n}^{2n_\gamma}}\left[ 1 + \mathcal{O}\left( (ML)^{-2} \right) \right],
   \end{split}\label{eq:pRes}
\end{equation}
where the last line includes an expansion in powers of $(ML)^{-1}$.
This expansion is legitimate provided that the sum over $\v{n}$ converges, which holds for $n_\gamma \geq 1$.
For $n_\gamma = 0$ there is a linear UV divergence that can be removed by adding a UV counterterm,
\begin{equation}
   \begin{split}
      \sum_{\v{n} \in \mathbb{Z}^3 } \frac{1}{\v{n}^2 + \v{r}^2} \rightarrow \lim_{\Lambda \rightarrow \infty} \sum_{\v{n} \in \mathbb{Z}^3 }^{|\v{n}| < \Lambda} \frac{1}{\v{n}^2 + \v{r}^2}- 4\pi \Lambda,
   \end{split}\label{eq:UV}
\end{equation}
and after making the replacement of Eq.~\eqref{eq:UV} the $1/(ML)$ expansion can be performed in Eq.~\eqref{eq:pRes}.
The photon pole residue at $-2\pi|\v{n}|/L$ involves energy denominators of order $1/L$ and is suppressed by $\mathcal{O}\left( (ML)^{-1} \right)$ compared to the particle-pole contributions in Eq.~\eqref{eq:pRes} where particle propagator energy denominators are of order $M/L^2$.
For all diagrams in Figs.~\ref{fig:radiation}-\ref{fig:radiation2}, particle poles only appear in either the upper or lower half of the complex plane and only photon poles contribute.
After performing the energy integrals these diagrams can be straightforwardly verified to be suppressed by $\mathcal{O}\left( (ML)^{-1} \right)$ or $\mathcal{O}\left( (ML)^{-2} \right)$ compared to Eq.~\eqref{eq:MNRNLO}.
The full NLO amplitude in NRQED$_L$ is therefore a sum of the amplitudes $\mathcal{M}^{\text{NRQED}_L}_{\rm NLO}(n_\gamma)$ associated with the diagrams in Fig.~\ref{fig:coulomb} obtained by substituting Eq.~\eqref{eq:pRes} into Eq.~\eqref{eq:MNRNLOdef},
\begin{equation}
   \begin{split}
      \mathcal{M}^{\text{NRQED}_L}_{\rm NLO}(n_\gamma) &= \frac{4\pi a}{M} \left( - \frac{\alpha M L}{2\pi^3} \right)^{n_\gamma} \left( \frac{a}{\pi L} \right)^{1-n_\gamma} \sum_{\v{n}}{}^{(\prime)} \frac{1}{(\v{n}^2 + \v{r}^2)\v{n}^{2n_\gamma}}\left[ 1 + \mathcal{O}\left( (ML)^{-1} \right) \right].
   \end{split}\label{eq:MNRNLO}
\end{equation}
Contributions with different $n_\gamma$ differ parametrically and can be matched independently between NRQED$_L$ and QED$_L$.

The QED$_L$ amplitudes associated with the NLO diagrams in Fig.~\ref{fig:coulomb} are given by
\begin{equation}
   \begin{split}
      &\mathcal{M}^{\text{QED}_L}_{\rm NLO}(n_\gamma) = -\frac{i 64 \pi^2 a^2}{L^3} \left( \frac{-\alpha}{2aM} \right)^{n_\gamma}  \int \frac{dk^0}{2\pi} \sum_{\v{n}}{}^{(\prime)} \frac{1}{(k^0 + M)^2 - \frac{4\pi^2(\v{r}+\v{n})^2}{L^2} - M^2 + i\epsilon}  \\
      &\hspace{20pt} \times \frac{1}{ (k^0 - M)^2 - \frac{4\pi^2(\v{r}-\v{n})^2}{L^2} - M^2 + i\epsilon } \left( \frac{ N_\gamma^\xi(k^0, \v{n})  }{(k^0)^2 - \frac{4\pi^2\v{n}^2}{L^2} + i \epsilon }  \right)^{n_\gamma},
   \end{split}
\end{equation}
where
\begin{equation}
   \begin{split}
      N_\gamma^\xi(k^0,\v{n}) &= (k^0)^2 - 4M^2 + \left( \frac{4\pi^2}{L^2} \right) (4\v{r}^2 - \v{n})^2 \\
      &\hspace{20pt} - (1-\xi)\left[ \frac{-(k^0)^4 + 4M^2 (k^0)^2 + \left( \frac{4\pi^2}{L^2} \right) (4\v{n}\cdot \v{r} - \v{n}^4)}{(k^0)^2 - \left(\frac{4\pi^2}{L^2}\right) \v{n}^2 + i\epsilon } \right].
   \end{split}\label{eq:Ngamma}
\end{equation}
The $k^0$ integral can be performed with contour integration.
Closing the contour in the upper half plane, the result includes contributions from a particle pole $k^0 = M - \sqrt{M^2 + 4\pi^2(\v{n}+\v{r})^2/L^2} = -2\pi^2(\v{n}+\v{r})^2/(ML^2)\left[1 + \mathcal{O}\left((ML)^{-2}\right)\right]$, a photon pole at $k^0 = 2\pi|\v{n}|/L$, and an antiparticle pole at $k^0 = -2M\left[1 + \mathcal{O}\left((ML)^{-2}\right)\right]$,
\begin{equation}
   \begin{split}
      &\mathcal{M}^{\text{QED}_L}_{\rm NLO}(n_\gamma) = \frac{64 \pi^2 a^2}{L^3} \left( \frac{-\alpha}{4aM} \right)^{n_\gamma}   \sum_{\v{n}}{}^{(\prime)} \left[ \text{Res}\left(M - \sqrt{M^2 + \frac{4\pi^2(\v{n}+\v{r})^2}{L^2} - i\epsilon}\right) \right. \\
      &\hspace{20pt} \left. + \text{Res}\left(\frac{-2\pi|\v{n}|}{L} + i\epsilon\right) + \text{Res}\left(-M - \sqrt{M^2 + \frac{4\pi^2}{L^2}(\v{n}-\v{r})^2 + i\epsilon} \right) \right].
   \end{split}\label{eq:MNLOdef}
\end{equation}
Taking the $\epsilon \rightarrow 0$ limit and expanding to leading order in $(ML)^{-1}$, the residue at the particle pole is given by
\begin{equation}
   \begin{split}
      \text{Res}\left(M - \sqrt{M^2 + \frac{4\pi^2(\v{n}+\v{r})^2}{L^2} - i\epsilon} \right) &= \frac{L^2}{16\pi^2 M}\left( \frac{M^2L^2}{\pi^2} \right)^{n_\gamma} \frac{1}{(\v{n}^2+\v{r}^2)\v{n}^{2n_\gamma}}.
   \end{split}\label{eq:ResQEDL}
\end{equation}
As in the NRQED$_L$ case, the residue at the photon pole $-2\pi|\v{n}|/L$ is suppressed compared to the residue at the particle pole by $\mathcal{O}(ML)^{-1}$.

The residue at the antiparticle pole at $-2M(1+\mathcal{O}(ML)^{-1})$ does not appear in the corresponding NRQED$_L$ expression Eq.~\eqref{eq:MNRNLOdef} and is therefore associated with contributions that do not appear in loop diagrams in NRQED$_L$.
This residue is given by
\begin{equation}
   \begin{split}
      &\text{Res}\left(-M - \sqrt{M^2 + \frac{4\pi^2}{L^2} + i\epsilon} \right) = -\frac{1}{16M^3} \left( - \frac{\pi^2  }{M^2 L^2} \right)^{n_\gamma} \\
      &\hspace{150pt} \times \left[ 2\v{r}^2(-4 + \xi)  + \v{n}^2(-3+2\xi) - 4\v{r}\cdot \v{n}(-2+\xi)  \right]^{n_\gamma}.
   \end{split}\label{eq:ResQEDLanti}
\end{equation}
The term involving two contact interactions involves the UV divergent sum $\sum_{\v{n}}{}^{(\prime)} 1$, which for $n_\gamma = 0$ must be consistent with the infinite-volume result $\int d^Dk \; 1 = 0$ after subtracting UV counterterms,
\begin{equation}
   \begin{split}
      \sum_{\v{n}\in\mathbb{Z}^3} 1 \rightarrow 0.
   \end{split}\label{eq:sum1}
\end{equation}
This leads to a vanishing contribution from the $n_\gamma = 0$ diagram appearing in the absence of QED$_L$, which is consistent with the expectation that antiparticle-pole contributions from off-shell intermediate states that are absent non-relativistically do not lead to power-law FV effects in local field theories.
The terms with $n_\gamma \geq 1$ differ by zero mode subtraction.
Since the $\v{n} = 0$ contribution to $\sum_{\v{n}} 1$ is unity, it follows from Eq.~\eqref{eq:sum1} that
\begin{equation}
   \begin{split}
      \sum_{\v{n}\in\mathbb{Z}^3 \setminus \{\v{0}\} }  1 \rightarrow -1.
   \end{split}\label{eq:sum2}
\end{equation}
Sums with $\v{n}^{2k}$ with $k > 0$ similarly require UV counterterms and vanish after including them.
Zero-mode contributions to these sums vanish, and so
\begin{equation}
   \begin{split}
      \sum_{\v{n}\in\mathbb{Z}^3 \setminus \{\v{0}\}} \v{n}^{2k} \rightarrow 0,
   \end{split}\label{eq:sum2zero}
\end{equation}
for $k > 0$. After subtracting UV counterterms the antiparticle pole contribution becomes
\begin{equation}
   \begin{split}
      \sum_{\v{n} \in \mathbb{Z}^3 \setminus \{\v{0}\}} \; \text{Res}\left(-M - \sqrt{M^2 + \frac{4\pi^2}{L^2} + i\epsilon} \right) \rightarrow  \frac{1}{16M^3} \left( - \frac{\pi^2 2\v{r}^2(-4 + \xi) }{M^2 L^2} \right)^{n_\gamma} ,
   \end{split}\label{eq:ResQEDLanti2}
\end{equation}
for $n_\gamma \geq 1$.
This leads to a finite contribution to $\mathcal{M}^{\text{QED}_L}_{\rm NLO}$ that does not arise in the corresponding loop diagrams contributing to $\mathcal{M}^{\text{NRQED}_L}_{\rm NLO}$.
Local counterterms involving powers of $L^{-1}$ must be added to the NRQED$_L$ Lagrangian in order to reproduce this relativistic effect arising from zero-mode subtraction.
Inserting Eq.~\eqref{eq:ResQEDLanti2} into Eq.~\eqref{eq:MNLOdef} shows that the dominant contribution to $\mathcal{M}^{\text{QED}_L}_{\rm NLO}$ arising from the $n_\gamma = 1$ diagram antiparticle pole is
\begin{equation}
   \begin{split}
      -\frac{4\pi a}{M} \left(\frac{ \alpha \pi^3 (-4 + \xi) \v{r}^2}{4 M^5 L^5}\right) = -\frac{8\pi a}{M} \left(\frac{  \pi (-4 + \xi) }{16 }\right)  \frac{ \alpha}{(M L)^3}   \left(\frac{\v{p}^2}{M^2}\right).
   \end{split}\label{eq:nonlocalCT}
\end{equation}
This contribution vanishes for FV systems in the two-particle rest frame where $\v{p} = 0$.
For boosted systems with a non-zero center-of-mass velocity (assumed to be non-relativistic $\v{p}^2 \ll M^2$ in the $(ML)^{-1}$ expansion above), this contribution is proportional to the LO contact interaction times the velocity squared times a relativistic QED suppression factor of $\alpha / (ML)^{-3}$.
For boosted systems, the NRQED$_L$ contact interaction in Eq.~\eqref{eq:NRQEDLapp} must therefore be supplemented with a nonlocal counterterm suppressed by the same factor of $\alpha / (ML)^{-3}$.
This is analogous to the self-energy of a single scalar field, which is shown in Ref.~\cite{Davoudi:2018qpl} to require a nonlocal counterterm equal to the scalar mass times the velocity squared times a relativistic QED suppression factor of $\alpha / (ML)^{-3}$.
The LQCD+QED$_L$ calculations discussed in the main text are performed in the center-of-mass rest frame, and these nonlocal counterterms vanish.
Non-vanishing nonlocal counterterms appear for fermion masses in NRQED$_L$ even in the two-particle rest frame, but since effects proportional to $(ML)^{-1}$ are neglected throughout this work it is consistent to neglect all nonlocal counterterms suppressed by $\alpha / (ML)^{-3}$.

The NLO QED$_L$ amplitude is therefore given by inserting the particle-pole residue in Eq~\eqref{eq:ResQEDL} into Eq.~\eqref{eq:MNLOdef},
\begin{equation}
   \begin{split}
      \mathcal{M}^{\text{QED}_L}_{\rm NLO}(n_\gamma) &= \frac{4\pi a}{M} \left( -\frac{\alpha M L}{2\pi^3} \right)^{n_\gamma} \left( \frac{a}{\pi L} \right)^{1-n_\gamma} \sum_{\v{n}}{}^\prime \frac{1}{(\v{n}^2 + \v{r}^{2}) \v{n}^{2n_\gamma}}\left[  1 + \mathcal{O}\left( (ML)^{-1} \right) \right],
   \end{split}\label{eq:MNLO}
\end{equation}
which is identical to Eq.~\eqref{eq:MNRNLO}.
The $(ML)^{-1}$ suppression of diagrams in Figs.~\ref{fig:radiation} and \ref{fig:radiation2} arising from the absence of particle pole contributions is identical in QED$_L$ and NRQED$_L$.
Additional diagrams appearing in QED$_L$ but not NRQED$_L$ associated with the two-particle-two-photon vertex or particle-antiparticle pair creation can be similarly verified to be suppressed by powers of $(ML)^{-1}$.
The NRQED$_L$ Lagrangian in Eq.~\eqref{eq:NRQEDLapp} therefore reproduces QED$_L$ at NLO.

Matching at NNLO proceeds similarly.
The NNLO diagrams shown in Fig.~\ref{fig:coulomb} can all be expressed in terms of the amplitude 
\begin{equation}
   \begin{split}
      &\mathcal{M}^{\text{NRQED}_L}_{\rm NNLO}(n_1,n_2,n_3) \\
      &= \frac{64\pi^3 a^3}{M^3 L^6} \left( \frac{\alpha M}{a} \right)^{n_1+n_2+n_3}  \sum_{\v{n},\v{m}}{}^{(\prime)} \int \frac{dk^0}{2\pi} \frac{1}{\left(k^0 + \frac{2\pi^2(\v{r}-\v{n})^2}{ML^2} - i\epsilon\right) \left(k^0 - \frac{2\pi^2(\v{r}+\v{n})^2}{ML^2} + i\epsilon\right) }  \\
      &\hspace{10pt} \times \left(\frac{1 - (1-\xi)\frac{(k^0)^2}{(k^0)^2 - \frac{4\pi^2\v{n}^2}{L^2}}}{ (k^0)^2 - \frac{4\pi^2\v{n}^2}{L^2} + i\epsilon  }\right)^{n_1} \int \frac{dq^0}{2\pi} \frac{1}{  \left(q^0 + \frac{2\pi^2(\v{r}-\v{m})^2}{ML^2} - i\epsilon\right) \left(q^0 - \frac{2\pi^2(\v{r}+\v{m})^2}{ML^2} + i\epsilon\right)  } \\
      &\hspace{10pt} \times \left(\frac{1 - (1-\xi)\frac{(k^0 - q^0)^2}{(k^0 - q^0)^2 - \frac{4\pi^2(\v{n}-\v{m})^2}{L^2} }}{ (k^0 - q^0)^2 - \frac{4\pi^2(\v{n} - \v{m})^2}{L^2} + i\epsilon }  \right)^{n_2} \left( \frac{ 1 - (1-\xi)\frac{(q^0)^2}{(q^0)^2 - \frac{4\pi^2\v{m}^2}{L^2} }}{(q^0)^2 - \frac{4\pi^2\v{m}^2}{L^2} + i\epsilon  }\right)^{n_3},
   \end{split}\label{eq:MNRNNLOdef}
\end{equation}
where $n_i \in \{ 0,\ \}$ for $i\in\{1,2,3\}$ labels whether each interaction is a four-particle contact interaction or photon exchange and $\sum_{\v{n},\v{m}}{}^{(\prime)}$ excludes $\v{n}=0$ if $n_1 = 1$ or $n_2 =1$ and excludes $\v{m}=0$ if $n_2 = 1$ or $n_3 = 1$.
Both energy integrands include poles where a particle is on-shell as well as poles where a photon is on-shell, and, as above, photon-pole contributions are suppressed by $\mathcal{O}\left( (ML)^{-1} \right)$.
Evaluating the $q^0$ and $k^0$ integrals then gives
\begin{equation}
   \begin{split}
      \mathcal{M}^{\text{NRQED}_L}_{\rm NNLO}(n_1,n_2,n_3) &= - \frac{4\pi a}{M} \left( \frac{a}{\pi L} \right)^{2 - n_1 - n_2 - n_3} \left( - \frac{\alpha M L}{2\pi^3} \right)^{n_1+n_2+n_3} \\
      &\hspace{10pt} \times \sum_{\v{n},\v{m}}{}^{(\prime)} \frac{1}{(\v{n}^2 + \v{r}^2) (\v{m}^2+\v{r}^2) \v{n}^{2n_1} (\v{n} - \v{m})^{2n_2} \v{m}^{2n_3}} \left[ 1 + \mathcal{O}\left((ML)^{-1}\right) \right],
   \end{split}\label{eq:MNRNNLO}
\end{equation}
where UV counterterms should be included as in Eq.~\eqref{eq:UV}.
Similarly to the NLO case, the NNLO diagrams in Figs.~\ref{fig:radiation} and \ref{fig:radiation2} only include photon pole contributions and are suppressed by $\mathcal{O}\left( (ML)^{-1} \right)$ or $\mathcal{O}\left( (ML)^{-2} \right)$ compared to Eq.~\eqref{eq:MNRNNLO}.

The corresponding QED$_L$ NNLO amplitudes are given by
\begin{equation}
   \begin{split}
      &\mathcal{M}^{\text{QED}_L}_{\rm NNLO}(n_1,n_2,n_3) = \frac{1024 \pi^3 a^3 M}{L^6}  \left( -\frac{\alpha}{4 a M} \right)^{n_1+n_2+n_3}  \sum_{\v{n},\v{m}}{}^{(\prime)} \int \frac{dk^0}{2\pi} \left( \frac{ N_\gamma^\xi(k^0, \v{n})  }{(k^0)^2 - \frac{4\pi^2\v{n}^2}{L^2} + i \epsilon }  \right)^{n_1}   \\
      &\hspace{10pt} \times \frac{1}{ \left( (k^0 - M)^2 -  \frac{4\pi^2(\v{n}-\v{r})^2}{L^2} - M^2 + i\epsilon\right) \left((k^0 + M)^2 - \frac{4\pi^2(\v{n}+\v{r})^2}{L^2} - M^2 + i\epsilon\right) } \\
      &\hspace{10pt} \times \int \frac{dq^0}{2\pi} \left( \frac{ N_\gamma^\xi(k^0-q^0, \v{n}-\v{m}) }{(k^0-q^0)^2 - \frac{4\pi^2(\v{n}-\v{m})^2}{L^2} + i \epsilon }  \right)^{n_2}  \left( \frac{ N_\gamma^\xi(q^0, \v{m})  }{(q^0)^2 - \frac{4\pi^2\v{m}^2}{L^2} + i \epsilon }  \right)^{n_3}  \\
      &\hspace{10pt} \times \frac{1}{ \left( (q^0 - M)^2 - \frac{4\pi^2(\v{m}-\v{r})^2}{L^2} - M^2 - i\epsilon\right)   \left( (q^0 + M)^2 - \frac{4\pi^2(\v{m}+\v{r})^2}{L^2} - M^2 + i\epsilon\right)  }.
   \end{split}\label{eq:MNNLOdef}
\end{equation}
Contributions from photon and antiparticle poles are again suppressed by powers of $(ML)^{-1}$, and evaluating the $q^0$ and $k^0$ energy integrals gives 
\begin{equation}
   \begin{split}
      \mathcal{M}^{\text{QED}_L}_{\rm NNLO}(n_1,n_2,n_3) &=  - \frac{4\pi a }{M}  \left( \frac{a}{\pi L} \right)^{2 - n_1 - n_2 - n_3} \left( -\frac{\alpha ML}{2\pi^3} \right)^{n_1 + n_2 + n_3} \\
      &\hspace{10pt} \times \sum_{\v{n},\v{m}}{}^{(\prime)} \frac{1}{(\v{n}^2 + \v{p}^2) (\v{m}^2 + \v{p}^2)\v{n}^{2n_1}(\v{n}-\v{m})^{2n_2} \v{m}^{2n_3}}  \left[ 1 + \mathcal{O}\left( (ML)^{-1} \right) \right].
   \end{split}\label{eq:MNNLO}
\end{equation}
All other diagrams are again suppressed by powers of $(ML)^{-1}$, and agreement between Eq.~\eqref{eq:MNRNNLO} and Eq.~\eqref{eq:MNNLO} shows that nonlocal counterterms are suppressed by powers of $(ML)^{-1}$ and can be neglected to the accuracy considered here.

While the matching in this section has been explicitly performed for scalar QED$_L$, the $(ML)^{-1}$ suppression of photon and antiparticle poles only relies on the structure of QED$_L$ propagator denominators that are identical for scalars and fermions.
The $(ML)^{-3}$ suppression of antiparticle pole contributions leading to nonlocal two-body counterterms also arises from the denominator structure of the QED$_L$ propagators and is expected to be generic for bosons and fermions.
The numerator structure of the scalar QED$_L$ propagators above is relevant for $(ML)^{-1}$ suppressed relativistic effects, and in particular the scalar QED$_L$ antiparticle pole in Eq.~\eqref{eq:ResQEDLanti} includes vanishing numerator factors for a system at rest that lead to $\v{p}^2/M^2$ velocity suppression of nonlocal counterterms in scalar NRQED$_L$.
This cancellation might be absent for fermions, and nonlocal two-body counterterms might be relevant for QED$_L$ calculations of charged fermions in the center-of-mass rest frame.
This would parallel the situation for one-body nonlocal counterterms, which arise at $\mathcal{O}(\alpha/(ML)^3)$ for fermions with any center-of-mass velocity and for scalars with non-zero center-of-mass velocity.

\section{Fitting procedures}\label{app:fits}

\subsection{Energy level determination}

In general the spectral representation in Eq.~\eqref{eq:spectral} cannot be inverted to determine the full energy spectrum from finite samples of correlation functions over a finite range of source/sink separations.
Any fitting procedure to extract energies from correlation functions involves making several choices, in particular the range of $t$ to include in the fit, the number of excited states to include in a truncation of Eq.~\eqref{eq:spectral} to use as a fitting model, and how to estimate the covariance matrix from a finite statistical ensemble.
In order to assess the systematic uncertainties associated with these fitting choices and provide a reproducible procedure for extracting energy levels from correlation function results, we use an approach detailed here for making fitting choices based on well-defined statistical criteria and random sampling over the space of possible fitting choices.

The first step in this fitting procedure is choosing the maximum source/sink separation $t_{\rm max}$ included in the fit.
For (multi-)baryon correlation functions, the signal-to-noise (StN) problem implies that results with larger temporal separation $t$ make exponentially smaller contributions to $\chi^2$ when fitting energy levels from correlation functions.
For the nucleon, the scaling $\text{StN}(G) = G(t)/\sqrt{\text{Var}(G(t))} \sim e^{-(M_N - \frac{3}{2}m_\pi)t}$ predicted by Parisi~\cite{Parisi:1983ae} and Lepage~\cite{Lepage:1989hd} applies in the limit of a large statistical ensemble size $N \rightarrow \infty$ and shows that fit results are exponentially insensitive to the choice of $t_{\rm max}$.
Similar results apply for multi-nucleon systems with baryon number $A$ where $\text{StN}(G) = G(t)/\sqrt{\text{Var}(G(t))} \sim e^{-A(M_N - \frac{3}{2}m_\pi)t}$ for large $t$.
For fixed $N$ (and, since volume-averaging increases StN, for fixed $L$) it is important to choose a fixed $t_{\rm max}$ before the StN has degraded to the point where correlation function estimates are unreliable.\footnote{The complex phases of baryon correlation functions are circular random variables, and there is therefore a noise region at large $t$ where $\ln N$ is not much larger than the variance of the phase distribution and the sample mean is a systematically unreliable estimator of the average correlation function~\cite{Wagman:2016bam}.}
%complex phase fluctuations leads to a ``noise region'' where $\ln N$ is not much larger than the variance of the phase distribution and the sample mean is a systematically unreliable estimator of the average correlation function~\cite{Wagman:2016bam}.
The StN ratio decreases with $t$ at small and intermediate $t$ before saturating at an $\mathcal{O}(1)$ value in the noise region, and to avoid the noise region $t_{\rm max}$ should not exceed the smallest $t$ where $\text{StN}(G) = G(t) / \sqrt{\text{Var}G(t)}$ reaches a specified $\mathcal{O}(1)$ cutoff.
For positive-definite meson correlation functions this issue does not arise, and the maximum $t$ that can be reliably included in the fit is only limited by the accuracy by which finite-temperature effects are modeled.
In this work, finite-temperature excited-state effects are neglected, and finite-temperature effects on correlation functions with $t\sim T/2$, where $T$ is the length of the Euclidean time direction, are not reliably modeled.
%Finite-temperature effects are particular important for correlation functions describing large numbers of mesons because any subset of the multi-meson system can propagate around the boundary of the Euclidean time direction~\cite{Beane:2007es, Detmold:2008fn, Detmold:2010au, Detmold:2011kw, Detmold:2012wc}.
In our fitting procedure, $t_{\rm max}$ is therefore chosen to be the minimum $t$ satisfying either: the correlation function noise-to-signal ratio at $(t+a)$ is smaller than a specified tolerance (final results use $tol_{\rm noise} = 1.0$), the correlation function sample mean at $t + a$ is negative,\footnote{If the average correlation function is not expected to be positive definite, then this condition should not be enforced. In principle, the source and sink interpolating operators used in this work differ from one another, and it is possible for the sign of an average correlation function to fluctuate at small $t$ where excited-state contributions with opposite sign to the ground-state contribution can be significant. In practice, small $t$ fluctuations of the sign of the sample mean correlation function that could be attributed to excited-state effects are not observed in this work, and a negative sample mean correlation function is taken as an indicator of large statistical noise.} or $t+a$ is larger than a finite-temperature cutoff (final results use $tol_{\rm temp} = 3 T /8$).
The values of $tol_{\rm noise}$ and $tol_{\rm temp}$ are free parameters in our fitting procedure that must be varied to assess the sensitivity of fit results to these choices; for concreteness the parameter choices are presented here that lead to the final results quoted in the main text. 
Results are found to be relatively insensitive to the parameter choices controlling $t_{\rm max}$, and for example varying the noise tolerance in the range $tol_{\rm noise} \in [0.1,\ 1]$ leads to results with consistent central values at the $1\sigma$ level and few-percent variation of the corresponding uncertainties.

%Results are found to be relatively insensitive to the parameters controlling $t_{\rm max}$ with for examples variations in the noise cutoff between 0.05 - 0.5 found to give statistical consistent results for one- and two-baryon ground-state energies.

The next step in our fitting procedure is to choose $t_{\rm min}$, the minimum $t$ included in the fit.
The choice of $t_{\rm min}$ significantly impacts how well excited-state effects at small $t$ can be resolved and how many excited states should be included in fits.
Furthermore, the uncertainties of energy-level determinations are exponentially sensitive to the choice of $t_{\rm min}$ for baryons because of the StN problem.
Fit results are therefore more sensitive to the choice of $t_{\rm min}$ than to the choice of $t_{\rm max}$.
Rather than choose a single $t_{\rm min}$, it is preferable to sample from many possible choices of $t_{\rm min}$ and quantify the sensitivity to this choice as the systematic error.
The minimum permissible $t_{\rm min}$ is fixed by the temporal nonlocality in the lattice action, and for the improved action used in this work  the transfer matrix involves fields on two adjacent timeslices~\cite{Luscher:1984is} and $t_{\rm min} \geq 2$ is required.
%In order to insert a transfer matrix for Wilson fermions between the source and sink and derive the spectral representation in Eq.~\eqref{eq:spectral}, $t_{\rm min} \geq 2$ is required.
The largest allowed $t_{\rm min}$ is limited by $t_{\rm min} \leq t_{\rm max} - t_{\rm plateau}$, where $t_{\rm plateau}$ is a free parameter that is not found to significantly affect final results when varied over the range $2 \lesssim t_{\rm plateau} \lesssim 8$ (final results use $t_{\rm plateau} = 4$).
For each type of interpolating operator included in a combined fit, $t_{\rm min}$ is sampled randomly within this range until either all possible values of $t_{\rm min}$ have been chosen or a maximum of $N_{\rm fits} = 200$ fits have been performed.

With $t_{\rm min}$ and $t_{\rm max}$ specified, the covariance matrix $\mathcal{C}^{ij}_{tt^\prime}$ must be estimated for $t_{\rm min} \leq t,t^\prime \leq t_{\rm max}$ and interpolating operators $i,j \in \{1,\dots N_{\rm op}\}$.
Fits involving a large number $N_{\rm pts}$ of time separations and interpolating operator choices may not satisfy the condition $N \gg N_{\rm pts}^2$ needed to ensure that the $N$ terms contributing to the  $N_{\rm pts} \times N_{\rm pts}$ sample covariance can accurately estimate the true underlying covariance matrix, where $N_{\rm pts}$ is the total number of source/sink separations from all interpolating operators included in the fit.
Shrinkage techniques have been developed to provide more accurate estimates of the underlying covariance matrix than the sample covariance matrix when $N \gg N_{\rm pts}^2$ is not satisfied~\cite{stein1956,LEDOIT2004365}.
Shrinkage estimators $\mathcal{S}_{tt^\prime}^{ij}(\lambda)$ of the covariance matrix are constructed as mixtures of a well-conditioned target matrix $T_{tt^\prime}^{ij}$ and the covariance matrix $\mathcal{C}_{tt^\prime}^{ij}$ estimated using standard bootstrap techniques from $N_{\rm boot} = 200$ samples of $N$ correlation functions $G_{i,a}^b(t)$ with $a \in \{1,\ldots, N\}$ and $b\in \{1,\ldots,N_{\rm boot} \}$ drawn from the original correlation function ensemble with replacement,
\begin{equation}
   \begin{split}
      \mathcal{S}_{tt^\prime}^{ij}(\lambda) = \mathcal{C}_{tt^\prime}^{ij}(1-\lambda) + T_{tt^\prime}^{ij}\lambda,
   \end{split}\label{eq:shrinkage}
\end{equation}
where $0 \leq \lambda \leq 1$ is a shrinkage parameter.
A common choice of well-conditioned target matrix for many problems in statistics is the identity matrix; however, this does not accurately describe the underlying covariance matrix for correlation functions whose diagonal entries decrease exponentially with $t$.
Following applications of shrinkage to lattice QCD in Ref.~\cite{Rinaldi:2019thf}, we take $T_{tt^\prime}^{ij} = \text{diag}(\mathcal{C}_{tt^\prime}^{ij})$.
In this case, shrinkage corresponds to an interpolation between a fully correlated fit with $\lambda = 0$ and an uncorrelated fit with $\lambda = 1$.
Shrinkage gives an unbiased estimator of the underlying covariance matrix in the infinite-statistics limit provided $\lambda$ vanishes sufficiently quickly in this limit.
It can be shown~\cite{LEDOIT2004365} that for finite $N$ an optimal $\lambda^* \neq 0$ satisfying this restriction can be chosen in order to minimize the average mean-squared difference between $\mathcal{S}_{tt^\prime}^{ij}(\lambda)$ and the underlying covariance matrix, and that a sample estimate for $\lambda^*$ is given by
\begin{equation}
   \begin{split}
      \lambda^* &= \text{Max}\left(0, \text{Min}\left(1, \frac{\sum_{a=1}^N \sum_{t,t^\prime,i,j} \left[ \widetilde{G}_a^i(t) \widetilde{G}_a^j(t^\prime) - \widetilde{\overline{S}}^{ij}(t,t^\prime) \right]^2 }{N^2 \sum_{t,t^\prime,i,j} \left[ \widetilde{\overline{S}}^{ij}(t,t^\prime) - \delta^{ij} \delta_{t,t^\prime} \right]^2} \right) \right),
   \end{split}\label{eq:optshrinkage}
\end{equation}
where
\begin{equation}
   \begin{split}
      \widetilde{G}_a^i(t) &= \frac{G_a^i(t) - \overline{G}^i(t)}{\sqrt{\overline{S}^{ii}(t,t)}}, \\
      \widetilde{\overline{S}}^{ij}(t,t^\prime) &= \frac{\overline{S}^{ij}(t,t^\prime)}{\sqrt{\overline{S}^{ii}(t,t)\overline{S}^{jj}(t^\prime,t^\prime)}},
   \end{split}\label{eq:parts}
\end{equation}
are defined in terms of the sample mean correlation function and sample covariance as in Ref.~\cite{Rinaldi:2019thf}
\begin{equation}
   \begin{split}
      \overline{G}^i(t) &= \frac{1}{N} \sum_{a=1}^N G_a^i(t), \\
      \overline{S}^{ij}(t,t^\prime) &=  \frac{1}{N-1} \sum_{a=1}^N \left[ G_a^i(t) G_a^j(t^\prime) - \overline{G}^i(t) \overline{G}^j(t^\prime) \right],
   \end{split}
\end{equation}
such that shrinkage of $\widetilde{\overline{S}}^{ij}(t,t^\prime)$ with $T_{tt^\prime}^{ij} = \text{diag}(\widetilde{\overline{\mathcal{S}}}_{tt^\prime}^{ij})$ corresponds to shrinkage of  $\widetilde{\overline{S}}^{ij}(t,t^\prime)$ with the identity matrix as a target, and the results of Ref.~\cite{LEDOIT2004365} assuming an identity matrix target can be applied.
The covariance matrix estimate with optimal shrinkage is then given by $\mathcal{S}_{tt^\prime}^{ij}(\lambda^*)$.
Fits to truncations of Eq.~\eqref{eq:spectral} including $e$ excited states can then be performed by minimizing the corresponding $\chi^2$ function defined by
\begin{equation}
   \begin{split}
      \chi^2 = \sum_{t,t^\prime = t_{\rm min}}^{t_{\rm max}} \sum_{i,j = 1}^{N_{\rm op}} \left( \overline{G}_i^B(t) - f(t,\mathbf{E},\mathbf{Z}) \right) \left[ \mathcal{S}(\lambda^*)^{-1} \right]_{tt^\prime}^{ij}  \left( \overline{G}_j^B(t^\prime) - f(t^\prime,\mathbf{E},\mathbf{Z}) \right),
   \end{split}\label{eq:chisquared}
\end{equation}
where $\overline{G}_i^B = 2\overline{G}_i - \frac{1}{N_{\rm boot}}\sum_{b=1}^{N_{\rm boot}} \overline{G}_i^b$ includes a $1/N$ bias correction estimated using bootstrap techniques and $\mathbf{E}$ and $\mathbf{Z}$ denote the energies and overlap factors appearing in Eq.~\eqref{eq:spectral}, including excited-state and thermal effects.
Since the overlap factors enter $f_N$ linearly, the values of $\v{Z}$ minimizing $\chi^2$ for fixed $\v{E}$ can be determined by solving a system of linear equations analogous to variable projection techniques~\cite{varpro0,varpro1}.
$\chi^2$-minimization can therefore be efficiently performed by using a nonlinear optimization method to determine $\v{E}$ with the optimal $\v{Z}$ determined by solving a system of linear equations at each step of nonlinear optimization for $\v{E}$.
In order to ensure positivity of the spectrum and remove fitting degeneracies, the parameters used for nonlinear optimization are $\ln E_0$ and $\ln(E_k - E_{k-1})$ for $1\leq k \leq e$.

For each randomly sampled choice of $t_{\rm min}$, the next step in the fitting procedure is to determine the number of excited states to be included in the sum of exponentials used as a fit function.
This is done by first performing a fit including zero excited states and then adding successively more excited states until the addition of the next excited state does not improve the goodness of fit according to an information criterion.
This work employs the Akaike Information Criterion~\cite{1100705} (AIC) with a cutoff chosen to penalize overfitting in which a fit with $e$ excited states is only preferred over a fit with $e-1$ excited states if $\text{AIC}(e) - \text{AIC}(e-1) < -\mathcal{A} N_{\rm dof}(e)$ where $N_{\rm dof}(e) = N_{\rm pts} - N_{\rm params}(e)$ is the number of degrees of freedom of the fit, $N_{\rm params}(e)$ is the number of fit parameters for a fit with $e$ excited states, and $\text{AIC}(e) = 2N_{\rm params}(e) + \chi^2(e) + k$ with $\chi^2(e)$ the (unreduced) $\chi^2$ of the fit defined in Eq.~\eqref{eq:chisquared} and $k$ is an irrelevant $e$-independent constant.
This choice corresponds to a preference for an $e$ state fit only if it improves the $\chi^2/N_{\rm dof}$ by an $\mathcal{O}(1)$ value $\mathcal{A}$ compared to the $(e-1)$ state fit.
For baryon correlation functions a value of $\mathcal{A} = -0.5$ is used, while for multi-meson correlation functions a value of $\mathcal{A} = -0.1$ is used.\footnote{Optimal shrinkage values of $\lambda^* \gtrsim 0.1$ appear for $n$-meson correlation functions with $n \gtrsim 6$ and $\chi^2$ values are correspondingly lower than would be expected for fully correlated $\chi^2$ minimization, leading to smaller absolute changes in $\text{AIC}$ for multi-meson correlation functions than for multi-baryon correlation function. Increasing $-\mathcal{A}$ from $0.1$ over the range $-\mathcal{A} \in [0.1, 2]$ leads to consistent results with larger uncertainties because precise and accurate two-state fits with small $t_{\rm min}$ are rejected in favor of one-state fits more frequently.}

Bootstrap resampling techniques are then used to estimate the uncertainty on the ground-state energy extracted from the fit with the preferred number of excited states in each fit region~\cite{Young,davison_hinkley_1997}.
The same $\chi^2$-minimization procedure and estimated covariance matrix $\mathcal{S}(\lambda^*)$ are used to determine the spectrum for each of $N_{\rm boot}$ bootstrap resampled ensembles.
Results are found to be insensitive to the choice of $N_{\rm boot}$, and final results use  $N_{\rm boot} = 200$.
The $67\%$ confidence interval for the ground-state energy is then obtained from the quantiles of the distribution of differences between the $b$'th bootstrap sample result $E^{b,f}_0$ and the fit result $E_0^f$ obtained for fit range $f$,
\begin{equation}
   \begin{split}
      \delta E_0^f = \frac{1}{2} \left[ Q_{5/6}\left( E^{b,f}_0 - E_0^f \right)  - Q_{1/6}\left( E^{b,f}_0 - E_0^f \right) \right],
   \end{split}\label{eq:quantile}
\end{equation}
where $Q_{p}(x^f)$ is the $p$-th quantile of the set of fit results with elements $x^f$.
Using this definition $\delta E_0^f$ minimizes the impact of outlier bootstrap samples compared to a definition based on the standard deviation of $E^{b,f}_0 - E_0^f$~\cite{davison_hinkley_1997}.
An analogous procedure is used to estimate uncertainties for excited-state energies and overlap factors.

\begin{figure}[!t]
\resizebox{0.9\linewidth}{!}{\begin{tikzpicture}[scale=1.1,x=0.75pt,y=0.75pt,yscale=-1,xscale=1,>=latex, every picture/.style={line width=0.75pt}]

%uncomment if require: \path (0,877); %set diagram left start at 0, and has height of 877

%Shape: Parallelogram [id:dp8368751093472157] 
\draw   (239.9,1) -- (482,1) -- (462.1,27) -- (220,27) -- cycle ;
%Straight Lines [id:da8747561356567314] 
\draw[->]    (351,27) -- (351,50) ;
%Shape: Rectangle [id:dp19808797240136167] 
\draw   (265,108) -- (437,108) -- (437,136) -- (265,136) -- cycle ;
%Shape: Rectangle [id:dp5524066836033706] 
\draw   (129,49) -- (573,49) -- (573,83) -- (129,83) -- cycle ;
%Shape: Rectangle [id:dp6996835610268235] 
\draw   (143,161) -- (559,161) -- (559,189) -- (143,189) -- cycle ;
%Straight Lines [id:da5268394960164764] 
\draw[->]    (351,136) -- (351,161) ;
%Straight Lines [id:da697319523139351] 
\draw[->]    (351,189) -- (351,214) ;
%Shape: Rectangle [id:dp7645820520560072] 
\draw   (219,214) -- (483,214) -- (483,281) -- (219,281) -- cycle ;
%Straight Lines [id:da1694363185367309] 
\draw[->]    (351,281) -- (351,306) ;
%Shape: Rectangle [id:dp23254041538723424] 
\draw   (127,306) -- (575,306) -- (575,342) -- (127,342) -- cycle ;
%Straight Lines [id:da02848564869794845] 
\draw[->]    (156.5,342) -- (156.5,367) ;
%Shape: Diamond [id:dp12022271891251157] 
\draw   (156.5,367) -- (219.5,394) -- (156.5,421) -- (93.5,394) -- cycle ;
%Shape: Rectangle [id:dp08610647387180692] 
\draw   (119,447) -- (196,447) -- (196,473) -- (119,473) -- cycle ;
%Shape: Diamond [id:dp116357882106921] 
\draw   (156.44,498) -- (217.88,525.88) -- (156.44,553.76) -- (95,525.88) -- cycle ;
%Shape: Rectangle [id:dp7141469581111128] 
\draw   (247.26,363) -- (419,363) -- (419,439) -- (247.26,439) -- cycle ;
%Straight Lines [id:da6428374828754841] 
\draw[->]    (325.13,439) -- (325.13,464) ;
%Shape: Diamond [id:dp6455572173521081] 
\draw   (325.13,464) -- (401,496.5) -- (325.13,529) -- (249.26,496.5) -- cycle ;
%Shape: Rectangle [id:dp02577729971793985] 
\draw   (459,363) -- (637.76,363) -- (637.76,482) -- (459,482) -- cycle ;
%Straight Lines [id:da48649031342814764] 
\draw[->]    (548.13,482) -- (548.13,507) ;
%Shape: Rectangle [id:dp6428577355853148] 
\draw   (441.26,507) -- (655,507) -- (655,549) -- (441.26,549) -- cycle ;
%Straight Lines [id:da12439017513455897] 
\draw[->]    (548,549) -- (548,574) ;
%Shape: Rectangle [id:dp18384005726639296] 
\draw   (462,574) -- (634,574) -- (634,644.5) -- (462,644.5) -- cycle ;
%Shape: Diamond [id:dp5052373503126517] 
\draw   (547.75,669.75) -- (623.5,702.25) -- (547.75,734.75) -- (472,702.25) -- cycle ;
%Straight Lines [id:da15170798900044014] 
\draw[->]    (472,702.25) -- (449.5,702.25) ;
%Straight Lines [id:da35217374661939116] 
\draw[->]    (548,644.5) -- (548,669.5) ;
%Shape: Rectangle [id:dp0409357990418604] 
\draw  [color={rgb, 255:red, 208; green, 2; blue, 27 }  ,draw opacity=1 ][line width=2.25]  (508.5,764.79) .. controls (508.5,760.92) and (511.63,757.79) .. (515.5,757.79) -- (580.5,757.79) .. controls (584.37,757.79) and (587.5,760.92) .. (587.5,764.79) -- (587.5,774.75) .. controls (587.5,778.62) and (584.37,781.75) .. (580.5,781.75) -- (515.5,781.75) .. controls (511.63,781.75) and (508.5,778.62) .. (508.5,774.75) -- cycle ;

%Straight Lines [id:da33087771700236823] 
\draw[->]    (548,734.75) -- (548,757.79) ;

%Shape: Diamond [id:dp3997852885707376] 
\draw   (361.75,665.13) -- (449.5,702.25) -- (361.75,739.38) -- (274,702.25) -- cycle ;
%Shape: Rectangle [id:dp3405600798070888] 
\draw  [color={rgb, 255:red, 126; green, 211; blue, 33 }  ,draw opacity=1 ][line width=2.25]  (172.5,695.25) .. controls (172.5,691.94) and (175.19,689.25) .. (178.5,689.25) -- (245.5,689.25) .. controls (248.81,689.25) and (251.5,691.94) .. (251.5,695.25) -- (251.5,709.25) .. controls (251.5,712.56) and (248.81,715.25) .. (245.5,715.25) -- (178.5,715.25) .. controls (175.19,715.25) and (172.5,712.56) .. (172.5,709.25) -- cycle ;

%Shape: Rectangle [id:dp6204518939737291] 
\draw  [color={rgb, 255:red, 208; green, 2; blue, 27 }  ,draw opacity=1 ][line width=2.25]  (322.25,769.41) .. controls (322.25,765.55) and (325.38,762.41) .. (329.25,762.41) -- (394.25,762.41) .. controls (398.12,762.41) and (401.25,765.55) .. (401.25,769.41) -- (401.25,779.38) .. controls (401.25,783.24) and (398.12,786.38) .. (394.25,786.38) -- (329.25,786.38) .. controls (325.38,786.38) and (322.25,783.24) .. (322.25,779.38) -- cycle ;

%Straight Lines [id:da6481343190179693] 
\draw[->]    (361.75,739.38) -- (361.75,762.41) ;

%Straight Lines [id:da13265143928878254] 
\draw[->]    (274,702.25) -- (251.5,702.25) ;
%Shape: Rectangle [id:dp585176972952749] 
\draw   (110.5,242) -- (187.5,242) -- (187.5,268) -- (110.5,268) -- cycle ;
%Straight Lines [id:da5374498680879103] 
\draw[<-]    (219,255) -- (187.5,255) ;
%Straight Lines [id:da2491744560890342] 
\draw[<-]    (110.5,255) -- (79,255) -- (79,394) -- (93.5,394) ;
%Straight Lines [id:da3955298941937876] 
\draw[->]    (351,83) -- (351,108) ;
%Straight Lines [id:da44233007097453814] 
\draw[->]    (156.5,421) -- (156.5,446) ;
%Straight Lines [id:da4371238891676569] 
\draw[->]    (156.5,473) -- (156.5,498) ;
%Shape: Rectangle [id:dp12796701504749242] 
\draw  [color={rgb, 255:red, 208; green, 2; blue, 27 }  ,draw opacity=1 ][line width=2.25]  (117,583.04) .. controls (117,579.17) and (120.13,576.04) .. (124,576.04) -- (189,576.04) .. controls (192.87,576.04) and (196,579.17) .. (196,583.04) -- (196,593) .. controls (196,596.87) and (192.87,600) .. (189,600) -- (124,600) .. controls (120.13,600) and (117,596.87) .. (117,593) -- cycle ;

%Straight Lines [id:da8241575867441462] 
\draw[->]    (156.5,553) -- (156.5,576.04) ;

%Shape: Rectangle [id:dp1633097432439624] 
\draw  [color={rgb, 255:red, 208; green, 2; blue, 27 }  ,draw opacity=1 ][line width=2.25]  (285.63,559.04) .. controls (285.63,555.17) and (288.76,552.04) .. (292.63,552.04) -- (357.63,552.04) .. controls (361.5,552.04) and (364.63,555.17) .. (364.63,559.04) -- (364.63,569) .. controls (364.63,572.87) and (361.5,576) .. (357.63,576) -- (292.63,576) .. controls (288.76,576) and (285.63,572.87) .. (285.63,569) -- cycle ;

%Straight Lines [id:da4599551432765635] 
\draw[->]    (325.13,529) -- (325.13,552.04) ;

%Straight Lines [id:da6555029915372808] 
\draw[<-]    (247.26,401) -- (232.75,401) -- (232.75,525.88) -- (217.88,525.88) ;
%Straight Lines [id:da038635364444898124] 
\draw[<-]    (458.13,422.5) -- (437.25,422.5) -- (437.25,496.5) -- (401,496.5) ;

% Text Node
\draw (351,14) node  [font=\small]  {$\text{Correlation functions} \ G( t)$};
% Text Node
\draw (351,122) node  [font=\small]  {$t_{\text{min}} \in [ 2a,t_{\text{max}} -\textcolor[rgb]{0.29,0.41,0.89}{t_{\text{plateau}}}]$};
% Text Node
\draw (351,67) node  [font=\footnotesize]  {$t_{\text{max}} = \text{min}\{\ t\ |\ \left[ 1/\text{StN}( \overline{G}( t +a))  >\textcolor[rgb]{0.29,0.41,0.89}{tol_{\text{noise}}} \right] \lor \left[ \overline{G}( t +a) < 0 \right] \lor \left[ t + a  >\textcolor[rgb]{0.29,0.41,0.89}{tol_{\text{temp}}} \right] \}$ };
% Text Node
\draw (351,225) node  [font=\small] [align=left] {Excited-state model selection:};
% Text Node
\draw (351,255) node    {$f( t,\mathbf{E} ,\mathbf{Z}) =\sum\limits ^{e}_{n=0} Z_{n} e^{-E_{n} t} ,\ e=0$};
% Text Node
\draw (351,324) node  [font=\small]  {$\chi ^{2} \ \text{minimization with Nelder-Mead+VarPro using} \ \mathcal{S}\left( \lambda ^{*}\right)\rightarrow \{\mathbf{E}^{f} ,\mathbf{Z}^{f}\}$};
% Text Node
\draw (154,394) node  [font=\footnotesize]  {$\Delta AIC< -\mathcal{\textcolor[rgb]{0.29,0.41,0.89}{A}} N_{\text{dof}}$};
% Text Node
\draw (325.13,496.5) node  [font=\footnotesize]  {$|\mathbf{E}^{f'} -\mathbf{E}^{f} | >\textcolor[rgb]{0.29,0.41,0.89}{tol_{\text{sol}}}$};
% Text Node
\draw (81,404) node  [font=\small] [align=left] {yes};
% Text Node
\draw (149,255) node  [font=\footnotesize]  {$e\leftarrow e+1$};
% Text Node
\draw (157.5,460) node  [font=\footnotesize]  {$e\leftarrow e-1$};
% Text Node
\draw (156.5,588.02) node  [font=\small] [align=left] {Reject fit};
% Text Node
\draw (159.5,523.5) node  [font=\small]  {$\frac{\chi ^{2}}{N_{\text{dof}}}  >\textcolor[rgb]{0.29,0.41,0.89}{tol_{\chi ^{2}}}$};
% Text Node
\draw (212,702.25) node  [font=\small] [align=left] {Accept fit};
% Text Node
\draw (169.5,428) node  [font=\small] [align=left] {no};
% Text Node
\draw (333.13,401) node  [font=\small] [align=center] {$\displaystyle \chi ^{2}$ minimization with\\CG+VarPro using $\displaystyle \mathcal{S}\left( \lambda ^{*}\right)$\\$\displaystyle \rightarrow \{\mathbf{E}^{f'} ,\mathbf{Z}^{f'}\}$};
% Text Node
\draw (548.38,378) node  [font=\small] [align=left] {Confidence intervals:};
% Text Node
\draw (548.38,429) node  [font=\small] [align=center] {$\displaystyle \chi ^{2}$ minimization with\\NM+VarPro using $\displaystyle \mathcal{S}\left( \lambda ^{*}\right)$\\over bootstrap ensemble\\$\displaystyle \rightarrow \{\mathbf{E}^{b,f} ,\mathbf{Z}^{b,f}\}$};
% Text Node
\draw (325.13,564.02) node  [font=\small] [align=left] {Reject fit};
% Text Node
\draw (225,535) node  [font=\small] [align=left] {no};
% Text Node
\draw (175,557) node  [font=\small] [align=left] {yes};
% Text Node
\draw (345,533) node  [font=\small] [align=left] {yes};
% Text Node
\draw (413,505) node  [font=\small] [align=left] {no};
% Text Node
\draw (548,769.77) node  [font=\small] [align=left] {Reject fit};
% Text Node
\draw (361.75,774.39) node  [font=\small] [align=left] {Reject fit};
% Text Node
\draw (548.13,528) node  [font=\small]  {$\delta \mathbf{E}^f =\frac{Q_{5/6}\left( \mathbf{E}^{b,f} - \v{E}^{f}\right) -Q_{1/6}\left( \v{E}^{b,f} - \v{E}^{f}\right)}{2}$};
% Text Node
\draw (547.75,702.25) node  [font=\footnotesize]  {$|\mathbf{E}^{f''} -\mathbf{E}^{f} | >\textcolor[rgb]{0.29,0.41,0.89}{tol_{\text{corr}}}$};
% Text Node
\draw (548,609.25) node  [font=\small] [align=center] {$\displaystyle \chi ^{2}$ minimization with\\NM+VarPro using $\displaystyle \mathcal{S}( 1)$\\$\displaystyle \rightarrow \{\mathbf{E}^{f''} ,\mathbf{Z}^{f''}\}$};
% Text Node
\draw (566,739) node  [font=\small] [align=left] {yes};
% Text Node
\draw (467,692) node  [font=\small] [align=left] {no};
% Text Node
\draw (360.75,702.25) node  [font=\footnotesize]  {$Q_{1/2}\left(\mathbf{E}^{b,f} -\mathbf{E}^{f}\right)  >\textcolor[rgb]{0.29,0.41,0.89}{tol_{\text{med}}}$};
% Text Node
\draw (381,745) node  [font=\small] [align=left] {yes};
% Text Node
\draw (267.5,693) node  [font=\small] [align=left] {no};
% Text Node
\draw (351,175) node  [font=\small] [align=left] {Covariance matrix $\displaystyle \mathcal{S}( \lambda )$ with optimal shrinkage parameter $\displaystyle \lambda ^{*}$};

\end{tikzpicture}}
\caption{Flowchart representing the steps of the fitting procedure for one specific fitting range. Rectangular shapes represent process steps, while diamond shapes represent decision steps. Input parameters to the fitting procedure are shown in blue. As described in the text, the steps illustrated here are repeated $N_{\rm fits}$ times with different random choices of $t_{\rm min}$, and final results are obtained from weighted averages of fit results for the $t_{\rm min}$ choices leading to the ``Accept fit'' rectangle.    \label{fig:flowchart}}
\end{figure}
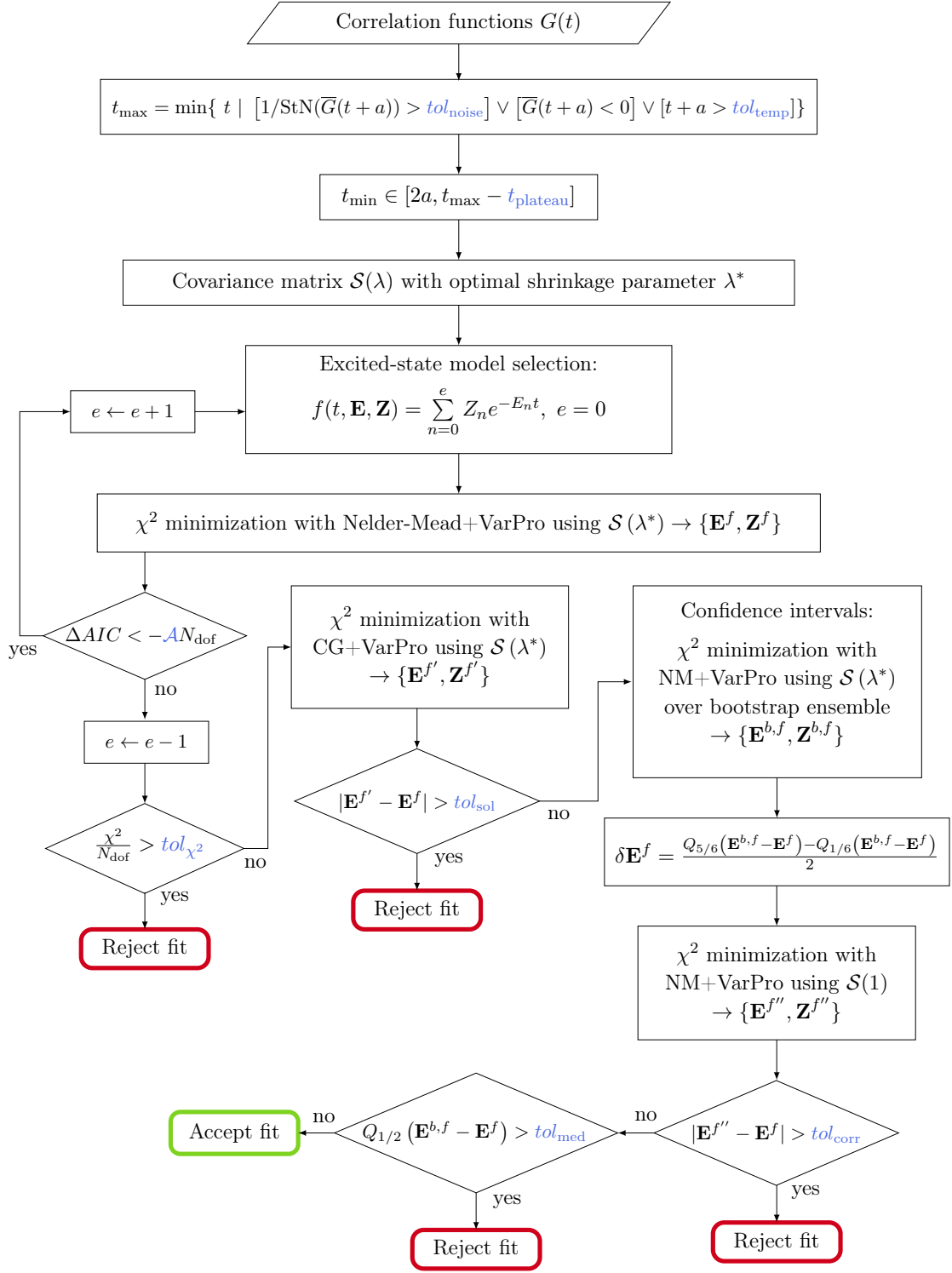

Several additional checks are used to ensure the robustness of $\chi^2$-minimization results: two different optimization algorithms, Nelder-Mead (NM) and conjugate gradient (CG), are used\footnote{Newton's method is used in place of Nelder-Mead if $N_{\rm states} = 1$, since Nelder-Mead does not work for a single fit parameter.} and are verified to give energies that differ by less than a specified tolerance\footnote{Fits resulting in ground-state energies less than $tol_{\rm sol}$, which appeared only for the $11\overline{K}^0$ system on the $L/a=32$ lattice volume, were also rejected.}  (final results use $tol_{\rm sol} = 10^{-5}$), median results for ground- and excited-state energies from the bootstrap samples are verified to agree with fit results from the average correlation functions for each energy level within a specified tolerance (final results use $tol_{\rm med} = 2\sigma$), uncorrelated fit results obtained by repeating the $\chi^2$-minimization procedure with $\mathcal{S}(\lambda=1)$ are verified to give consistent results for each energy level within a specified tolerance (final results use $tol_{\rm corr} = 5\sigma$), the $\chi^2/N_{\rm dof}$ is verified to be less than a specified tolerance (final results use $tol_{\chi^2}=2$).
This defines a reproducible and automatable procedure for fitting correlation functions, including sampling of possible fit ranges and excited-state model selection, in which fit results are functions of only the tolerances described above and the given correlation functions.
A graphical illustration of the fitting procedure is shown in Fig.~\ref{fig:flowchart}.
This fitting procedure was implemented in the \verb Julia ~language~\cite{Julia-2017} using the \verb Optim ~optimization package~\cite{mogensen2018optim} to obtain the results of this work.

Fits that pass all of the checks above are considered reliable estimates of the energy spectrum, and the final estimate of the ground-state energy $\overline{E}_0$ and its uncertainty $\delta \overline{E}_0$ are obtained by taking a weighted average of the $N_{\rm success}$ successful fit results $E_0^f$,
\begin{equation}
   \begin{split}
      \overline{E}_0 &= \sum_{f=1}^{N_{\rm success}} w^f E_0^f,\\
      \delta_{\rm stat} \overline{E}_0^2 &= \sum_{f=1}^{N_{\rm success}} w^f (\delta E_0^f)^2,\\
      \delta_{\rm sys} \overline{E}_0^2 &= \sum_{f=1}^{N_{\rm success}} w^f \left( E_0^f - \overline{E}_0 \right)^2,\\
      \delta \overline{E}_0 &= \sqrt{\delta_{\rm stat} \overline{E}_0^2 + \delta_{\rm sys} \overline{E}_0^2},
   \end{split}\label{eq:weightedave}
\end{equation}
where $f$ labels the choice of fit range specified by $t_{\rm min}$ for each interpolating operator\footnote{The total error $\delta \overline{E}_0$ describes the combined statistical uncertainty on $\overline{E}_0$ plus systematic uncertainty arising from the choice of fit range and fit model. The partitioning of this error into $\delta_{\rm stat} \overline{E}_0$ and $\delta_{\rm sys} \overline{E}_0$ only partially separates statistical and systematic uncertainties because $\delta_{\rm stat} E_0$ includes statistical errors plus systematic uncertainties related to fluctuations among the $\delta E^f_0$.}.
Each fit result provides an unbiased estimate of the ground-state energy.
The relative weights $w_f$ of each fit in the weighted average can therefore be chosen arbitrarily in the limit of large statistics; in practice it is advantageous to choose weights that penalize poor fits with larger $\chi^2/N_{\rm dof}$ and unconstraining fits with larger uncertainties $\delta E_0^f$.
Following Ref.~\cite{Rinaldi:2019thf}, we use the weights
\begin{equation}
   \begin{split}
      \widetilde{w}^f &= \frac{p_f \left( \delta E_0^f\right)^{-2} }{ \sum_{f^\prime = 1}^{N_{\rm success}} p_{f^\prime} \left( \delta E_0^{f^\prime} \right)^{-2}  },
   \end{split}\label{eq:weights}
\end{equation}
where $p_f = \Gamma(N_{\rm dof}/2, \chi_f^2/2)/\Gamma(N_{\rm dof}/2)$ is the $p$-value assuming $\chi^2$-distributed goodness-of-fit parameters with $\chi^2_f$ obtained by inserting $E_0^f$ into Eq.~\eqref{eq:chisquared}\footnote{For large $\lambda^*$, the $\chi^2$ function being minimized approaches an uncorrelated $\chi^2$ and the values of $\chi^2$ will not be distributed as $\chi^2$-distributed random variables with $N_{\rm dof}$ degrees-of-freedom. In this regime where finite $N$ artifacts are not negligible, the weights in Eq.~\eqref{eq:weights} still serve the purpose of penalizing comparatively less accurate descriptions of the results being fit and their correlations as estimated by $\mathcal{S}_{tt^\prime}^{ij}(\lambda^*)$, but the absolute sizes of the $p_f$ should not be interpreted as $p$-values for each fit. %In future work it might be advantageous to consider using $p$-value estimates calculated using bootstrap techniques as in Ref.~RBC?? in place of the $p_f$ in Eq.~\eqref{eq:weights}.
}.
Variation to the particular choices of specified tolerances have been studied, and the subsequent variation in the ensemble of successful fits is found to have little impact on the results of this weighted averaging.
The results $\overline{E}_0$ and $\delta \overline{E}_0$ obtained with this procedure are shown as the central values and uncertainties for single-particle energy results $E_{\pi^+}(L)$, $E_{\overline{K}^0}(L)$, $E_n(L)$, and $E_p(L)$ in Tables~\ref{tab:mesonM} and \ref{tab:baryon}.
Effective mass plots showing the smallest $t_{\rm min}$ fit with weight over $1/2$ of the maximum weight fit as well as $E_0^f$ and $\widetilde{w}^f / \text{Max}(\widetilde{w}^f)$ are shown in Appendix~\ref{sec:fitresults}.

\subsection{Multi-meson correlation functions}

To determine results for multi-meson ground-state energies with thermal effects taken into account, fits are performed iteratively starting with fits for $n=1$ mesons and then moving to fits with increasing $n$.
The excited-state fit form in Eq.~\eqref{eq:spectral} includes thermal effects describing $k$ forwards-propagating mesons and $n-k$ backwards-propagating mesons for $k < n/2$ that are included by using the central values $\overline{E}_0$ calculated for $E_k$ and $E_{n-k}$ in Eq.~\eqref{eq:spectral}.
Uncertainties in $E_k$ are found to be significantly smaller than uncertainties in $E_n$ for $k < n/2$, and for simplicity are not incorporated into the description of thermal effects.
The overlap factors for these thermal states are determined using linear algebra techniques~\cite{varpro0,varpro1} during each step of nonlinear optimization for the $N$-particle energy spectrum analogously to the procedure described above for other overlap factors.
This fit function is found to provide acceptable fits to multi-meson correlation functions without the need for additional free parameters describing excited-state thermal effects.

Before beginning the fitting procedure described above, correlation function results from all quark propagator sources on a given configuration are averaged and meson correlation functions are further blocked along the Markov chain to form $N_{\rm block} = 200$ approximately independent samples from the $N_{\rm cfg}$ configurations for each volume shown in Table~\ref{tab:ensem}.
Further averaging is found to give statistically consistent results, suggesting that autocorrelations can be neglected after this blocking.
To determine correlated differences of ground-state energies for different hadron type ($\pi^+,\ \overline{K}^0$) and hadron number, fits are performed independently to determine $E_0^f$ for each hadron type and number for each fit range sampled.
A fit range is considered to give a successful fit only if the checks on fit robustness described above are passed for each hadron type and number involved in the correlated difference.
For each successful fit range, bootstrap resampling is used to determine the uncertainties on correlated differences of the resulting $E_0^f$.
During bootstrap resampling, the same elements of these $N_{\rm block}$ samples are used to construct bootstrap ensembles for each hadron type and number.
Correlated differences of the bootstrap results $E_0^{f,b}$ are then formed, and confidence intervals are computed by applying Eq.~\eqref{eq:quantile} to the these correlated differences.
Finally, weighted averages of the resulting correlated differences and their associated uncertainties are taken using Eq.~\eqref{eq:weightedave}-\eqref{eq:weights}.
The results of this procedure are used to determine the FV energy shifts and differences between FV energy shifts for charged and uncharged hadrons shown in Tables~\ref{tab:mesonFV}-\ref{tab:baryon}.

\subsection{Multi-nucleon correlation functions}

Differences between multi-nucleon ground-state energies and the corresponding sums of their constituent nucleon masses are computed using correlated  differences of bootstrap results $E_0^{f,b}$ for multi-nucleon and single-nucleon correlation functions analogously to the multi-meson case described above. 
Correlated differences between one-nucleon and multi-nucleon energies can be determined much more precisely than multi-nucleon energies alone, and differences of one-nucleon and multi-nucleon fit results with different values of $N_{\rm states}$ are found to describe correlated differences of LQCD+QED$_L$ effective energy results poorly. 
$N_{\rm states}$ is therefore restricted to be identical between single-nucleon and multi-nucleon systems.
Otherwise, fits for multi-nucleon energy shifts are performed identically to fits for multi-meson energy shifts not including thermal effects.

\subsection{Fit results}\label{sec:fitresults}

\begin{figure}
  \centering
     \begin{minipage}[c][150pt][t]{70pt} $1\ \pi^+\newline L/a=32$ \end{minipage}
     \includegraphics[width=.40\textwidth]{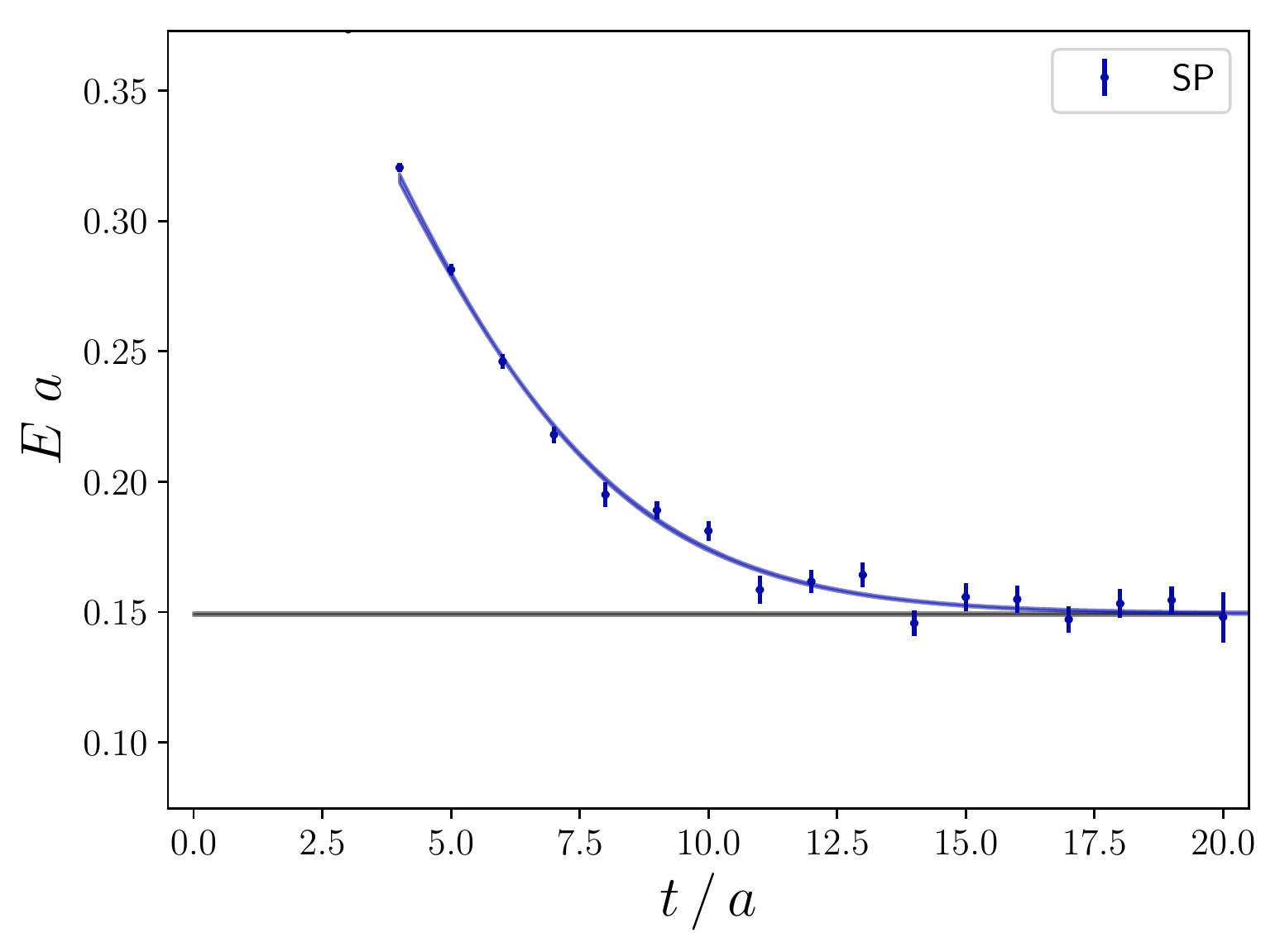}\hspace{10pt}
     \includegraphics[width=.40\textwidth]{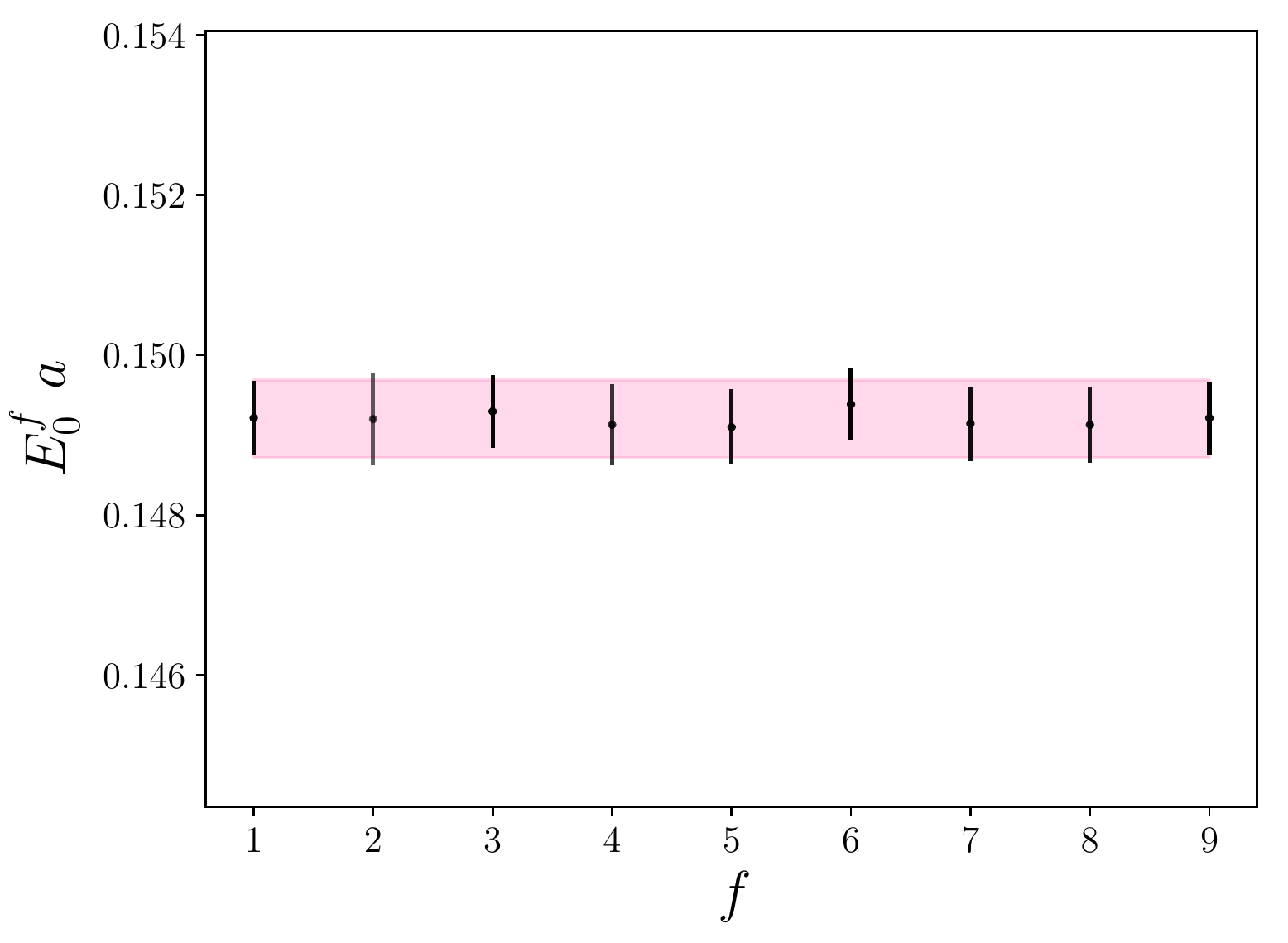} \\\vspace{-65pt}
     \begin{minipage}[c][150pt][t]{70pt} $2\ \pi^+\newline L/a=32$ \end{minipage}
     \includegraphics[width=.40\textwidth]{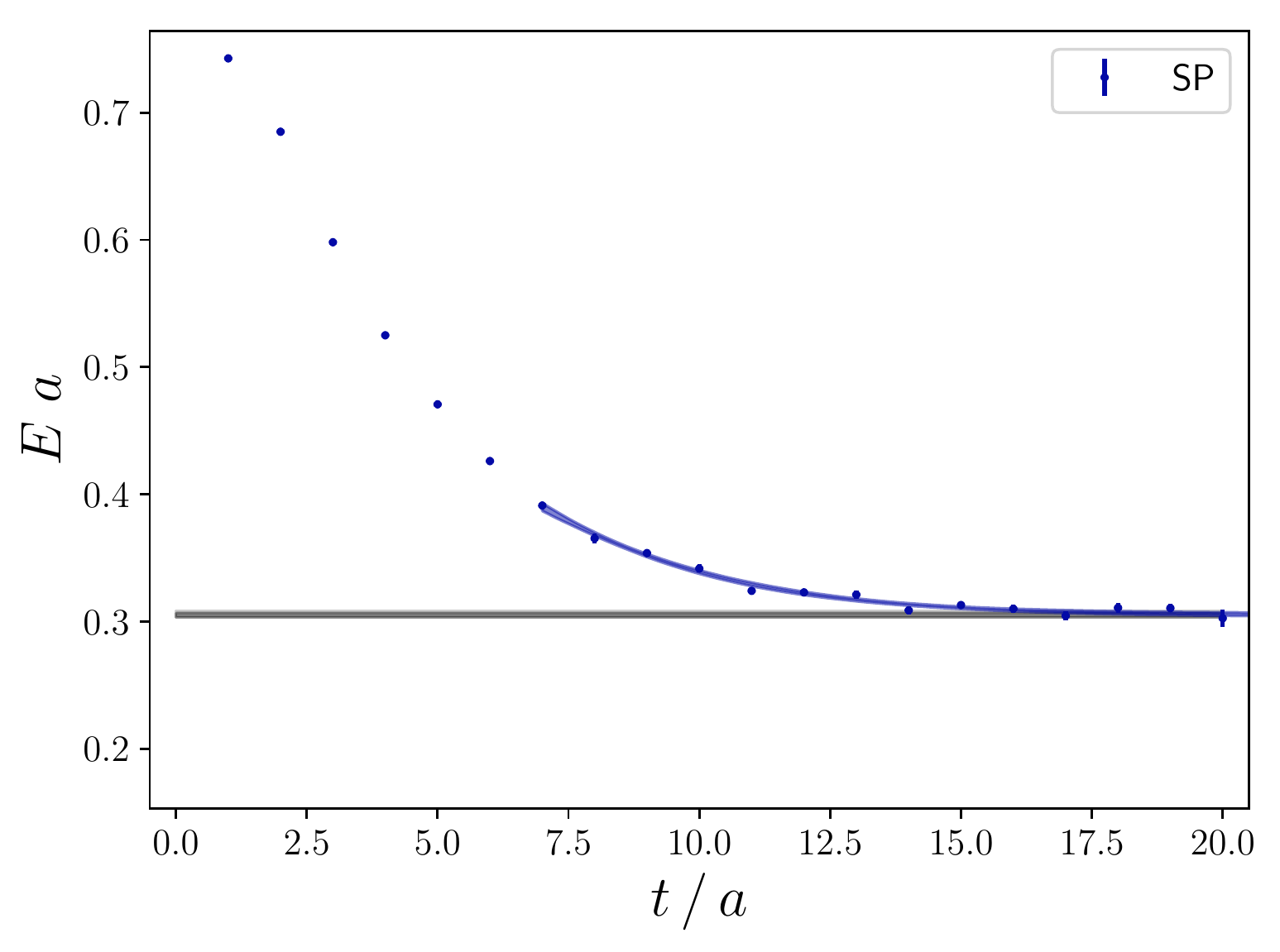}\hspace{10pt}
     \includegraphics[width=.40\textwidth]{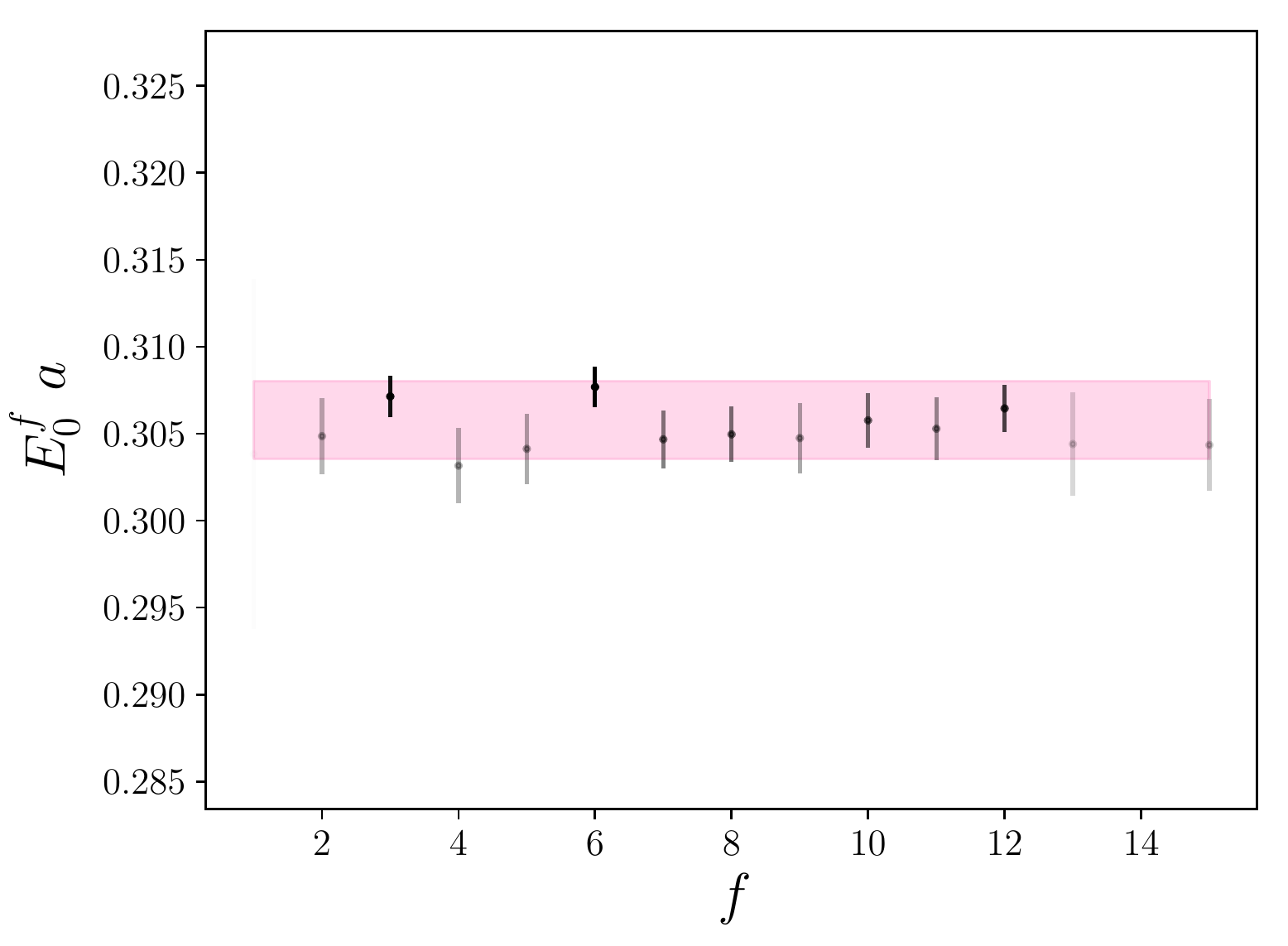} \\\vspace{-65pt}
     \begin{minipage}[c][150pt][t]{70pt} $3\ \pi^+\newline L/a=32$ \end{minipage}
     \includegraphics[width=.40\textwidth]{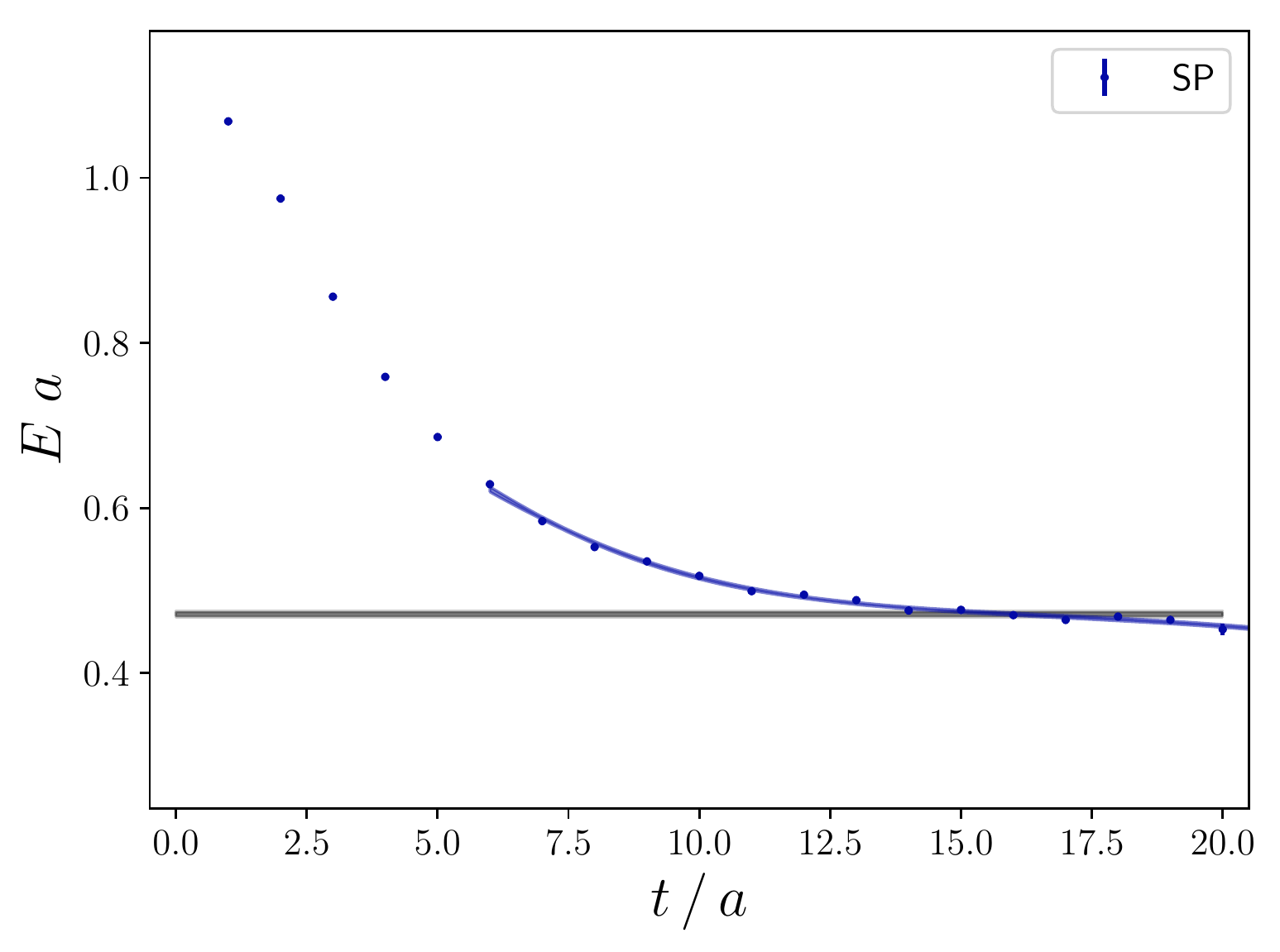}\hspace{10pt}
     \includegraphics[width=.40\textwidth]{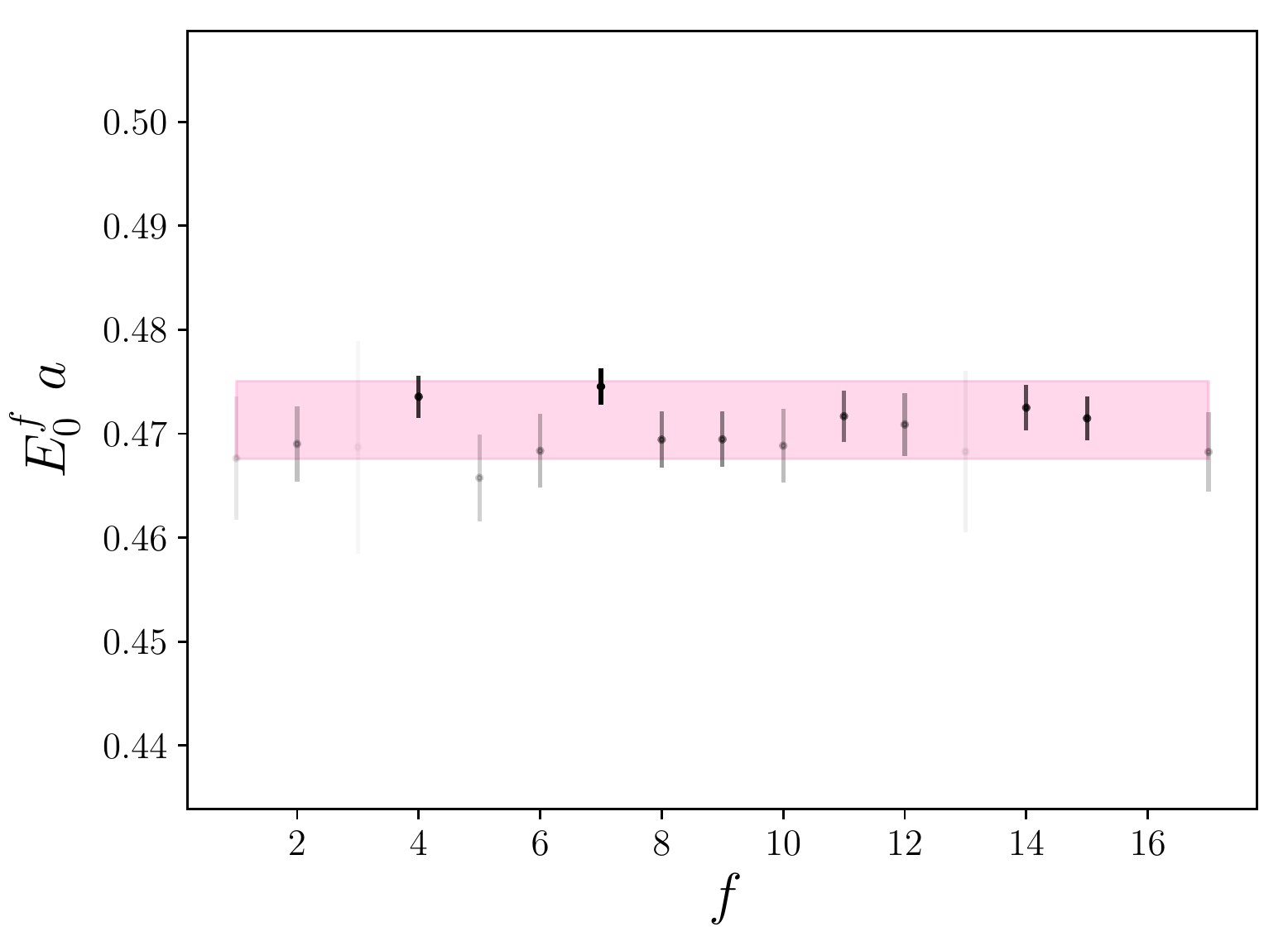} \\\vspace{-65pt}
     \begin{minipage}[c][150pt][t]{70pt} $4\ \pi^+\newline L/a=32$ \end{minipage}
     \includegraphics[width=.40\textwidth]{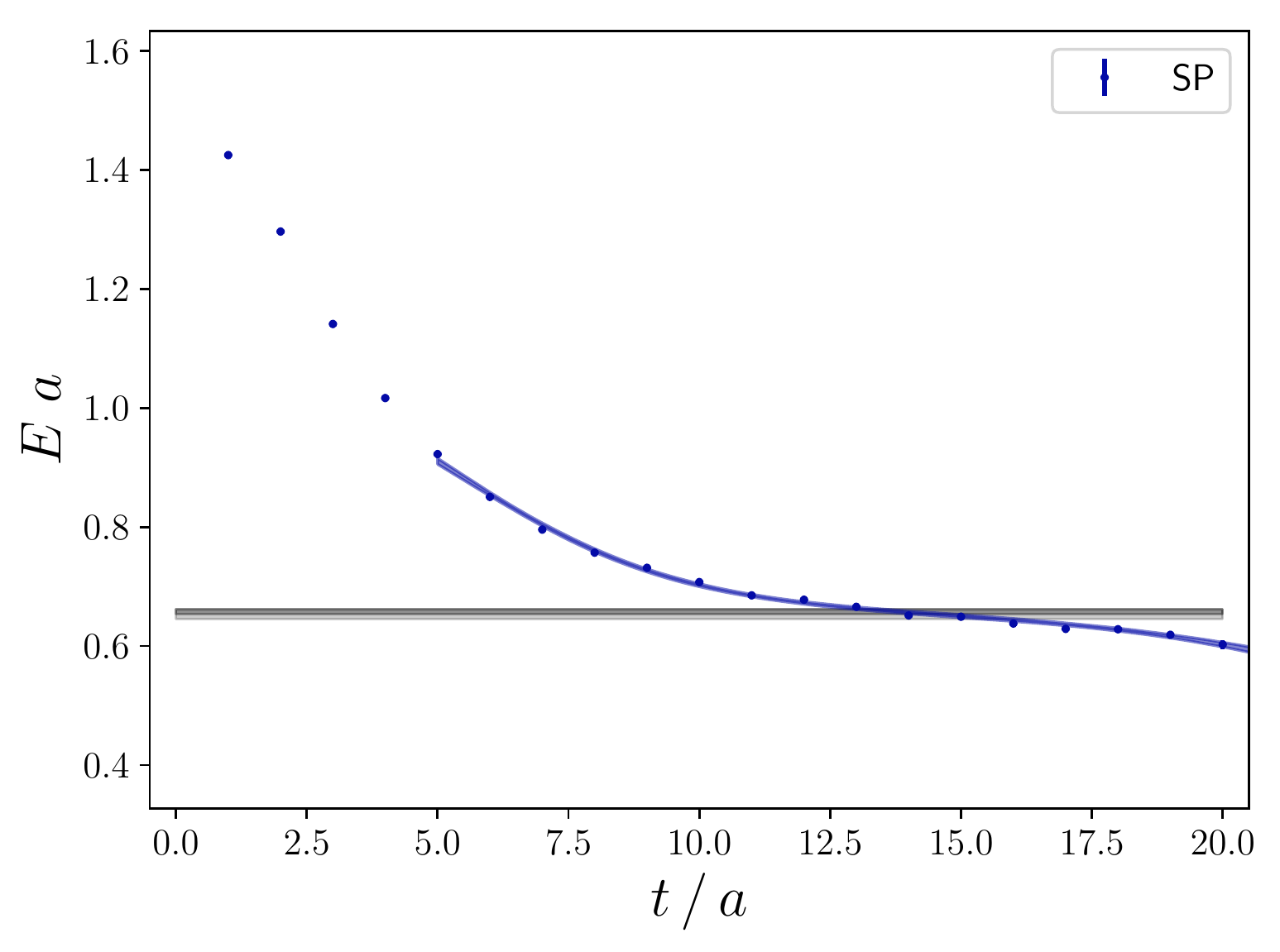}\hspace{10pt}
     \includegraphics[width=.40\textwidth]{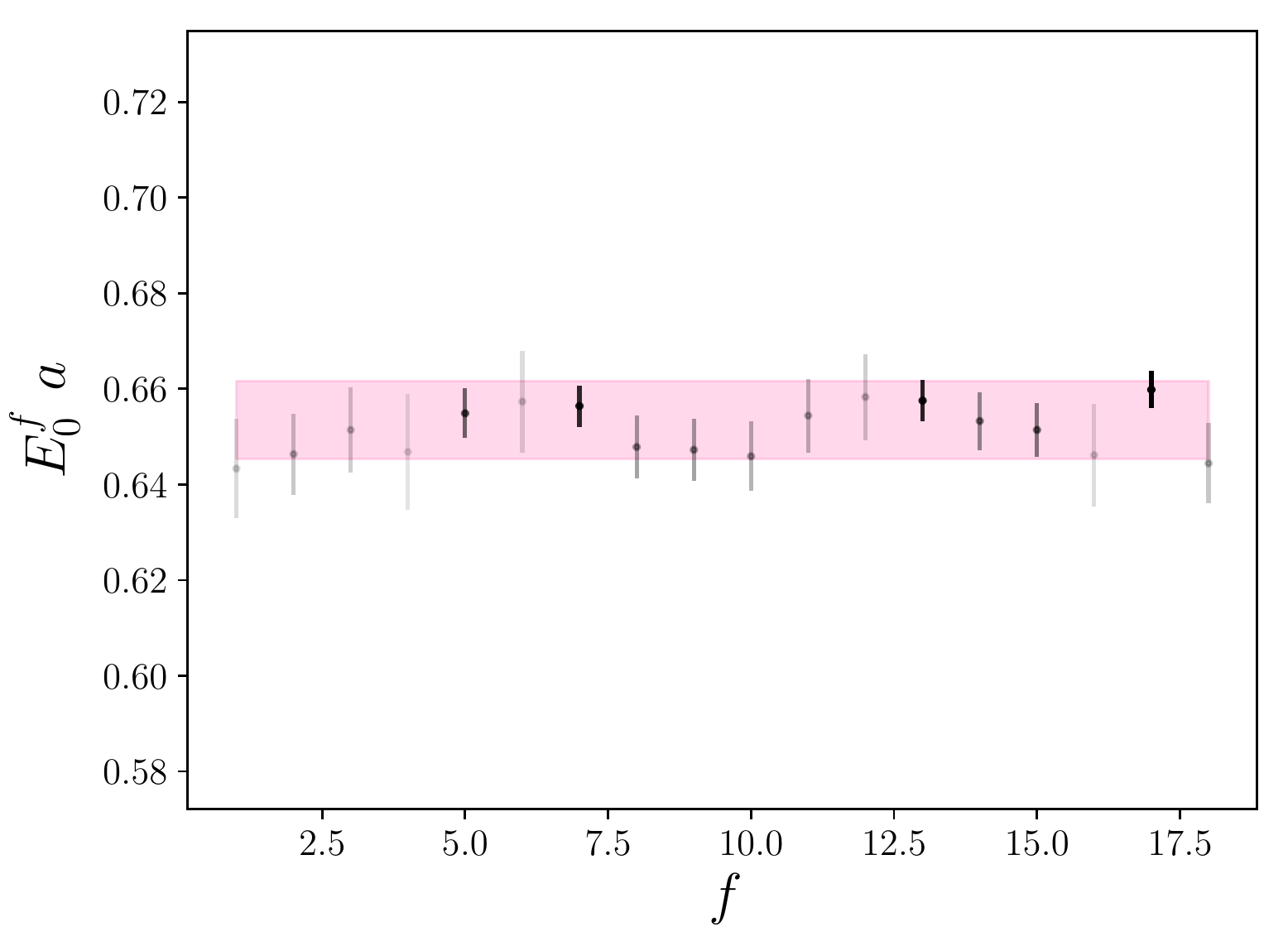} \\\vspace{-65pt}
   \caption{Fit results for systems of $n\in\{1,\dots,4\}$ $\pi^+$ mesons for the $L/a = 32$ lattice volume. The figures are analogous to Fig.~\ref{fig:pion48a}, see Appendix~\ref{app:fits} for a definition of the fitting procedure employed.
    \label{fig:pion32a}}
\end{figure}

\begin{figure}
  \centering
     \begin{minipage}[c][150pt][t]{70pt} $5\ \pi^+\newline L/a=32$ \end{minipage}
     \includegraphics[width=.40\textwidth]{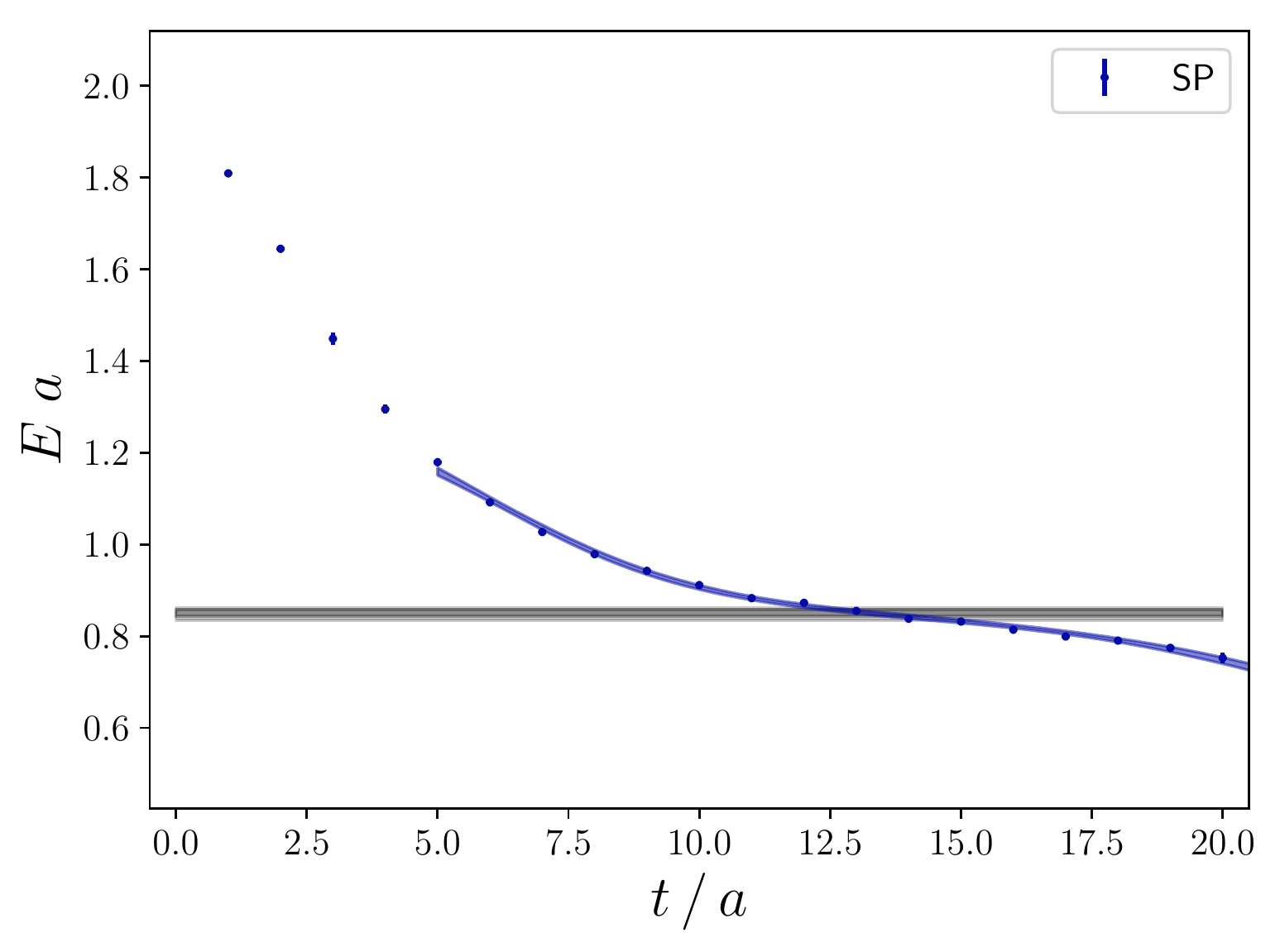}\hspace{10pt}
     \includegraphics[width=.40\textwidth]{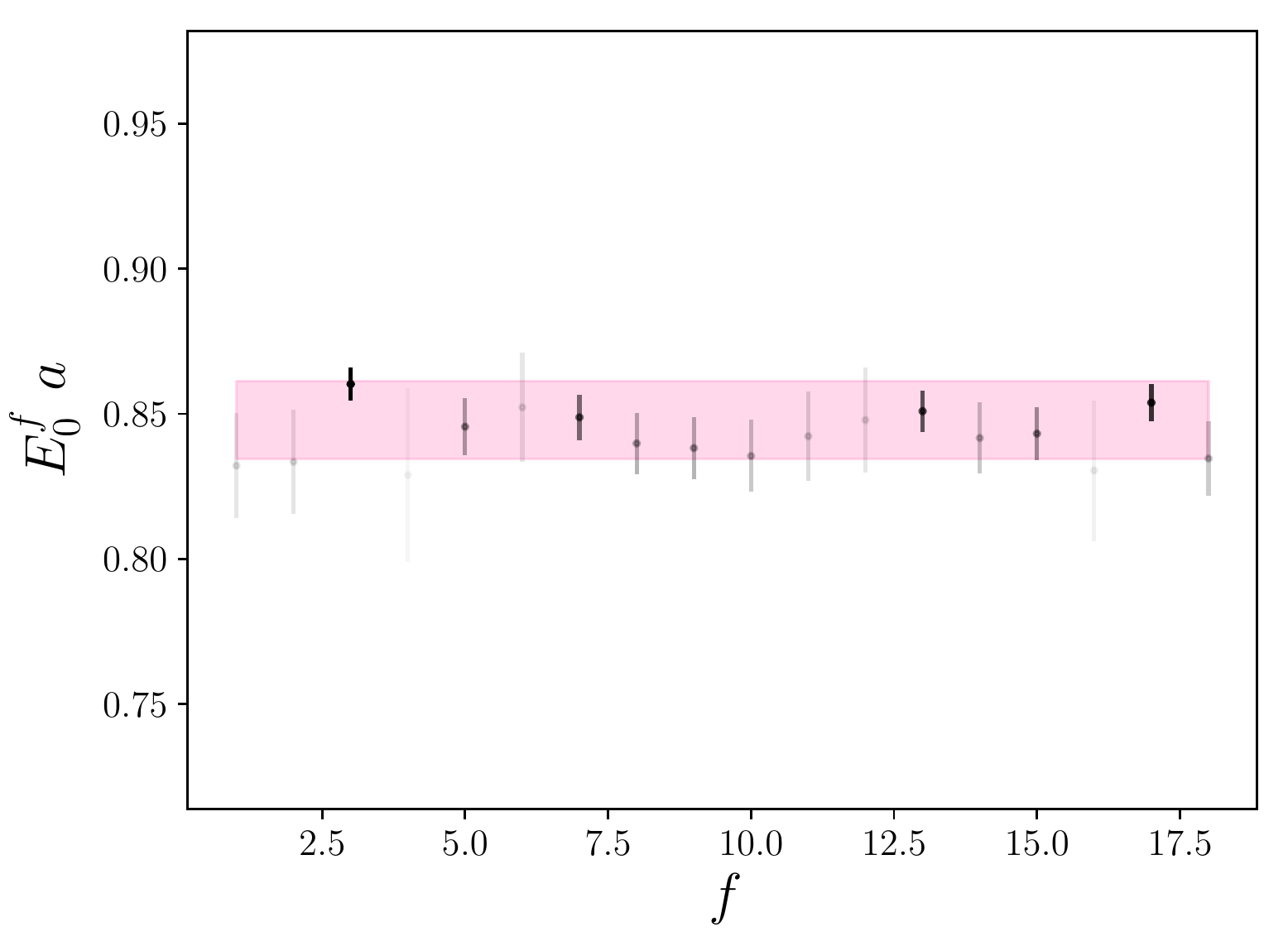} \\\vspace{-65pt}
     \begin{minipage}[c][150pt][t]{70pt} $6\ \pi^+\newline L/a=32$ \end{minipage}
     \includegraphics[width=.40\textwidth]{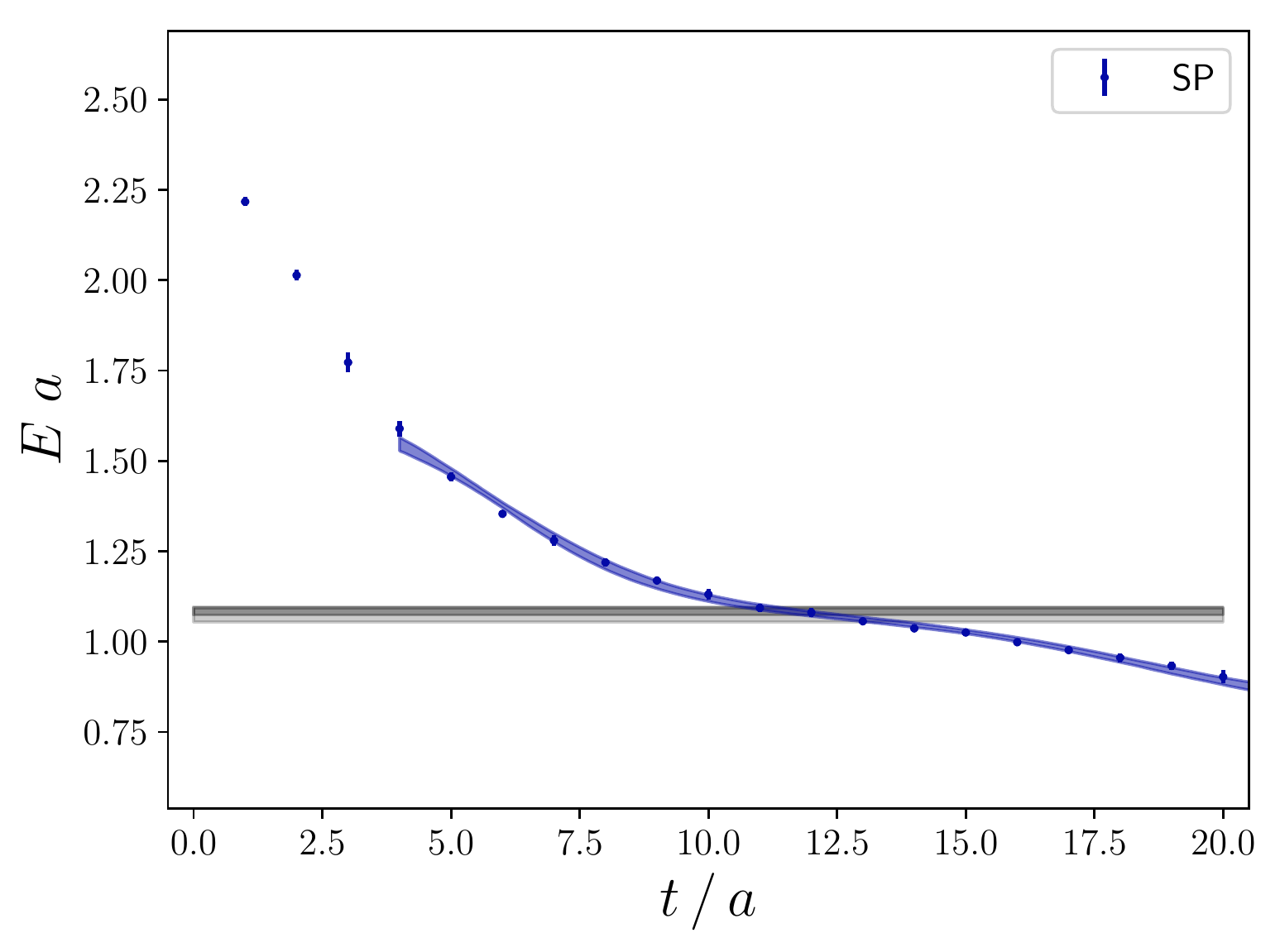}\hspace{10pt}
     \includegraphics[width=.40\textwidth]{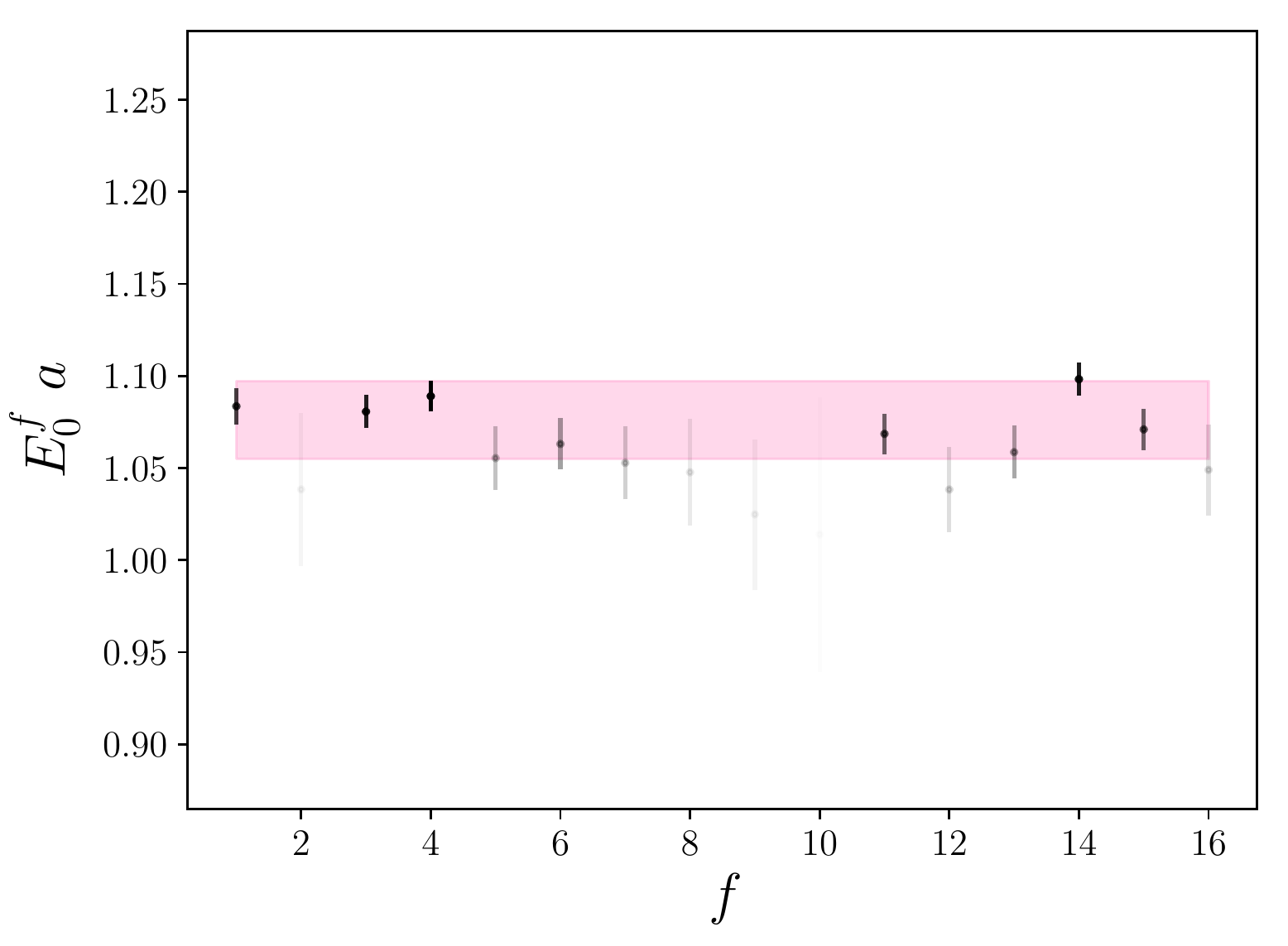} \\\vspace{-65pt}
     \begin{minipage}[c][150pt][t]{70pt} $7\ \pi^+\newline L/a=32$ \end{minipage}
     \includegraphics[width=.40\textwidth]{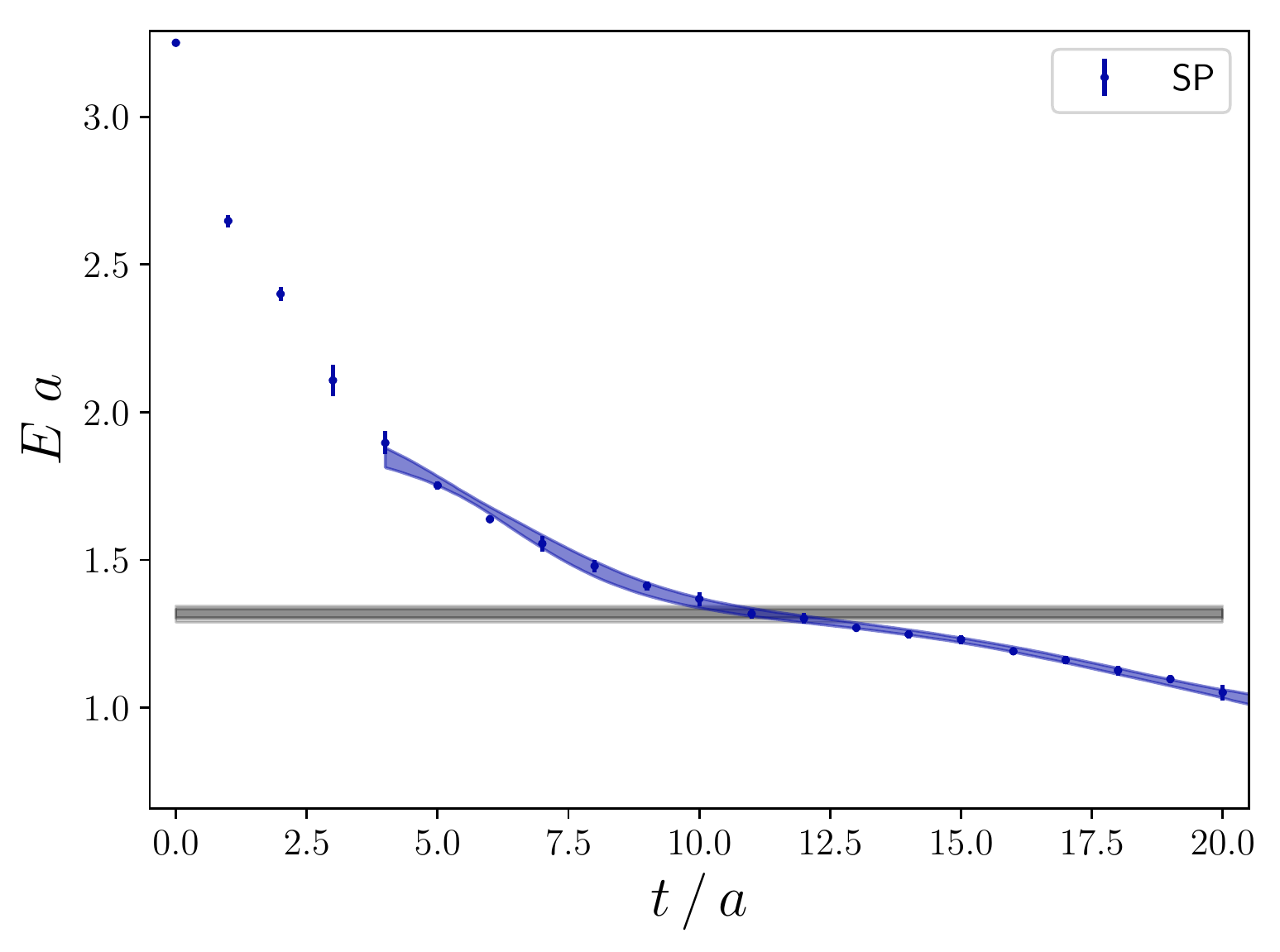}\hspace{10pt}
     \includegraphics[width=.40\textwidth]{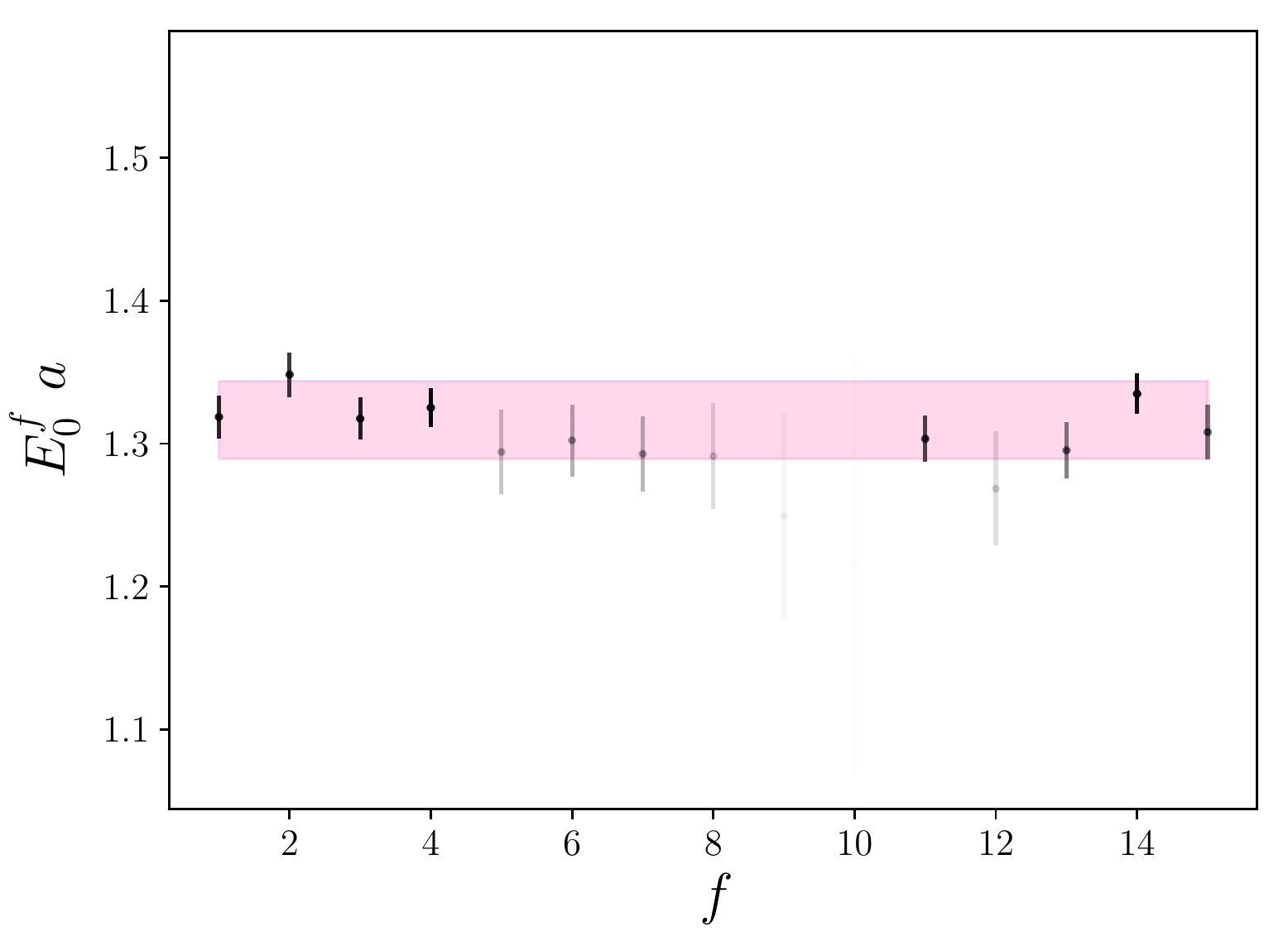} \\\vspace{-65pt}
     \begin{minipage}[c][150pt][t]{70pt} $8\ \pi^+\newline L/a=32$ \end{minipage}
     \includegraphics[width=.40\textwidth]{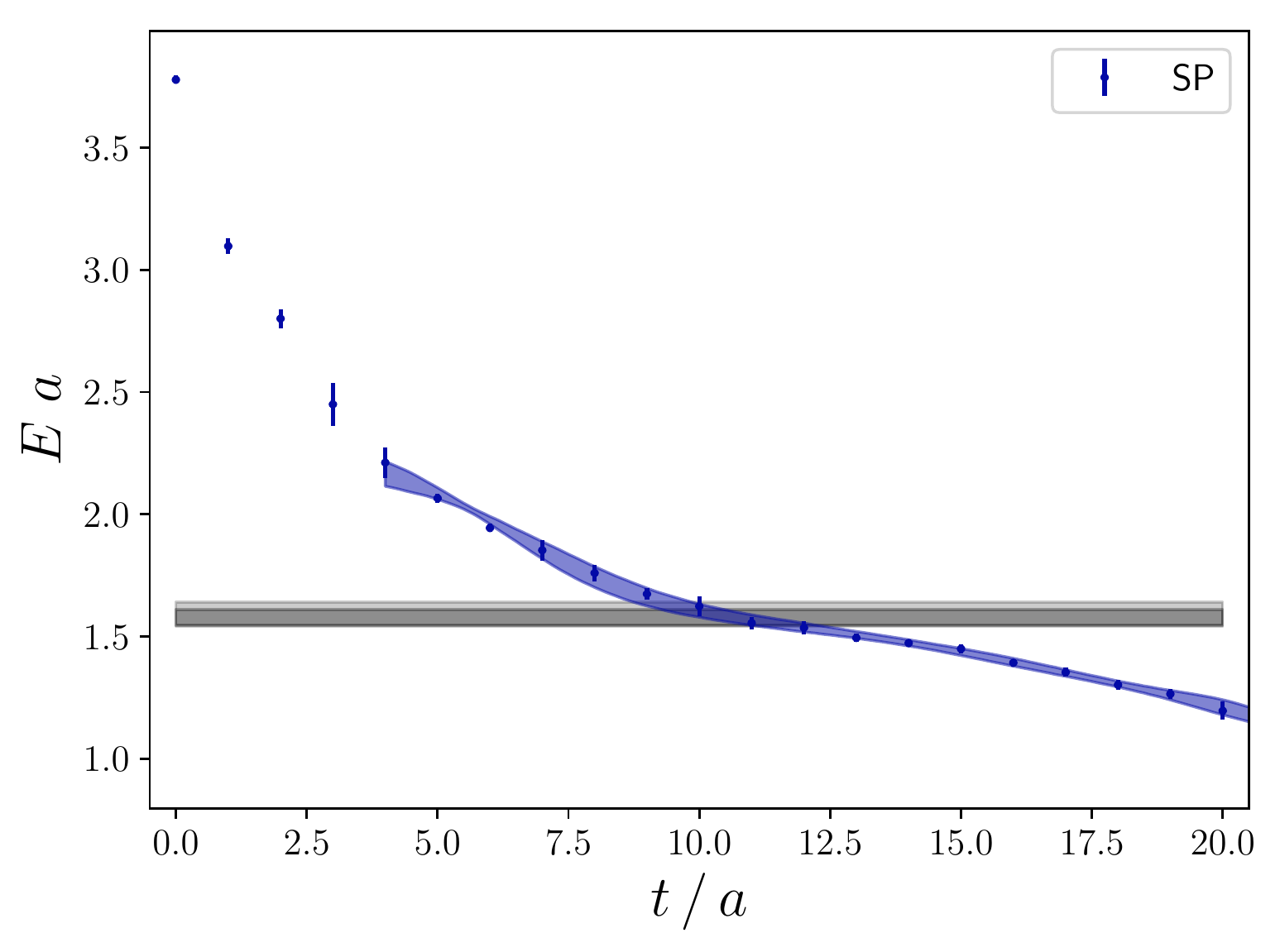}\hspace{10pt}
     \includegraphics[width=.40\textwidth]{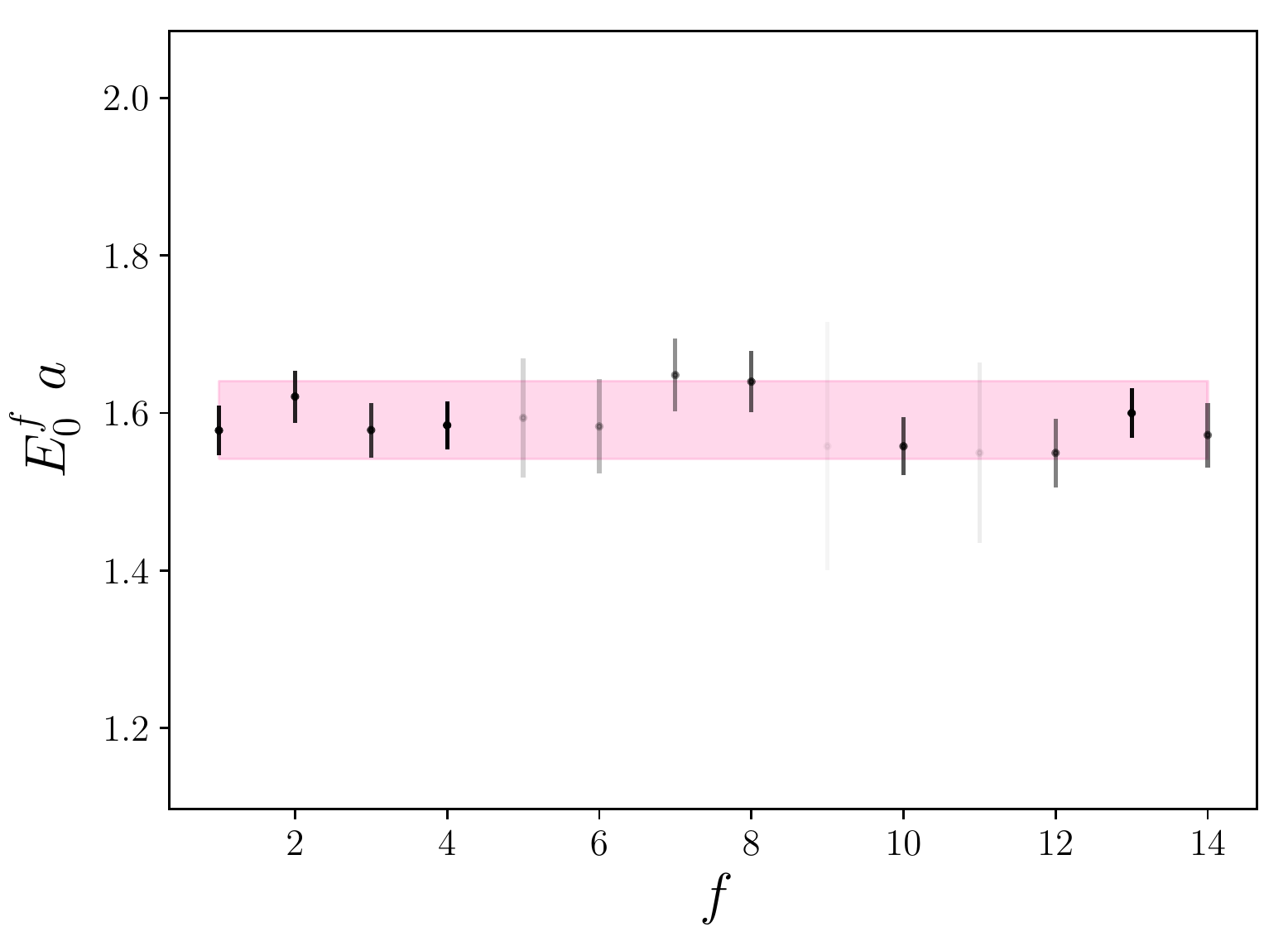} \\\vspace{-65pt}
   \caption{Fit results for systems of $n\in\{5,\dots,8\}$ $\pi^+$ mesons for the $L/a = 32$ lattice volume. The figures are analogous to Fig.~\ref{fig:pion48a}, see Appendix~\ref{app:fits} for a definition of the fitting procedure employed.
    \label{fig:pion32b}}
\end{figure}

\begin{figure}
  \centering
     \begin{minipage}[c][150pt][t]{70pt} $9\ \pi^+\newline L/a=32$ \end{minipage}
     \includegraphics[width=.40\textwidth]{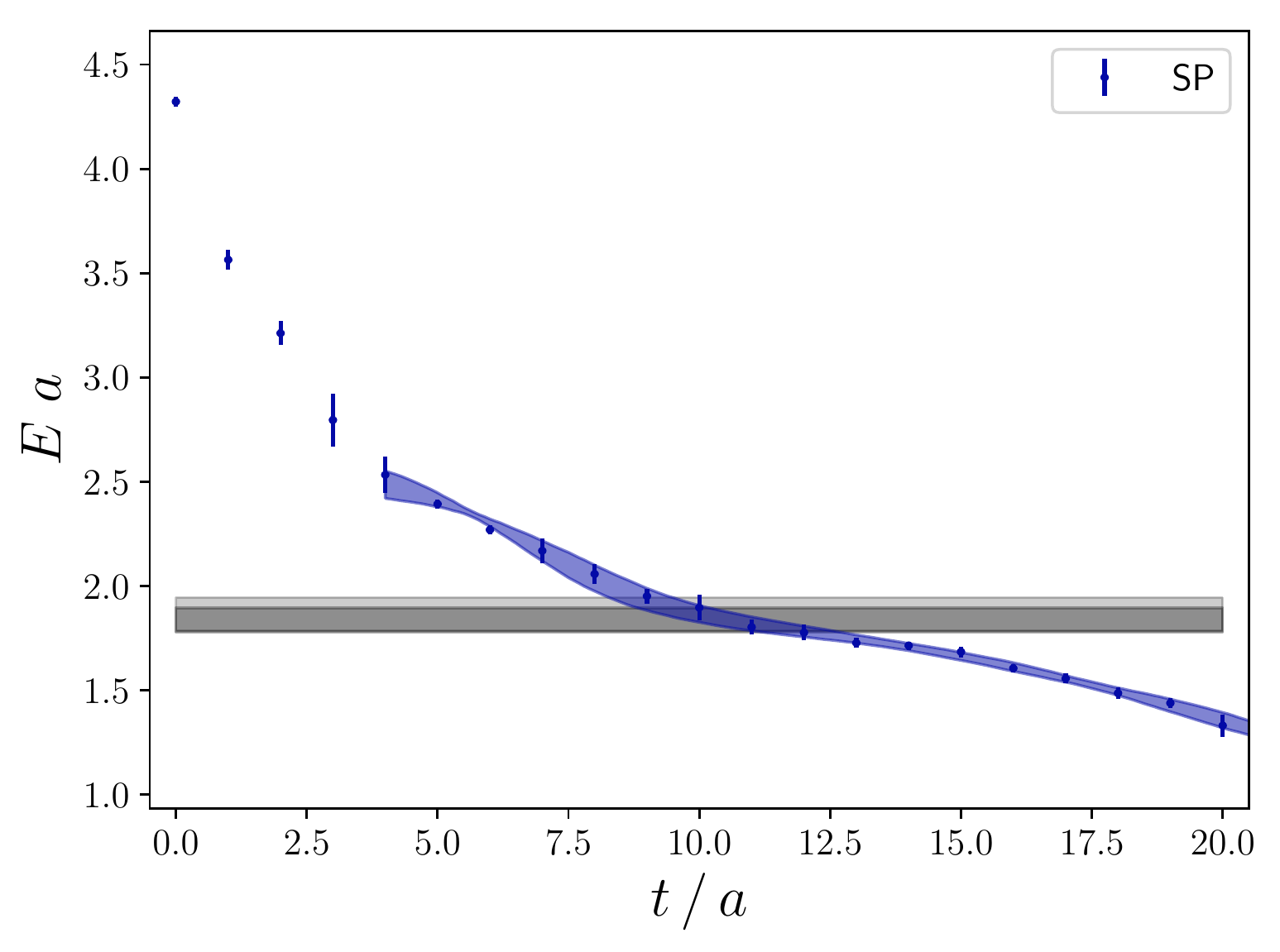}\hspace{10pt}
     \includegraphics[width=.40\textwidth]{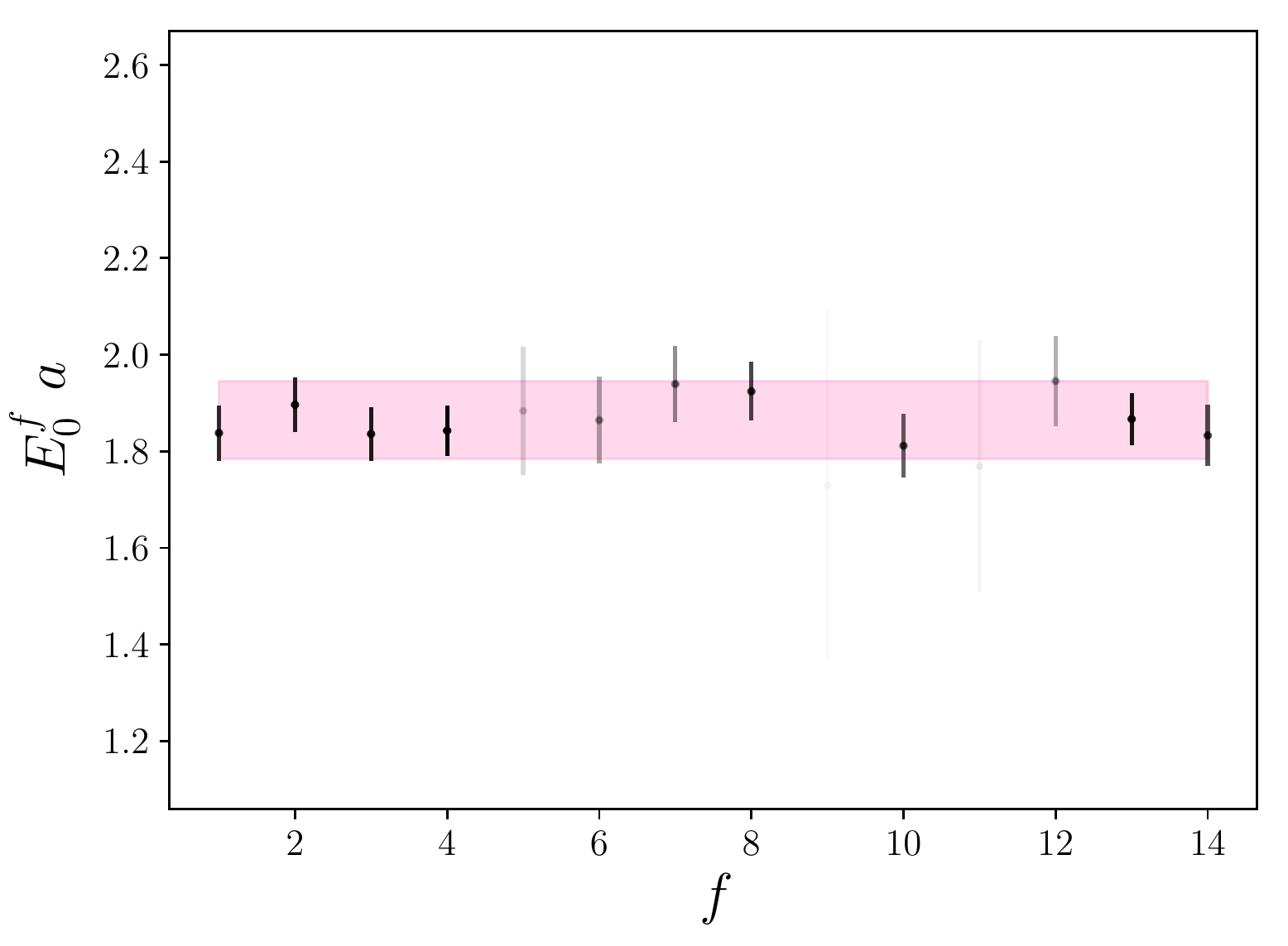} \\\vspace{-65pt}
     \begin{minipage}[c][150pt][t]{70pt} $10\ \pi^+\newline L/a=32$ \end{minipage}
     \includegraphics[width=.40\textwidth]{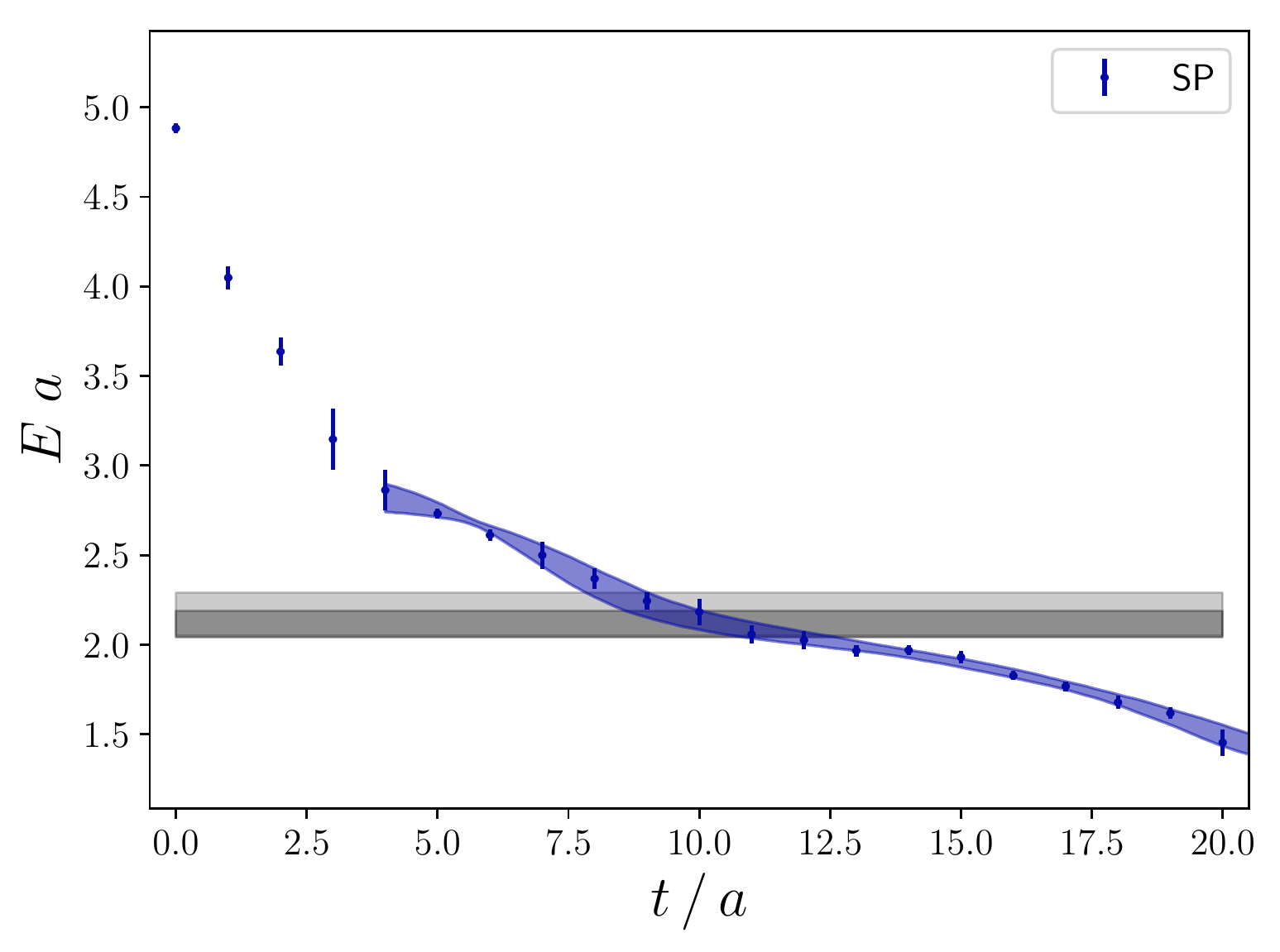}\hspace{10pt}
     \includegraphics[width=.40\textwidth]{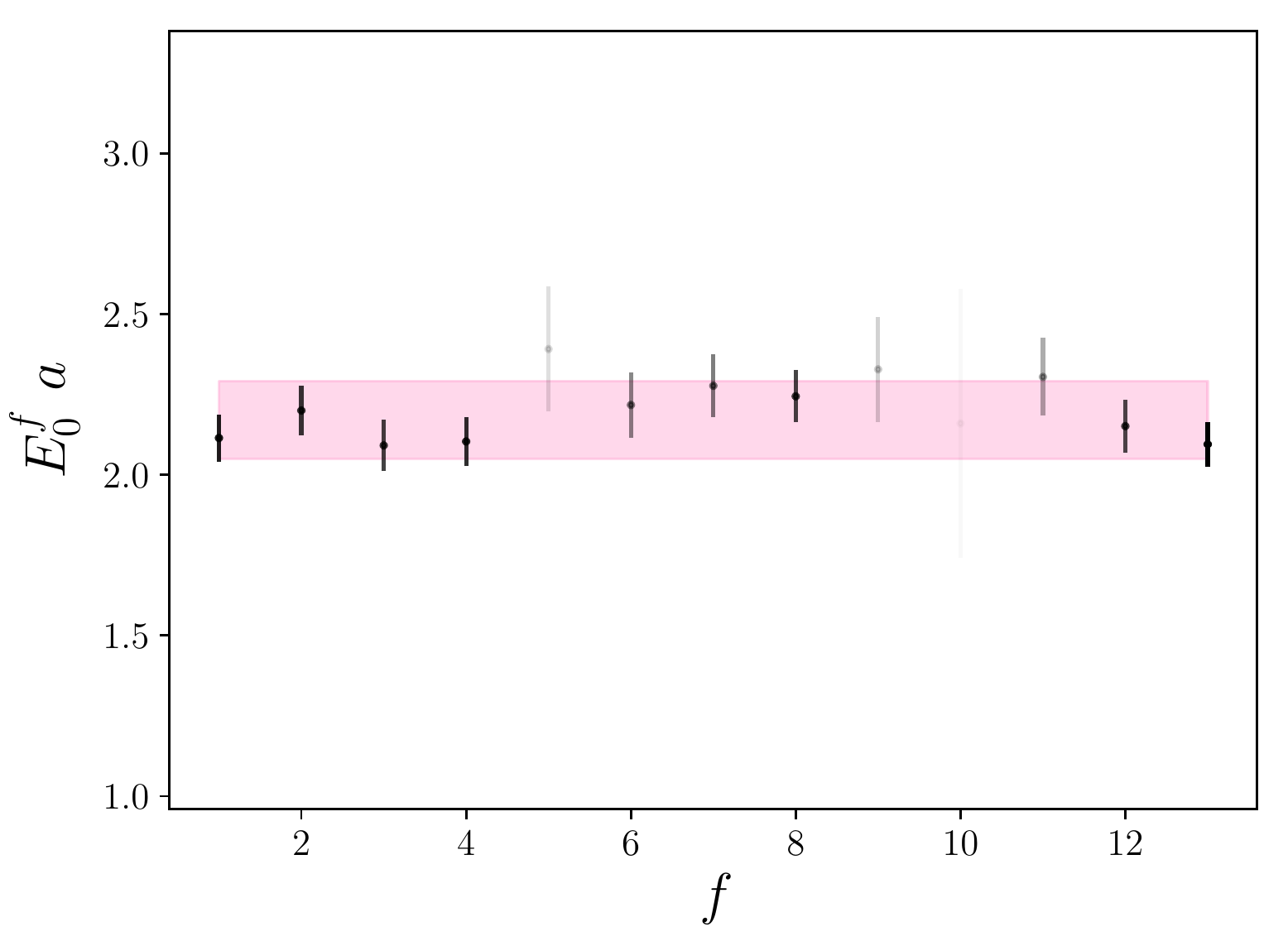} \\\vspace{-65pt}
     \begin{minipage}[c][150pt][t]{70pt} $11\ \pi^+\newline L/a=32$ \end{minipage}
     \includegraphics[width=.40\textwidth]{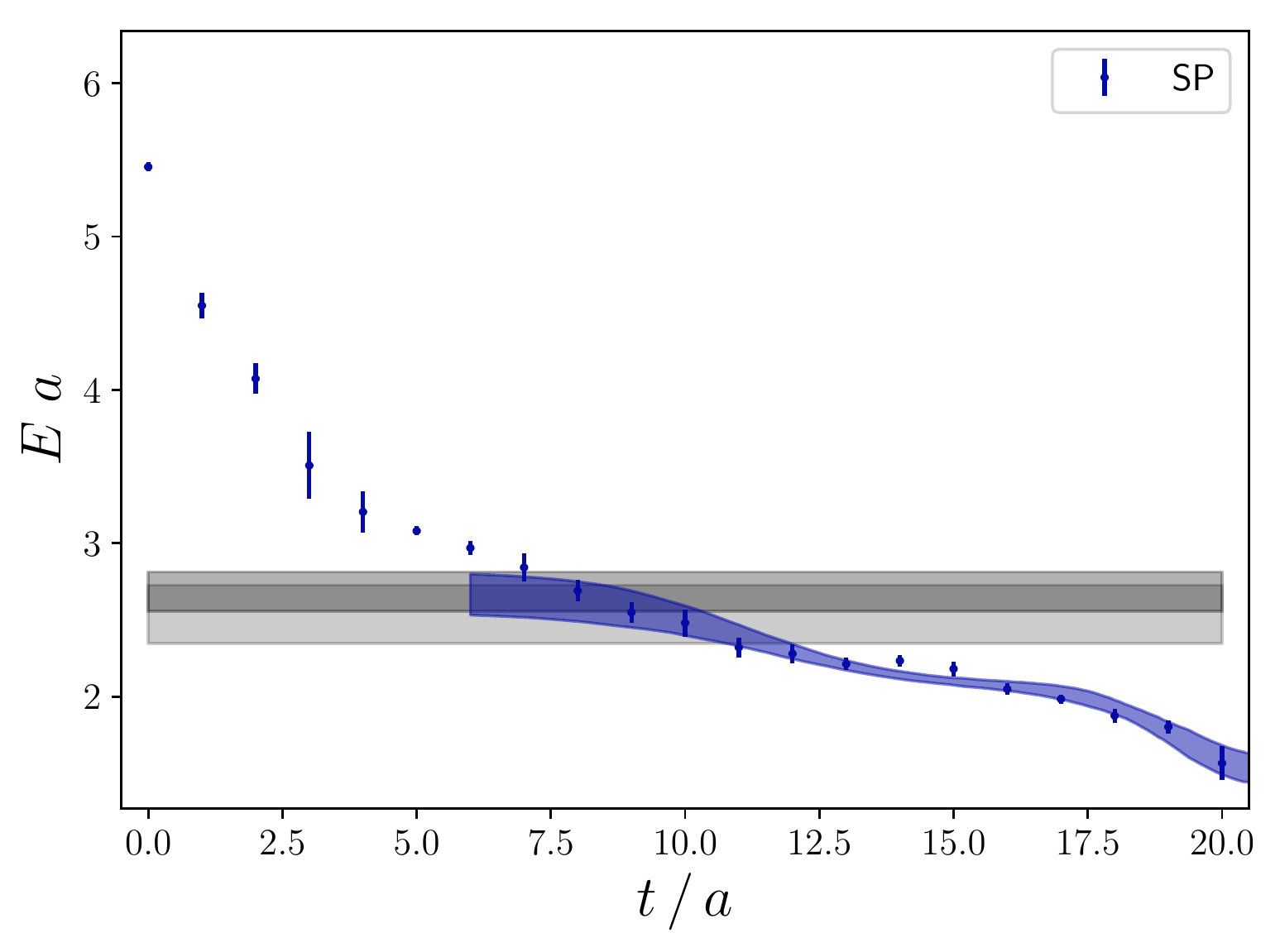}\hspace{10pt}
     \includegraphics[width=.40\textwidth]{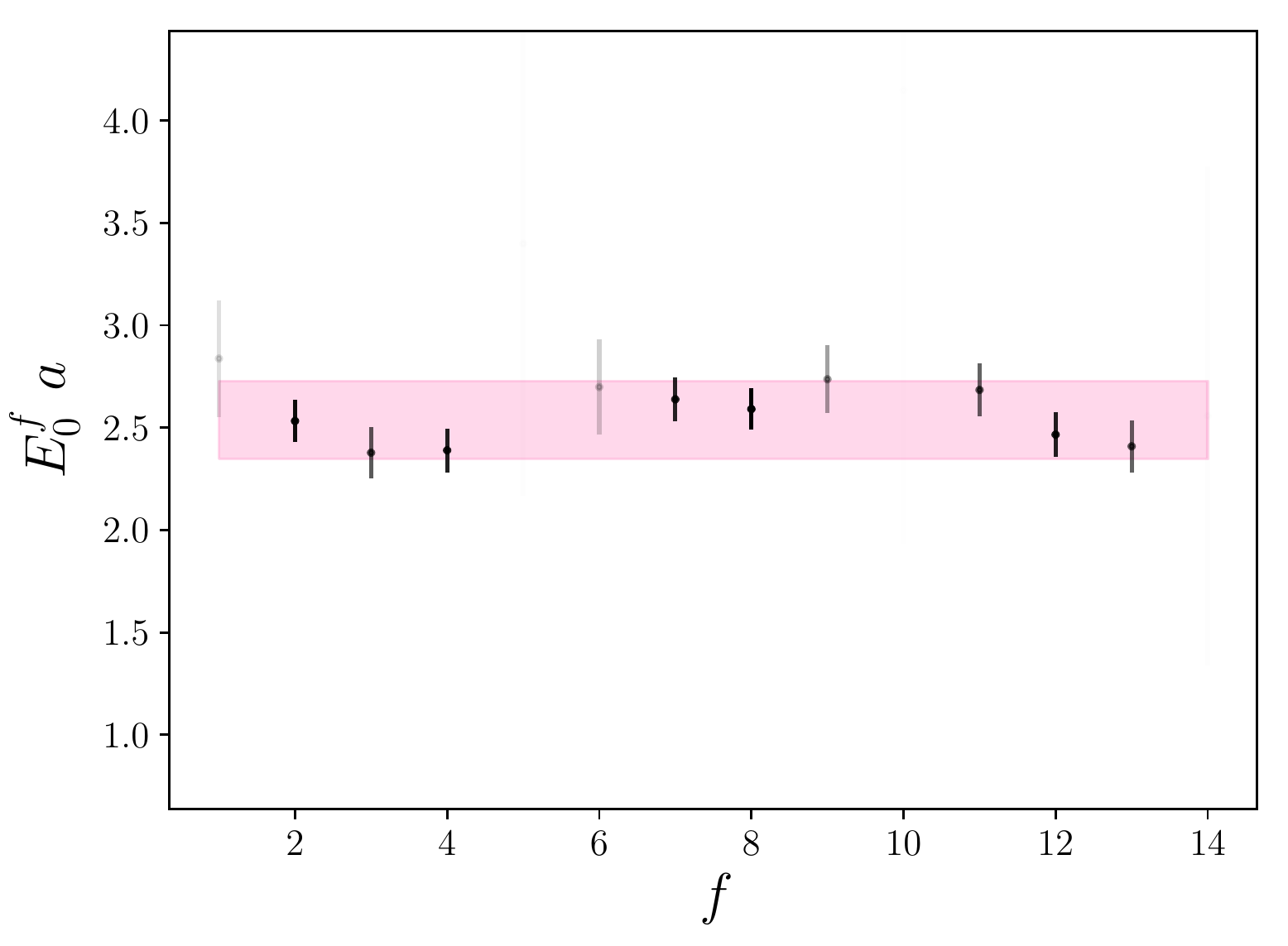} \\\vspace{-65pt}
     \begin{minipage}[c][150pt][t]{70pt} $12\ \pi^+\newline L/a=32$ \end{minipage}
     \includegraphics[width=.40\textwidth]{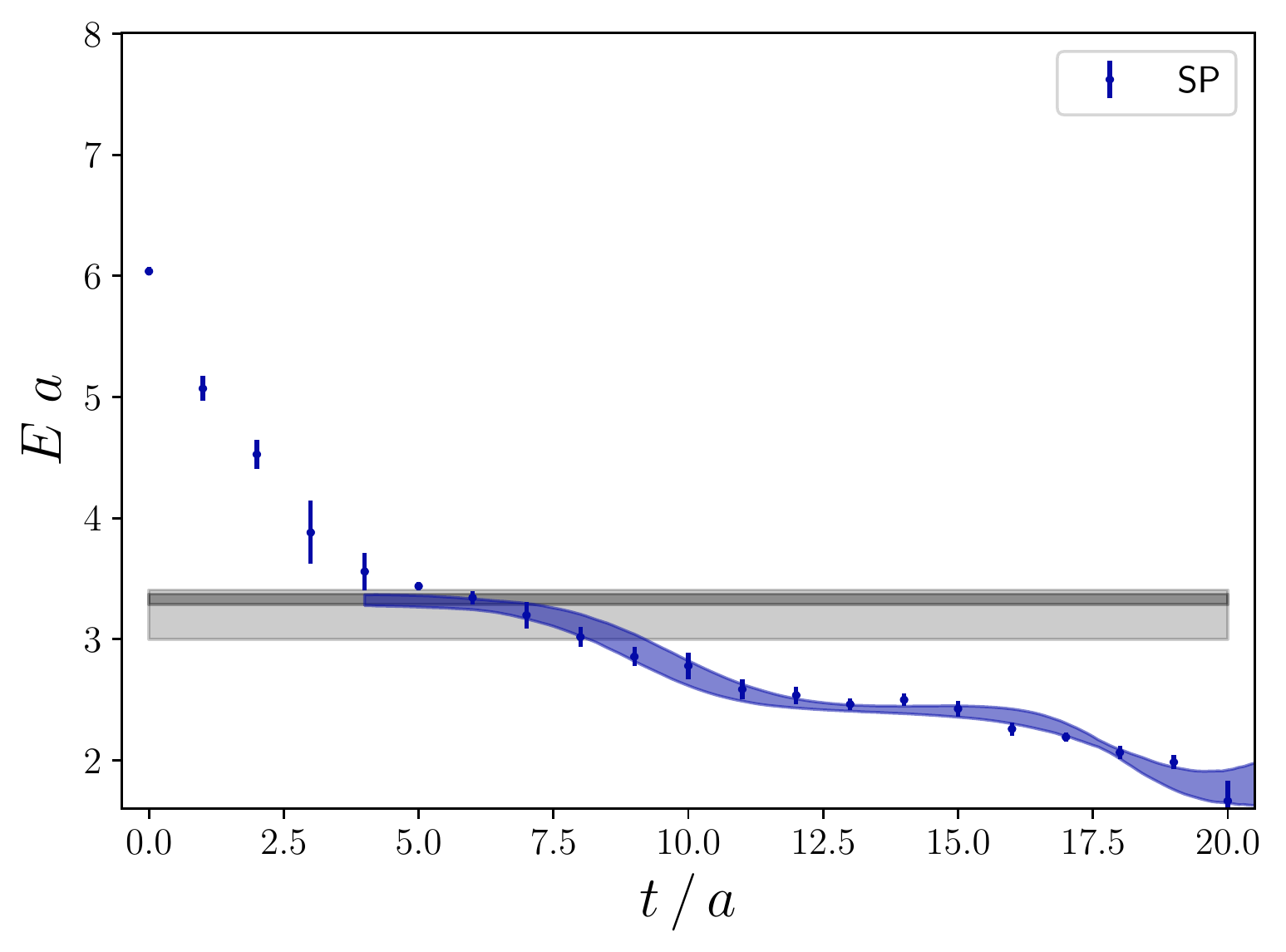}\hspace{10pt}
     \includegraphics[width=.40\textwidth]{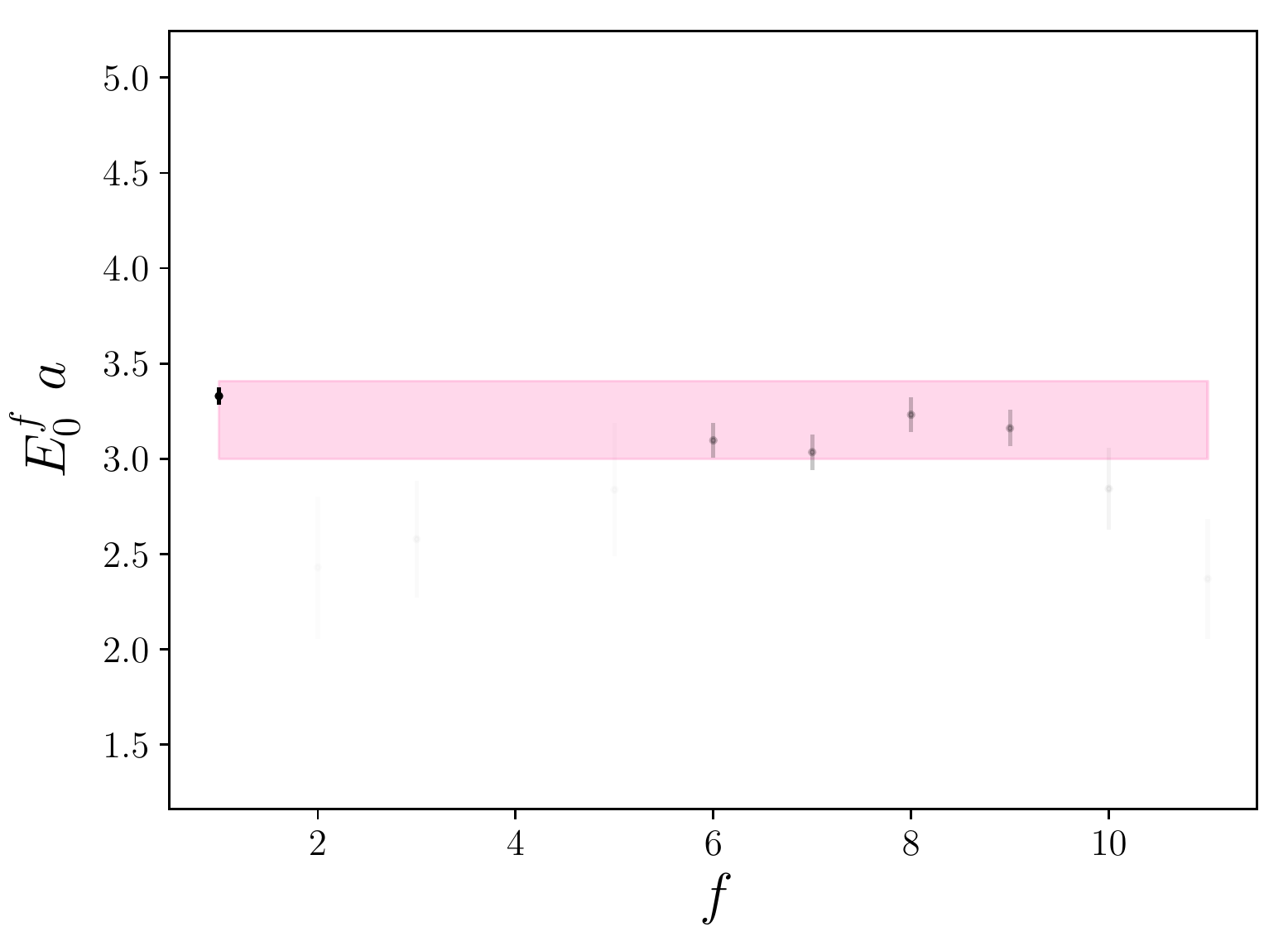} \\\vspace{-65pt}
   \caption{Fit results for systems of $n\in\{9,\dots,12\}$ $\pi^+$ mesons for the $L/a = 32$ lattice volume. The figures are analogous to Fig.~\ref{fig:pion48a}, see Appendix~\ref{app:fits} for a definition of the fitting procedure employed.
    \label{fig:pion32c}}
\end{figure}

\begin{figure}
  \centering
     \begin{minipage}[c][150pt][t]{70pt} $1\ \overline{K^0}$ \newline $L/a=48$ \end{minipage}
     \includegraphics[width=.40\textwidth]{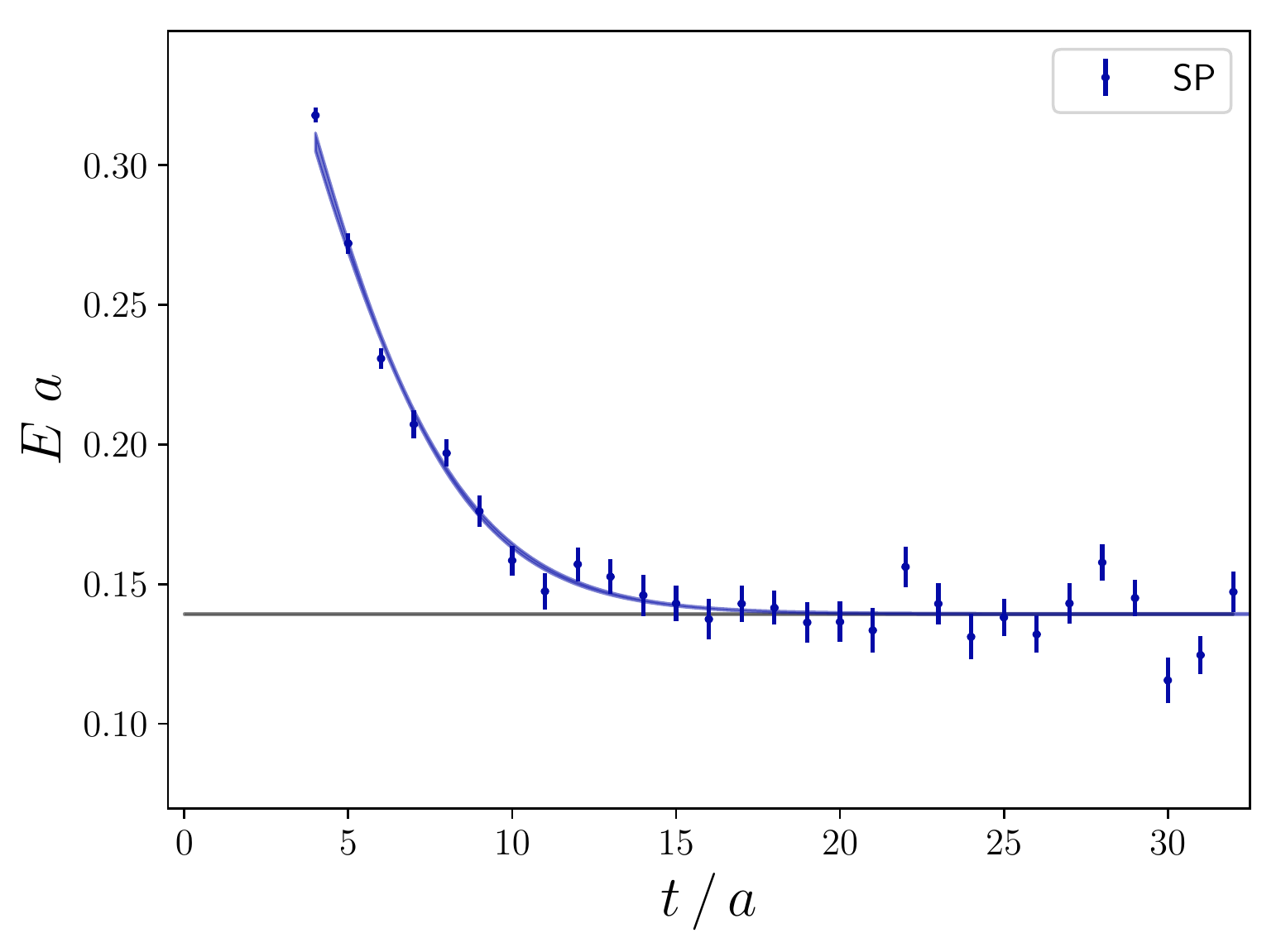}\hspace{10pt}
     \includegraphics[width=.40\textwidth]{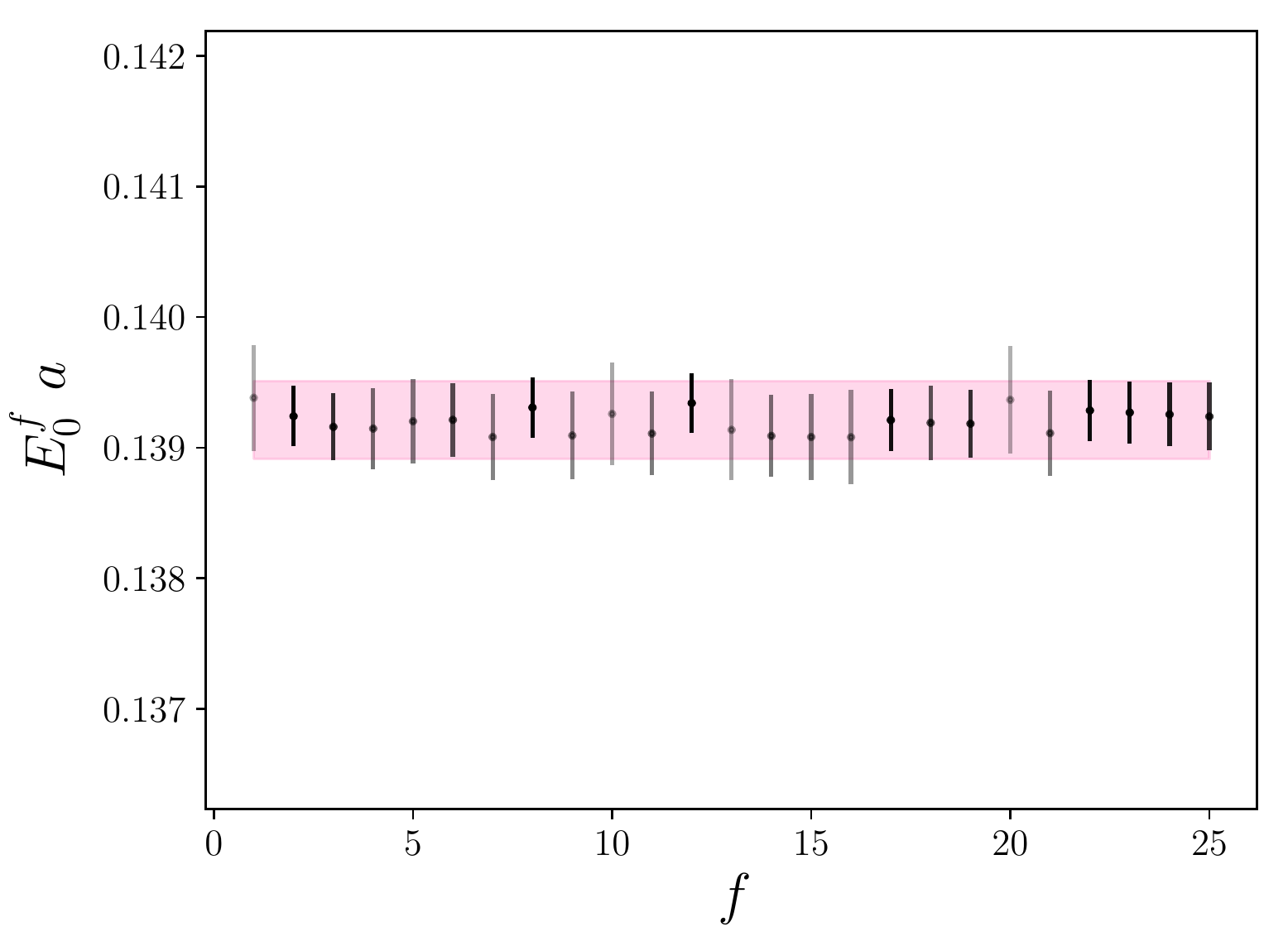} \\\vspace{-65pt}
     \begin{minipage}[c][150pt][t]{70pt} $2\ \overline{K^0}\newline L/a=48$ \end{minipage}
     \includegraphics[width=.40\textwidth]{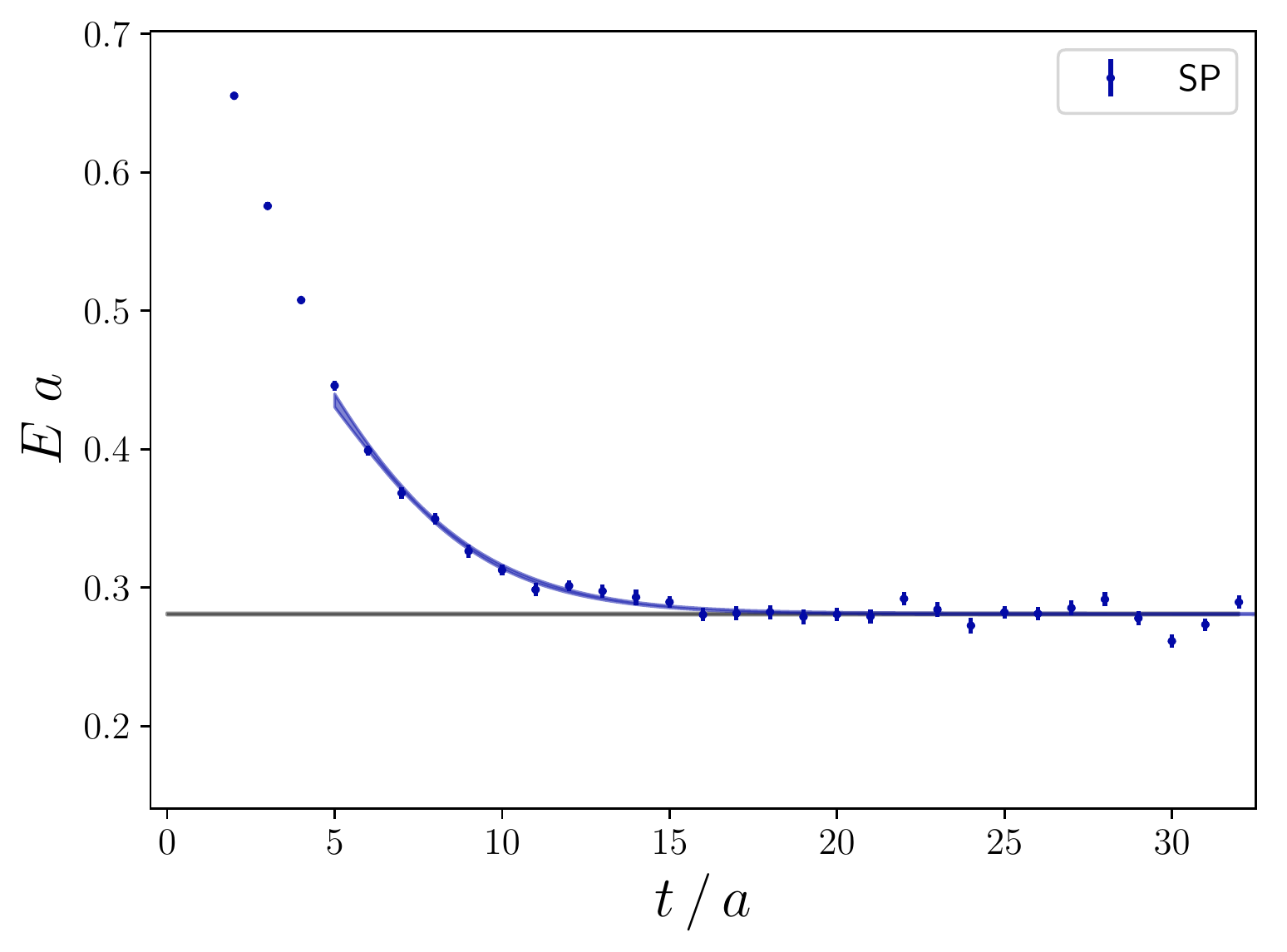}\hspace{10pt}
     \includegraphics[width=.40\textwidth]{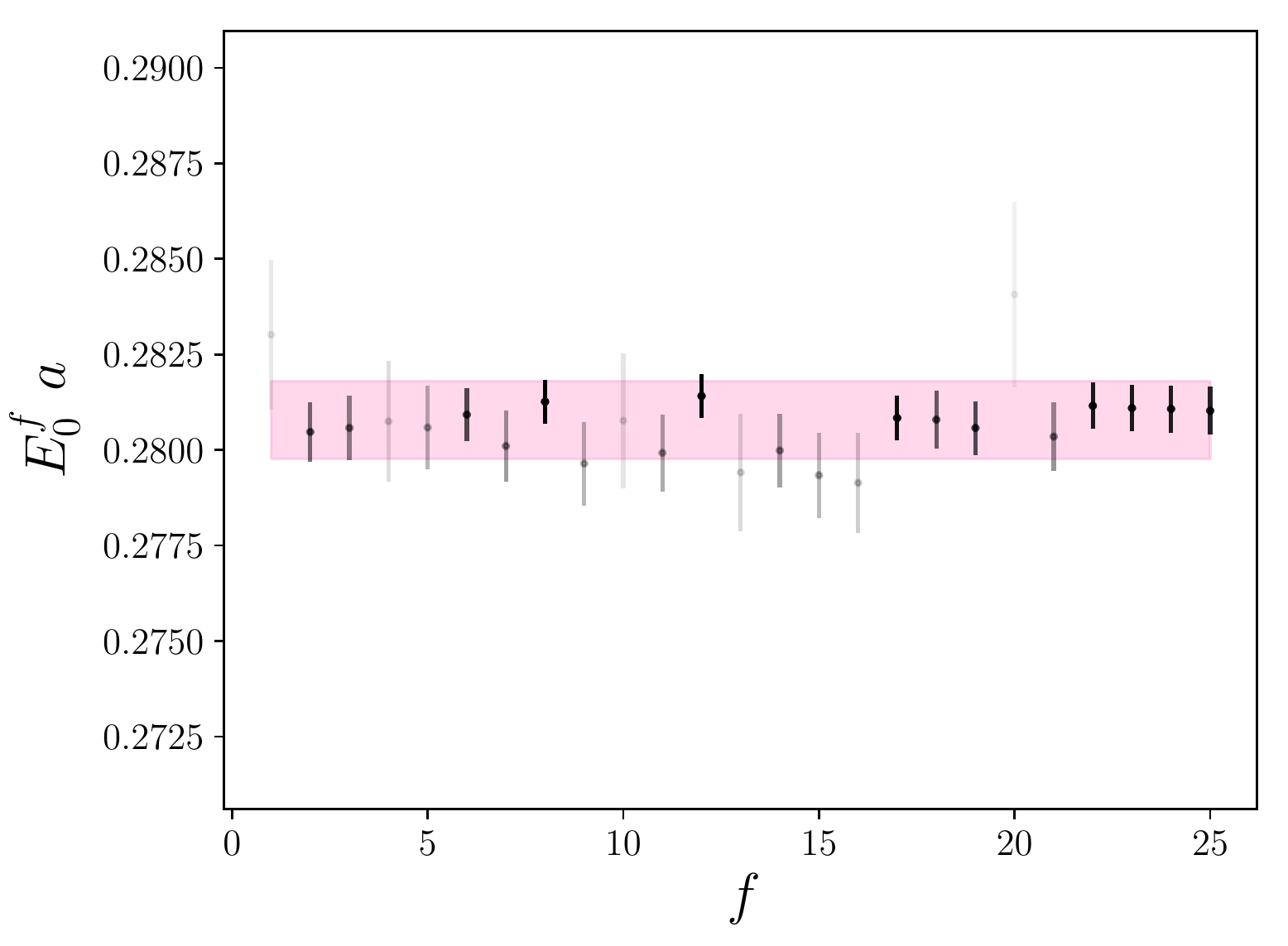} \\\vspace{-65pt}
     \begin{minipage}[c][150pt][t]{70pt} $3\ \overline{K^0}\newline L/a=48$ \end{minipage}
     \includegraphics[width=.40\textwidth]{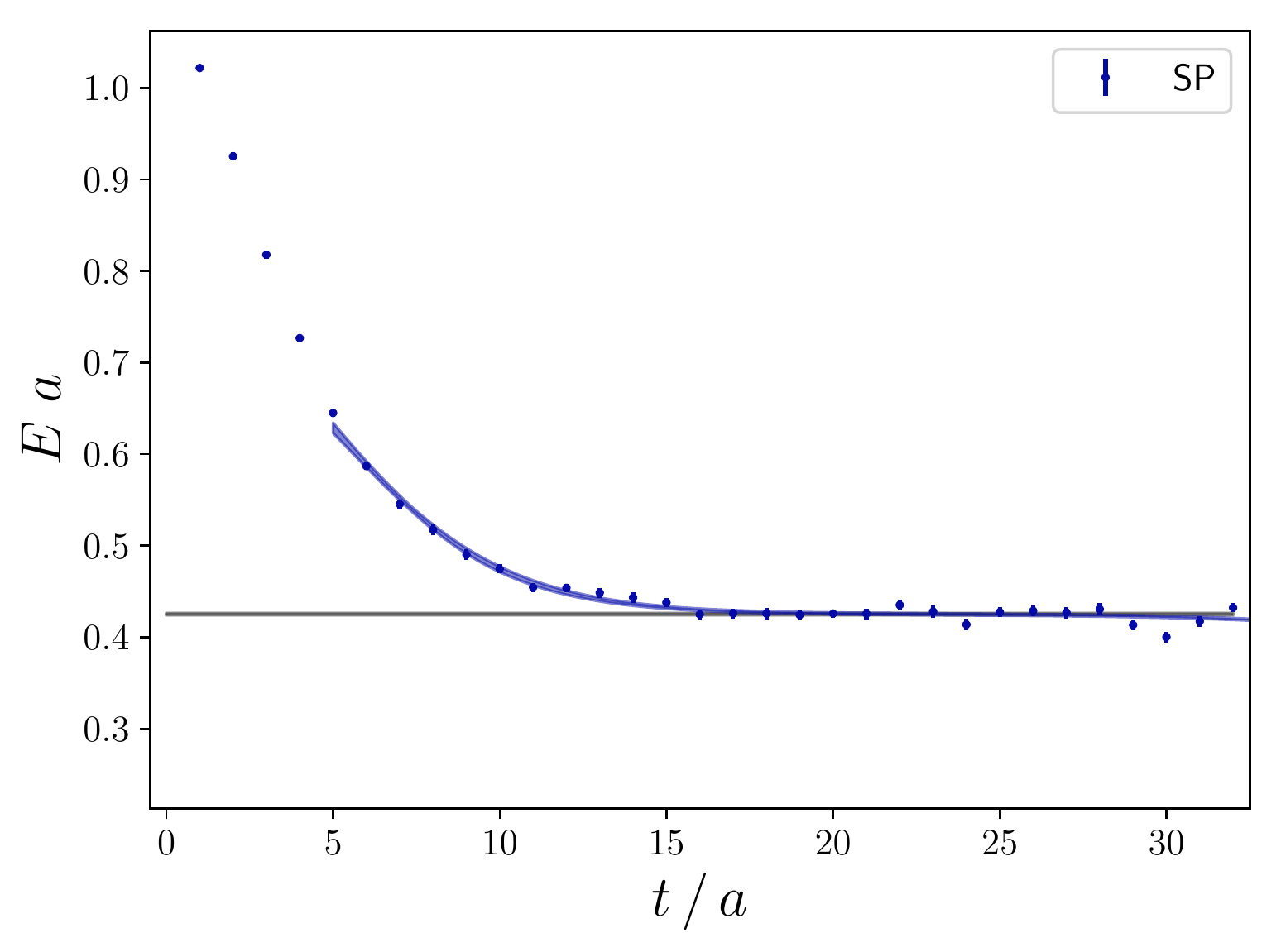}\hspace{10pt}
     \includegraphics[width=.40\textwidth]{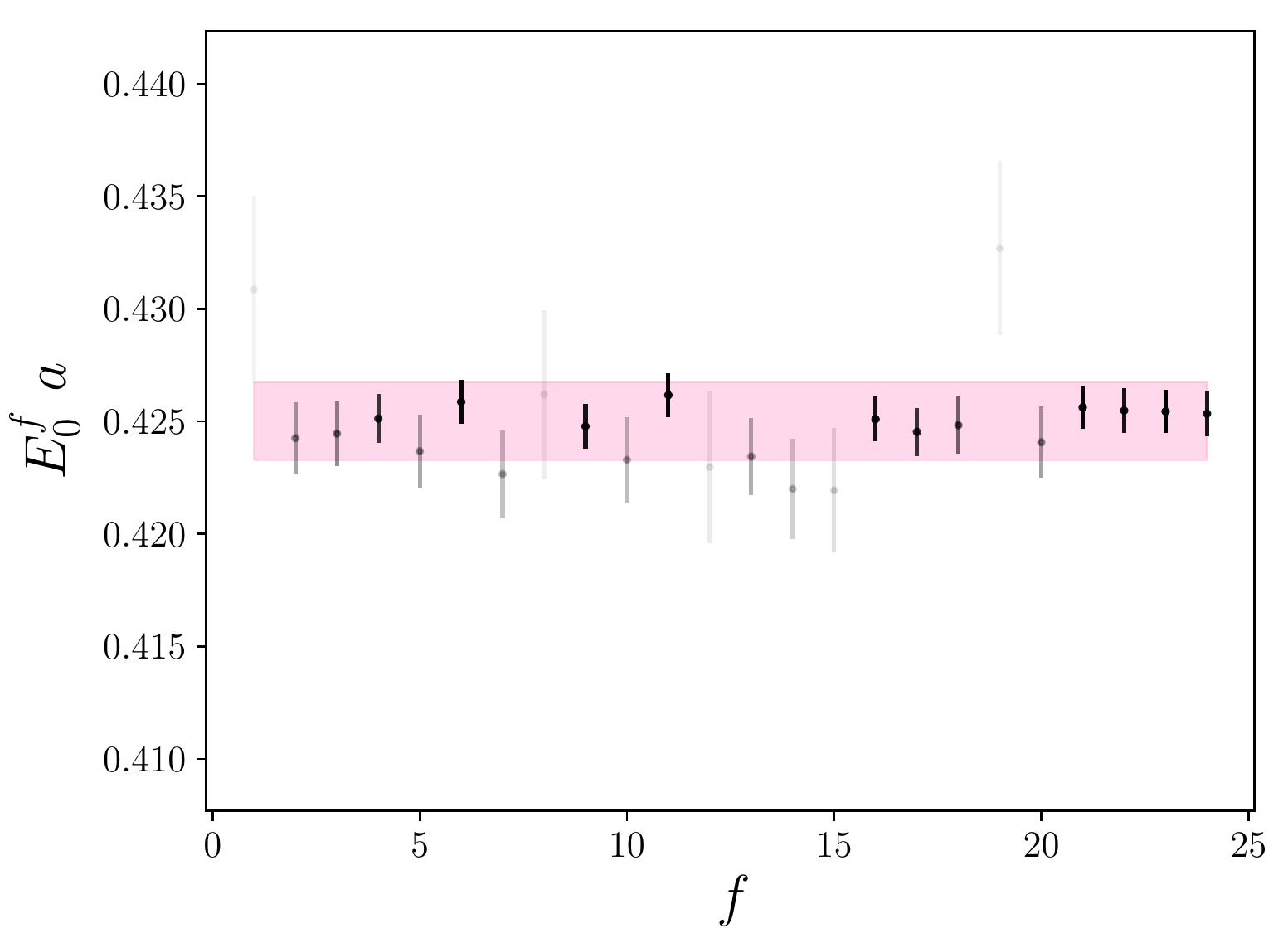} \\\vspace{-65pt}
     \begin{minipage}[c][150pt][t]{70pt} $4\ \overline{K^0}\newline L/a=48$ \end{minipage}
     \includegraphics[width=.40\textwidth]{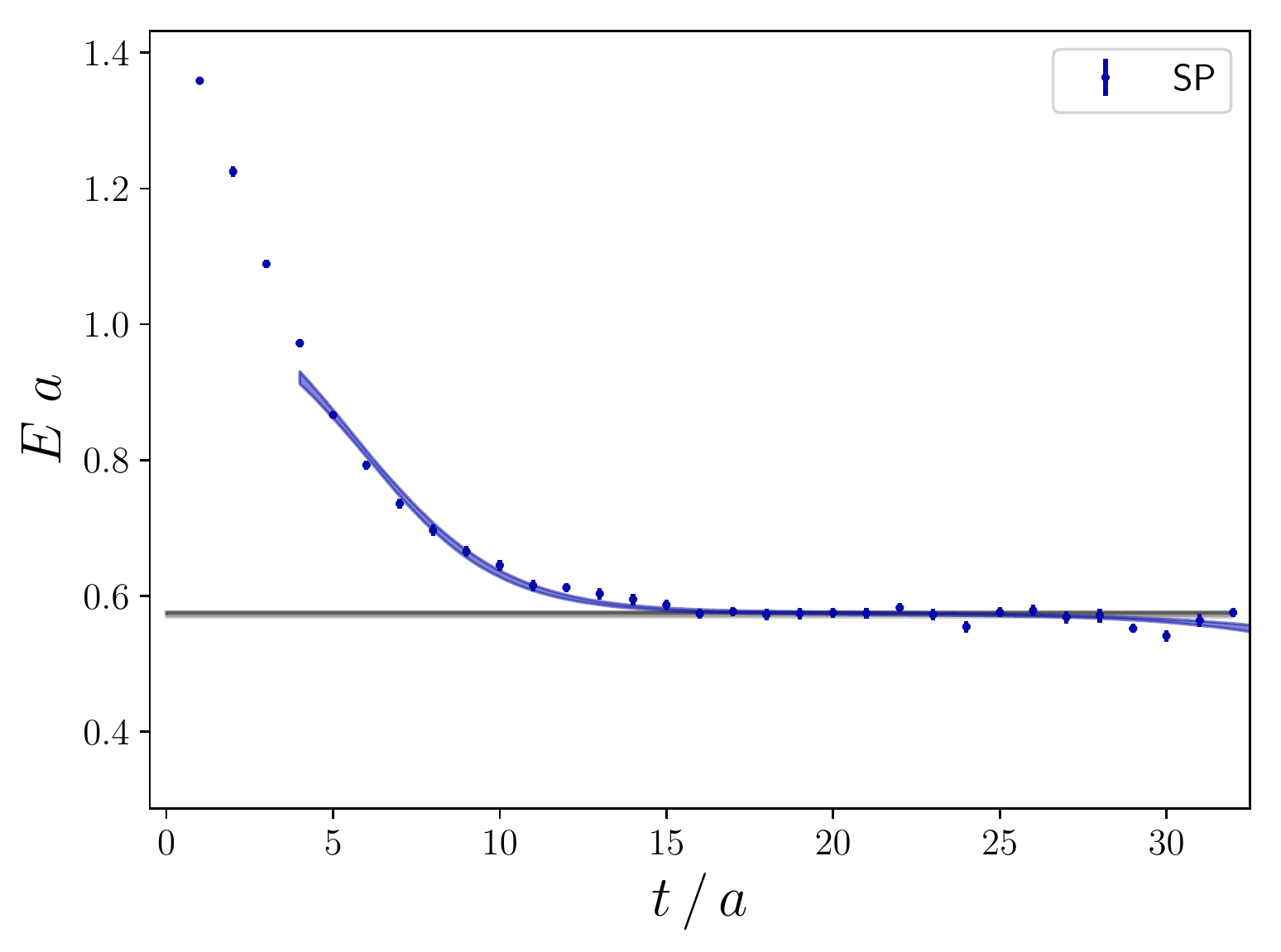}\hspace{10pt}
     \includegraphics[width=.40\textwidth]{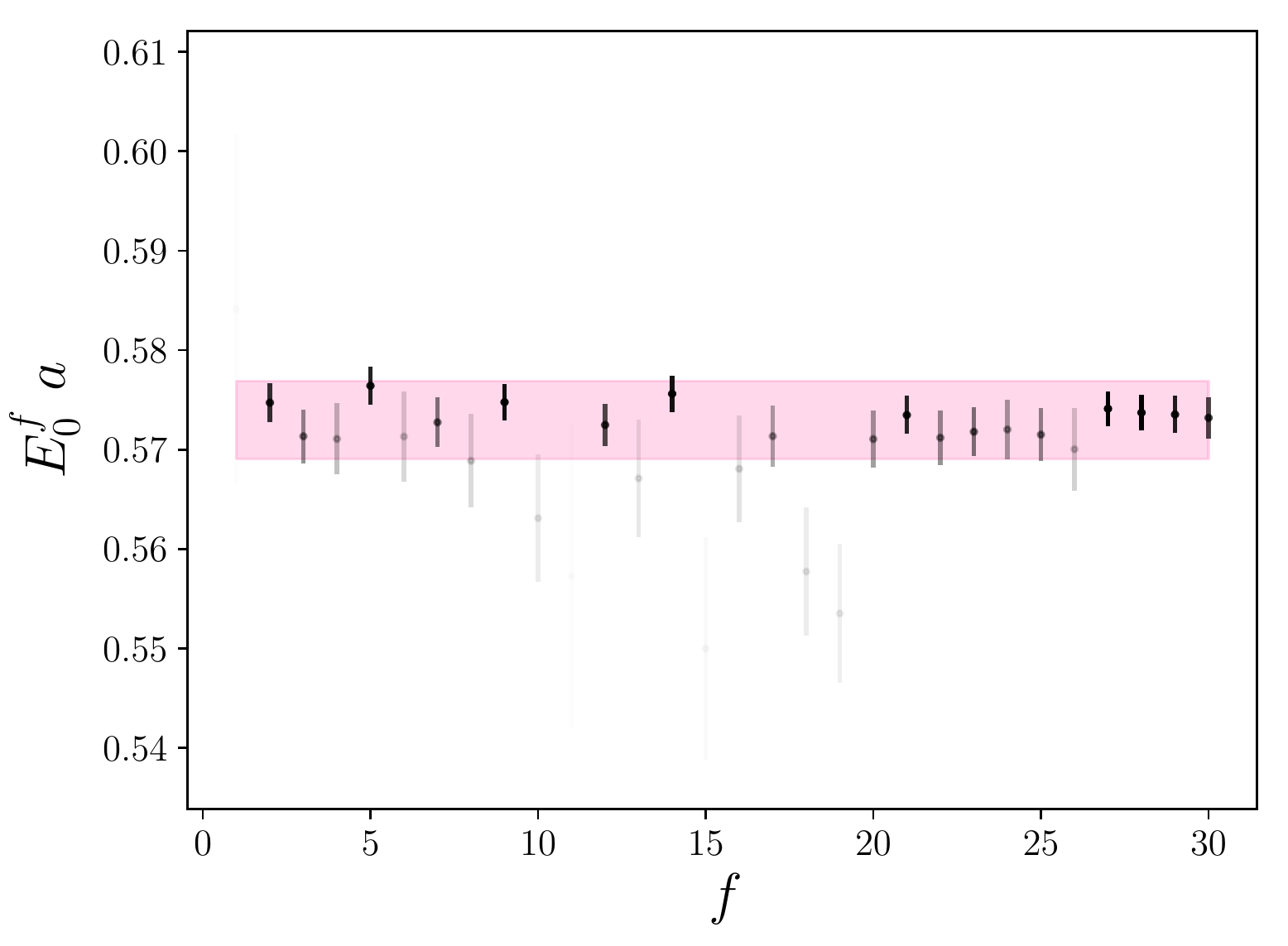} \\\vspace{-65pt}
   \caption{Fit results for systems of $n\in\{1,\dots,4\}$ $\overline{K^0}$ mesons for the $L/a = 48$ lattice volume. The figures are analogous to Fig.~\ref{fig:pion48a}, see Appendix~\ref{app:fits} for a definition of the fitting procedure employed.
    \label{fig:kaon48a}}
\end{figure}

\begin{figure}
  \centering
     \begin{minipage}[c][150pt][t]{70pt} $5\ \overline{K^0}\newline L/a=48$ \end{minipage}
     \includegraphics[width=.40\textwidth]{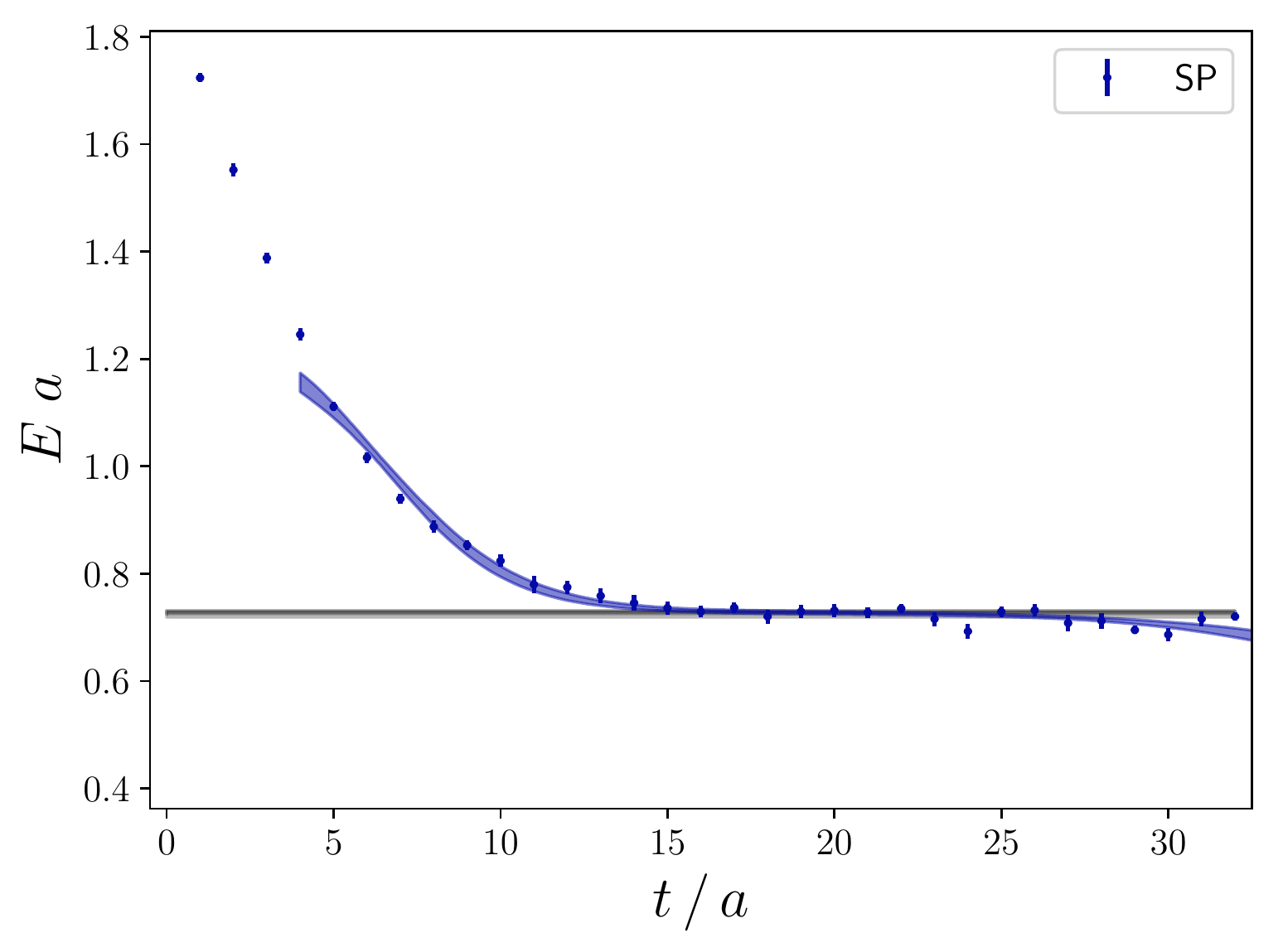}\hspace{10pt}
     \includegraphics[width=.40\textwidth]{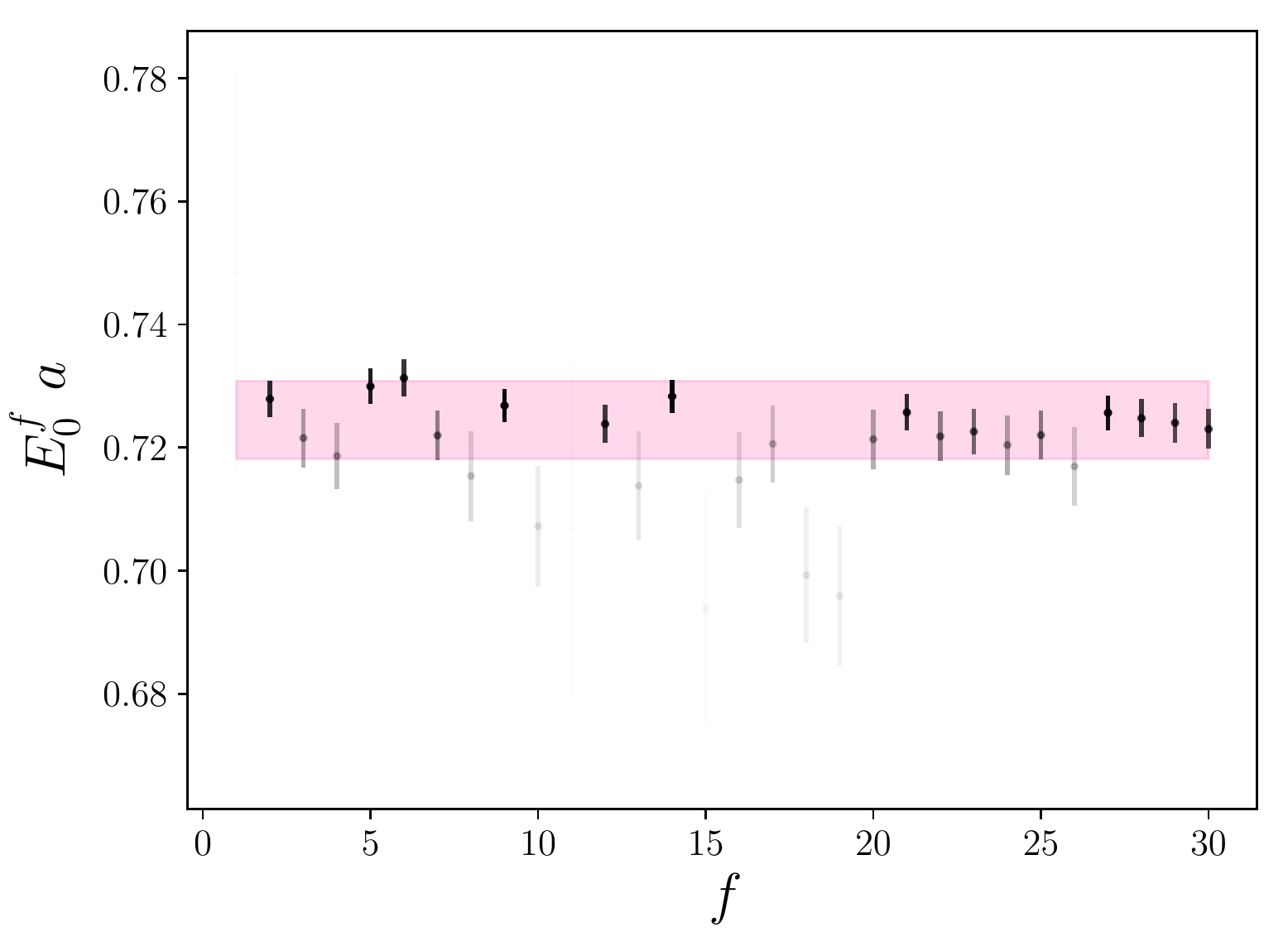} \\\vspace{-65pt}
     \begin{minipage}[c][150pt][t]{70pt} $6\ \overline{K^0}\newline L/a=48$ \end{minipage}
     \includegraphics[width=.40\textwidth]{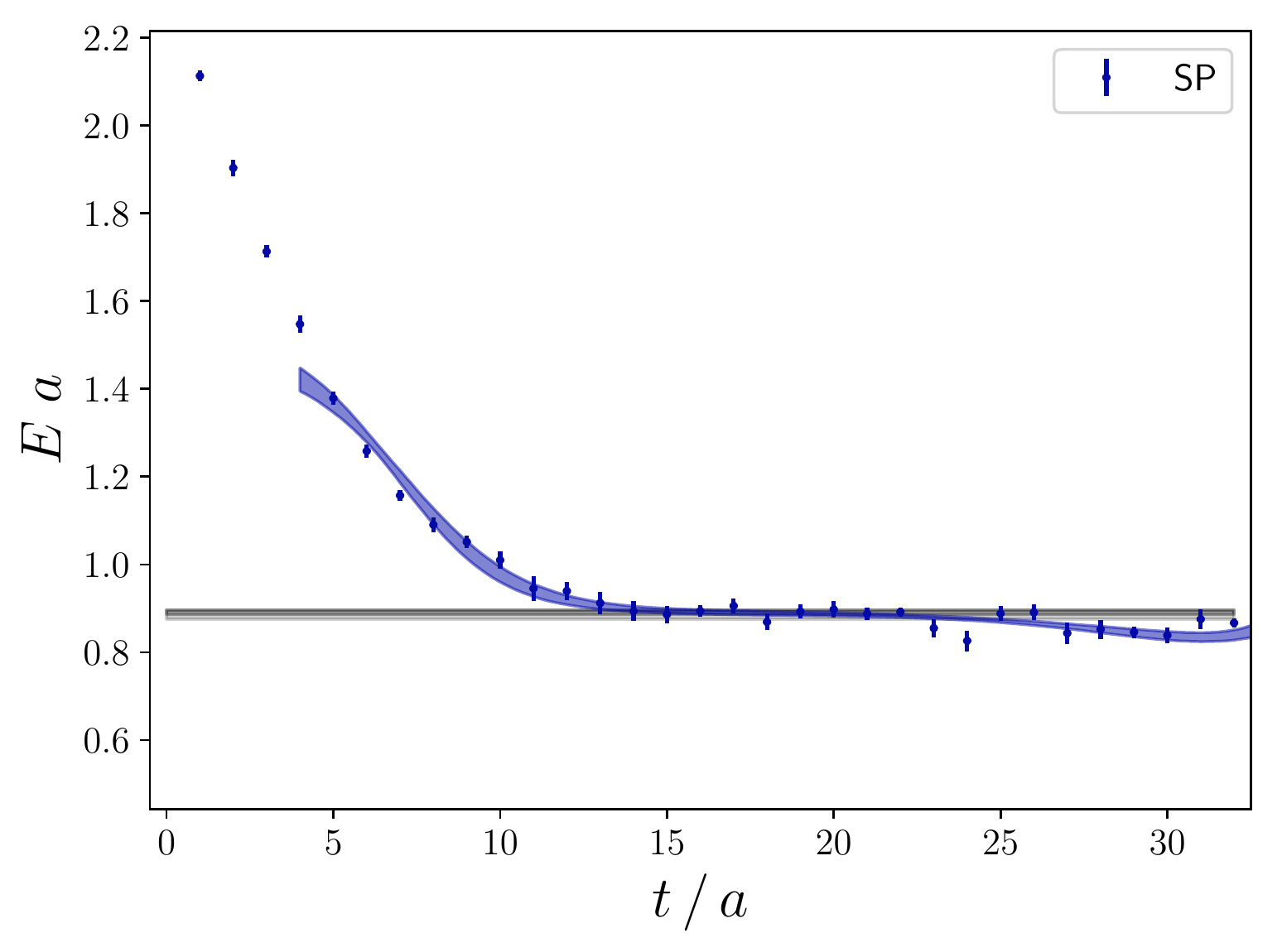}\hspace{10pt}
     \includegraphics[width=.40\textwidth]{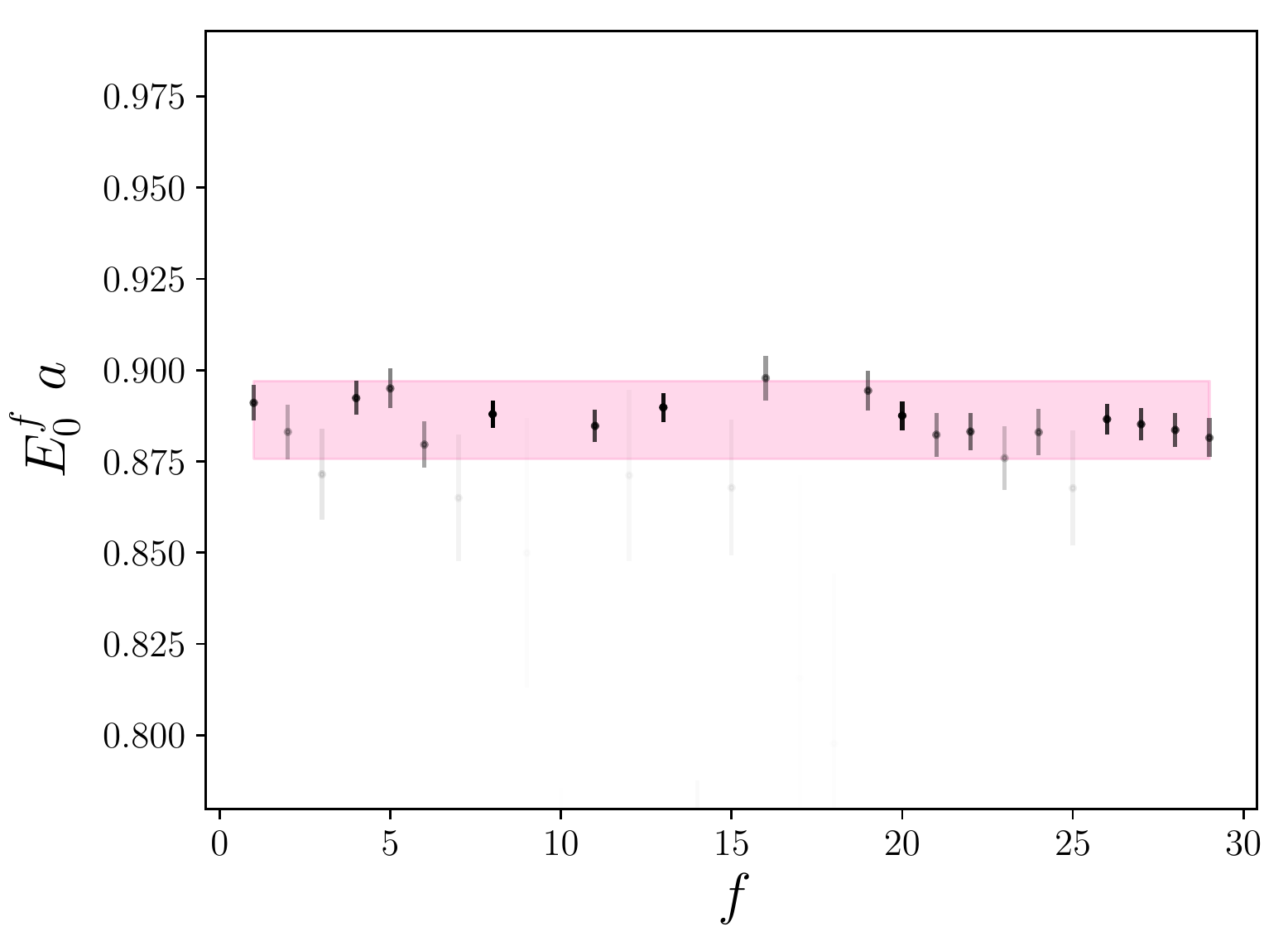} \\\vspace{-65pt}
     \begin{minipage}[c][150pt][t]{70pt} $7\ \overline{K^0}\newline L/a=48$ \end{minipage}
     \includegraphics[width=.40\textwidth]{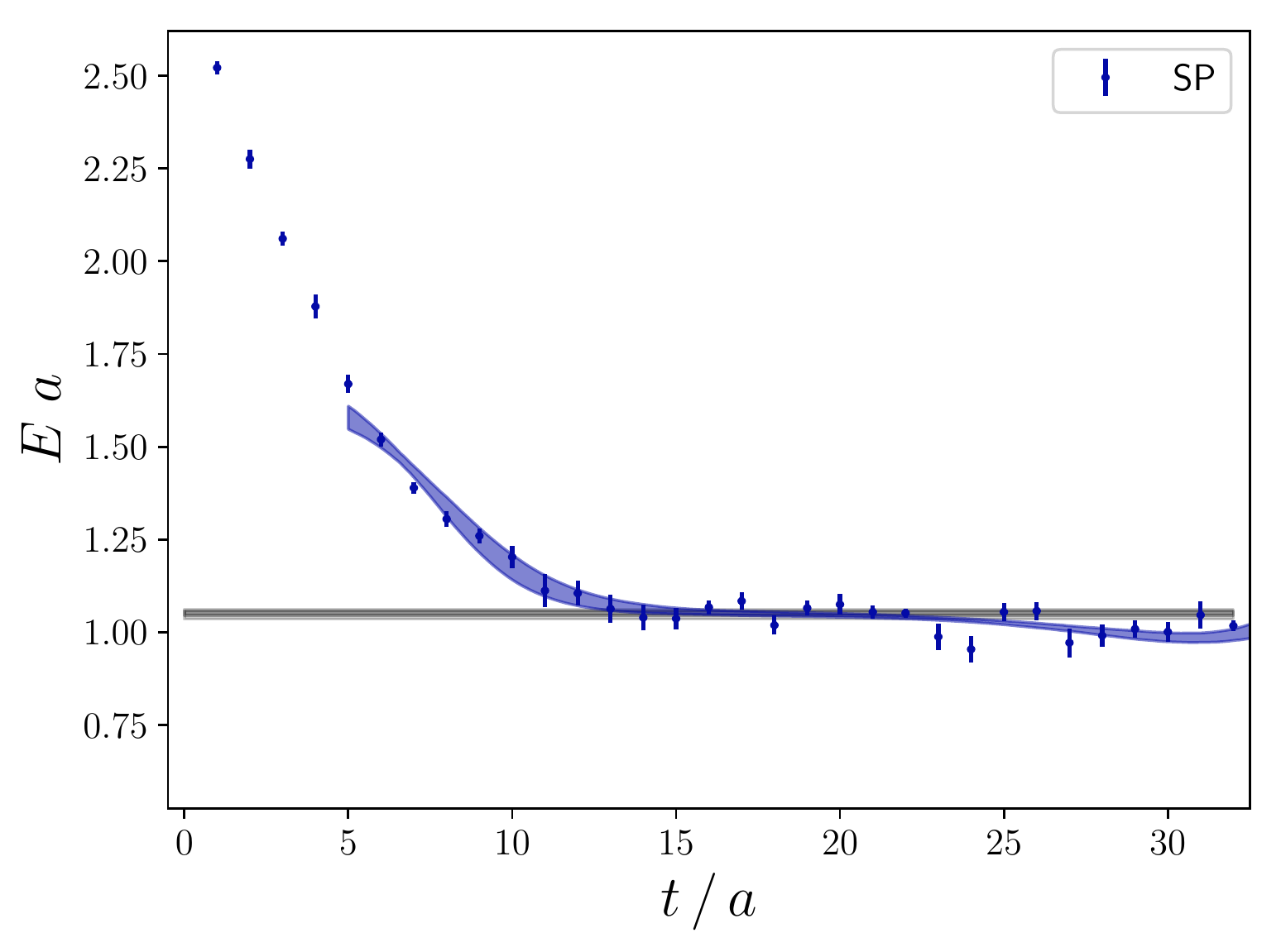}\hspace{10pt}
     \includegraphics[width=.40\textwidth]{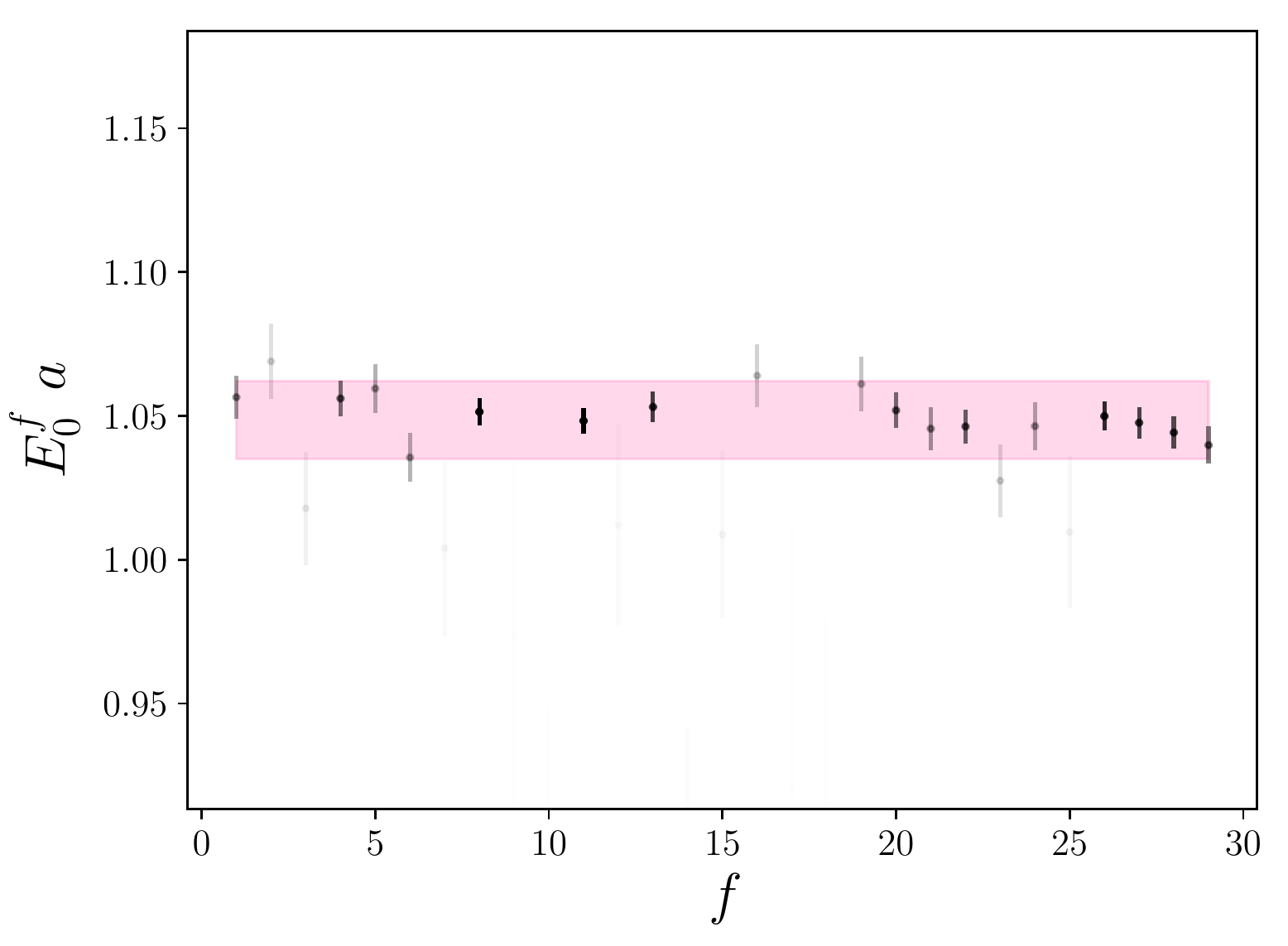} \\\vspace{-65pt}
     \begin{minipage}[c][150pt][t]{70pt} $8\ \overline{K^0}\newline L/a=48$ \end{minipage}
     \includegraphics[width=.40\textwidth]{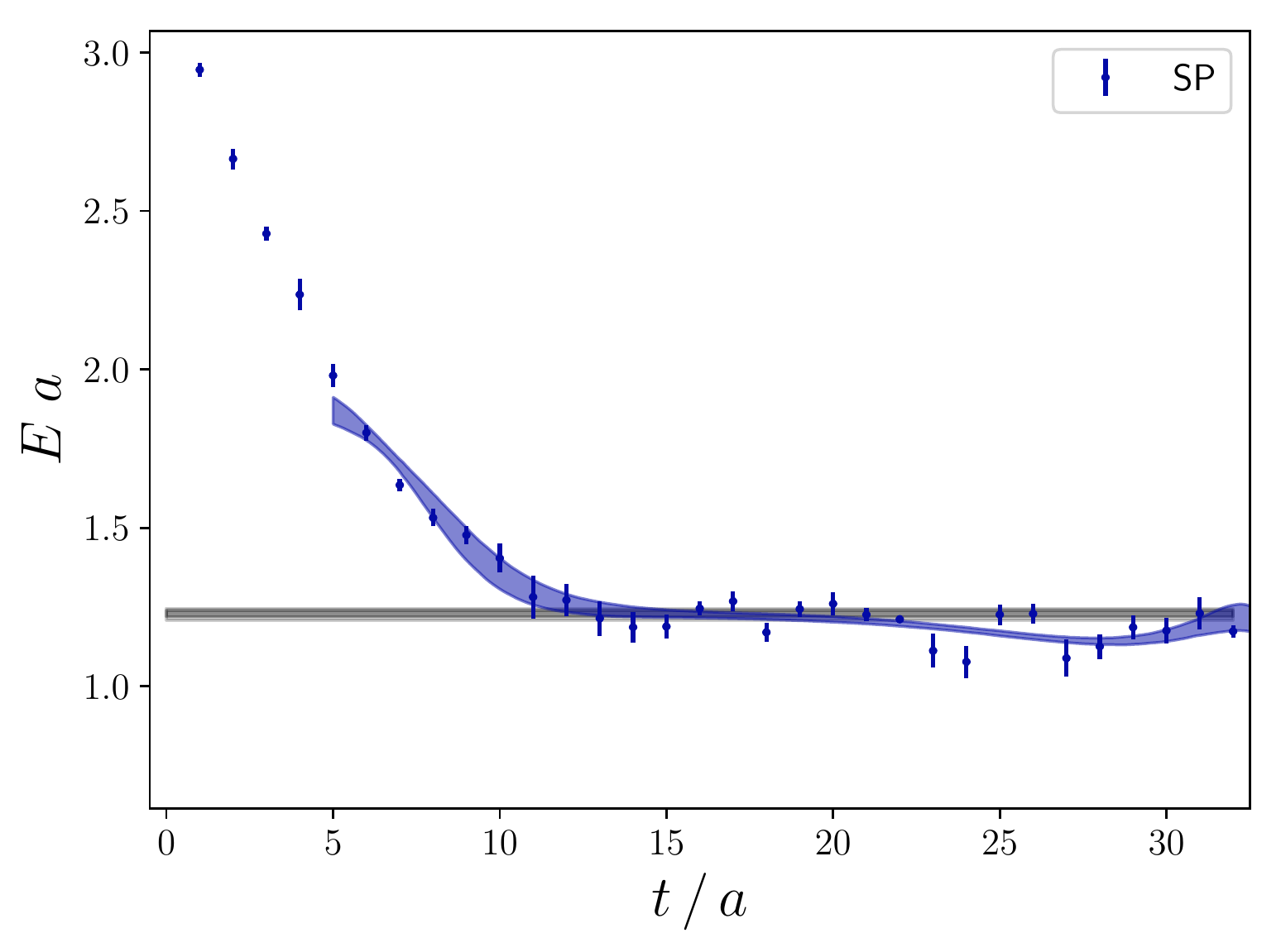}\hspace{10pt}
     \includegraphics[width=.40\textwidth]{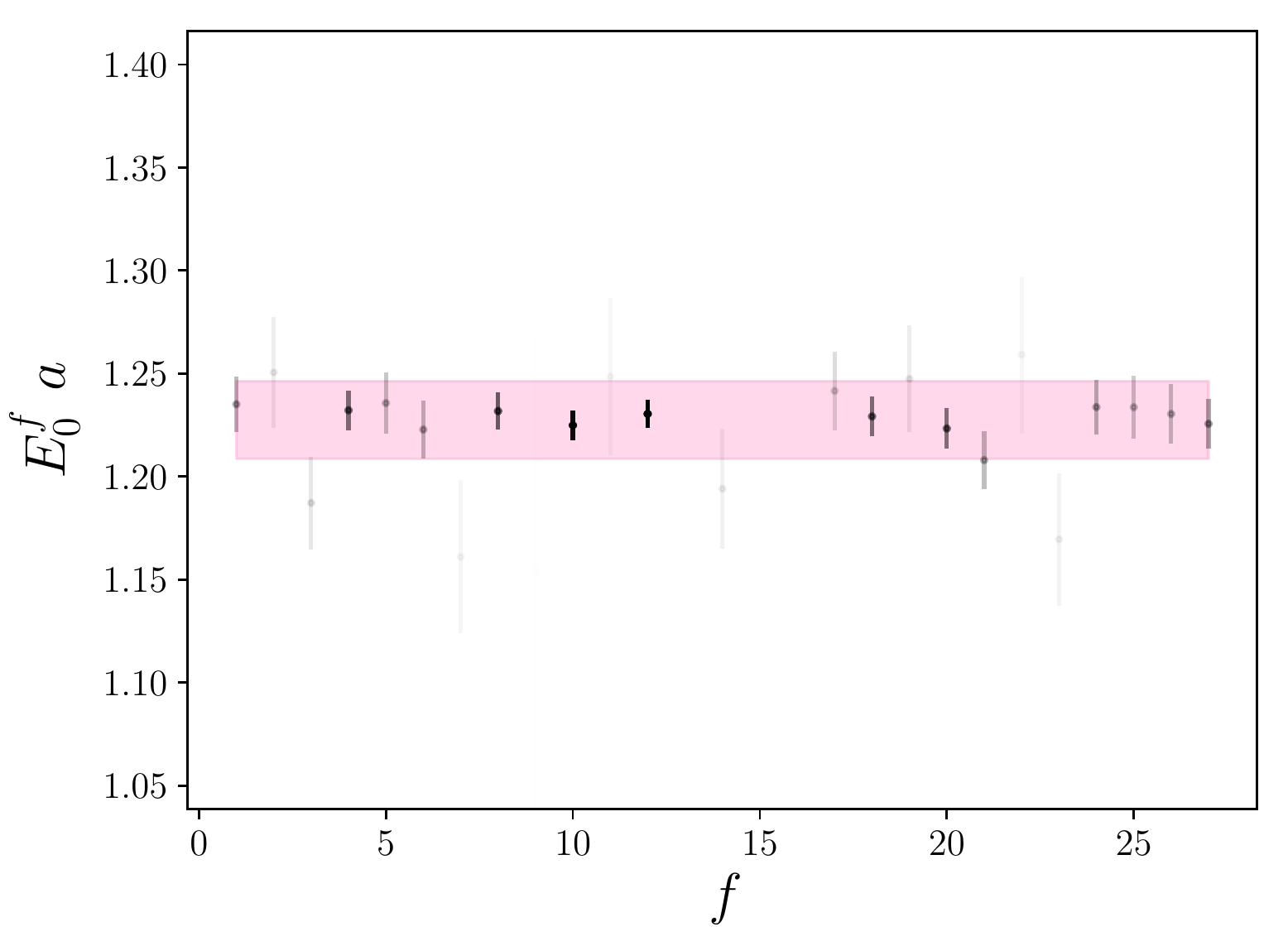} \\\vspace{-65pt}
   \caption{Fit results for systems of $n\in\{5,\dots,8\}$ $\overline{K^0}$ mesons for the $L/a = 48$ lattice volume. The figures are analogous to Fig.~\ref{fig:pion48a}, see Appendix~\ref{app:fits} for a definition of the fitting procedure employed.
    \label{fig:kaon48b}}
\end{figure}

\begin{figure}
  \centering
     \begin{minipage}[c][150pt][t]{70pt} $9\ \overline{K^0}\newline L/a=48$ \end{minipage}
     \includegraphics[width=.40\textwidth]{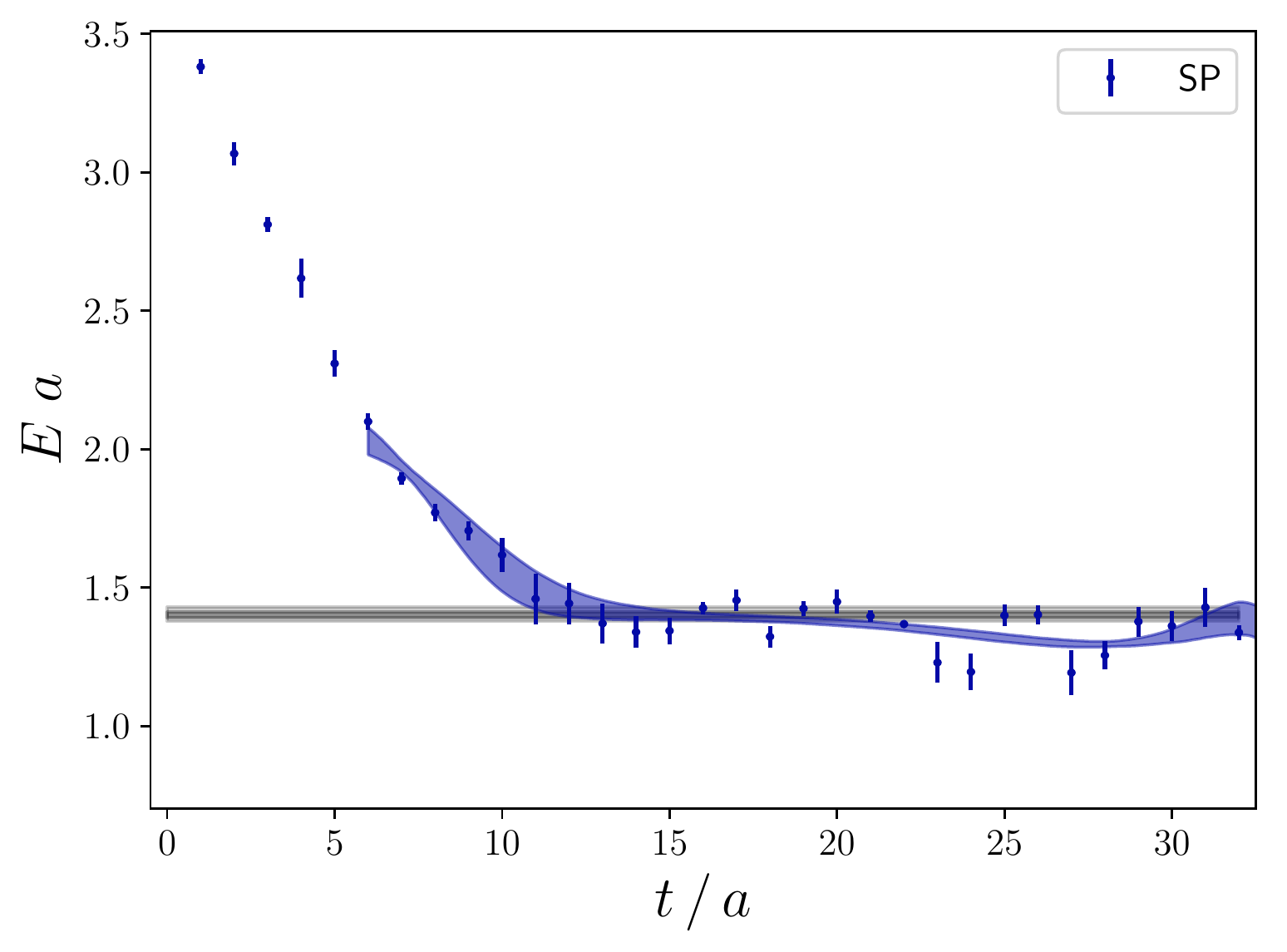}\hspace{10pt}
     \includegraphics[width=.40\textwidth]{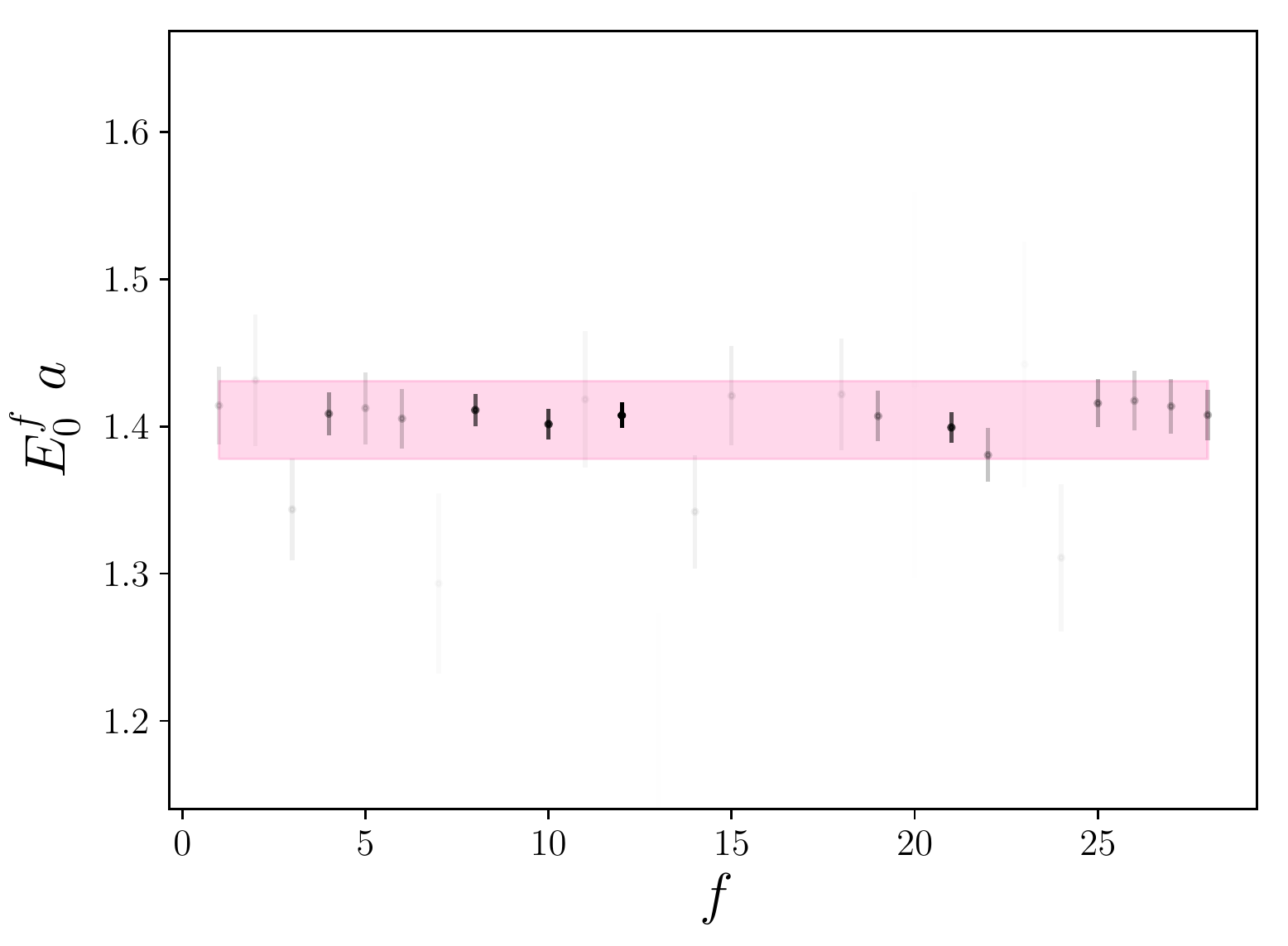} \\\vspace{-65pt}
     \begin{minipage}[c][150pt][t]{70pt} $10\ \overline{K^0}\newline L/a=48$ \end{minipage}
     \includegraphics[width=.40\textwidth]{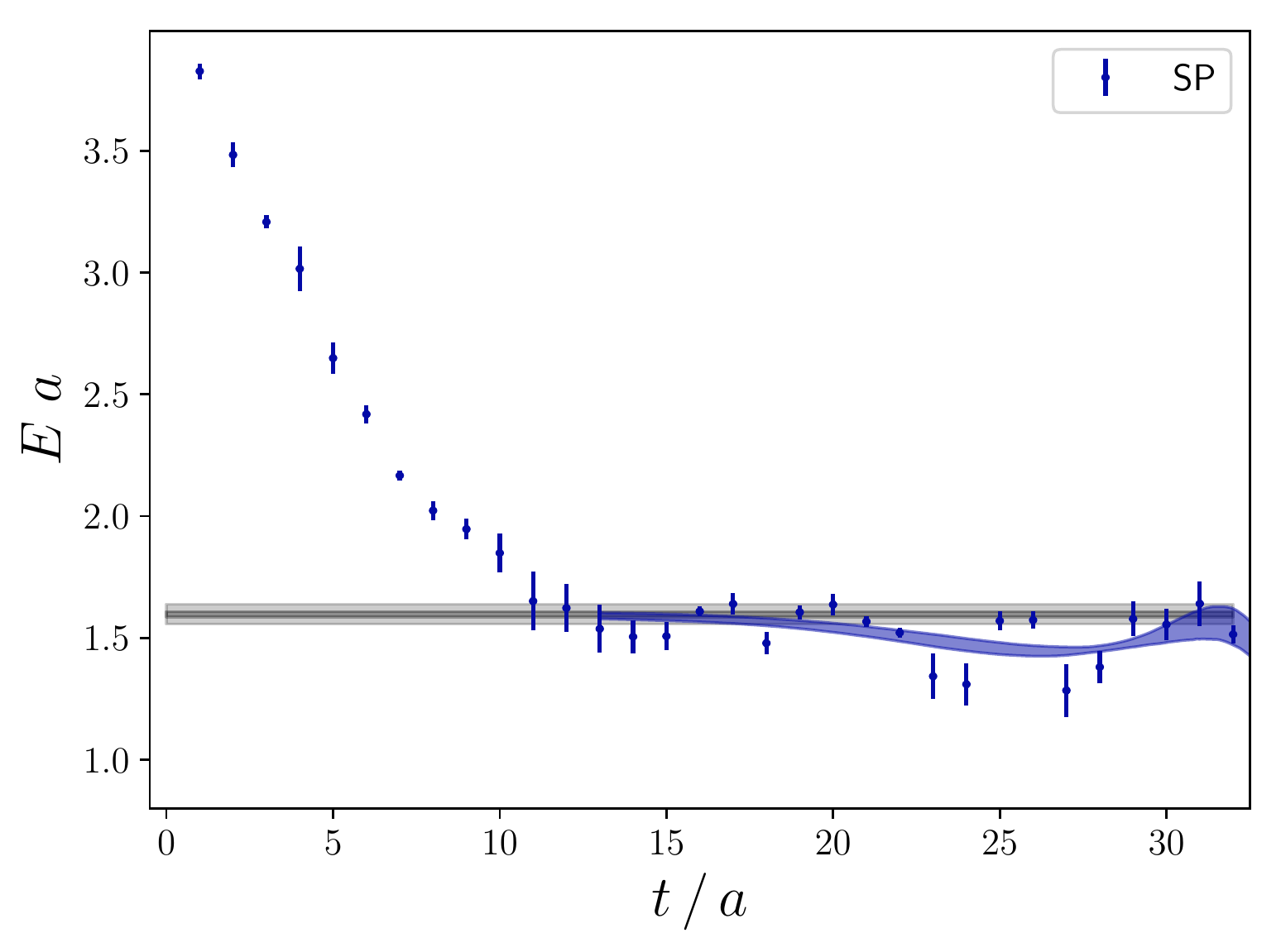}\hspace{10pt}
     \includegraphics[width=.40\textwidth]{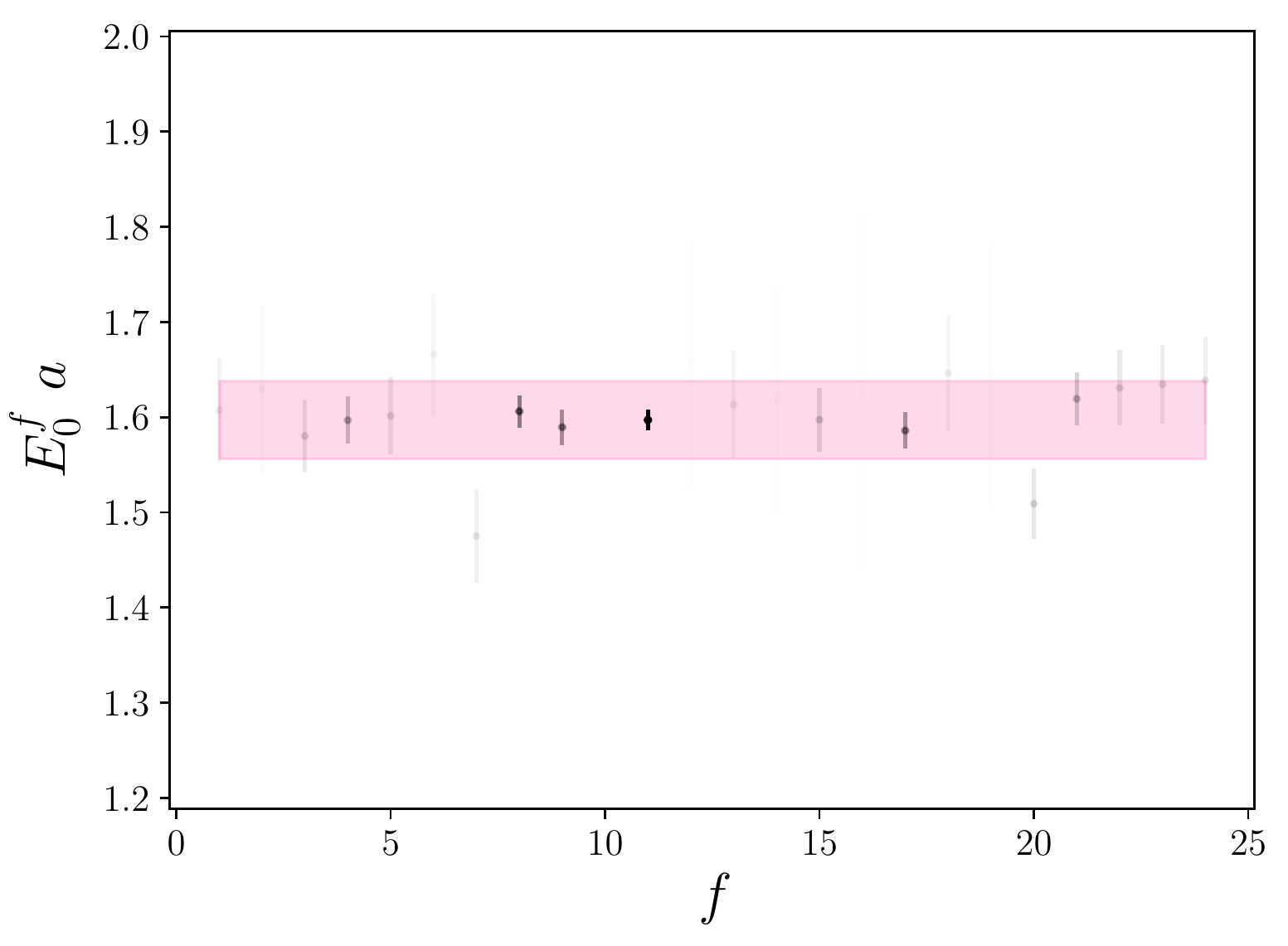} \\\vspace{-65pt}
     \begin{minipage}[c][150pt][t]{70pt} $11\ \overline{K^0}\newline L/a=48$ \end{minipage}
     \includegraphics[width=.40\textwidth]{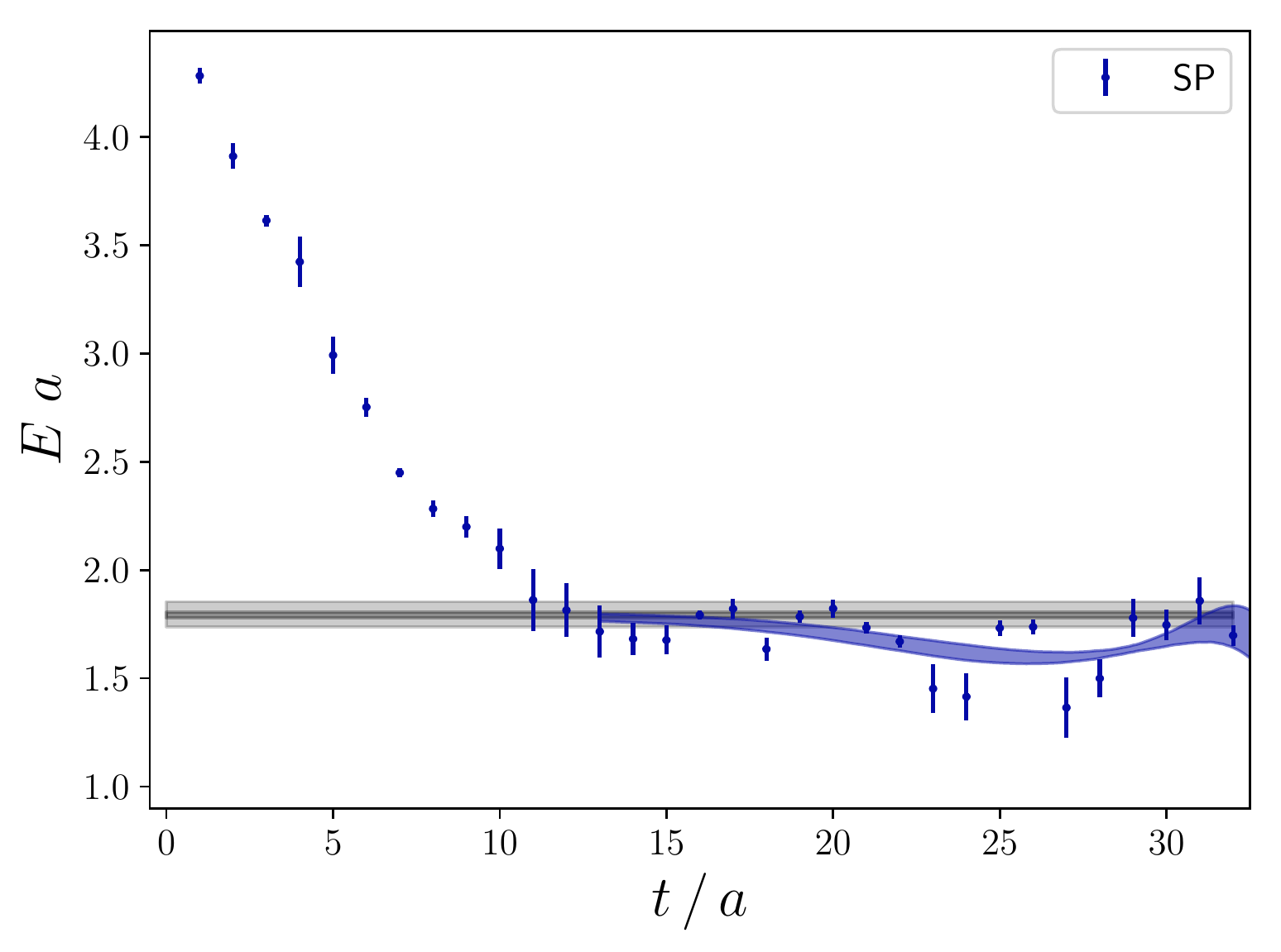}\hspace{10pt}
     \includegraphics[width=.40\textwidth]{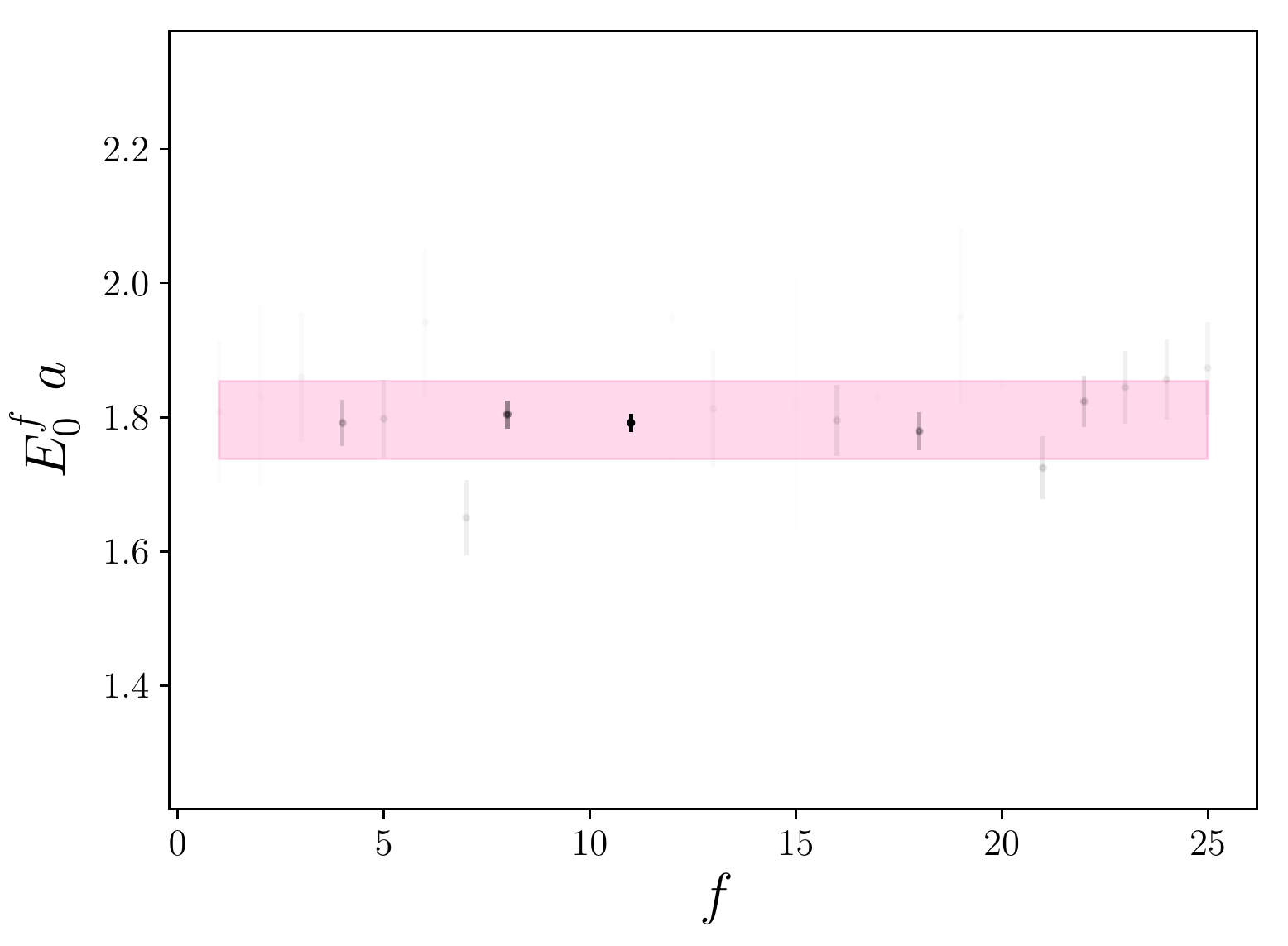} \\\vspace{-65pt}
     \begin{minipage}[c][150pt][t]{70pt} $12\ \overline{K^0}\newline L/a=48$ \end{minipage}
     \includegraphics[width=.40\textwidth]{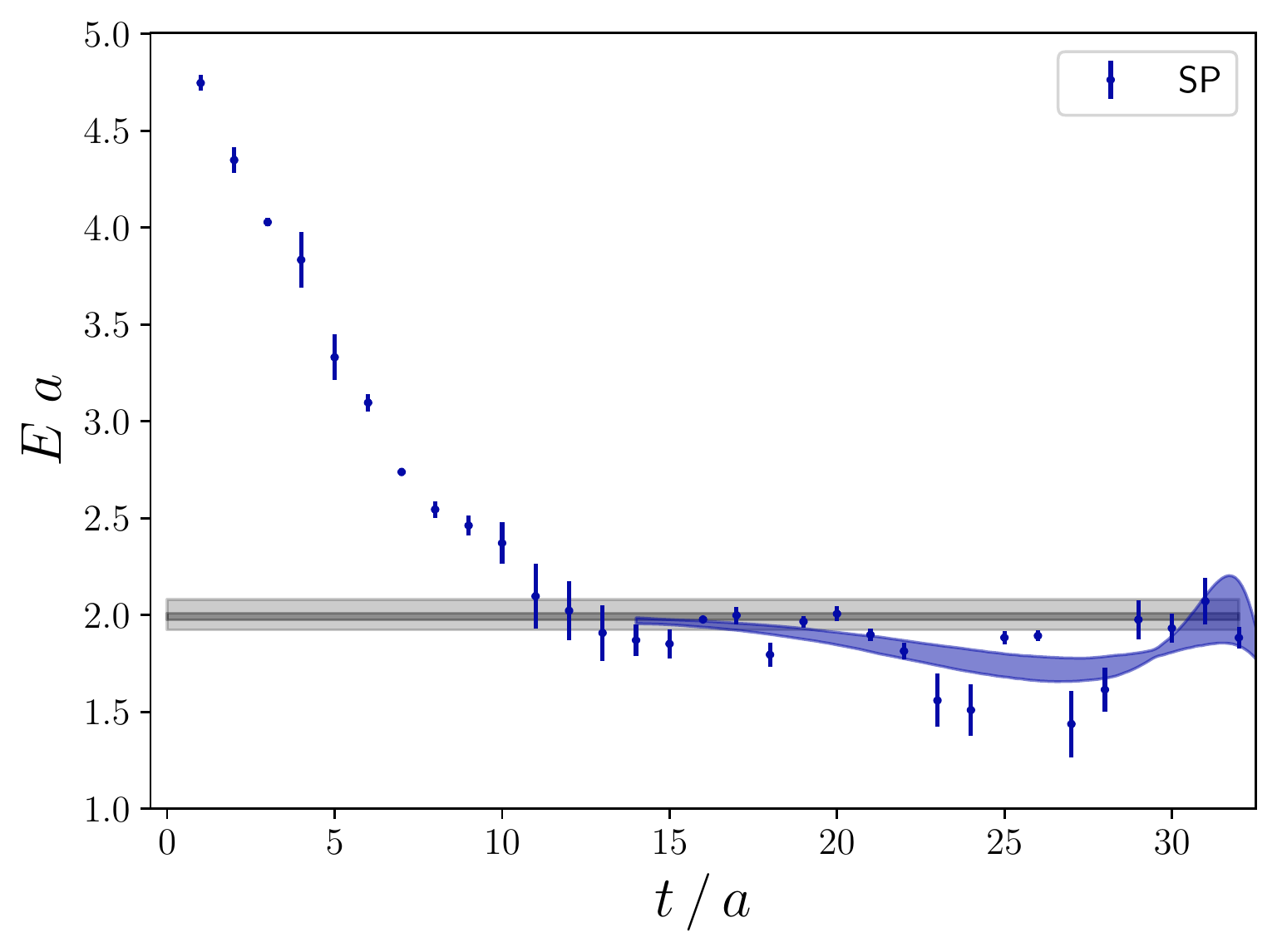}\hspace{10pt}
     \includegraphics[width=.40\textwidth]{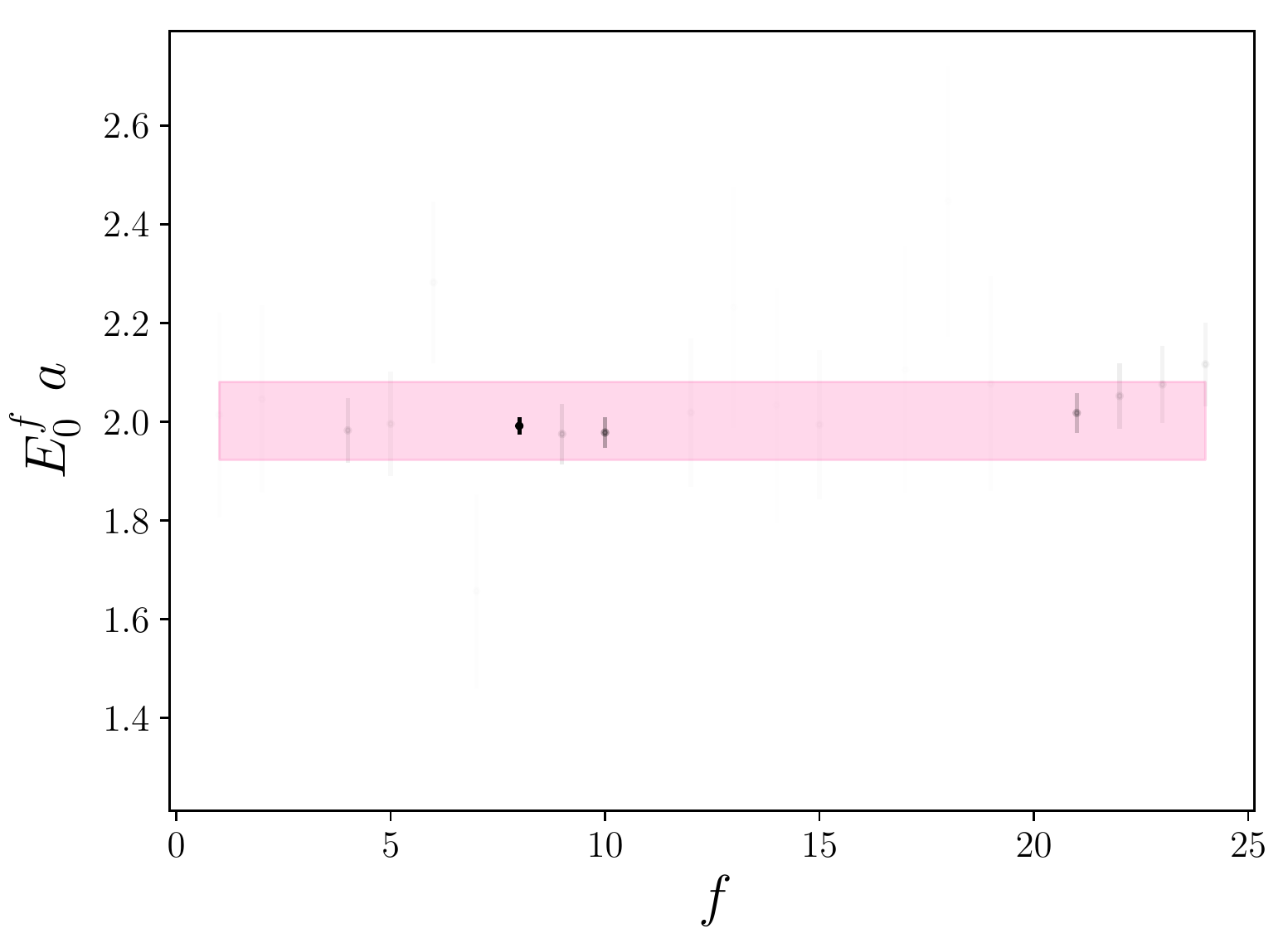} \\\vspace{-65pt}
   \caption{Fit results for systems of $n\in\{9,\dots,12\}$ $\overline{K^0}$ mesons for the $L/a = 48$ lattice volume. The figures are analogous to Fig.~\ref{fig:pion48a}, see Appendix~\ref{app:fits} for a definition of the fitting procedure employed.
    \label{fig:kaon48c}}
\end{figure}

\begin{figure}
  \centering
     \begin{minipage}[c][150pt][t]{70pt} $1\ \overline{K^0}\newline L/a=32$ \end{minipage}
     \includegraphics[width=.40\textwidth]{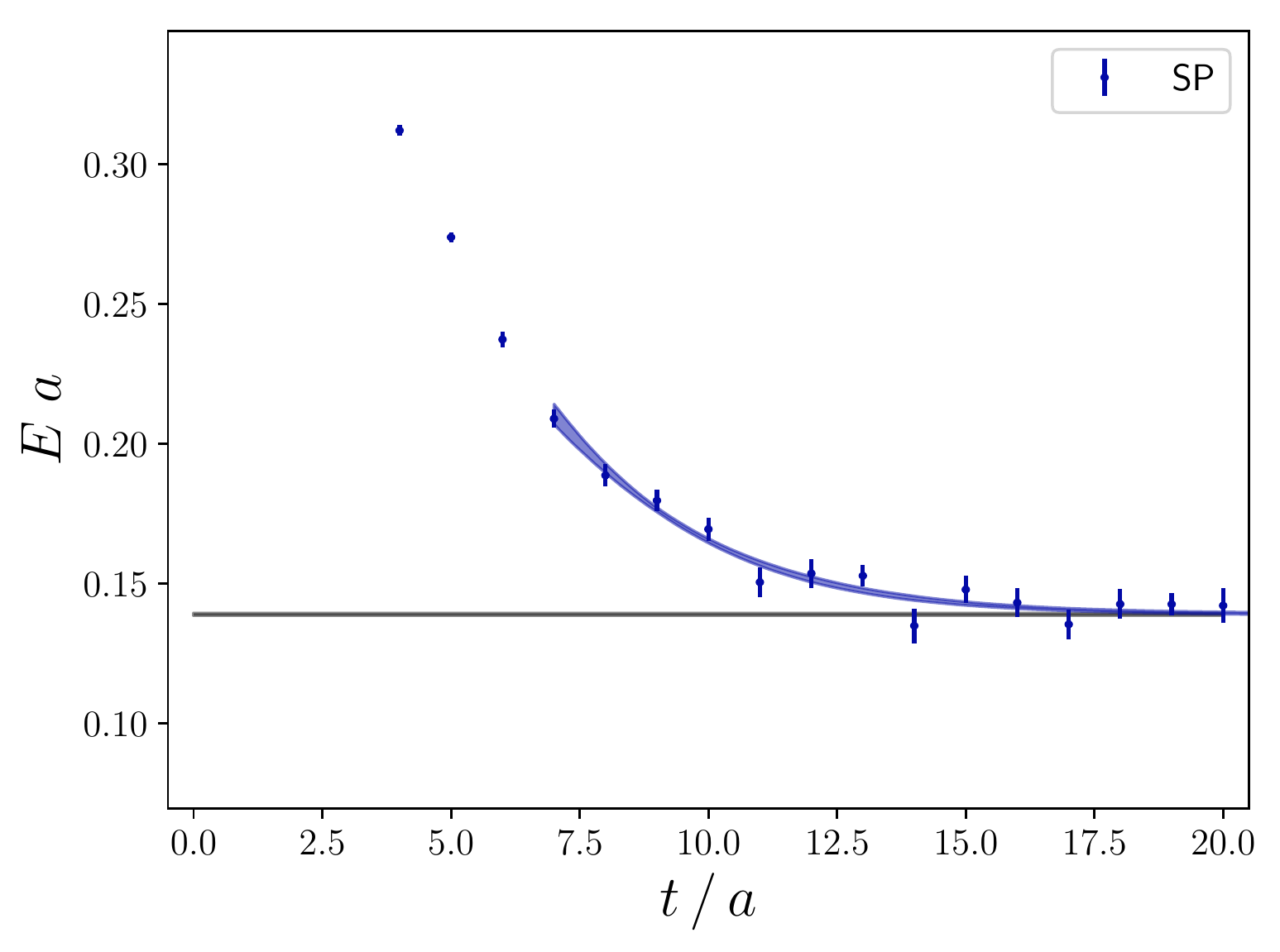}\hspace{10pt}
     \includegraphics[width=.40\textwidth]{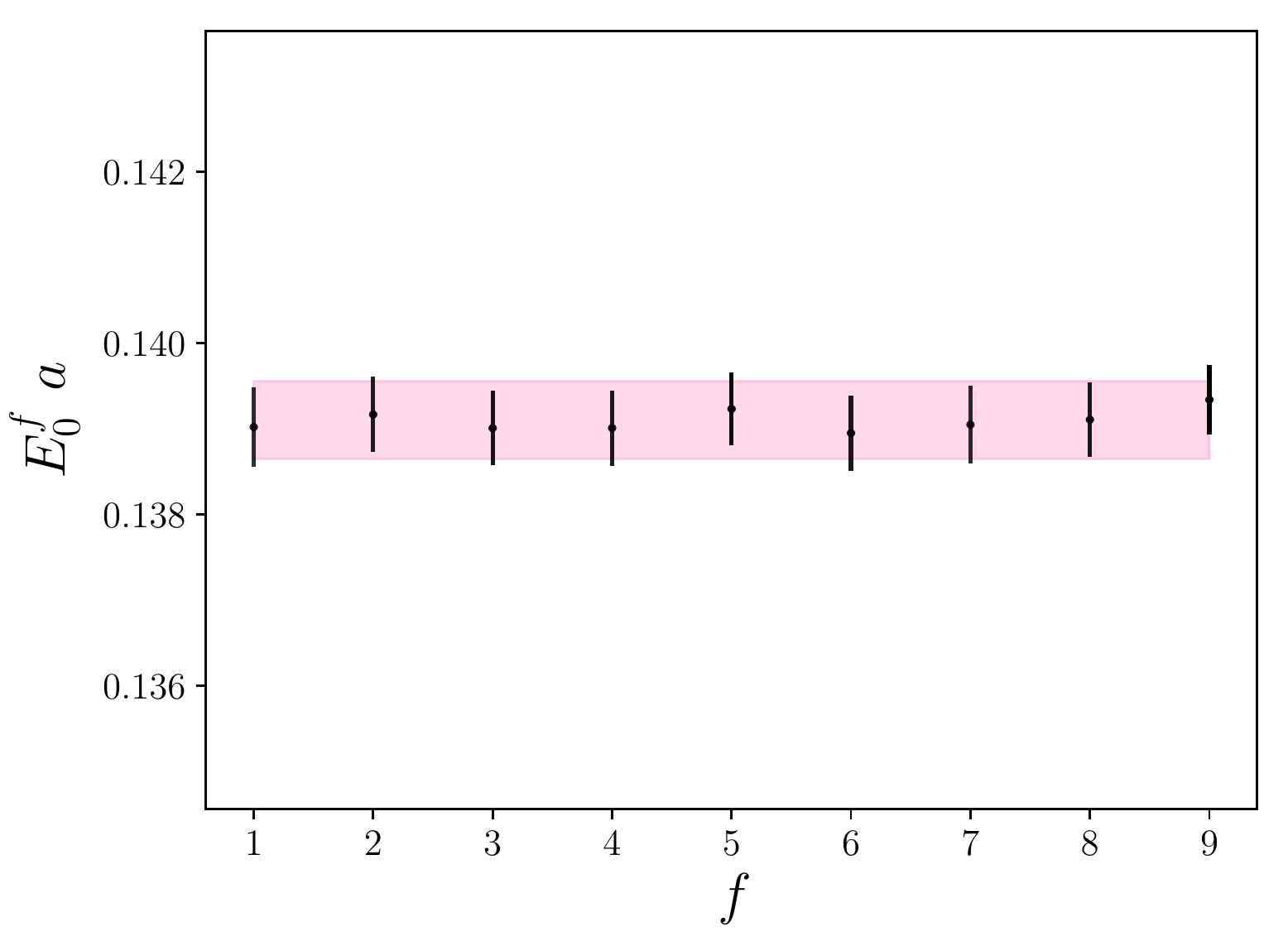} \\\vspace{-65pt}
     \begin{minipage}[c][150pt][t]{70pt} $2\ \overline{K^0}\newline L/a=32$ \end{minipage}
     \includegraphics[width=.40\textwidth]{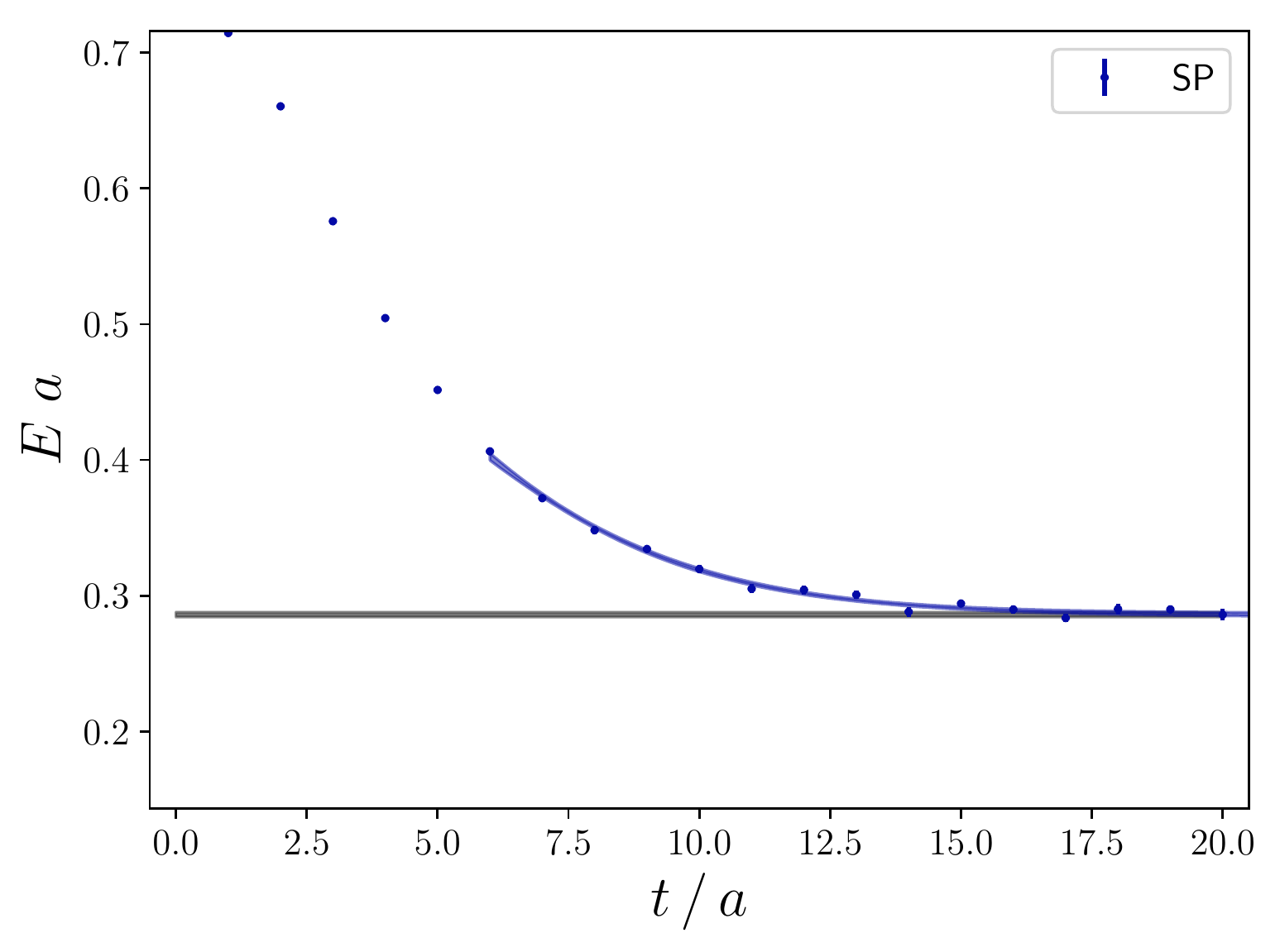}\hspace{10pt}
     \includegraphics[width=.40\textwidth]{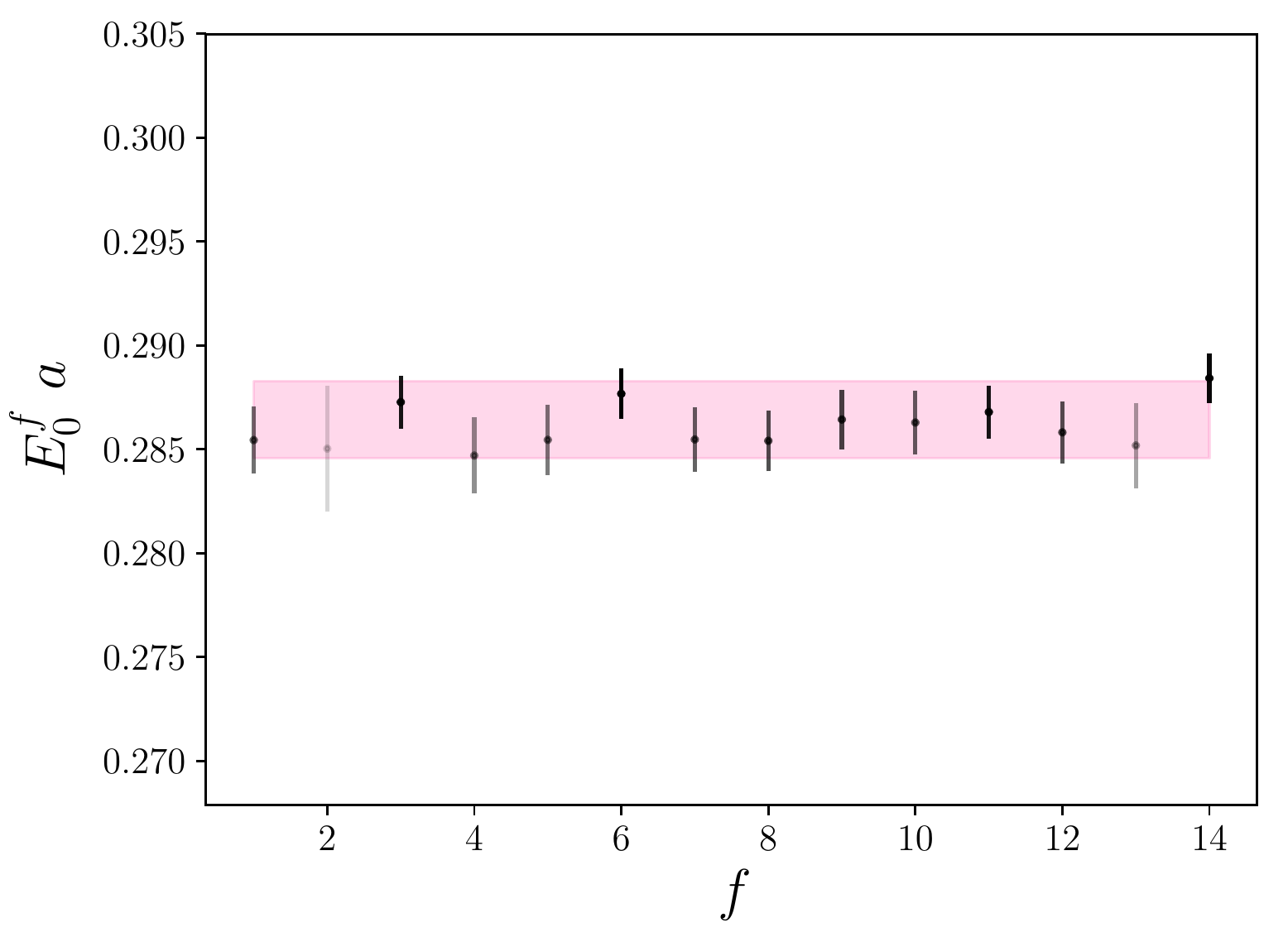} \\\vspace{-65pt}
     \begin{minipage}[c][150pt][t]{70pt} $3\ \overline{K^0}\newline L/a=32$ \end{minipage}
     \includegraphics[width=.40\textwidth]{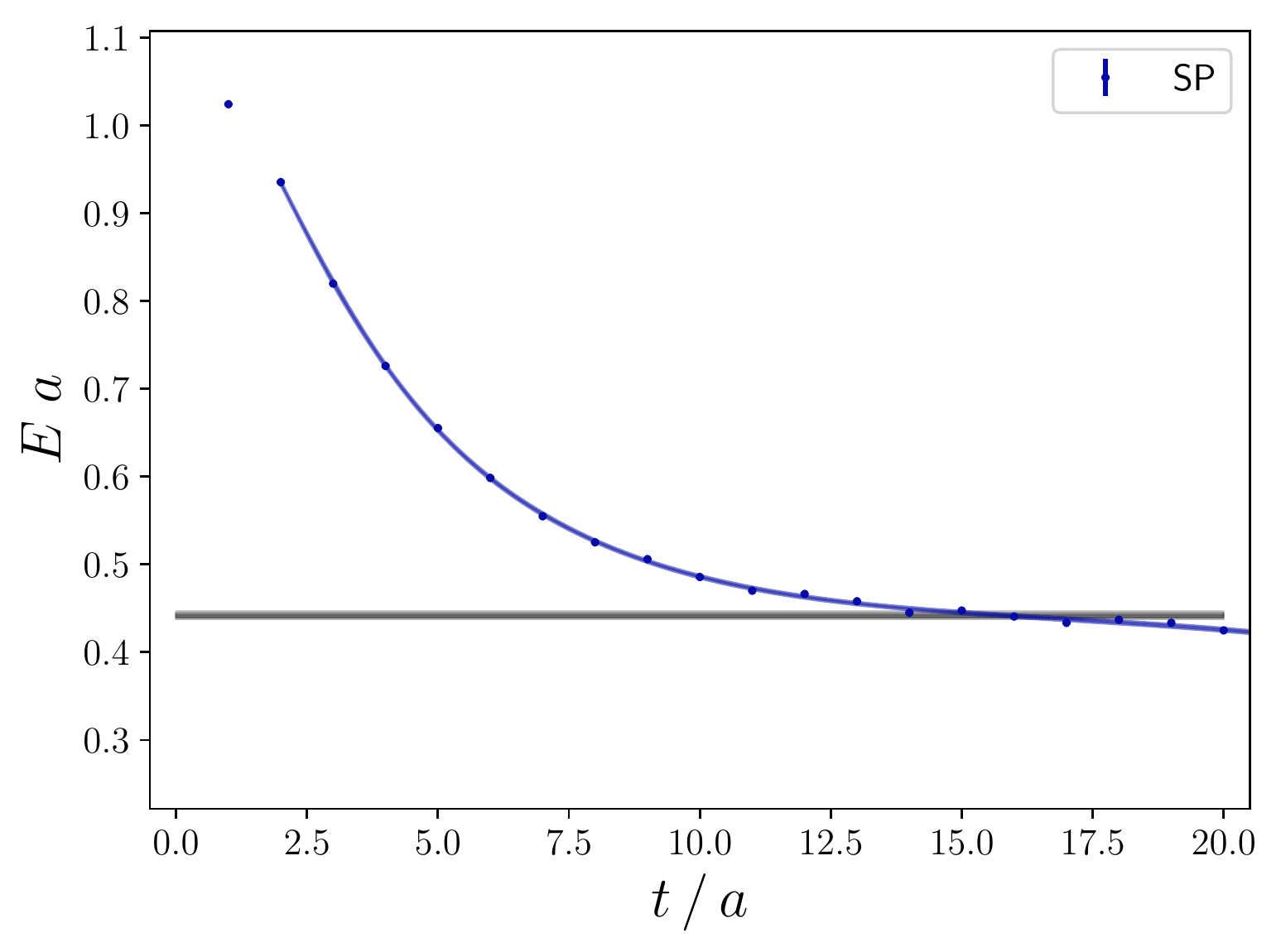}\hspace{10pt}
     \includegraphics[width=.40\textwidth]{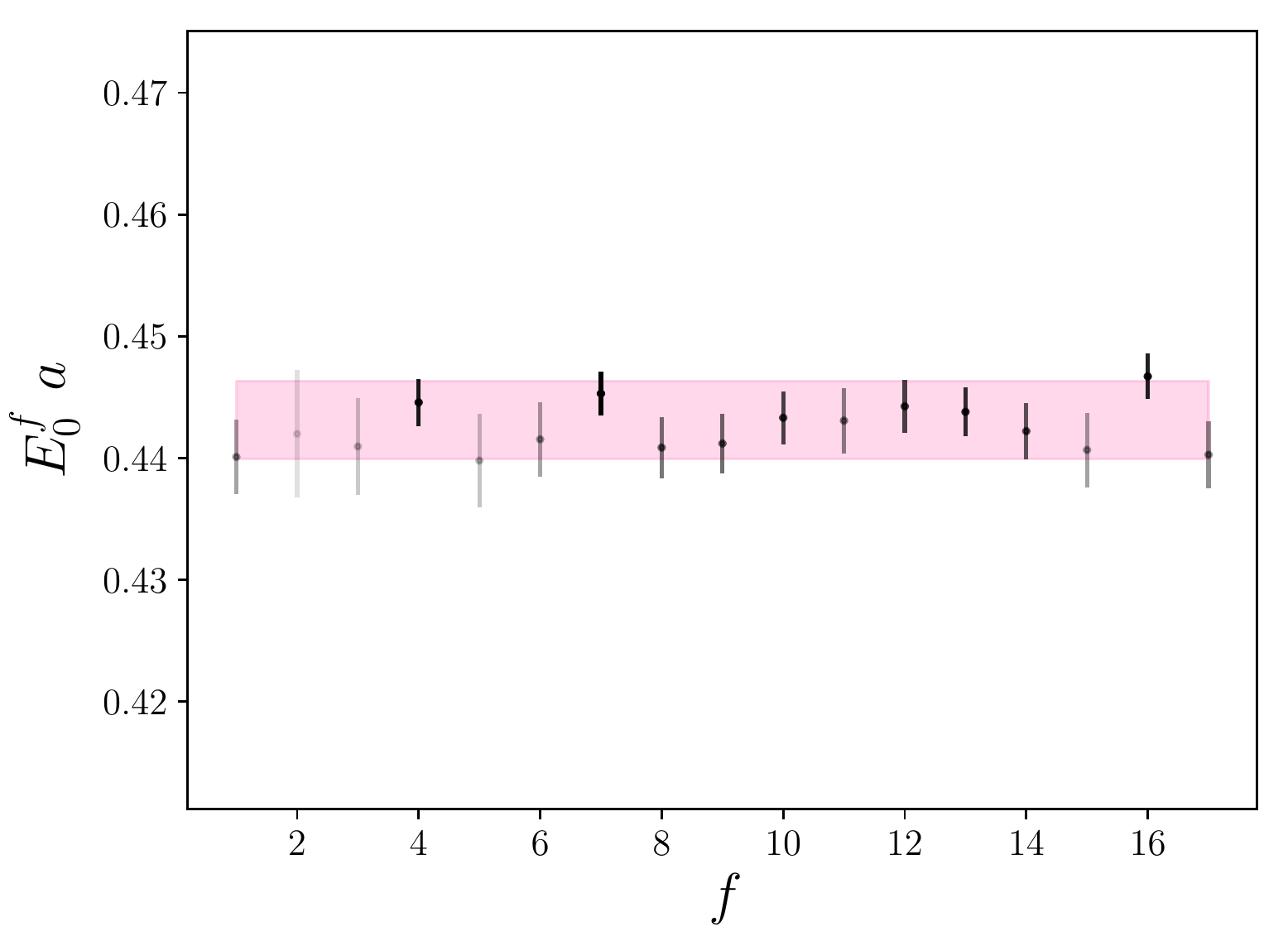} \\\vspace{-65pt}
     \begin{minipage}[c][150pt][t]{70pt} $4\ \overline{K^0}\newline L/a=32$ \end{minipage}
     \includegraphics[width=.40\textwidth]{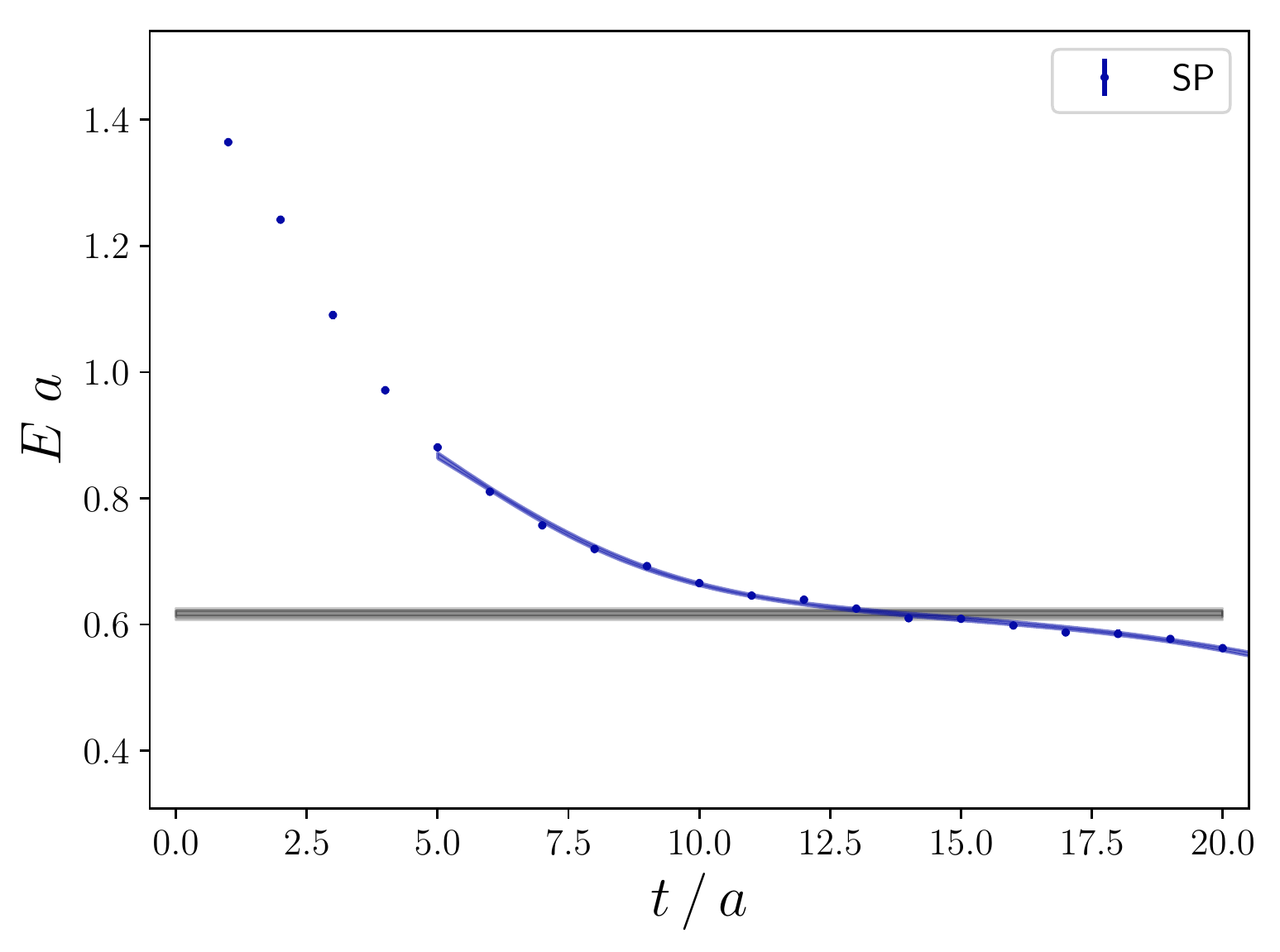}\hspace{10pt}
     \includegraphics[width=.40\textwidth]{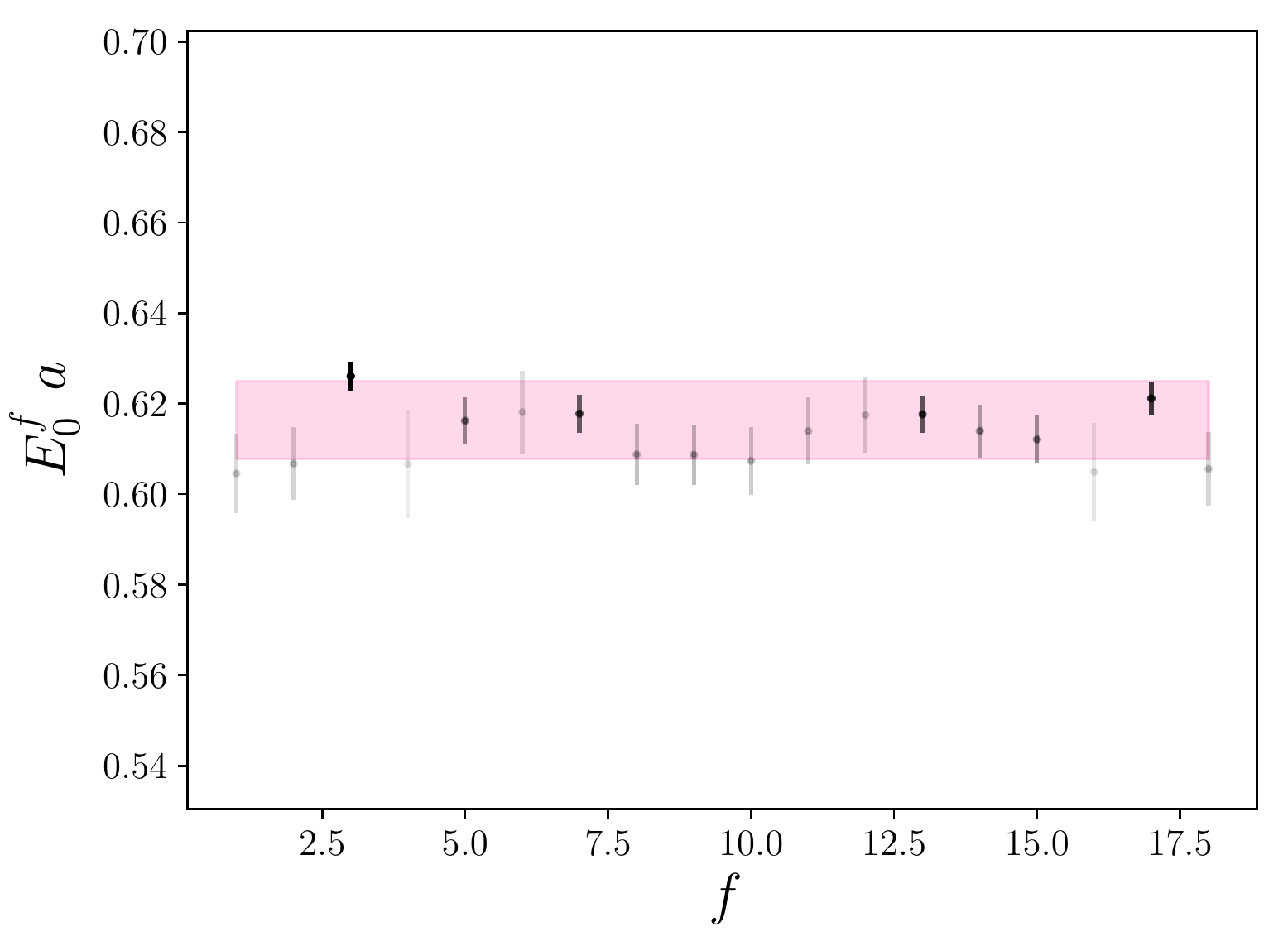} \\\vspace{-65pt}
   \caption{Fit results for systems of $n\in\{1,\dots,4\}$ $\overline{K^0}$ mesons for the $L/a = 32$ lattice volume. The figures are analogous to Fig.~\ref{fig:pion48a}, see Appendix~\ref{app:fits} for a definition of the fitting procedure employed.
    \label{fig:kaon32a}}
\end{figure}

\begin{figure}
  \centering
     \begin{minipage}[c][150pt][t]{70pt} $5\ \overline{K^0}\newline L/a=32$ \end{minipage}
     \includegraphics[width=.40\textwidth]{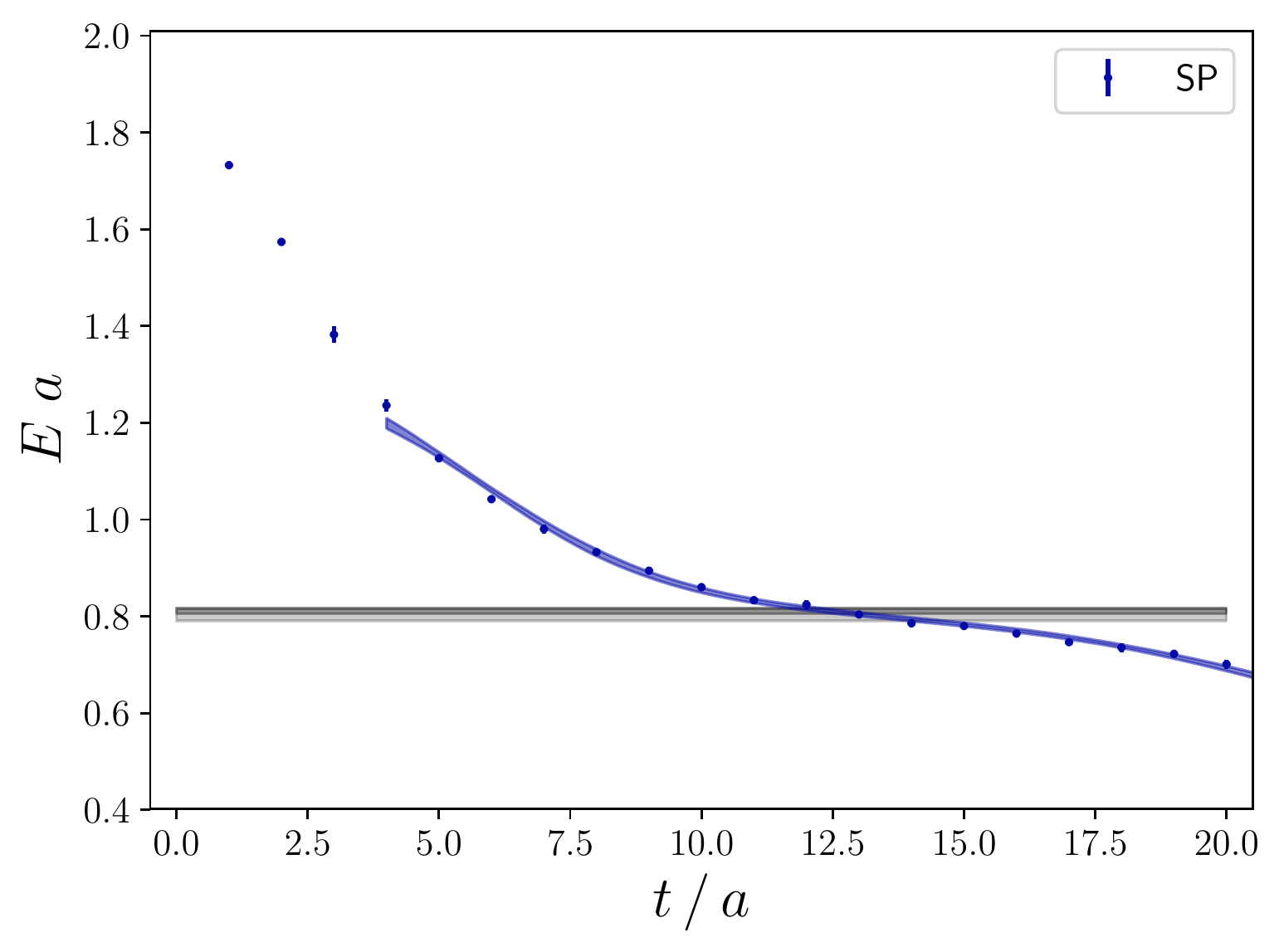}\hspace{10pt}
     \includegraphics[width=.40\textwidth]{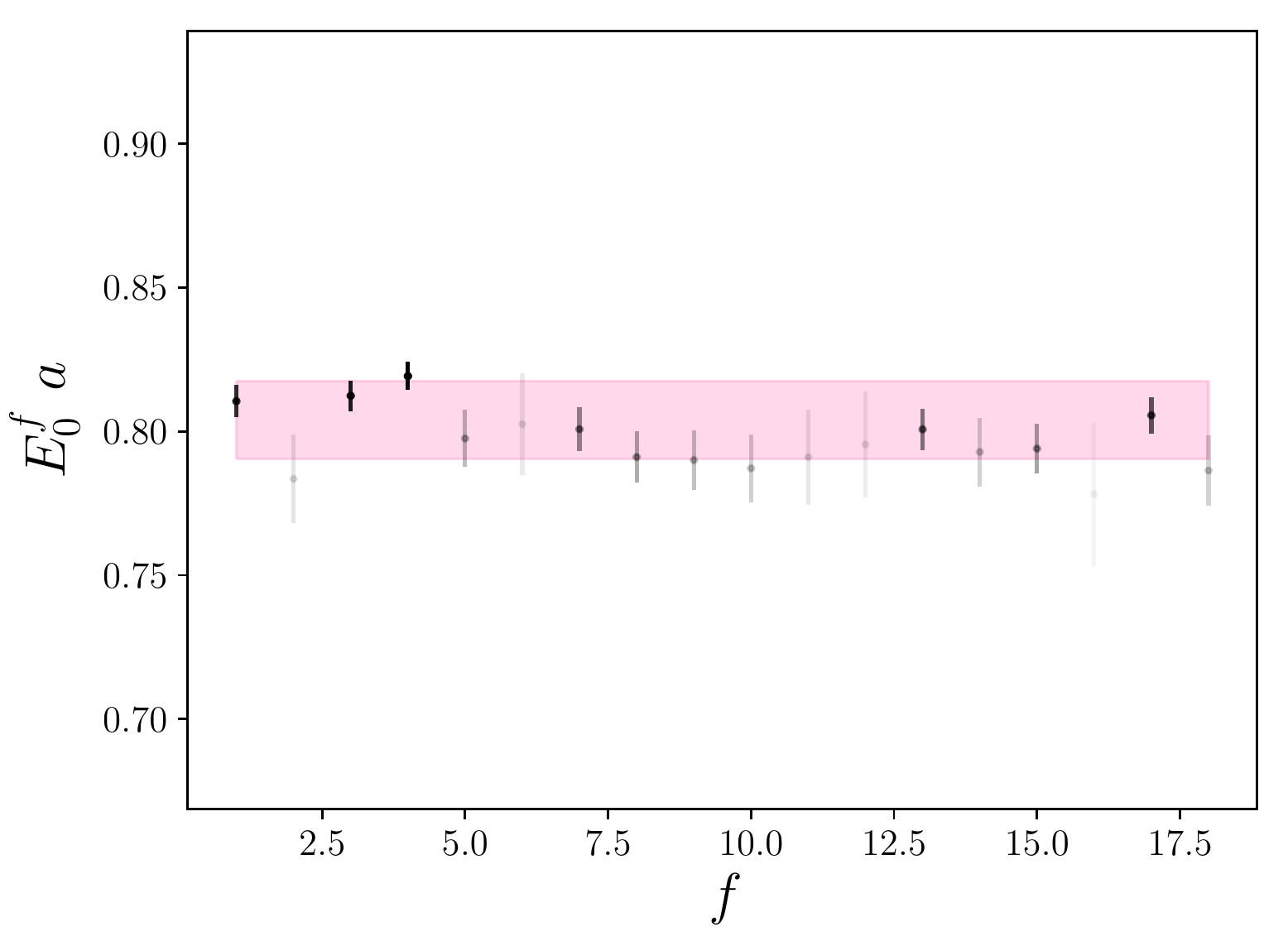} \\\vspace{-65pt}
     \begin{minipage}[c][150pt][t]{70pt} $6\ \overline{K^0}\newline L/a=32$ \end{minipage}
     \includegraphics[width=.40\textwidth]{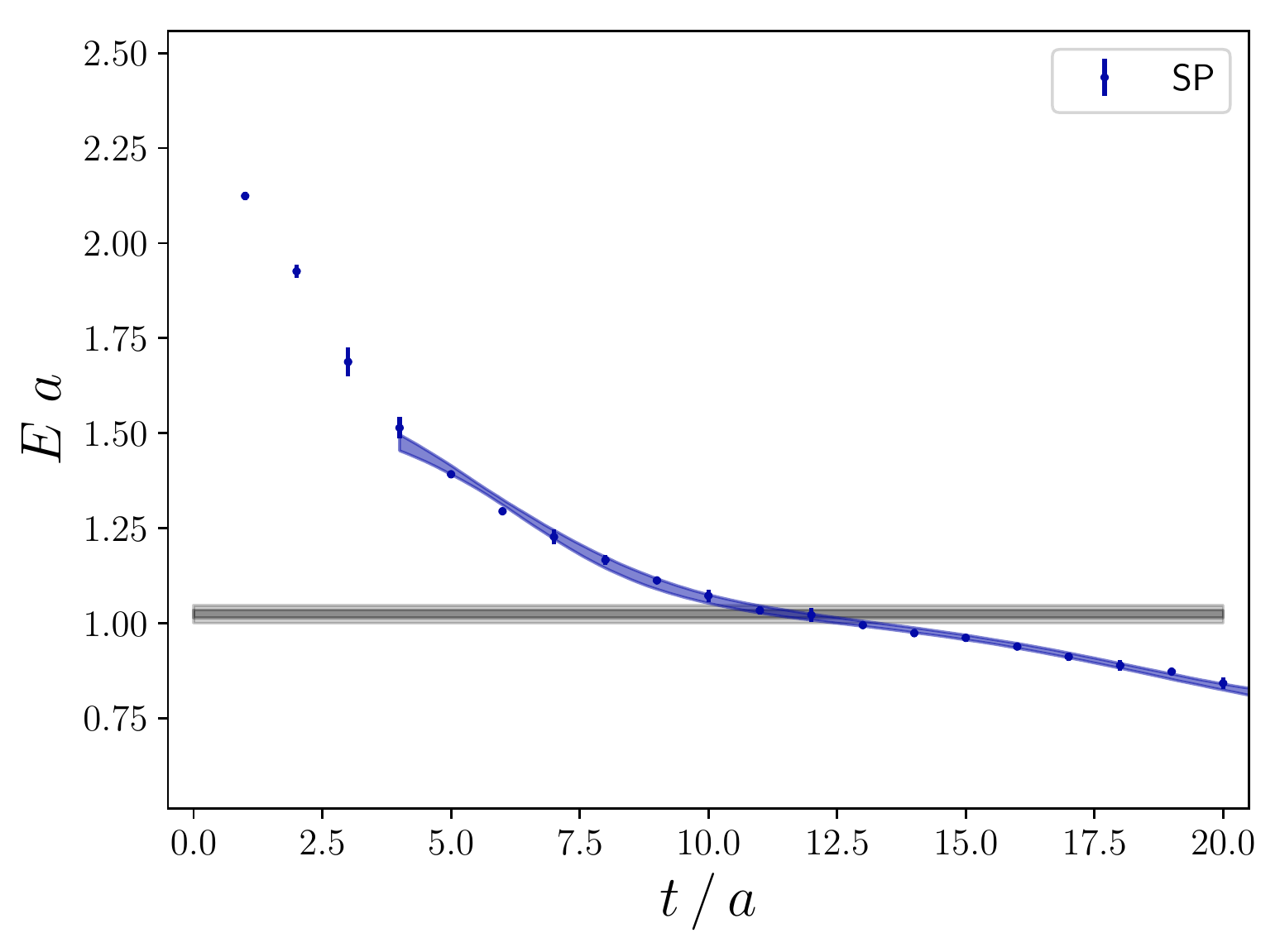}\hspace{10pt}
     \includegraphics[width=.40\textwidth]{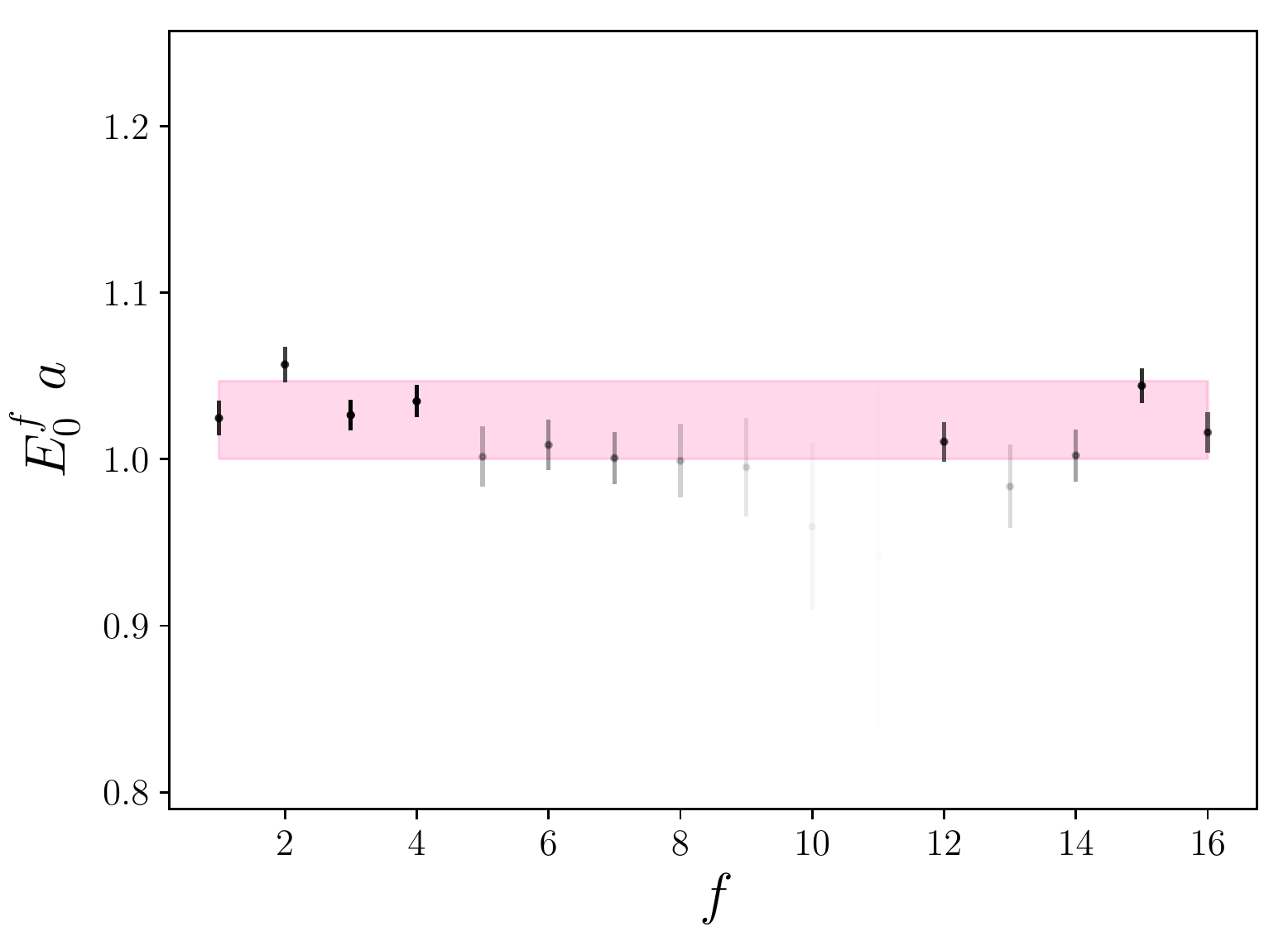} \\\vspace{-65pt}
     \begin{minipage}[c][150pt][t]{70pt} $7\ \overline{K^0}\newline L/a=32$ \end{minipage}
     \includegraphics[width=.40\textwidth]{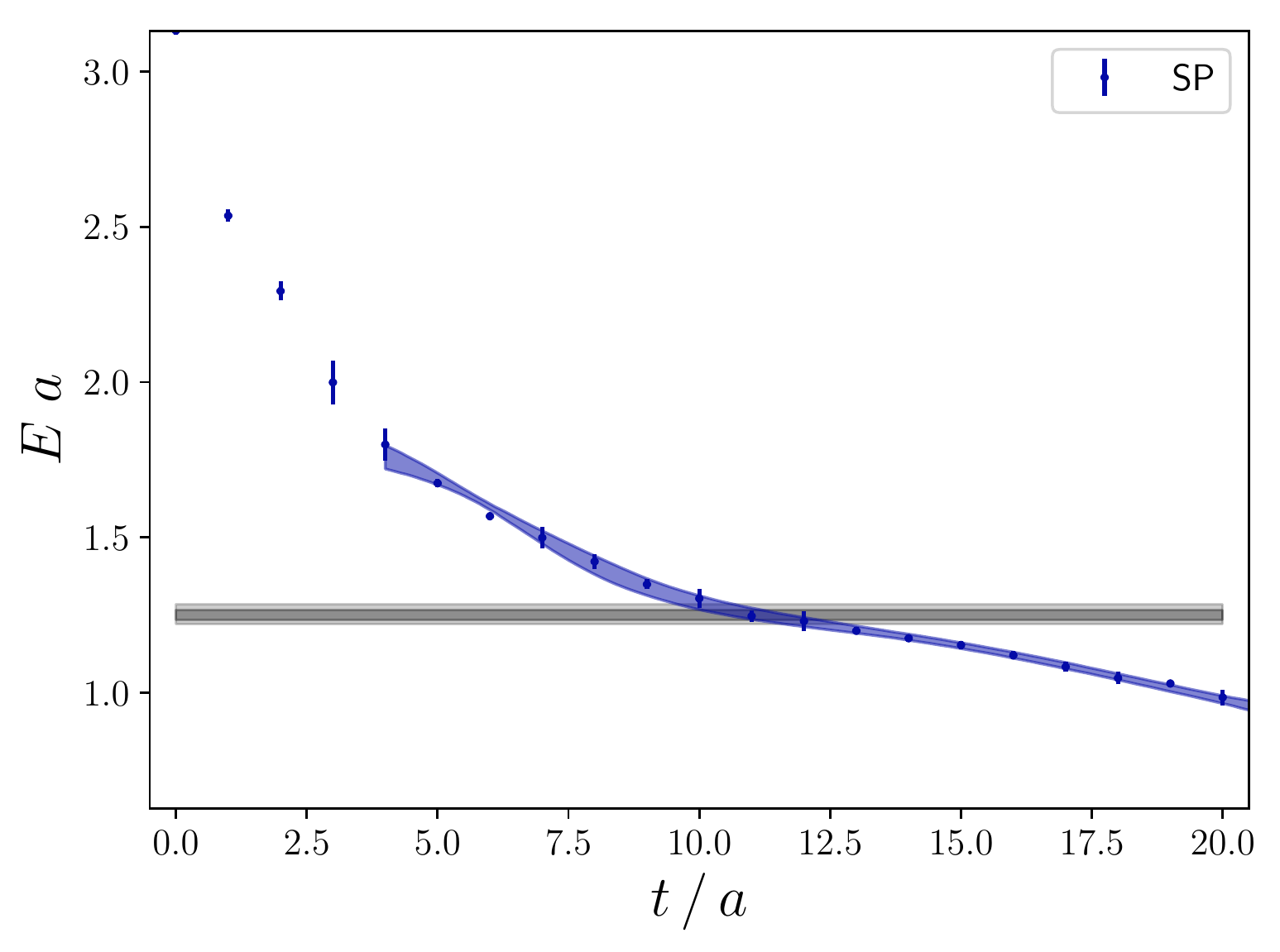}\hspace{10pt}
     \includegraphics[width=.40\textwidth]{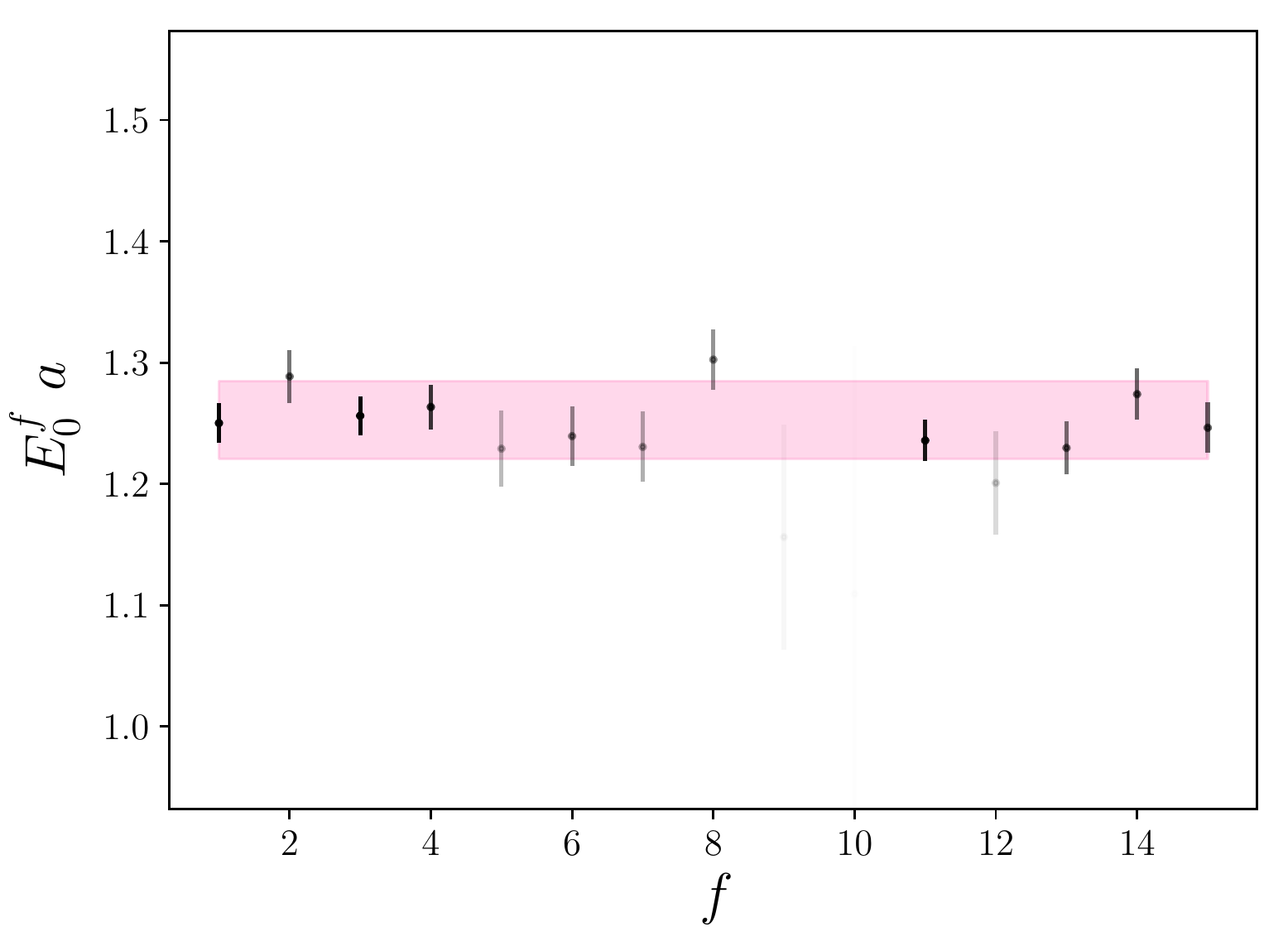} \\\vspace{-65pt}
     \begin{minipage}[c][150pt][t]{70pt} $8\ \overline{K^0}\newline L/a=32$ \end{minipage}
     \includegraphics[width=.40\textwidth]{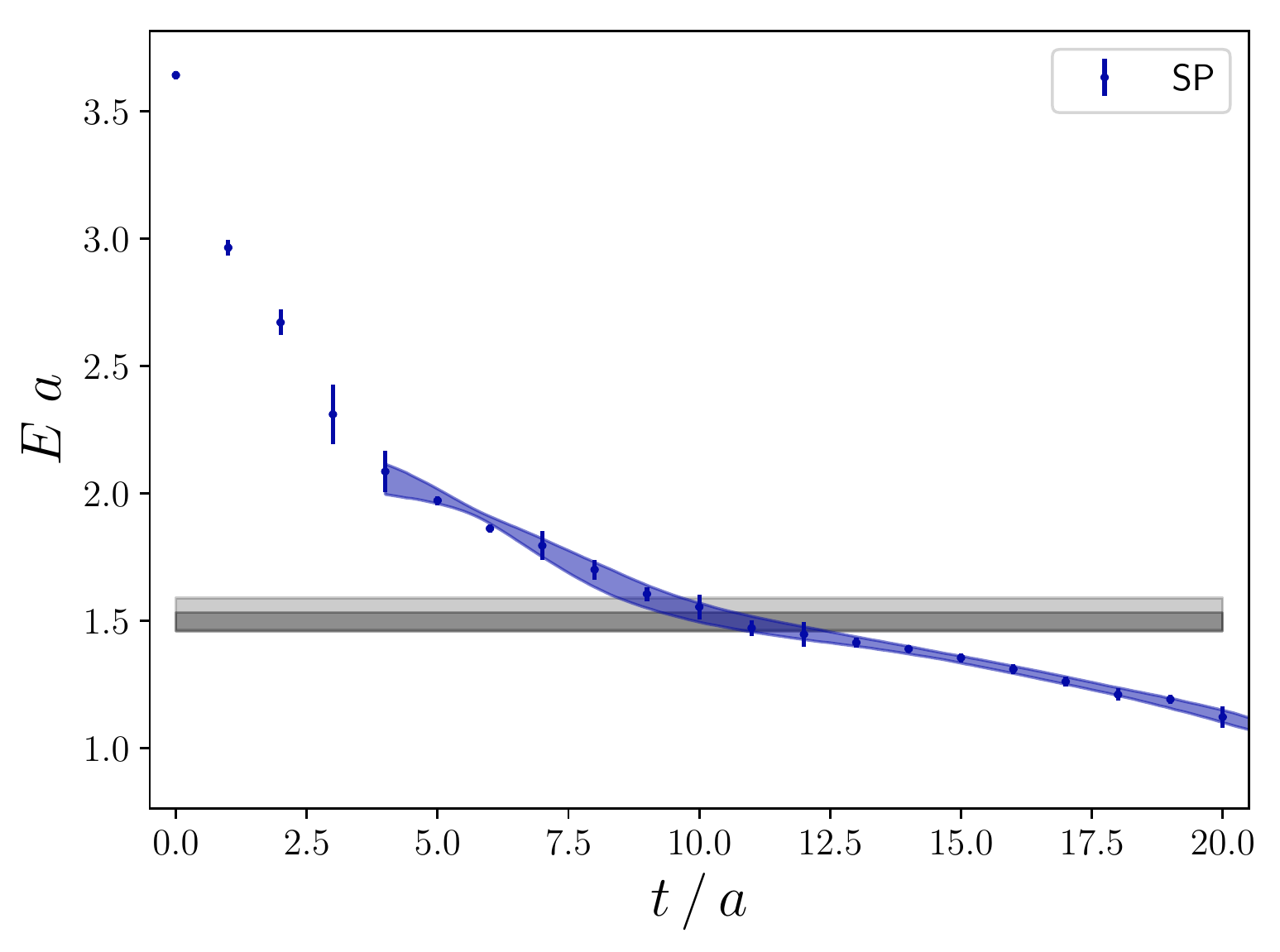}\hspace{10pt}
     \includegraphics[width=.40\textwidth]{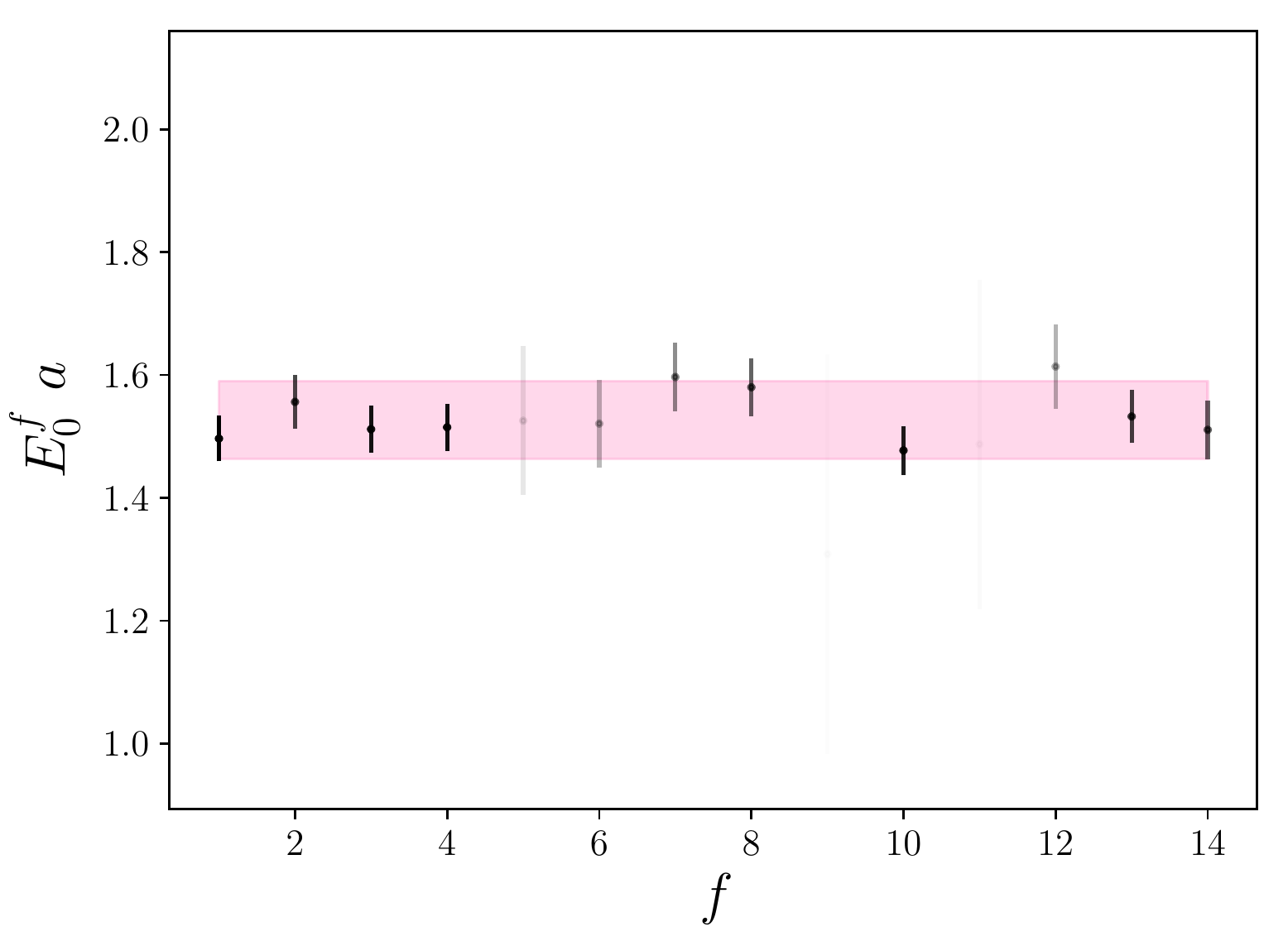} \\\vspace{-65pt}
   \caption{Fit results for systems of $n\in\{5,\dots,8\}$ $\overline{K^0}$ mesons for the $L/a = 32$ lattice volume. The figures are analogous to Fig.~\ref{fig:pion48a}, see Appendix~\ref{app:fits} for a definition of the fitting procedure employed.
    \label{fig:kaon32b}}
\end{figure}

\begin{figure}
  \centering
     \begin{minipage}[c][150pt][t]{70pt} $9\ \overline{K^0}\newline L/a=32$ \end{minipage}
     \includegraphics[width=.40\textwidth]{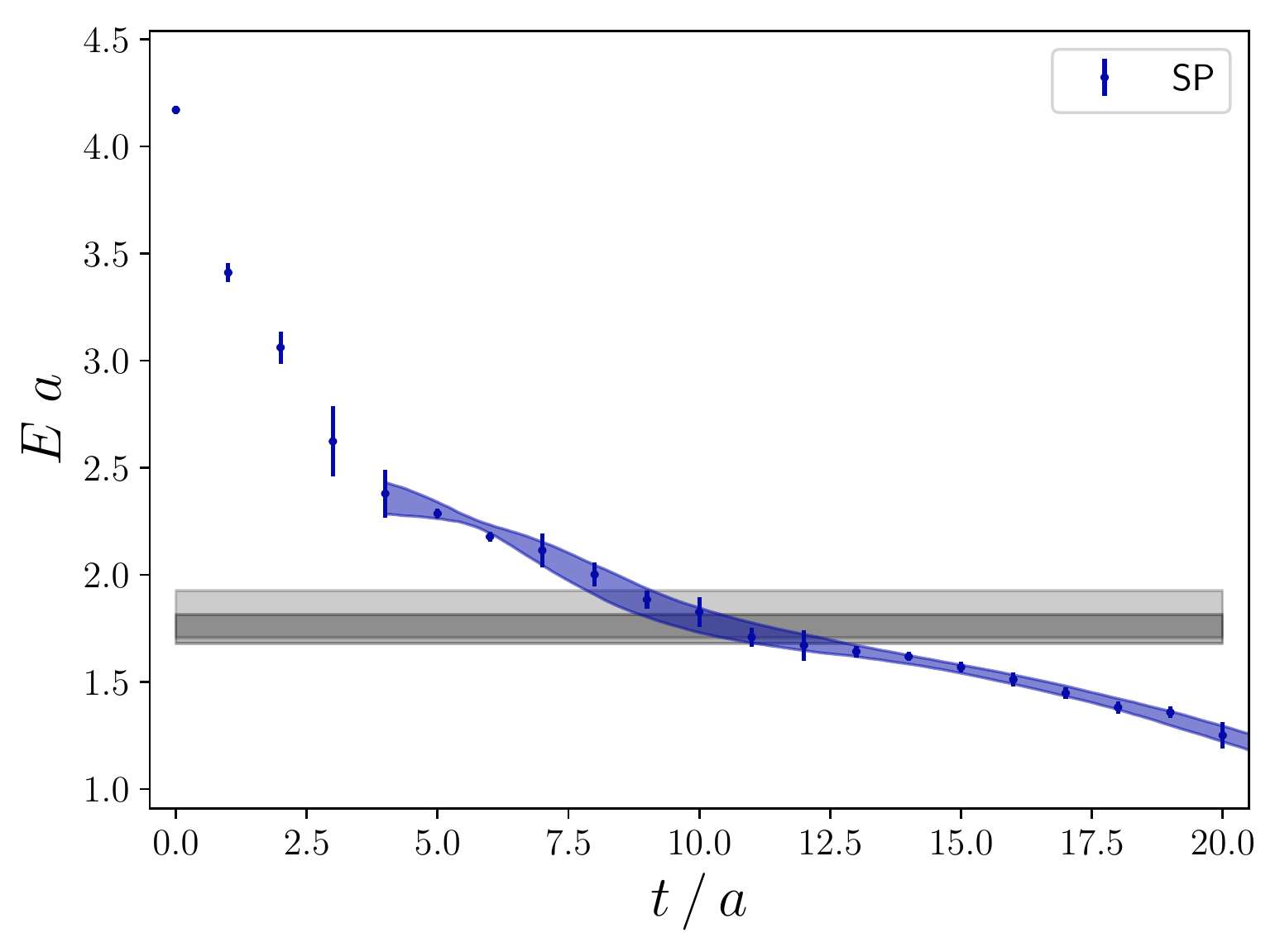}\hspace{10pt}
     \includegraphics[width=.40\textwidth]{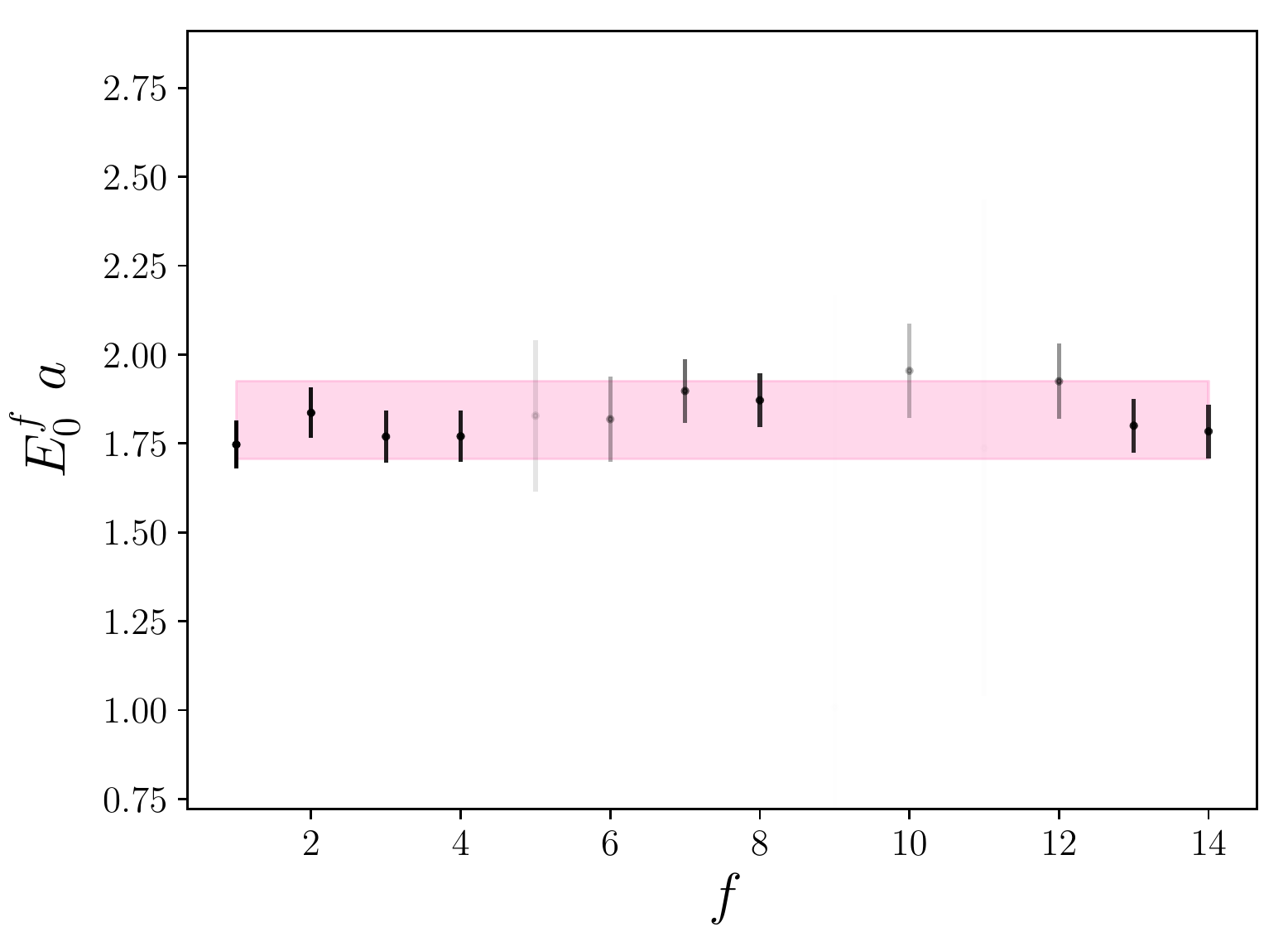} \\\vspace{-65pt}
     \begin{minipage}[c][150pt][t]{70pt} $10\ \overline{K^0}\newline L/a=32$ \end{minipage}
     \includegraphics[width=.40\textwidth]{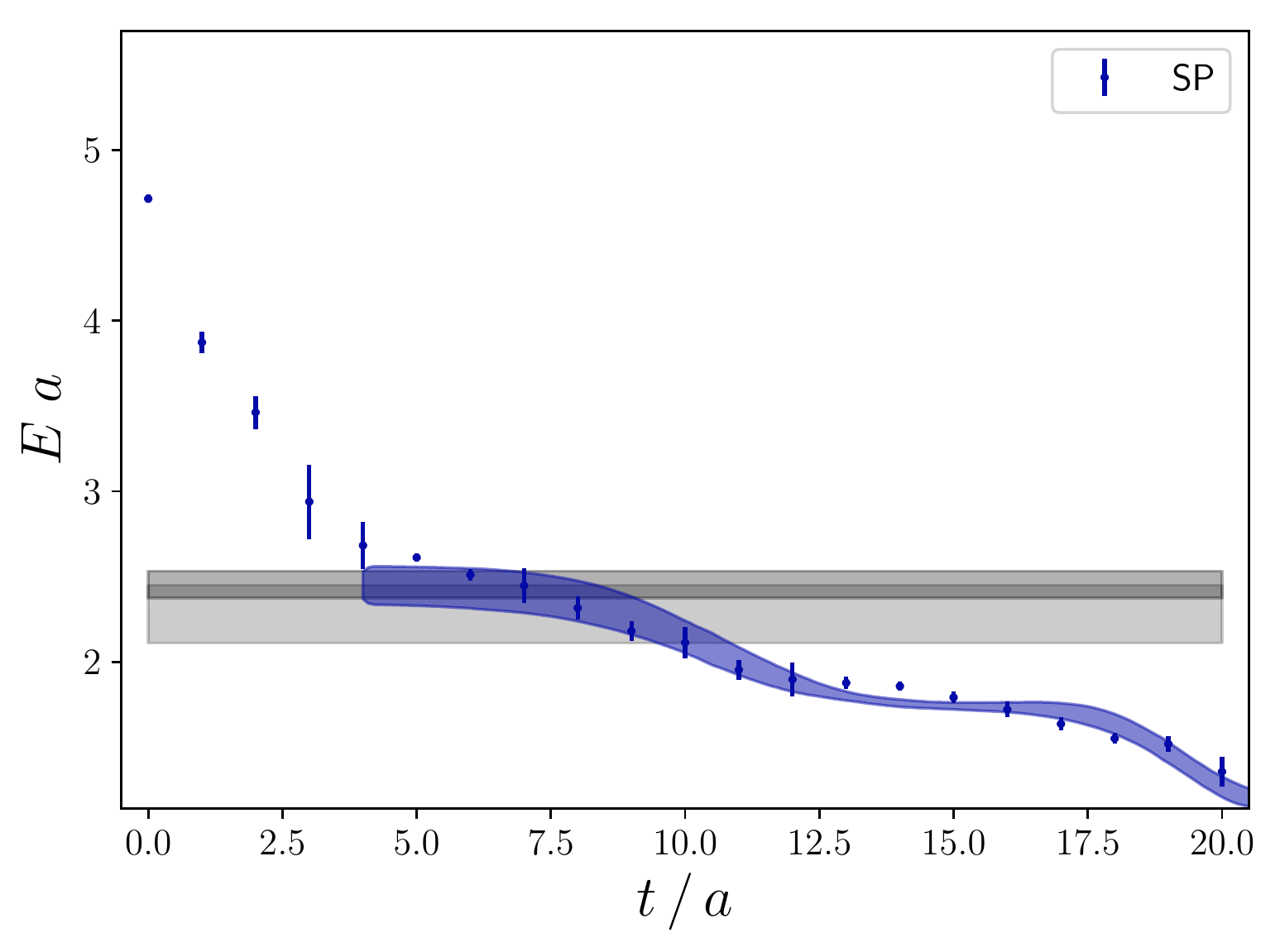}\hspace{10pt}
     \includegraphics[width=.40\textwidth]{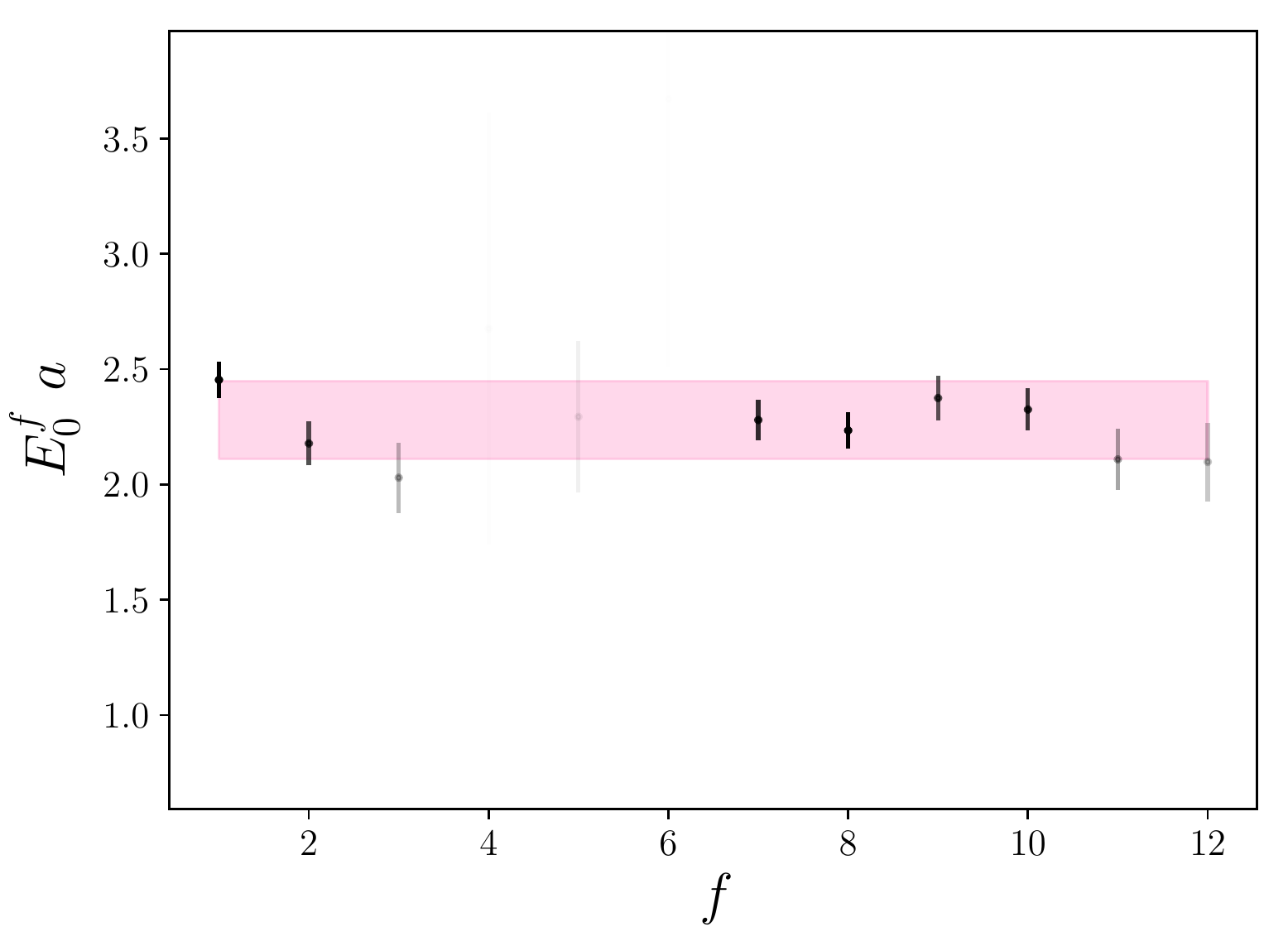} \\\vspace{-65pt}
     \begin{minipage}[c][150pt][t]{70pt} $11\ \overline{K^0}\newline L/a=32$ \end{minipage}
     \includegraphics[width=.40\textwidth]{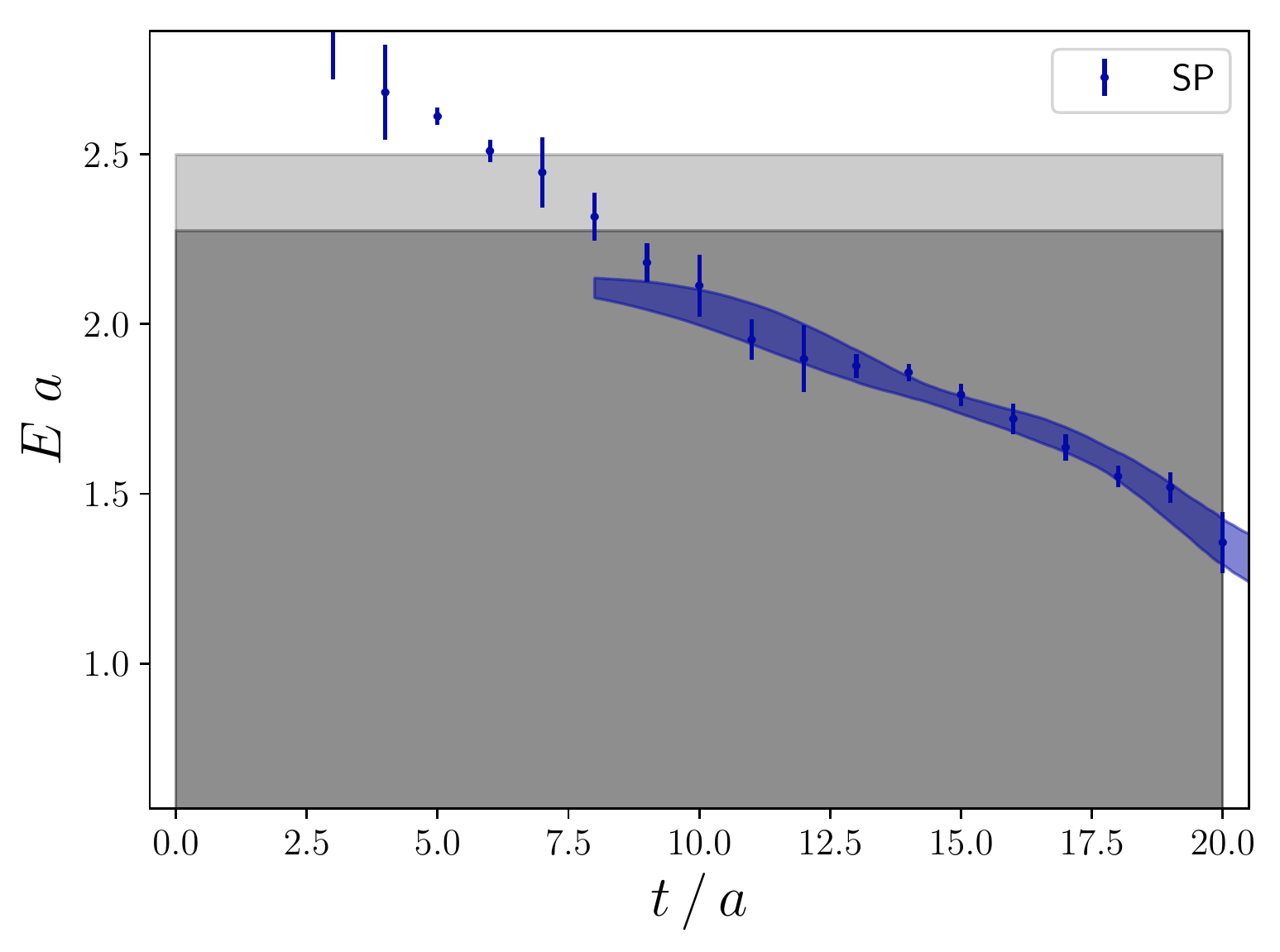}\hspace{10pt}
     \includegraphics[width=.40\textwidth]{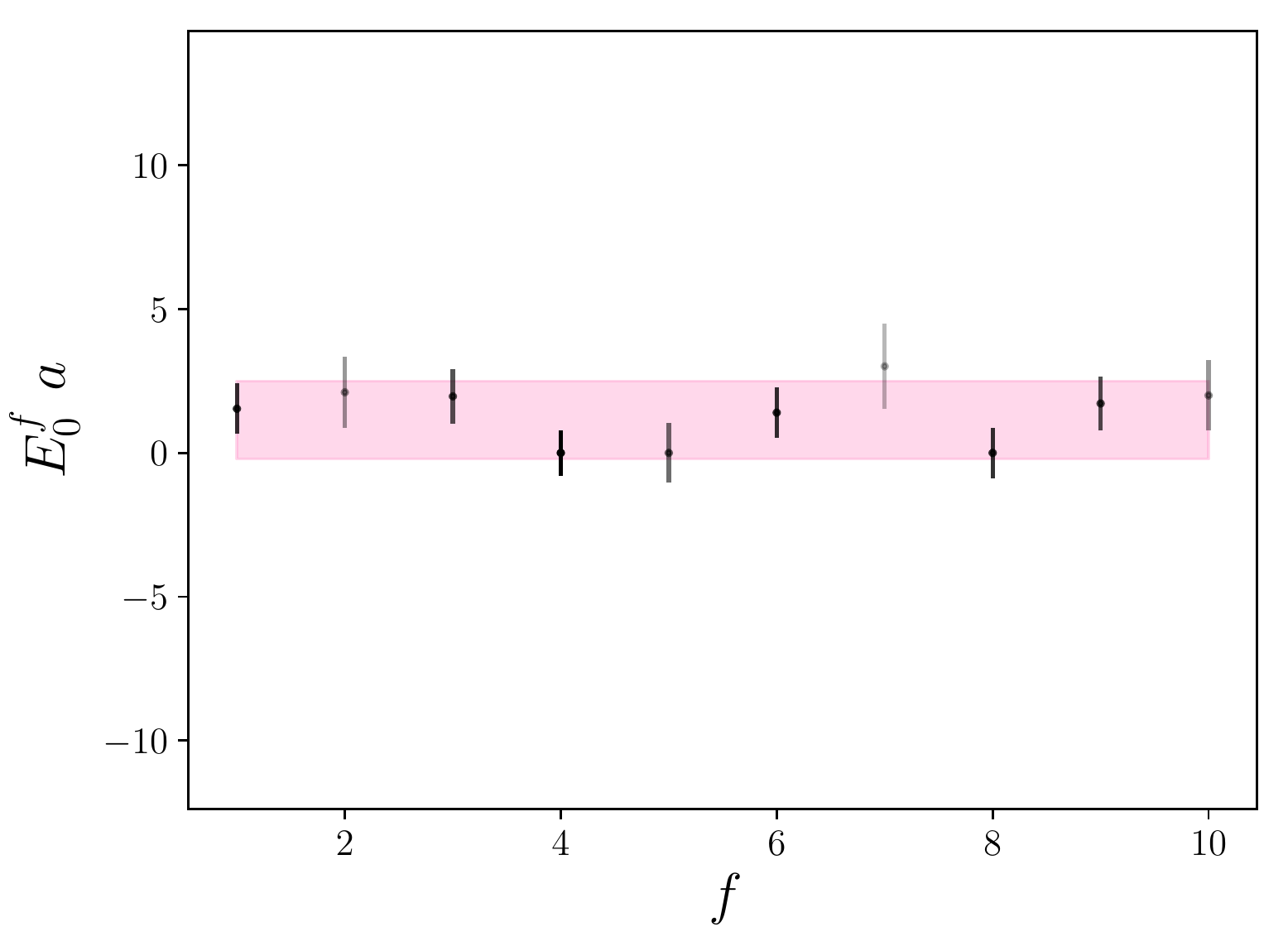} \\\vspace{-65pt}
     \begin{minipage}[c][150pt][t]{70pt} $12\ \overline{K^0}\newline L/a=32$ \end{minipage}
     \includegraphics[width=.40\textwidth]{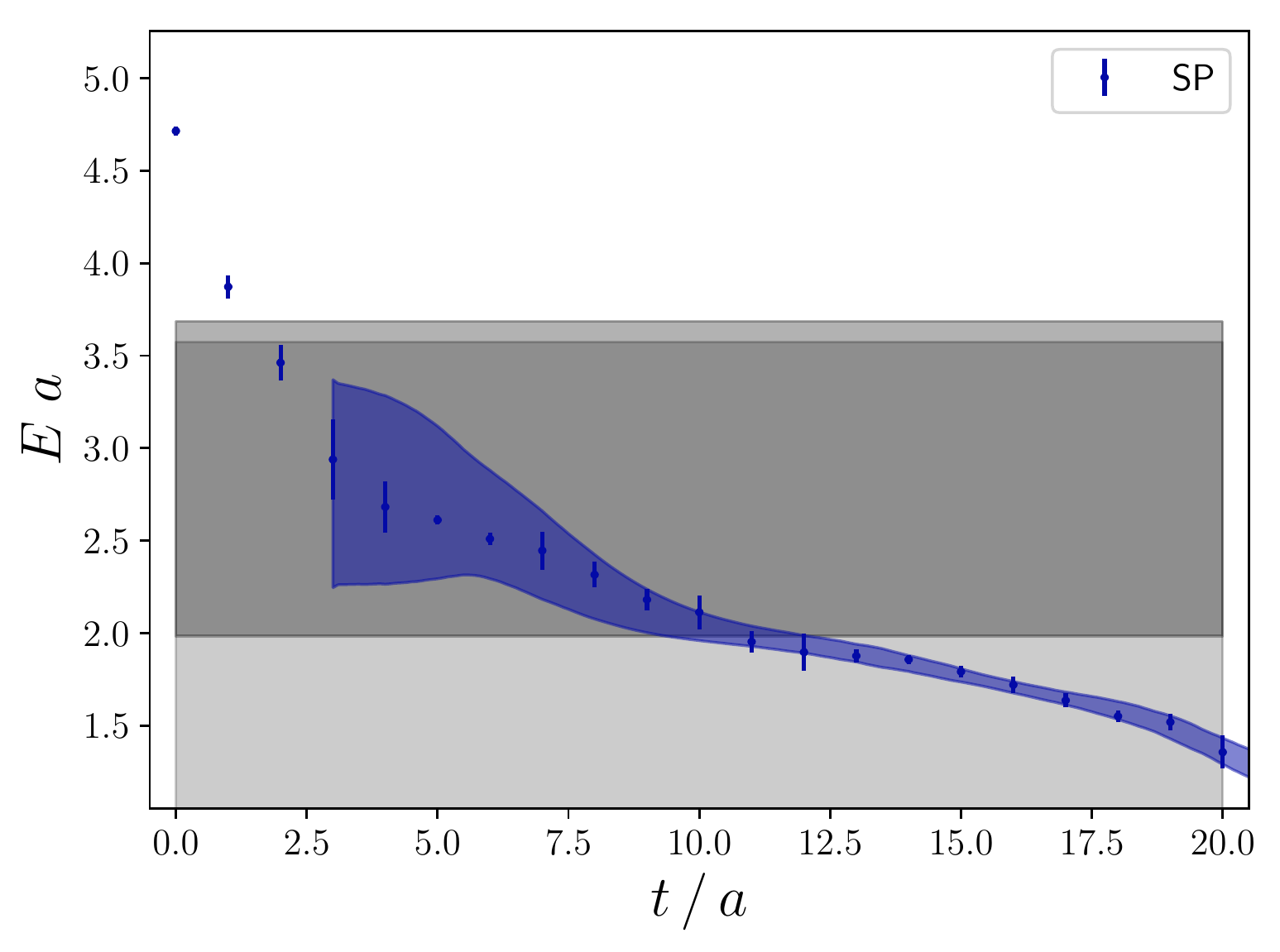}\hspace{10pt}
     \includegraphics[width=.40\textwidth]{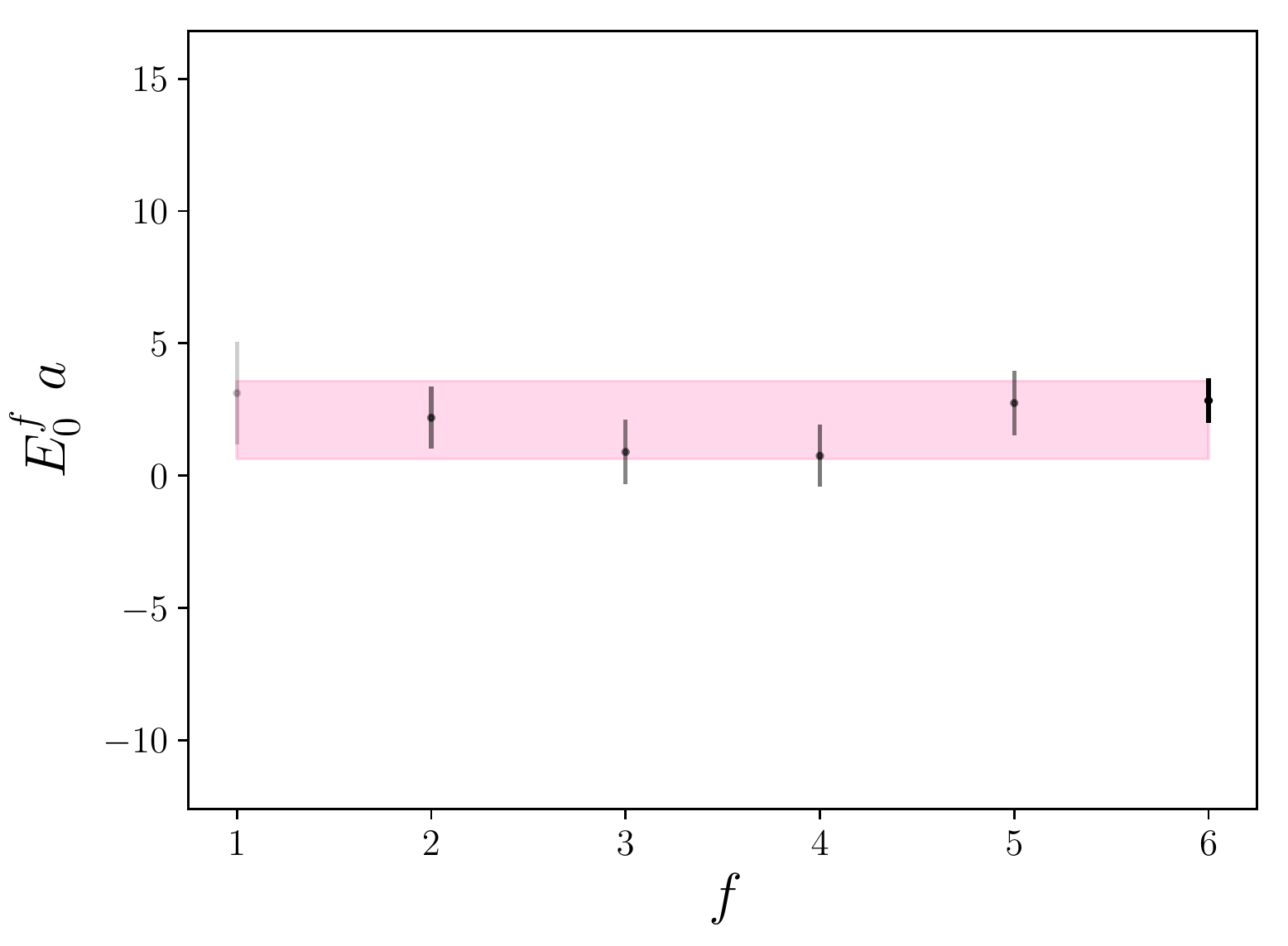} \\\vspace{-65pt}
   \caption{Fit results for systems of $n\in\{9,\dots,12\}$ $\overline{K^0}$ mesons for the $L/a = 32$ lattice volume. The figures are analogous to Fig.~\ref{fig:pion48a}, see Appendix~\ref{app:fits} for a definition of the fitting procedure employed.
    \label{fig:kaon32c}}
\end{figure}

\begin{figure}
  \centering
     \begin{minipage}[c][150pt][t]{70pt} $p \newline L/a=32$ \end{minipage}
     \includegraphics[width=.40\textwidth]{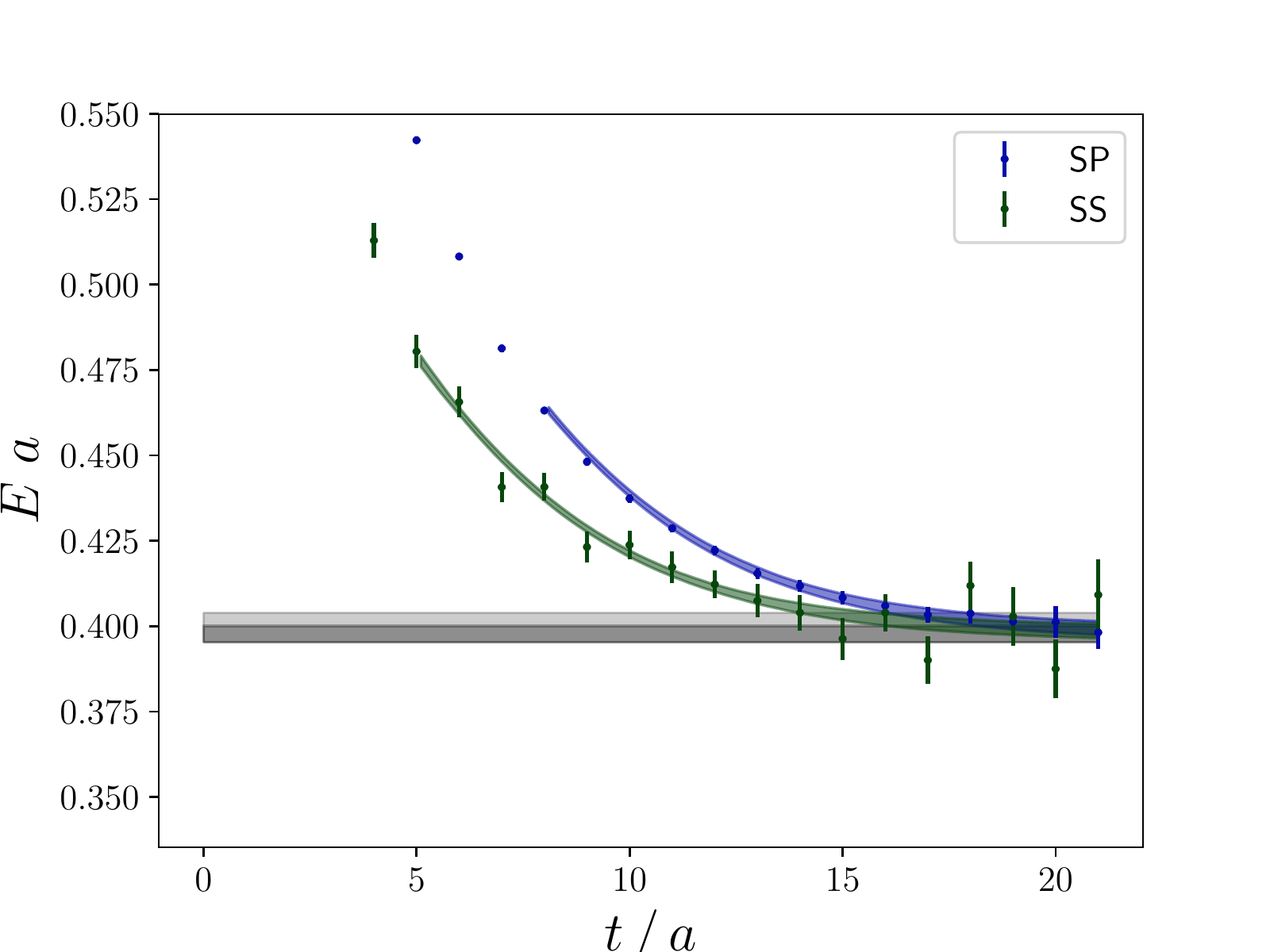} \hspace{10pt}
     \includegraphics[width=.40\textwidth]{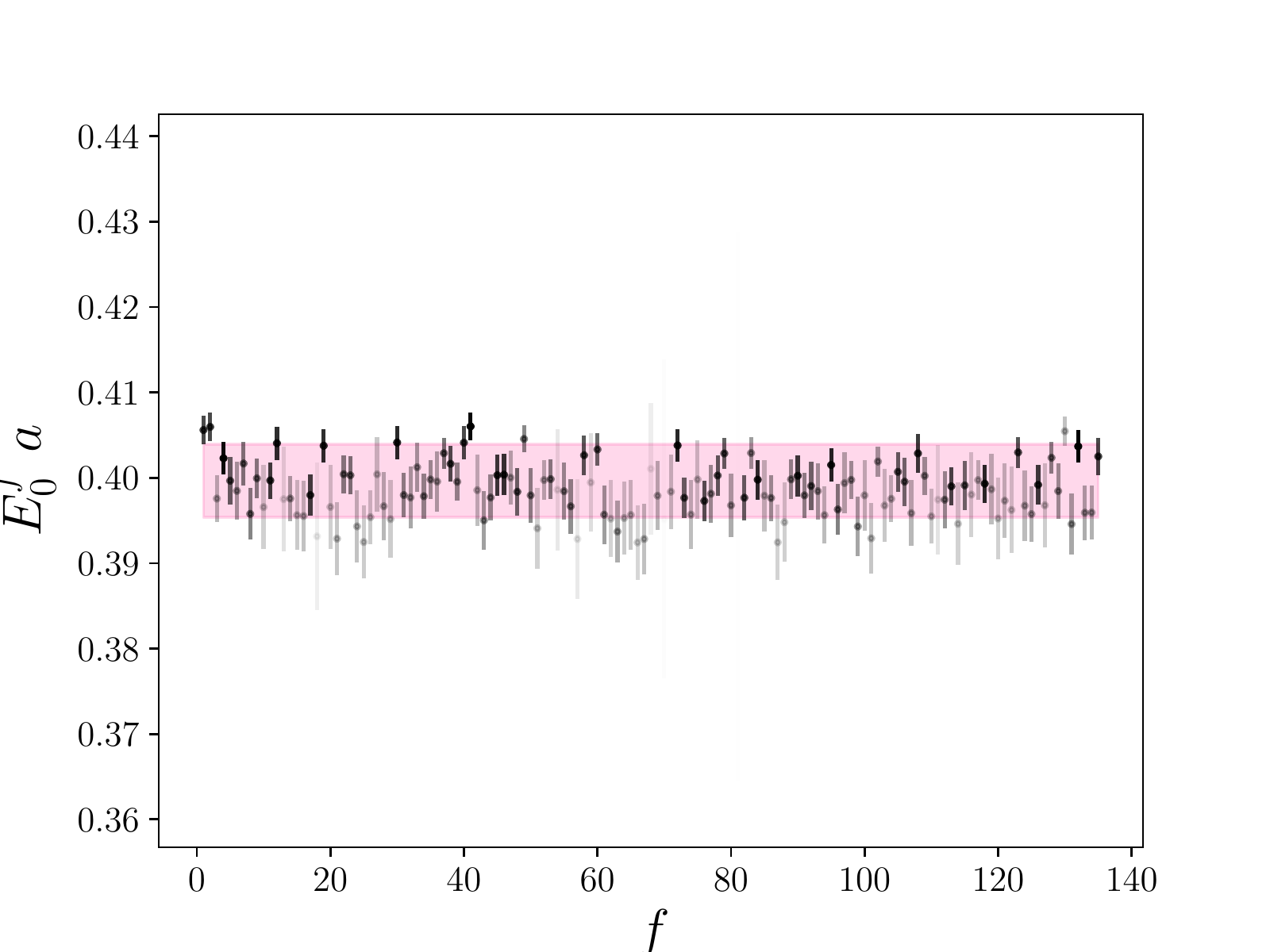} \\\vspace{-65pt}
     \begin{minipage}[c][150pt][t]{70pt} $p \newline L/a=48$ \end{minipage}
     \includegraphics[width=.40\textwidth]{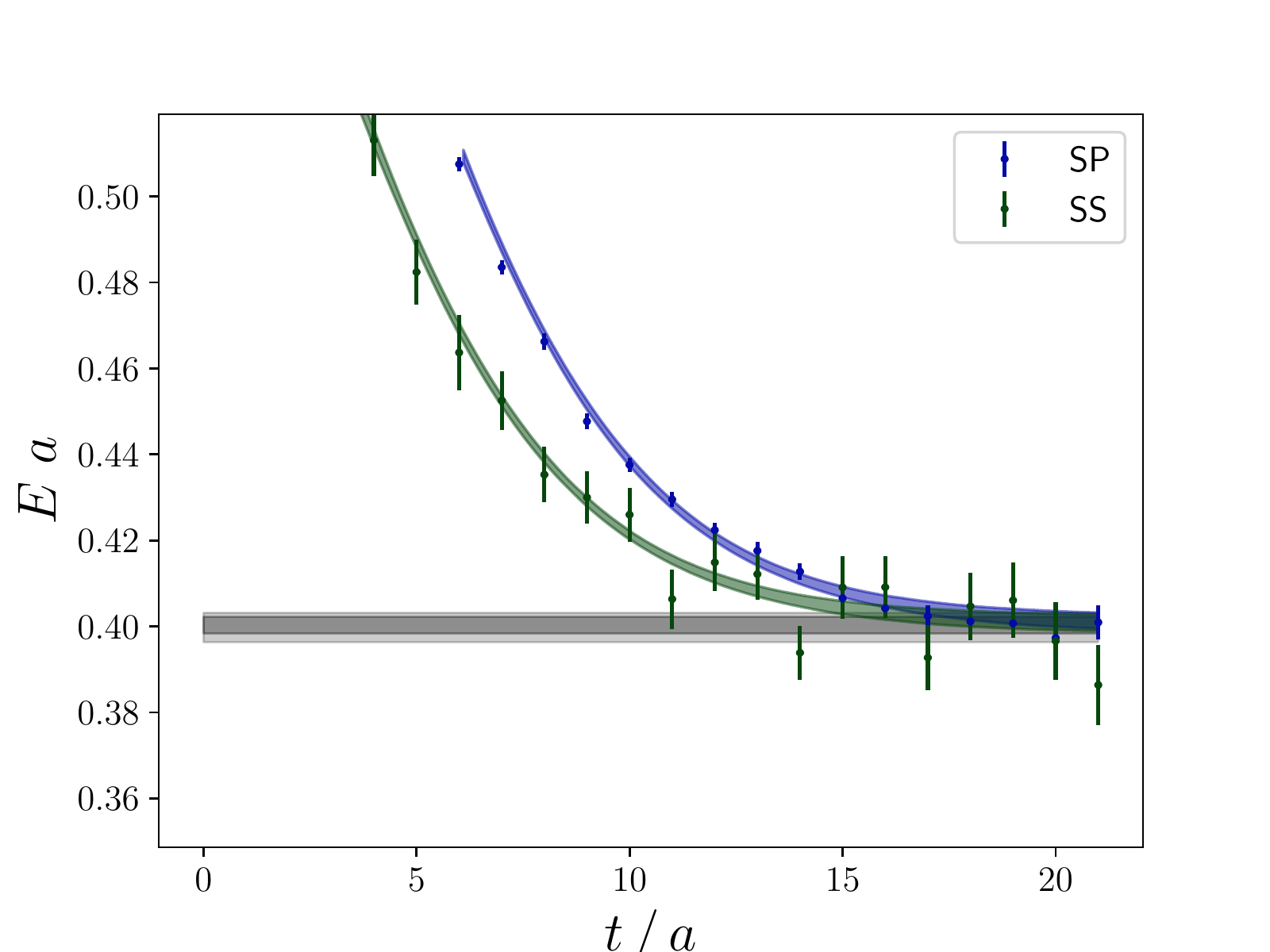} \hspace{10pt}
     \includegraphics[width=.40\textwidth]{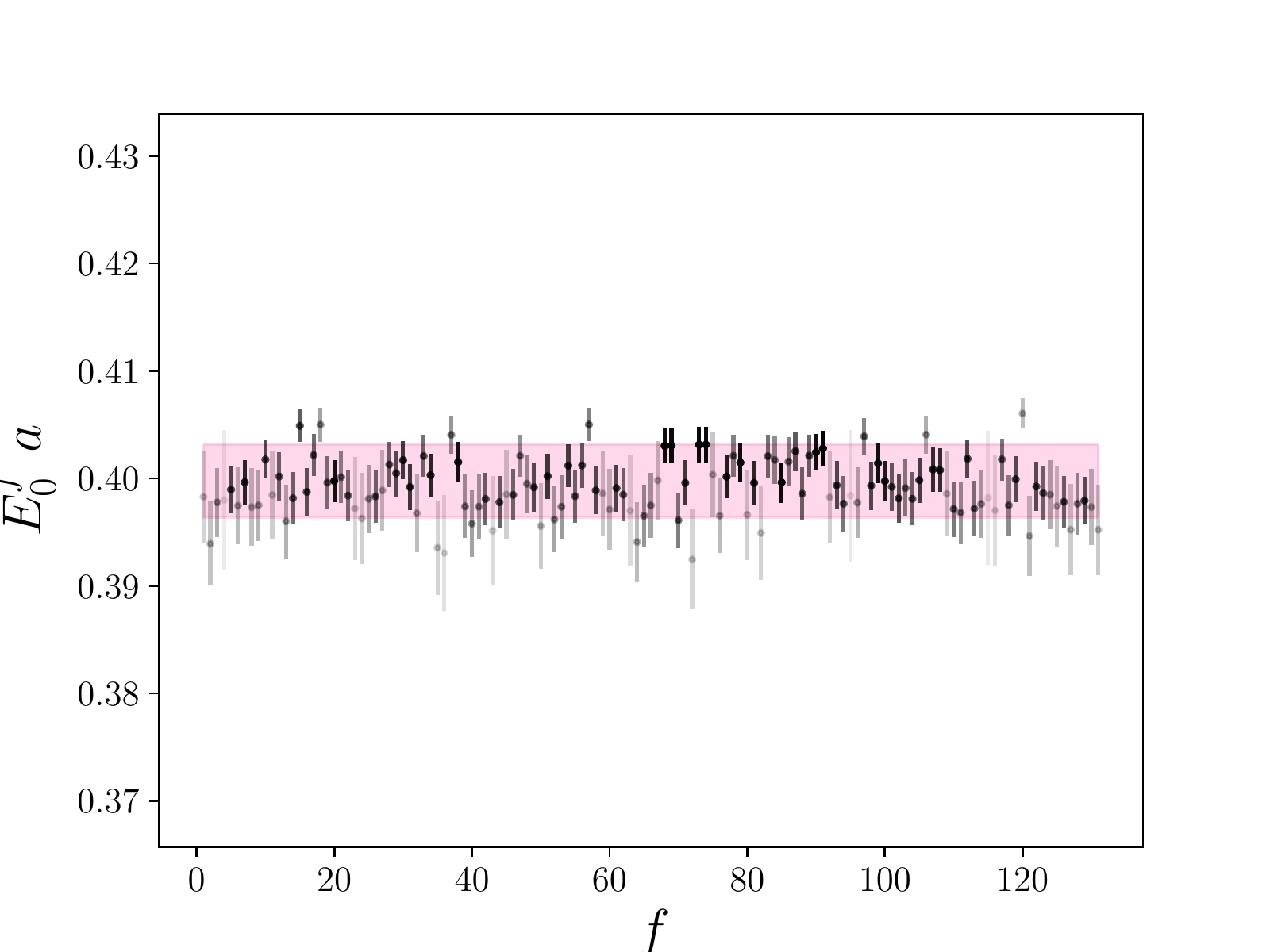} \\\vspace{-65pt}
     \begin{minipage}[c][150pt][t]{70pt} $n \newline L/a=32$ \end{minipage}
     \includegraphics[width=.40\textwidth]{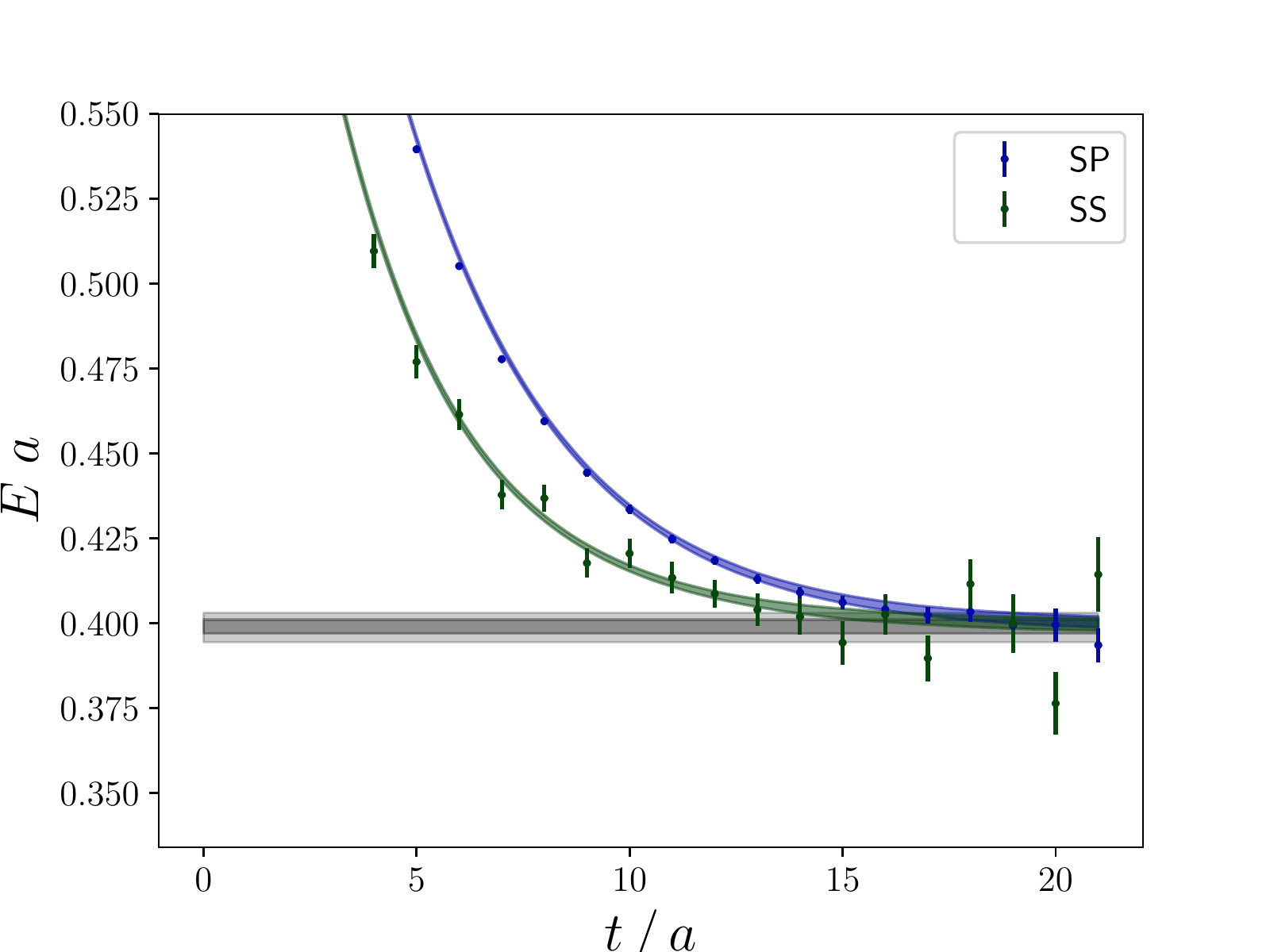} \hspace{10pt}
     \includegraphics[width=.40\textwidth]{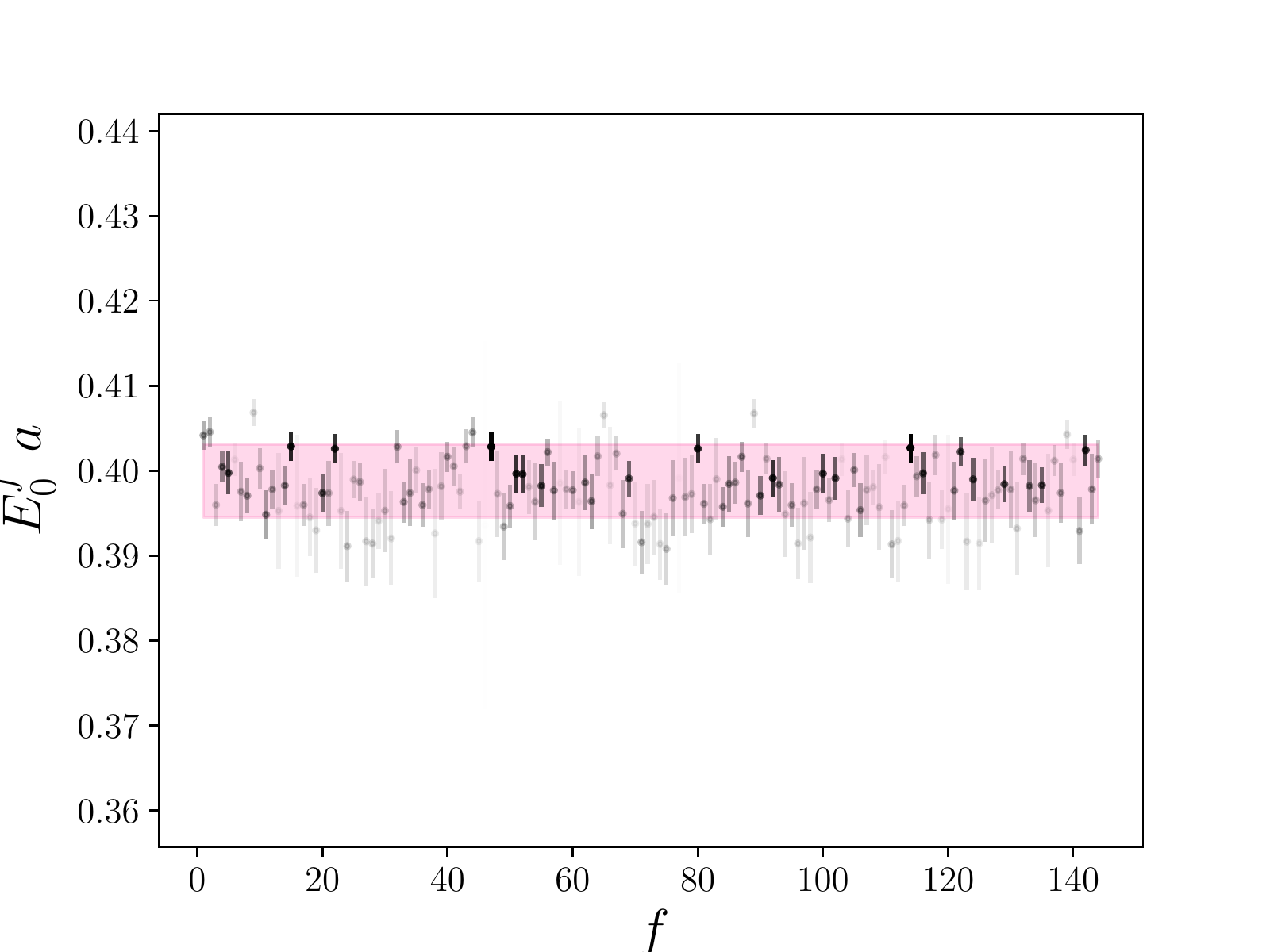} \\\vspace{-65pt}
     \begin{minipage}[c][150pt][t]{70pt} $n \newline L/a=48$ \end{minipage}
     \includegraphics[width=.40\textwidth]{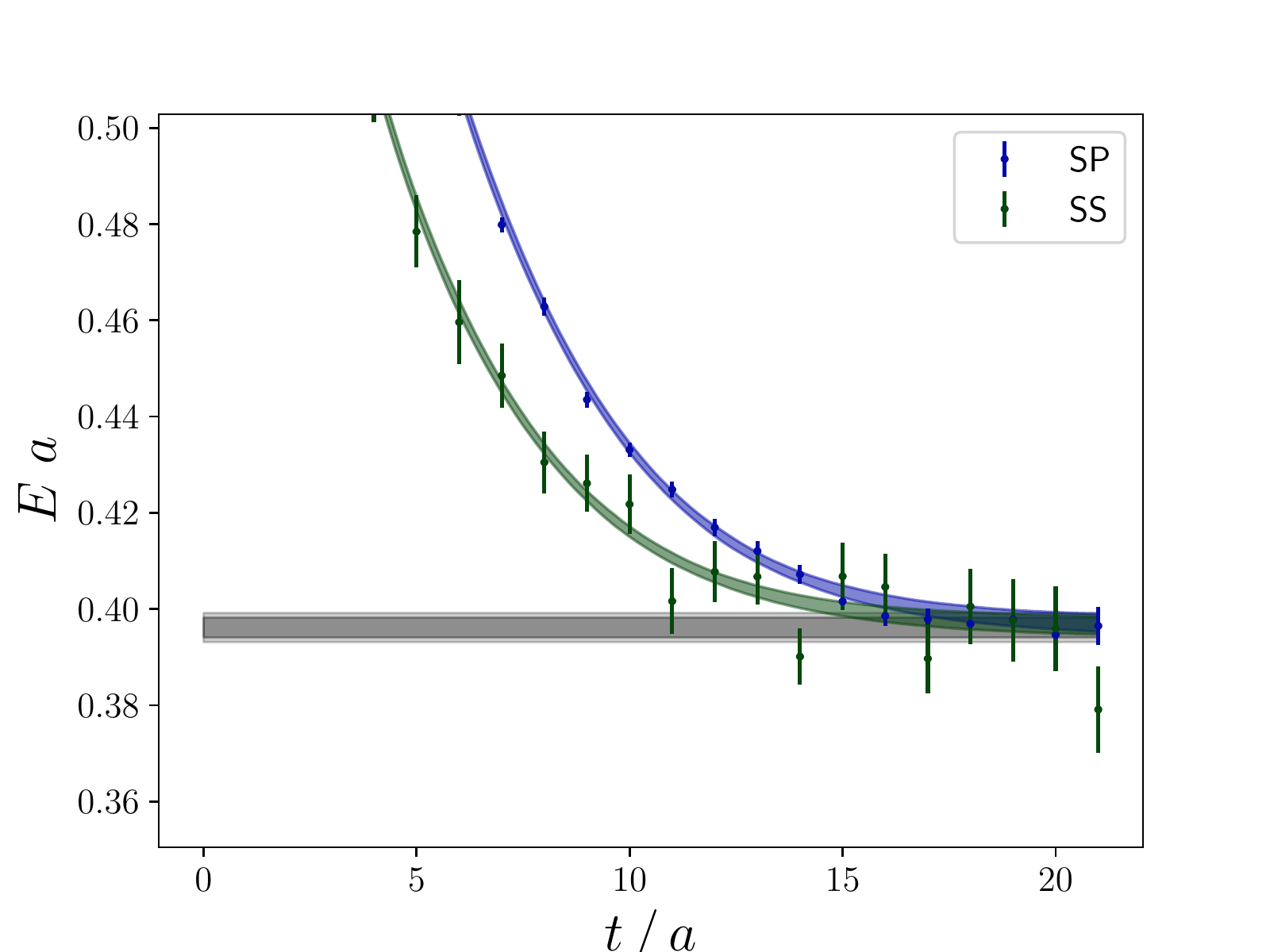} \hspace{10pt}
     \includegraphics[width=.40\textwidth]{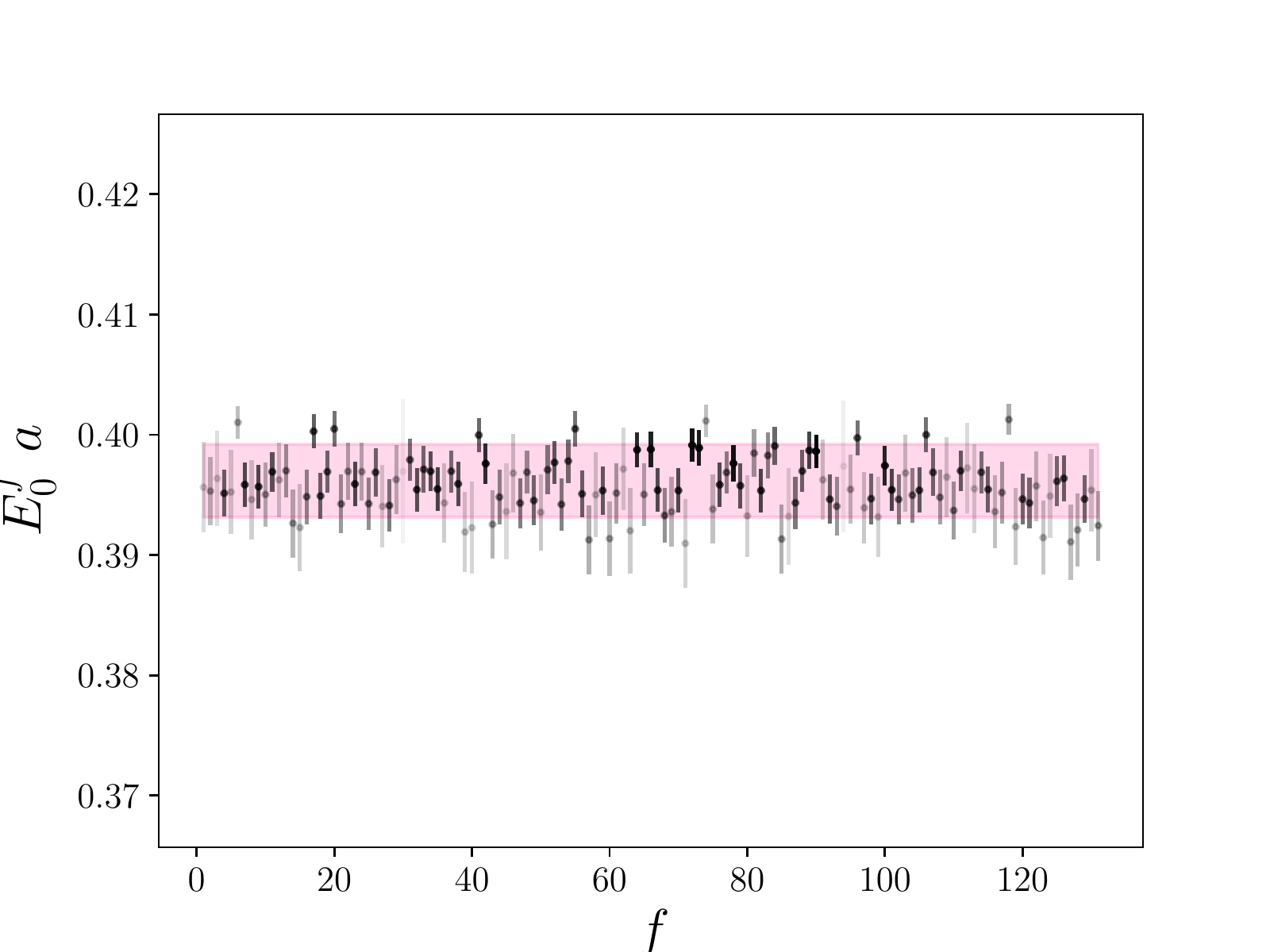} \\\vspace{-65pt}
   \caption{Fit results for proton and neutron systems. The figures are analogous to Fig.~\ref{fig:pion48a}, see Appendix~\ref{app:fits} for a definition of the fitting procedure employed.
    \label{fig:baryon1}}
\end{figure}

\begin{figure}
  \centering
     \begin{minipage}[c][150pt][t]{70pt} $pp \newline L/a=32$ \end{minipage}
     \includegraphics[width=.40\textwidth]{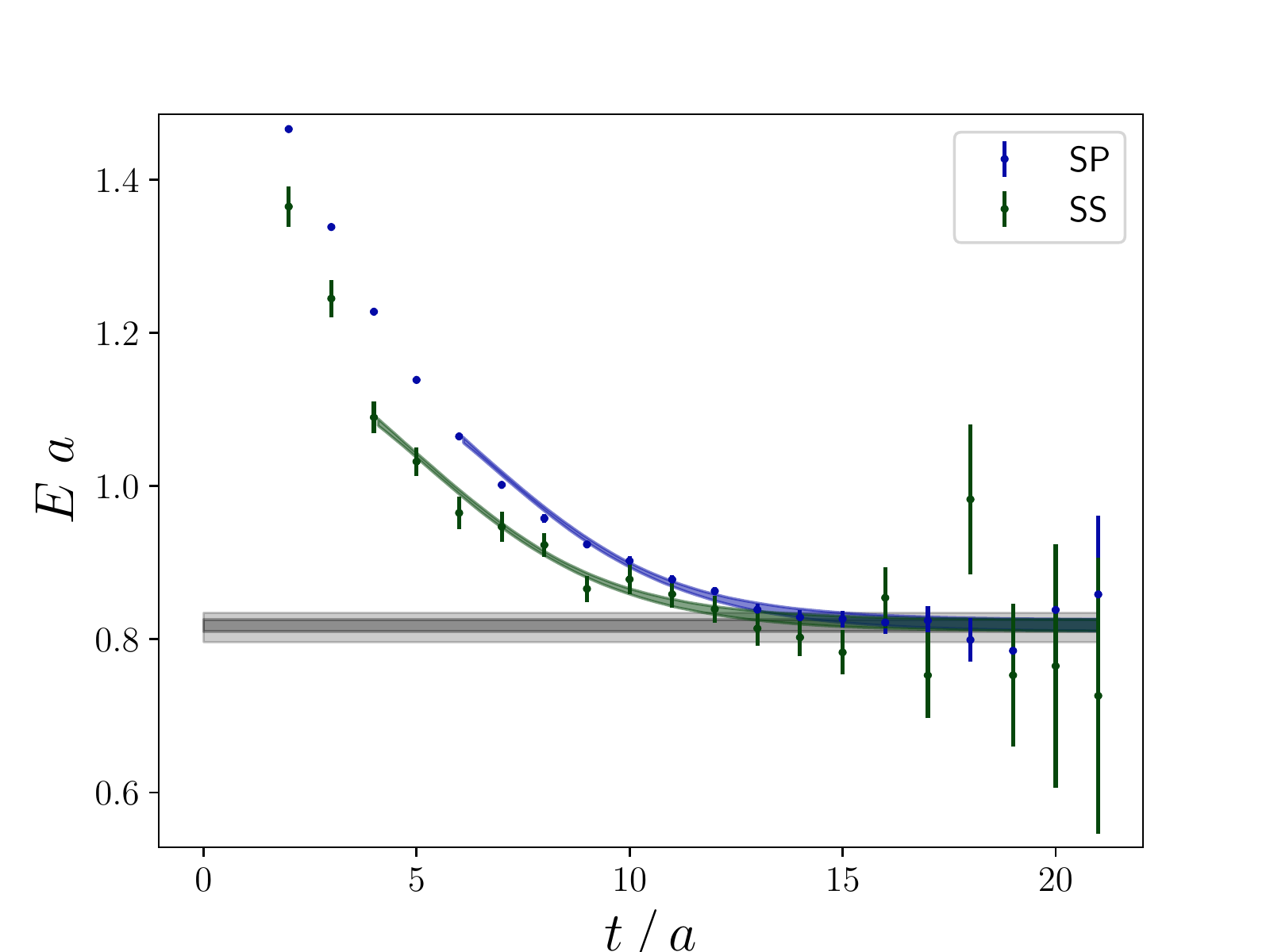} \hspace{10pt}
     \includegraphics[width=.40\textwidth]{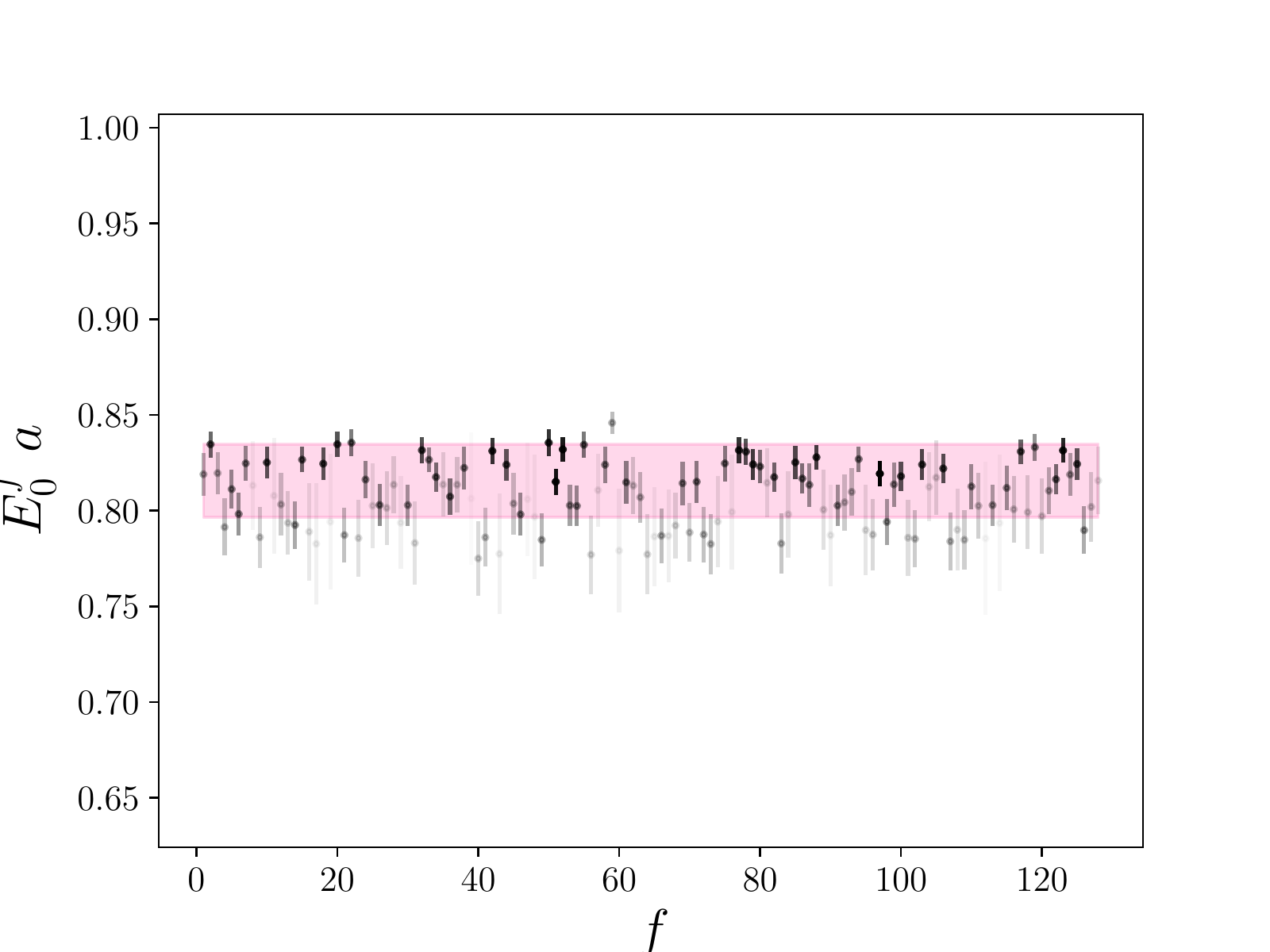} \\\vspace{-65pt}
     \begin{minipage}[c][150pt][t]{70pt} $pp \newline L/a=48$ \end{minipage}
     \includegraphics[width=.40\textwidth]{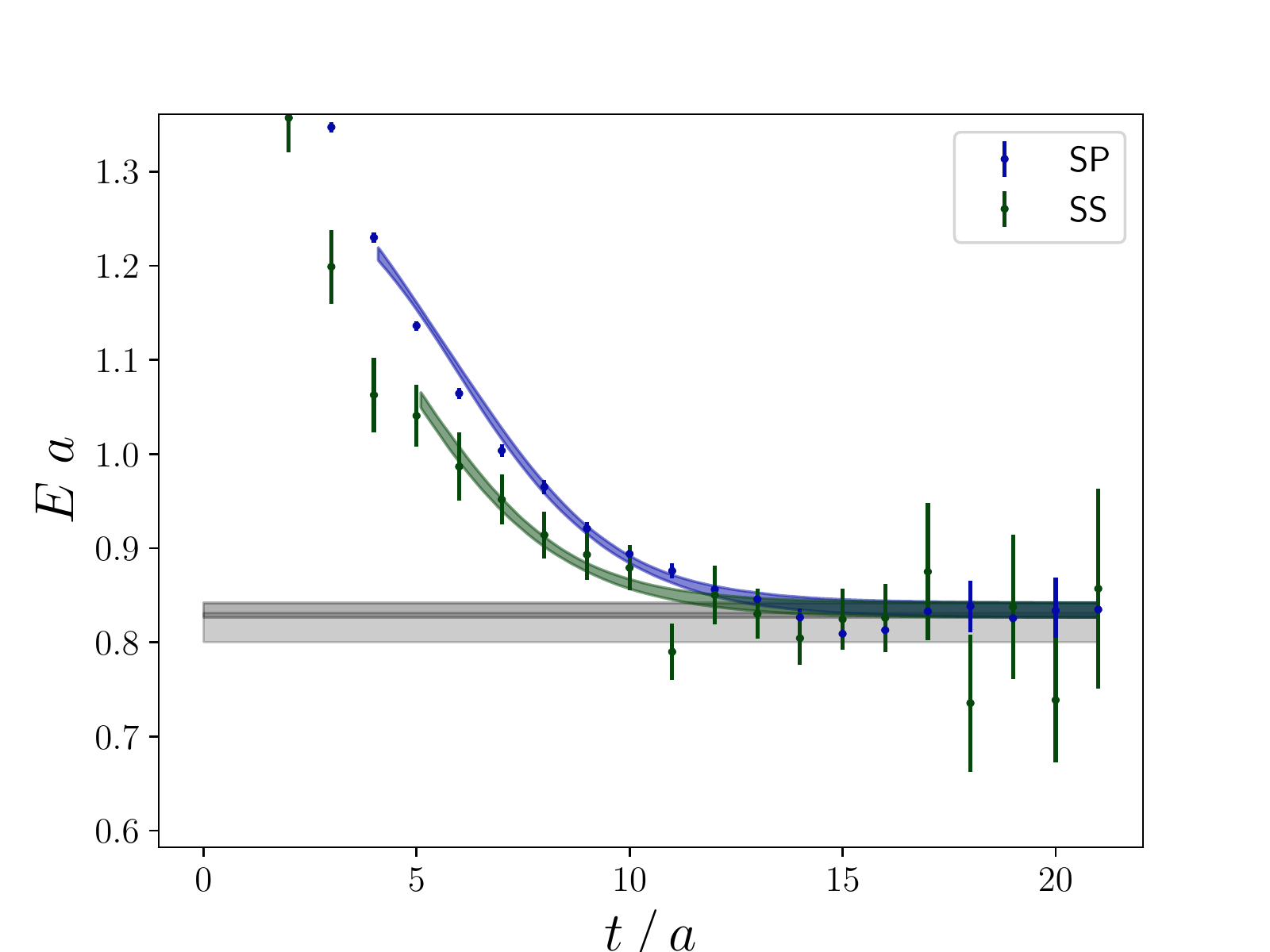} \hspace{10pt}
     \includegraphics[width=.40\textwidth]{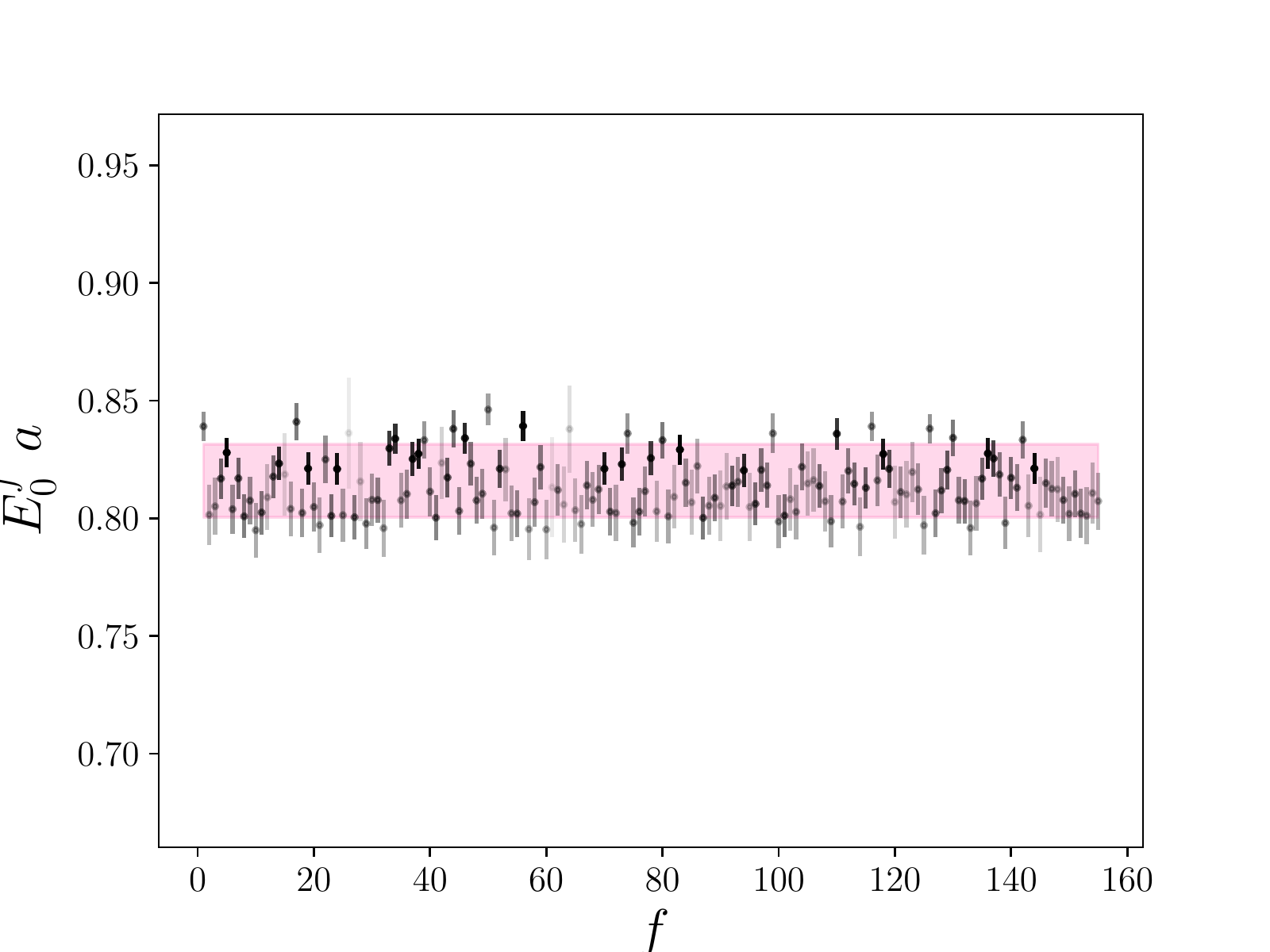} \\\vspace{-65pt}
     \begin{minipage}[c][150pt][t]{70pt} $nn \newline L/a=32$ \end{minipage}
     \includegraphics[width=.40\textwidth]{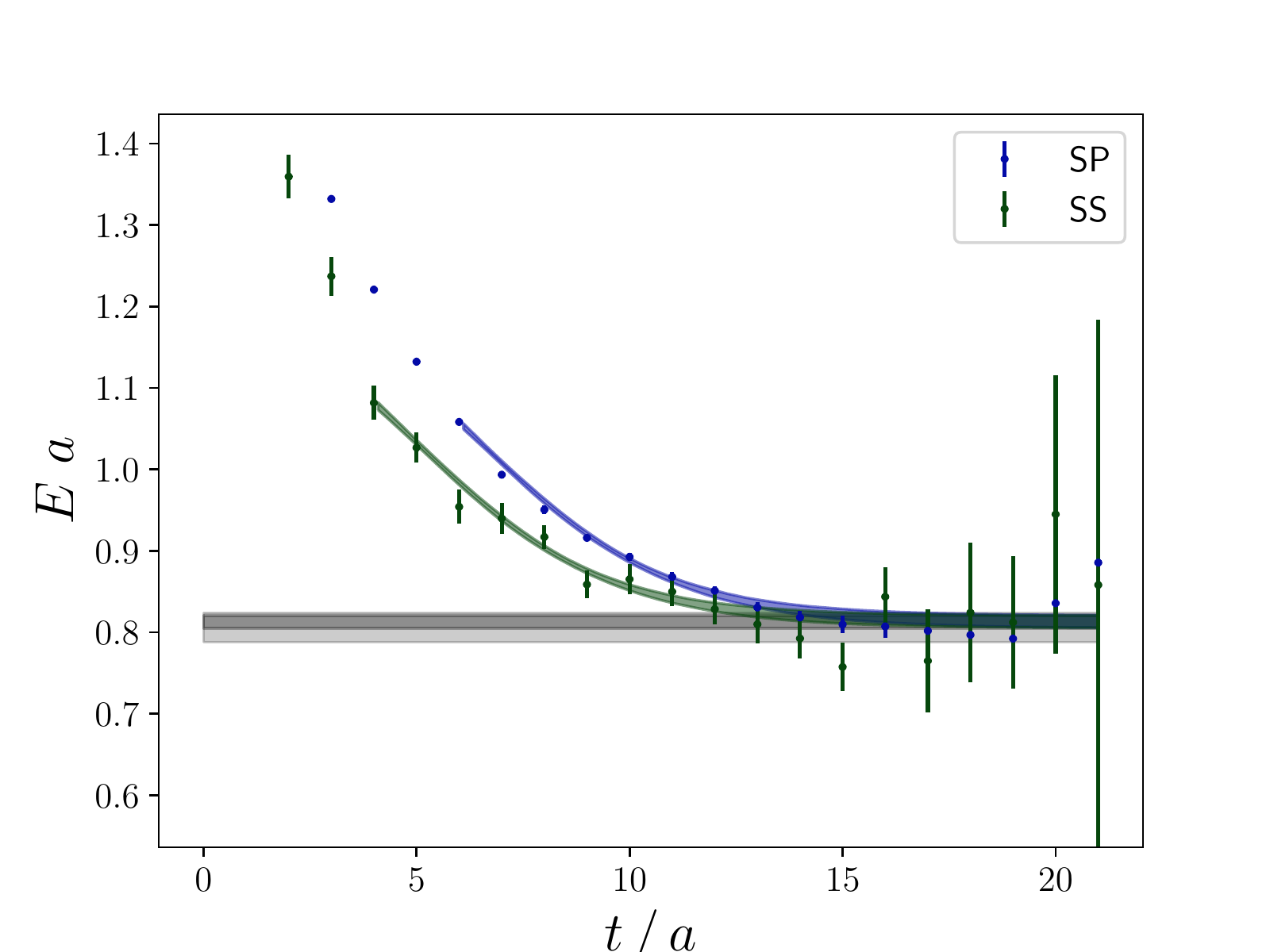} \hspace{10pt}
     \includegraphics[width=.40\textwidth]{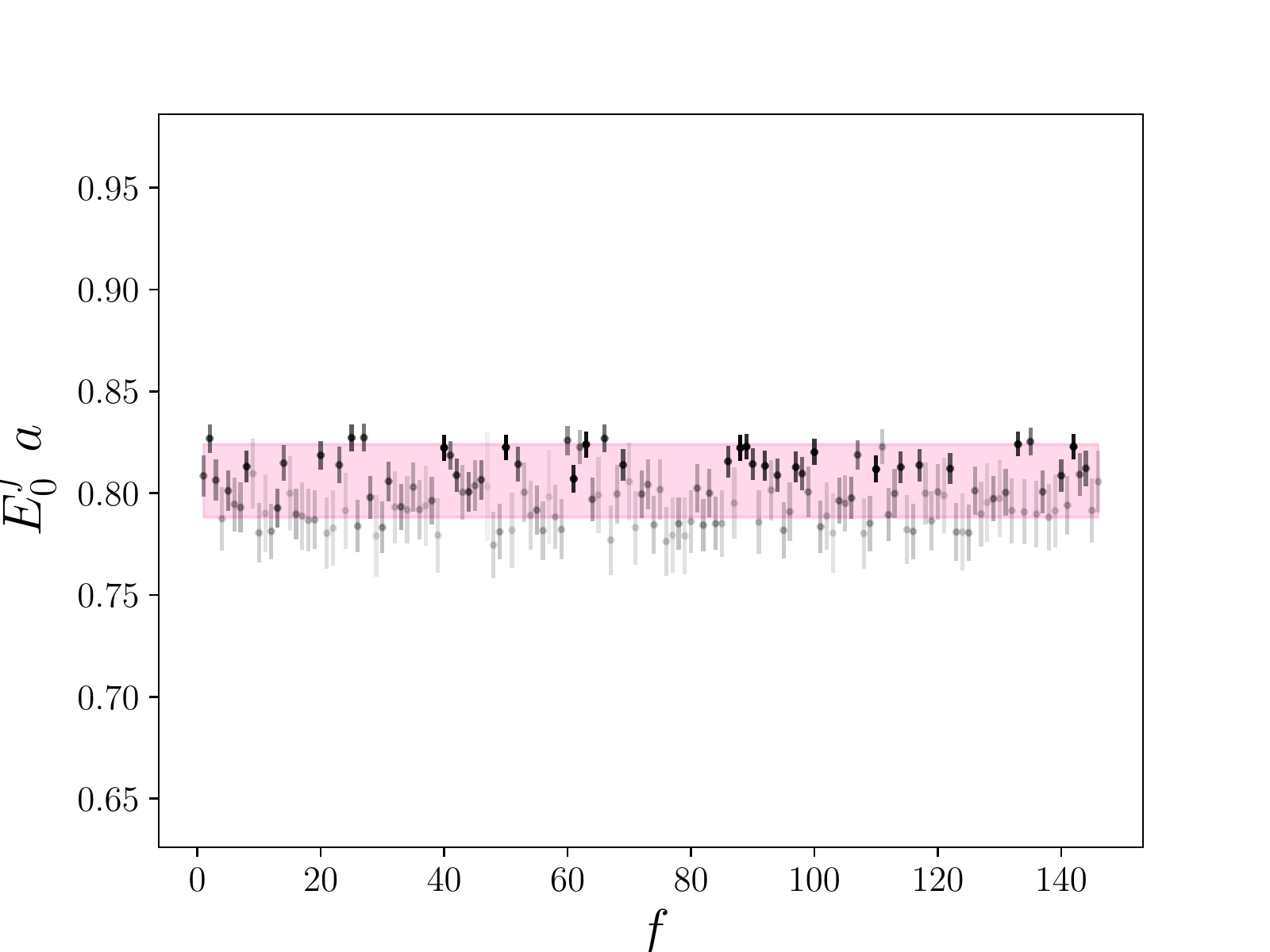} \\\vspace{-65pt}
     \begin{minipage}[c][150pt][t]{70pt} $nn \newline L/a=48$ \end{minipage}
     \includegraphics[width=.40\textwidth]{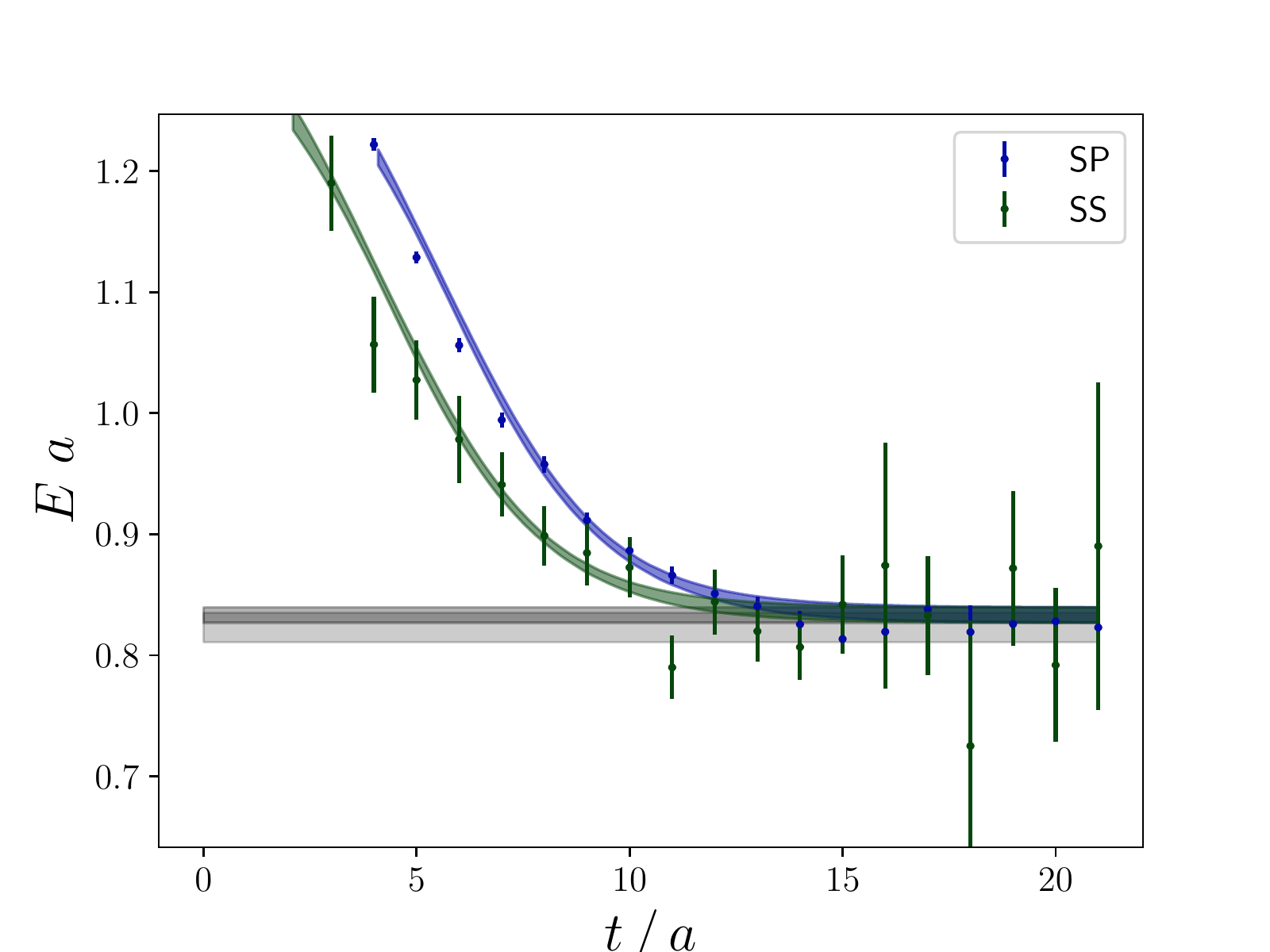} \hspace{10pt}
     \includegraphics[width=.40\textwidth]{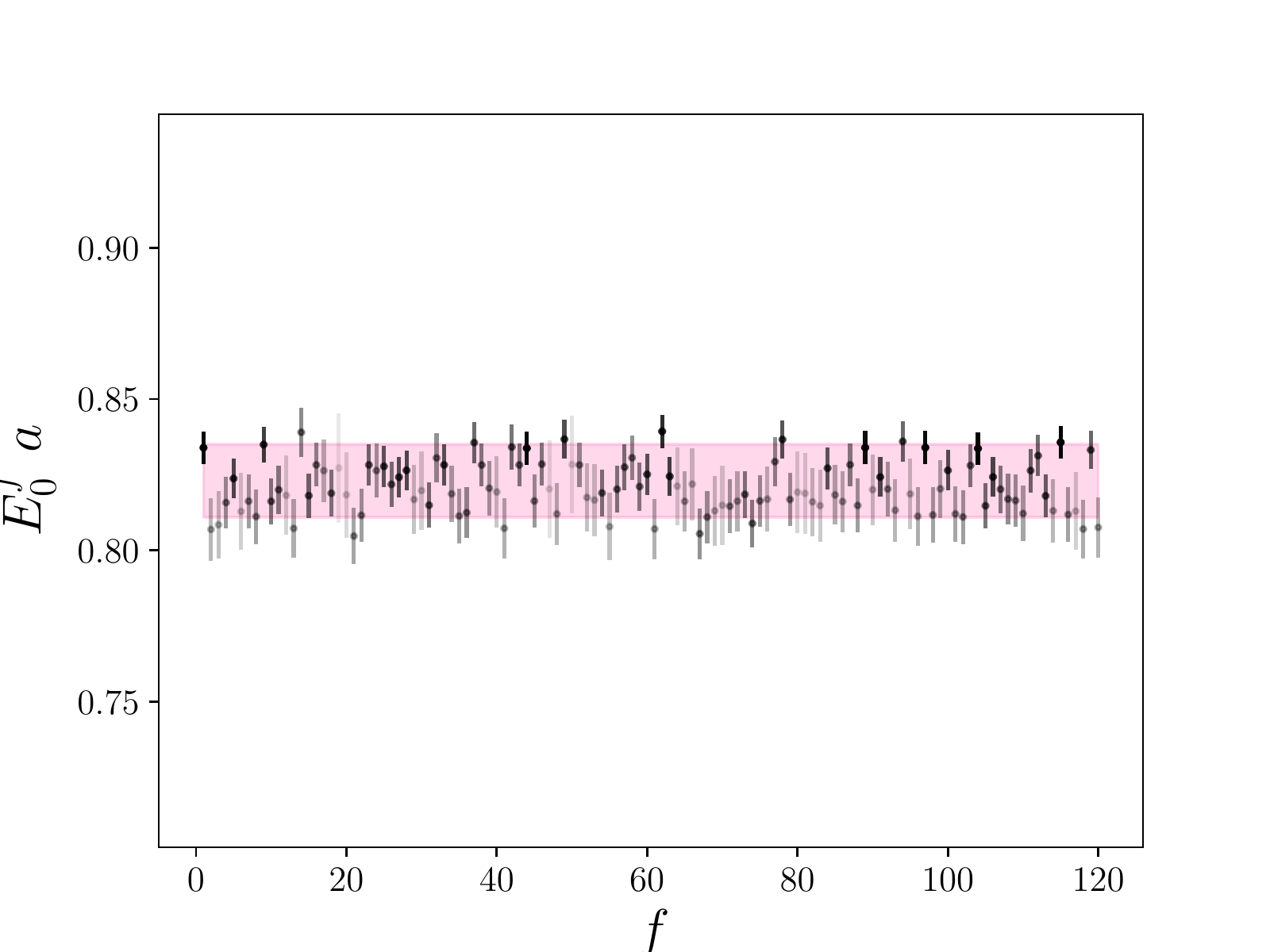} \\\vspace{-65pt}
   \caption{Fit results for systems of two protons and systems of two neutrons. The figures are analogous to Fig.~\ref{fig:pion48a}, see Appendix~\ref{app:fits} for a definition of the fitting procedure employed.
    \label{fig:baryon2a}}
\end{figure}

\begin{figure}
  \centering
  \begin{minipage}[c][150pt][t]{70pt} $pn({}^1S_0) \newline L/a=32$ \end{minipage}
     \includegraphics[width=.40\textwidth]{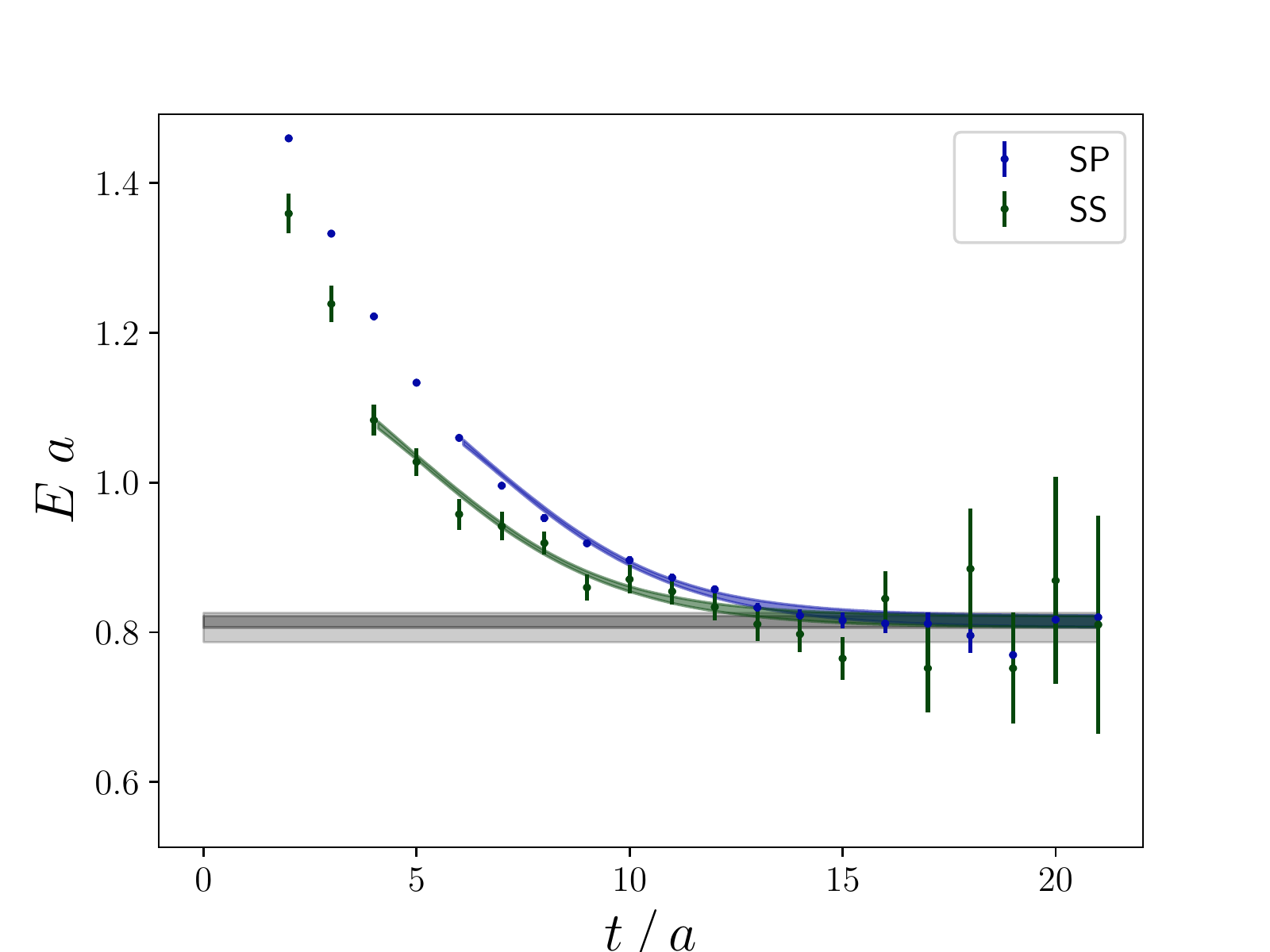} \hspace{10pt}
     \includegraphics[width=.40\textwidth]{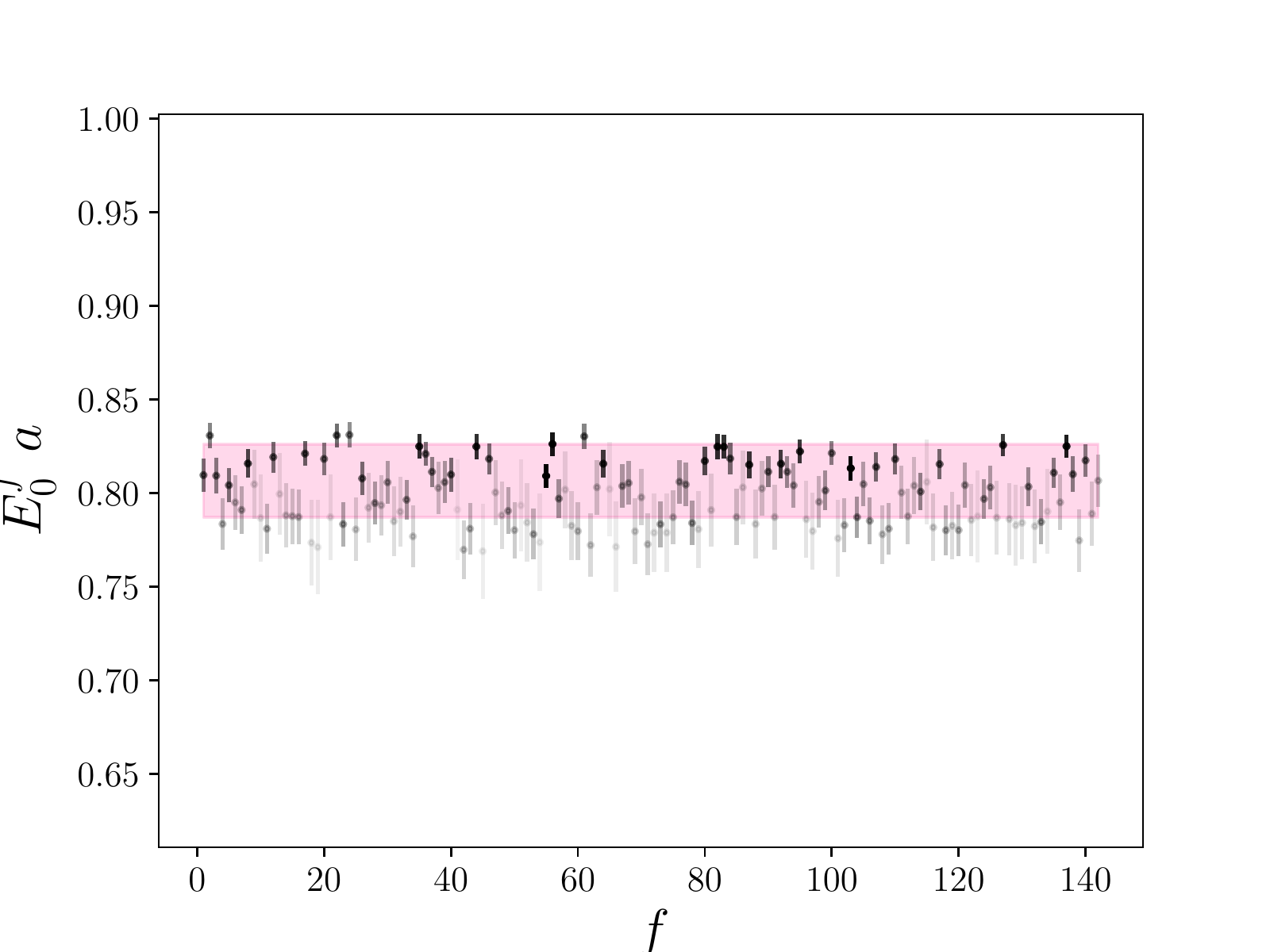} \\\vspace{-65pt}
     \begin{minipage}[c][150pt][t]{70pt} $pn({}^1S_0) \newline L/a=48$ \end{minipage}
     \includegraphics[width=.40\textwidth]{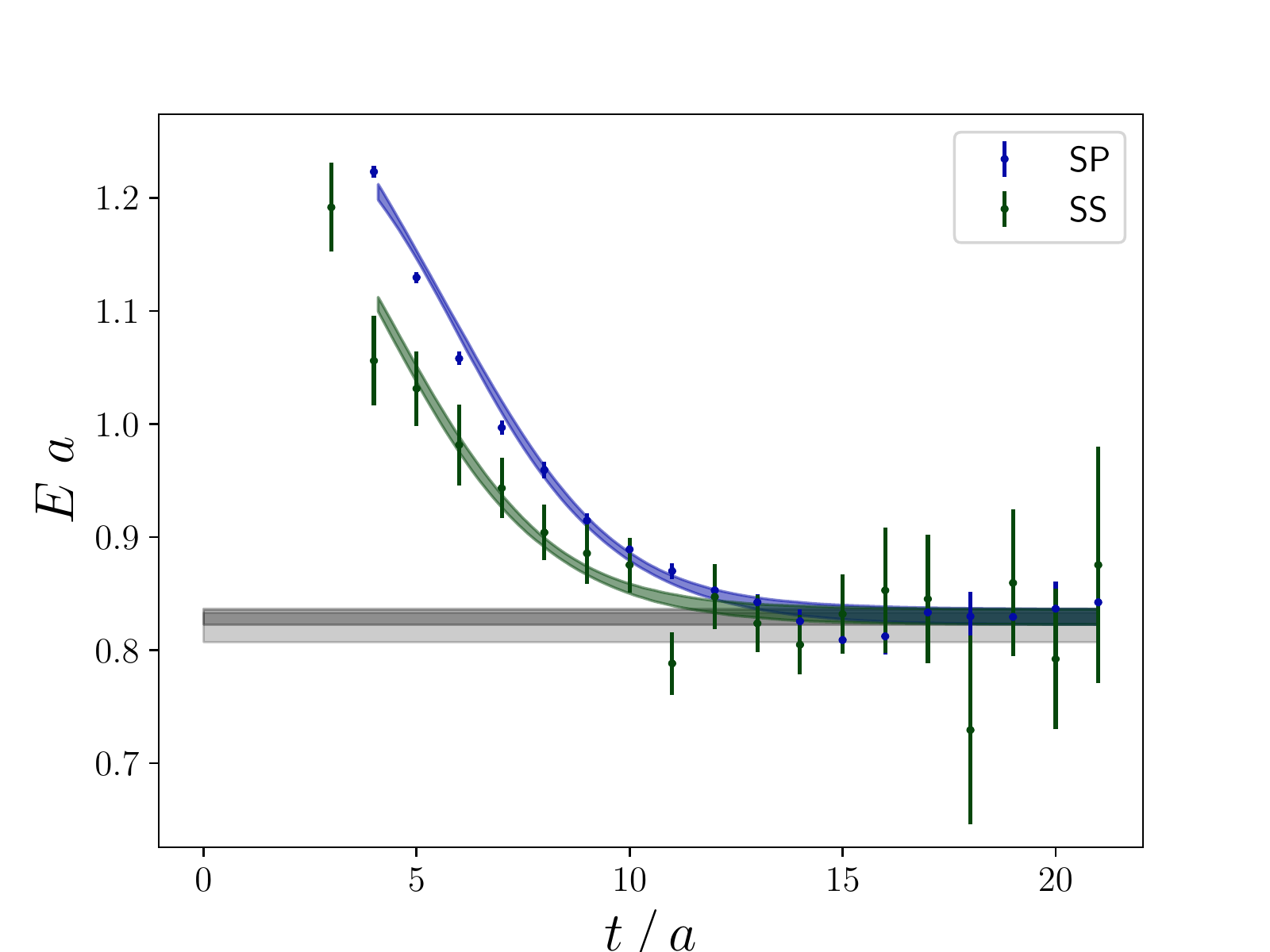} \hspace{10pt}
     \includegraphics[width=.40\textwidth]{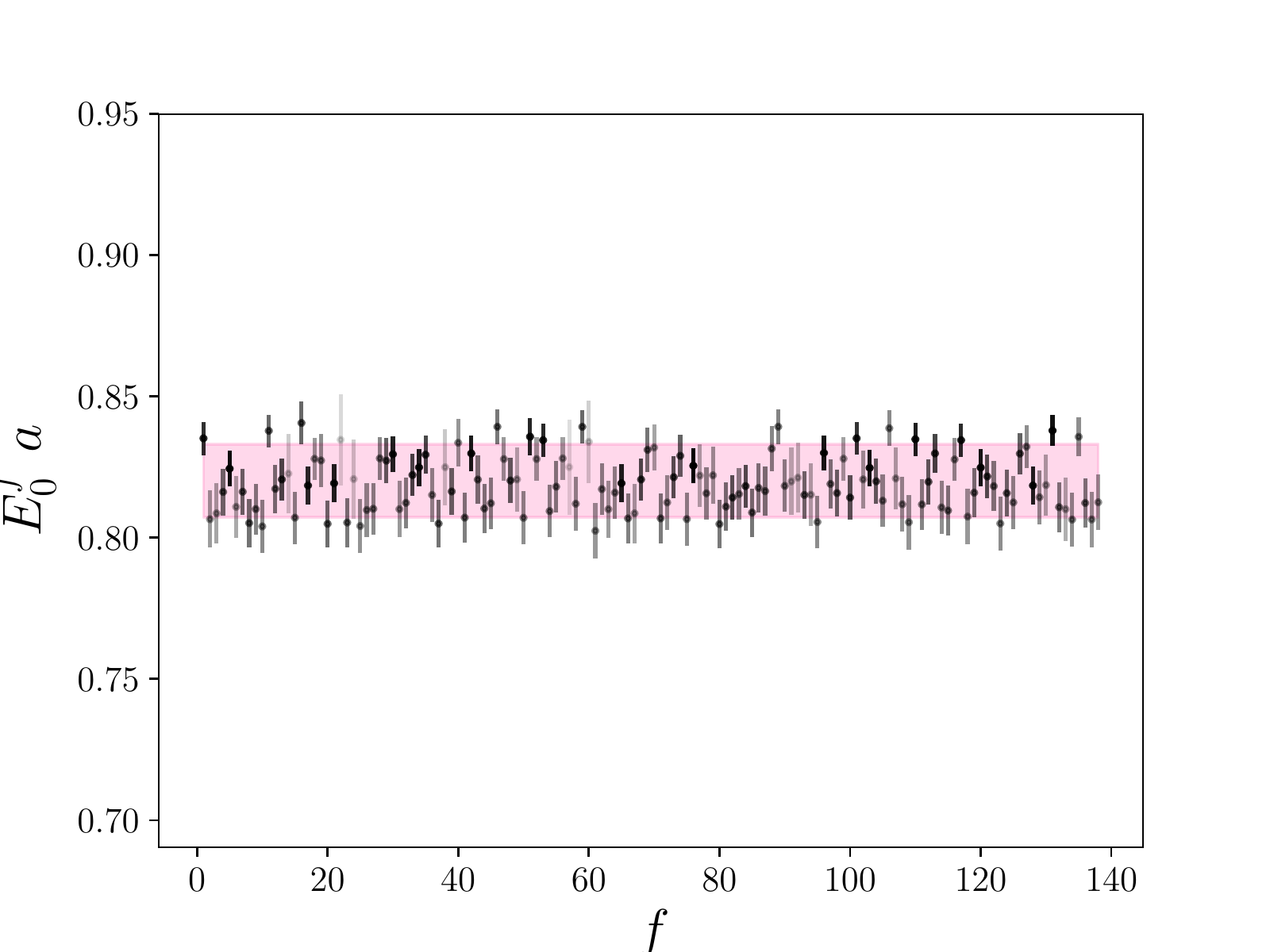} \\\vspace{-65pt}
     \begin{minipage}[c][150pt][t]{70pt} $pn({}^3S_1) \newline L/a=32$ \end{minipage}
     \includegraphics[width=.40\textwidth]{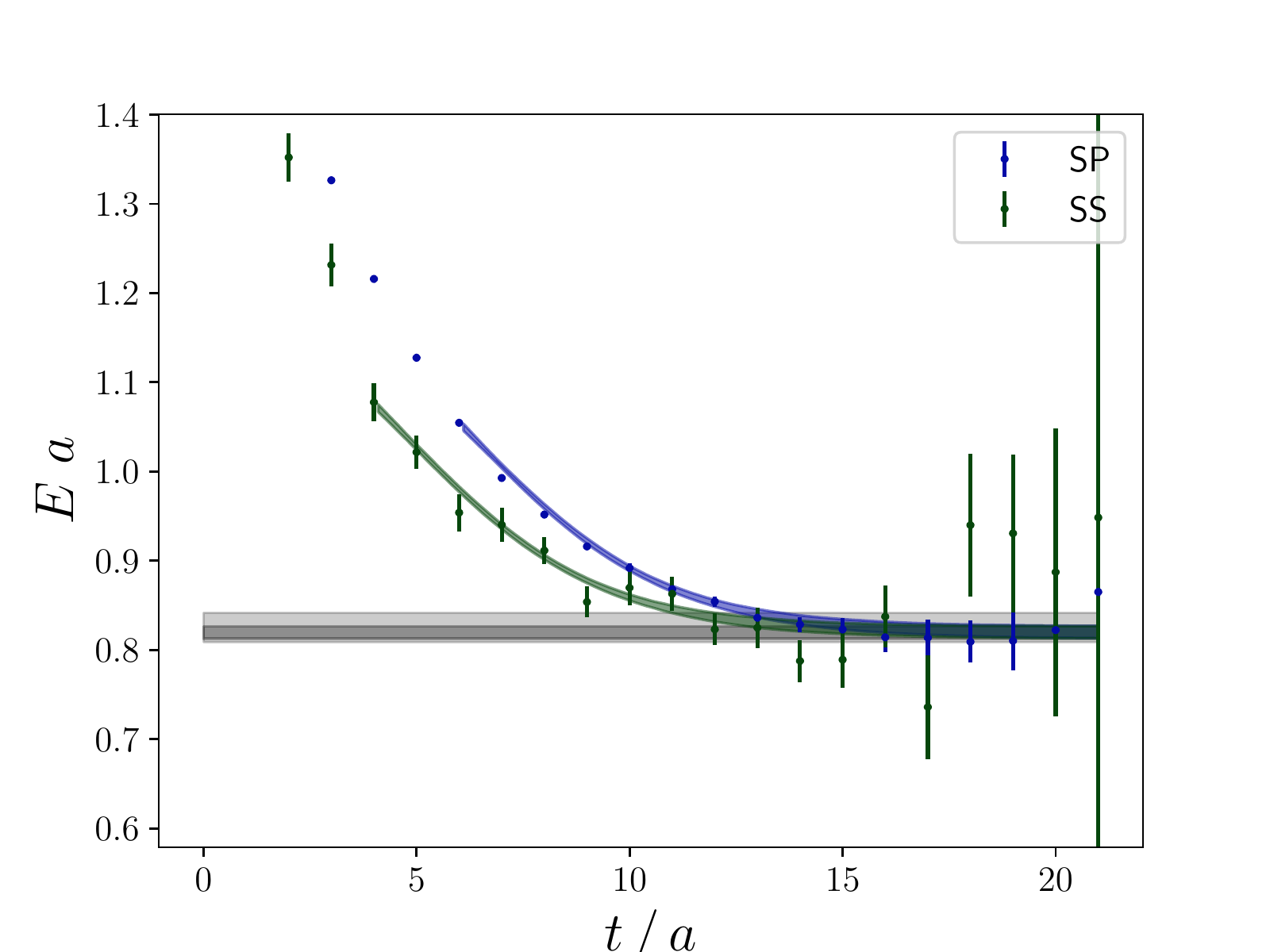} \hspace{10pt}
     \includegraphics[width=.40\textwidth]{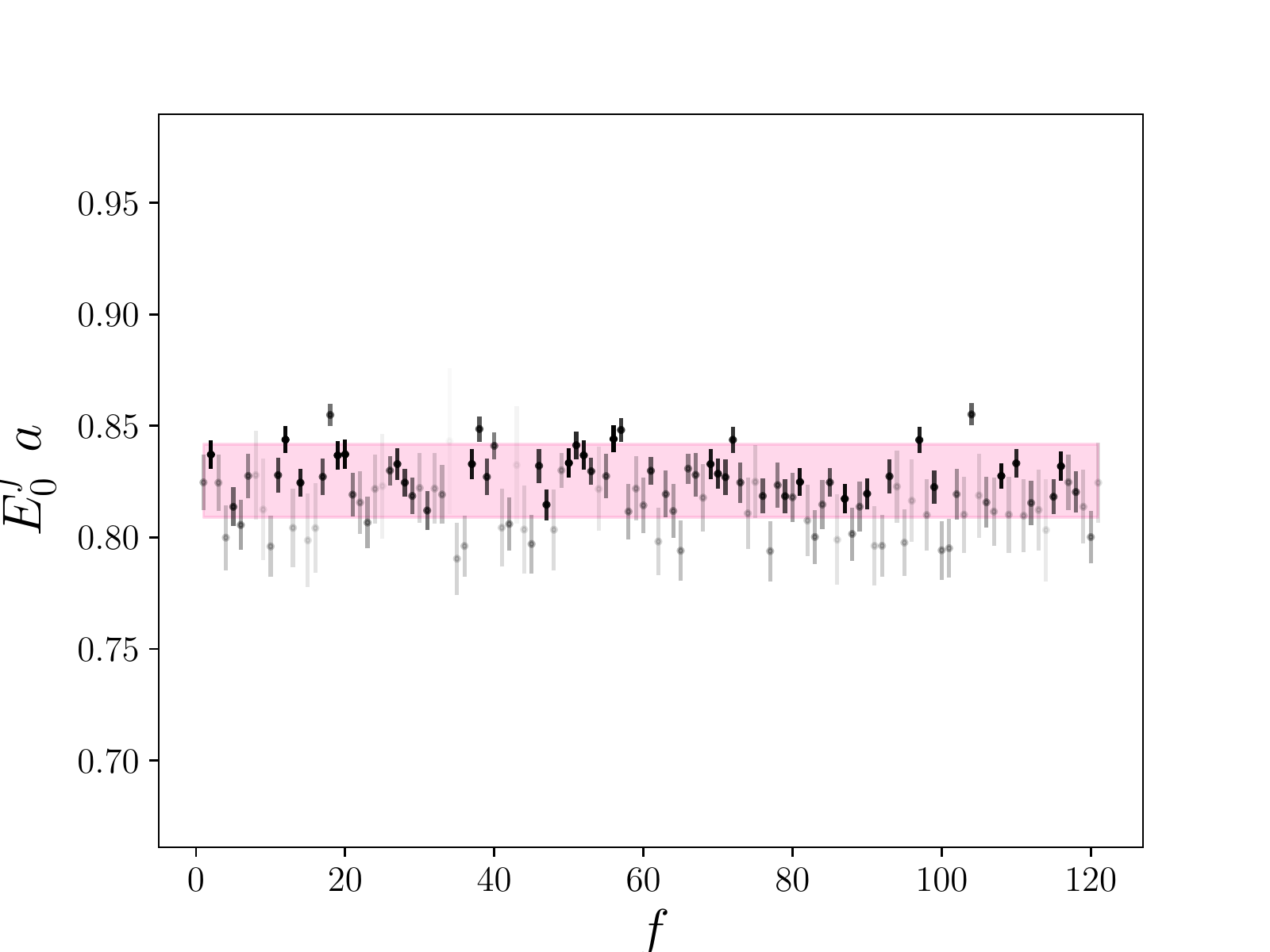} \\\vspace{-65pt}
     \begin{minipage}[c][150pt][t]{70pt} $pn({}^3S_1) \newline L/a=48$ \end{minipage}
     \includegraphics[width=.40\textwidth]{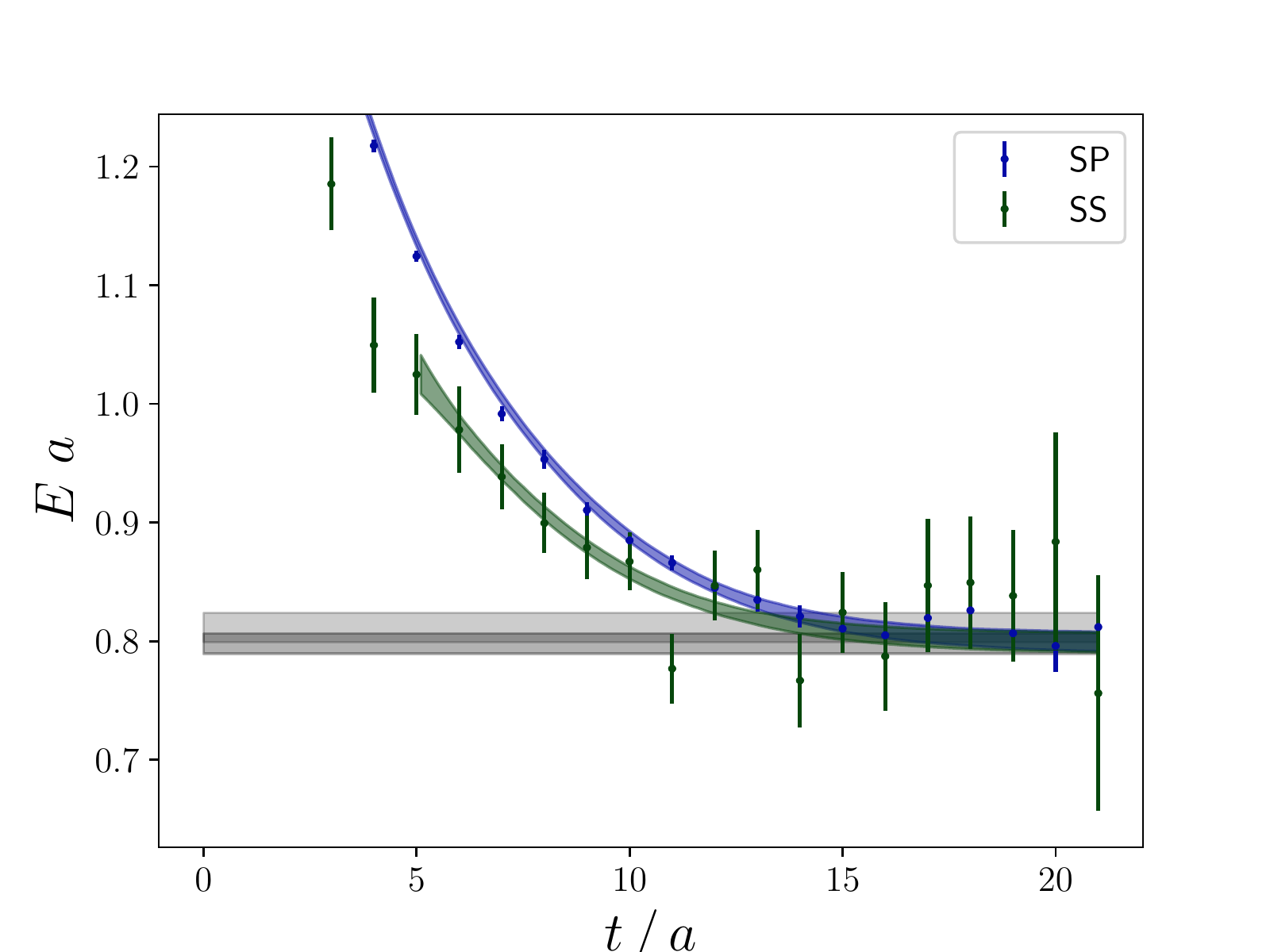} \hspace{10pt}
     \includegraphics[width=.40\textwidth]{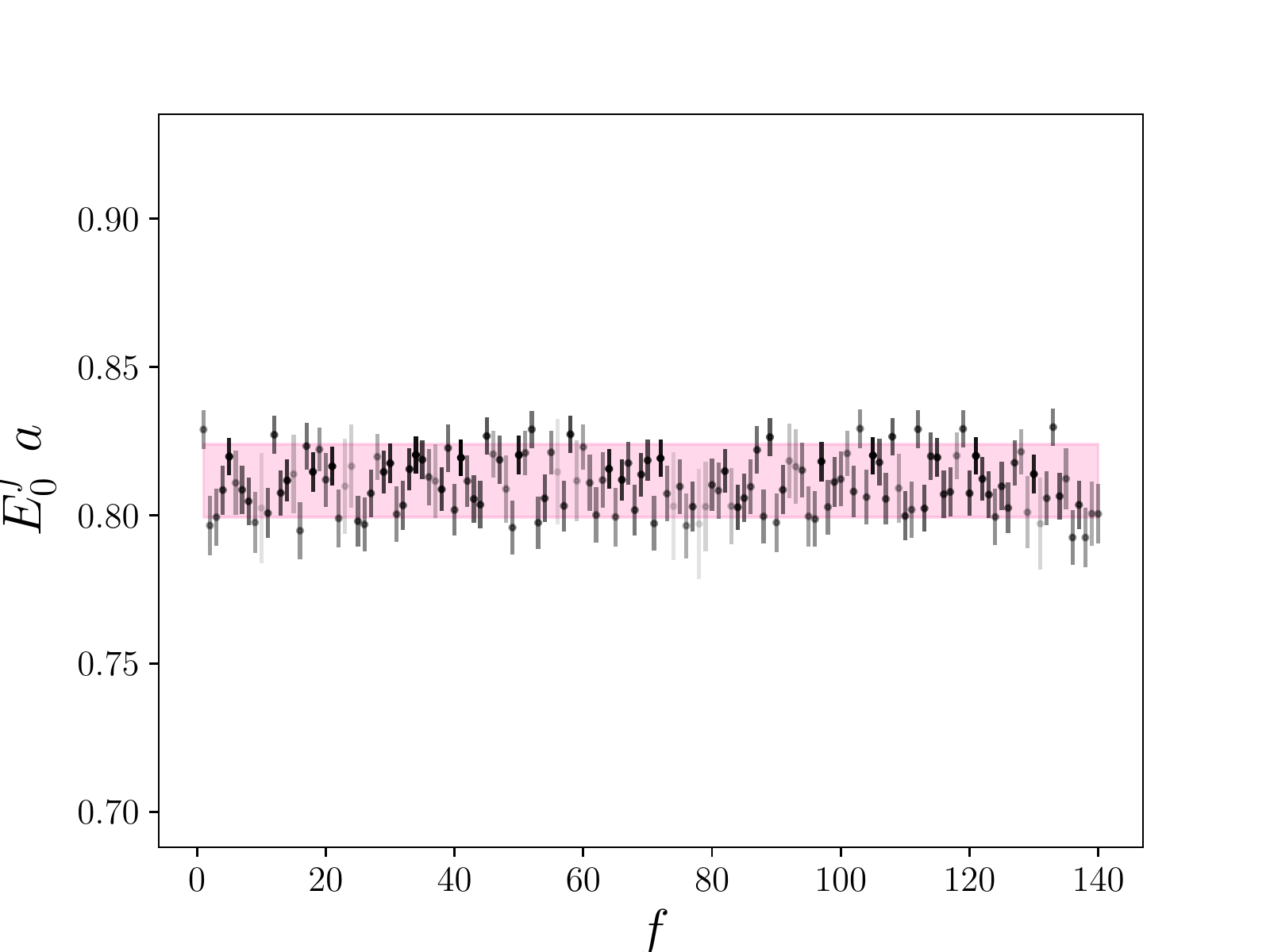} \\\vspace{-65pt}
   \caption{Fit results for systems of one neutron and one proton in the cubic irreps associated with ${}^3S_1$ and ${}^1S_0$ systems in the infinite-volume limit. The figures are analogous to Fig.~\ref{fig:pion48a}, see Appendix~\ref{app:fits} for a definition of the fitting procedure employed.
    \label{fig:baryon2b}}
\end{figure}

\begin{figure}
  \centering
   \begin{minipage}[c][150pt][t]{70pt} ${}^3\text{H} \newline L/a=32$ \end{minipage}
     \includegraphics[width=.40\textwidth]{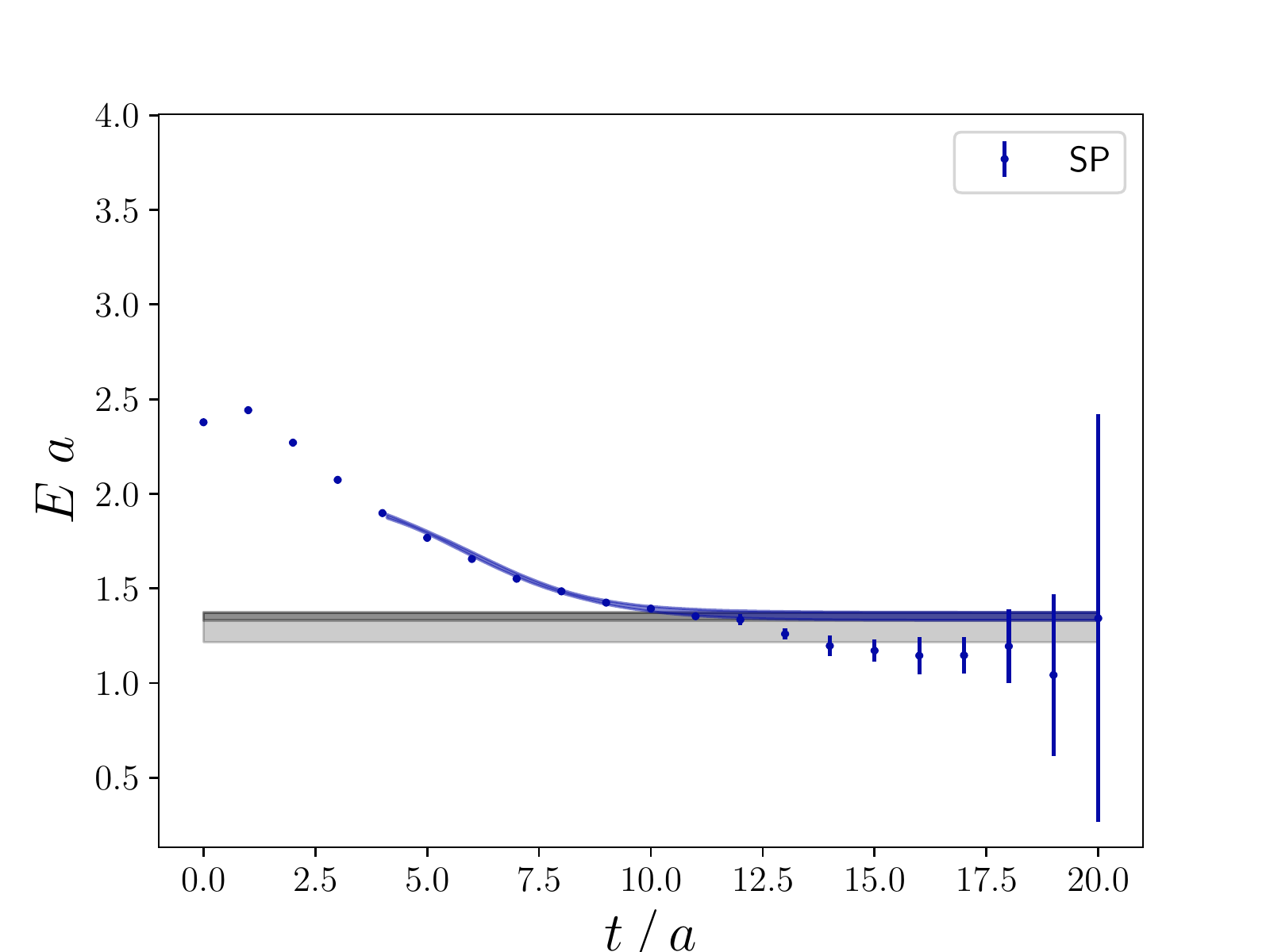} \hspace{10pt}
     \includegraphics[width=.40\textwidth]{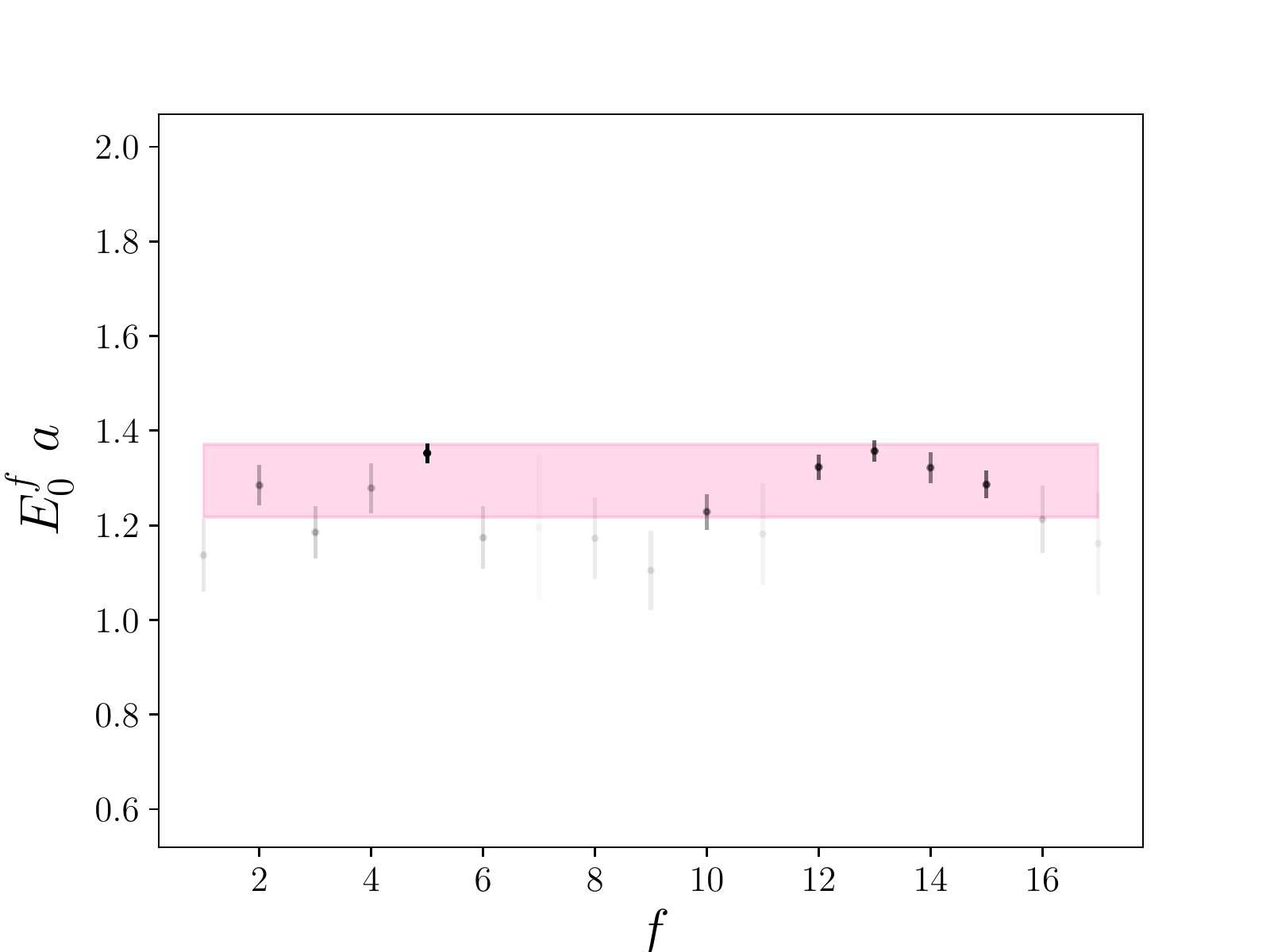} \\\vspace{-65pt}
     \begin{minipage}[c][150pt][t]{70pt} ${}^3\text{H} \newline L/a=48$ \end{minipage}
     \includegraphics[width=.40\textwidth]{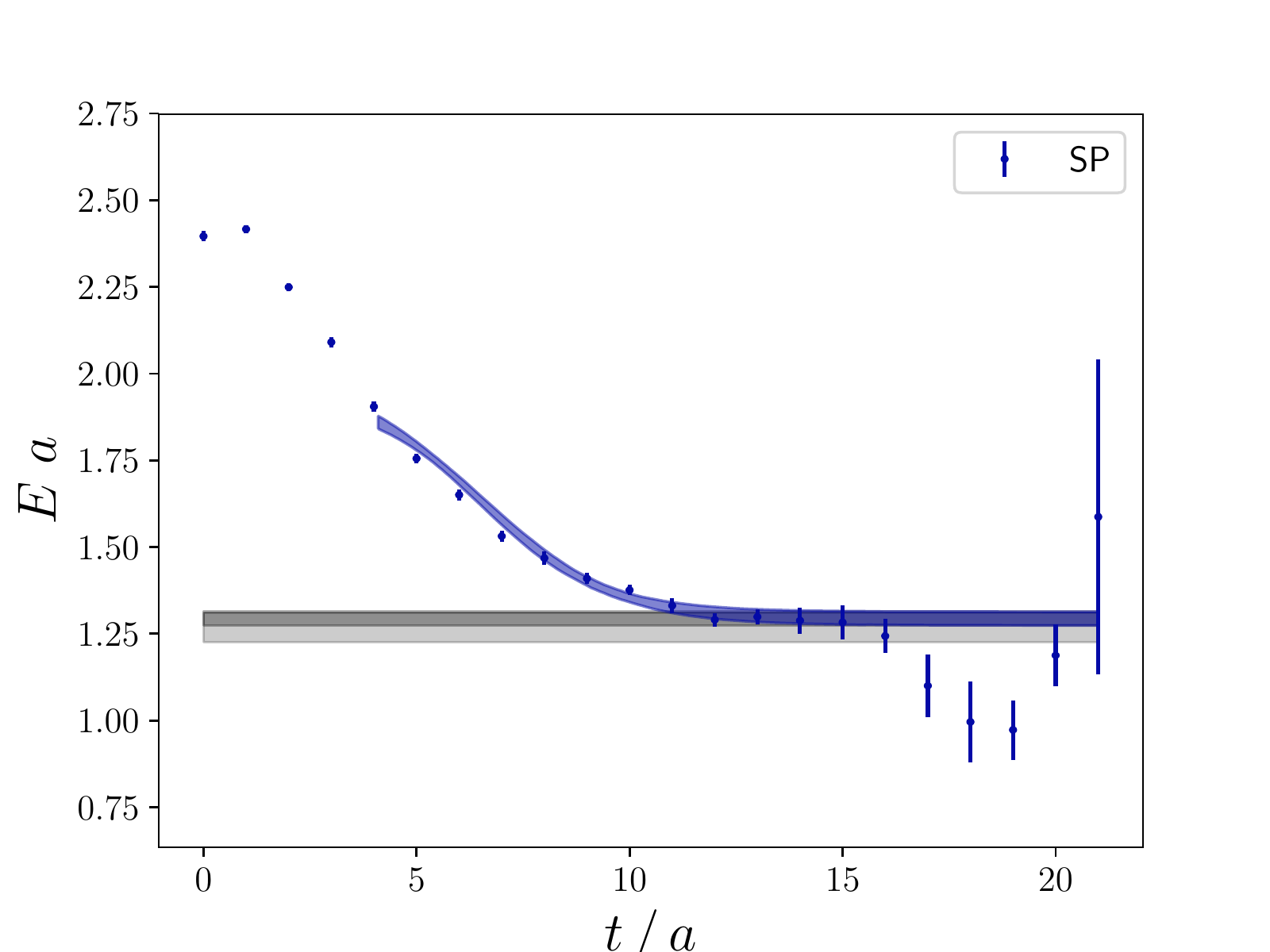} \hspace{10pt}
     \includegraphics[width=.40\textwidth]{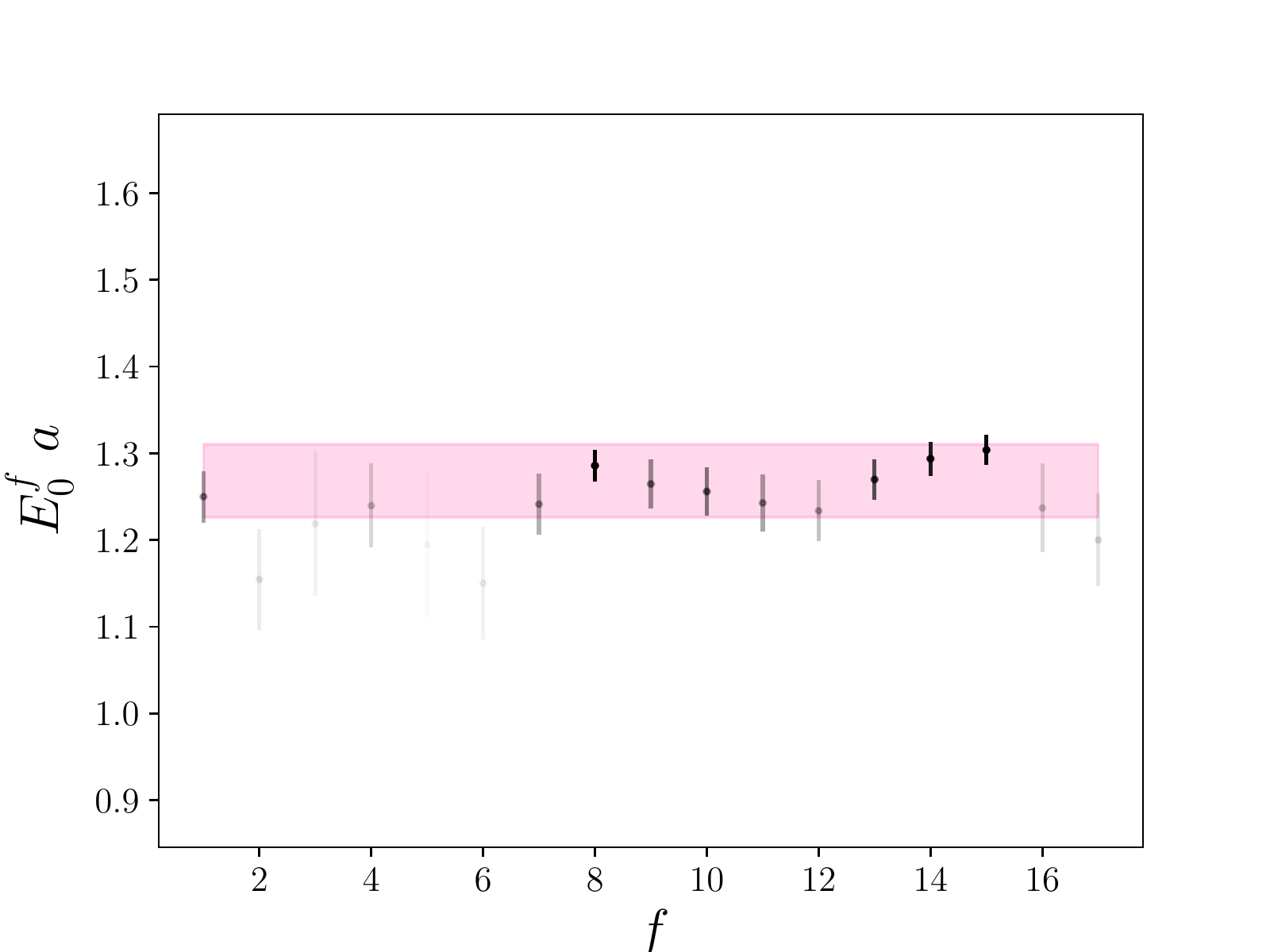} \\\vspace{-65pt}
     \begin{minipage}[c][150pt][t]{70pt} ${}^3\text{He} \newline L/a=32$ \end{minipage}
     \includegraphics[width=.40\textwidth]{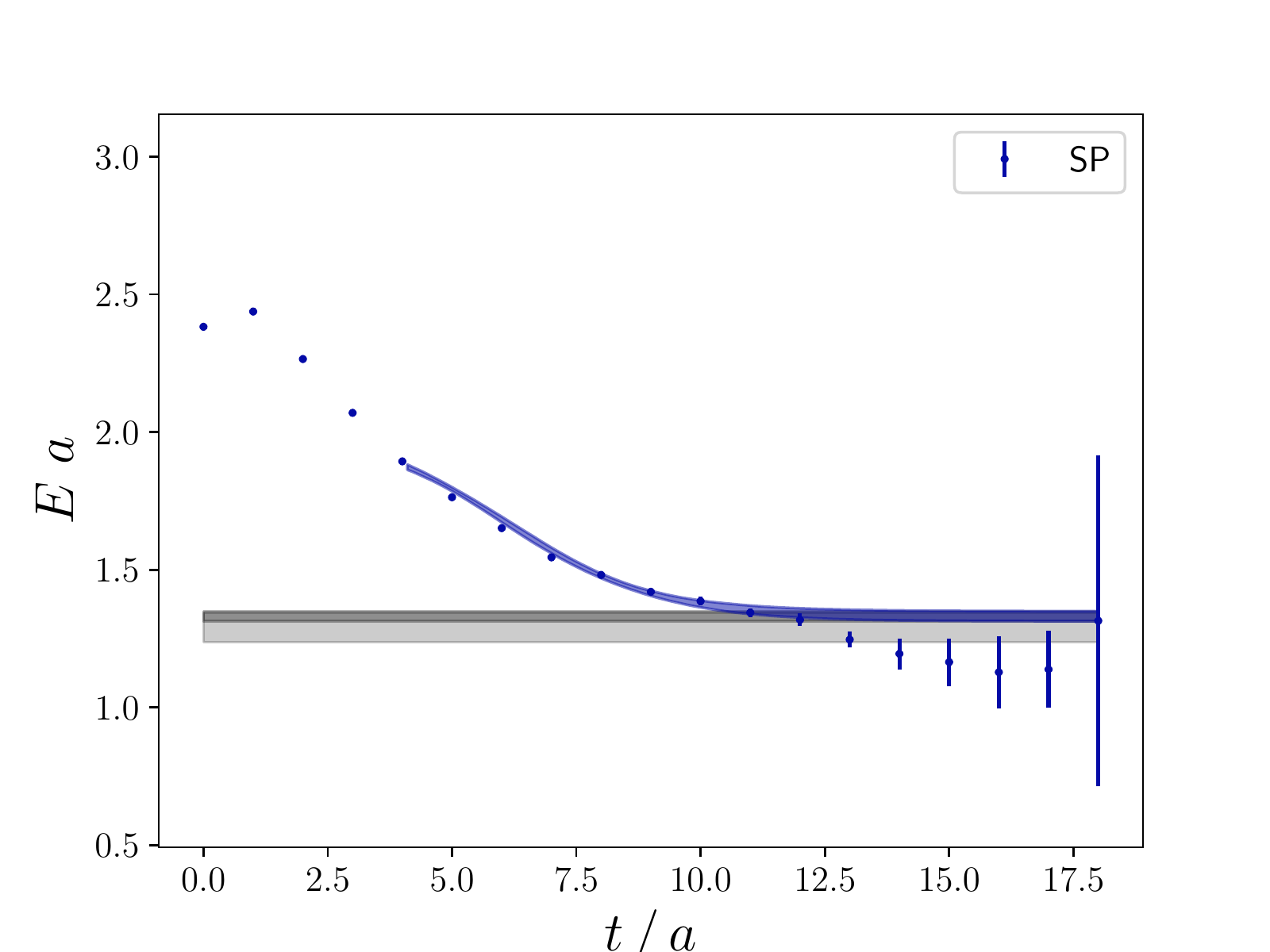} \hspace{10pt}
     \includegraphics[width=.40\textwidth]{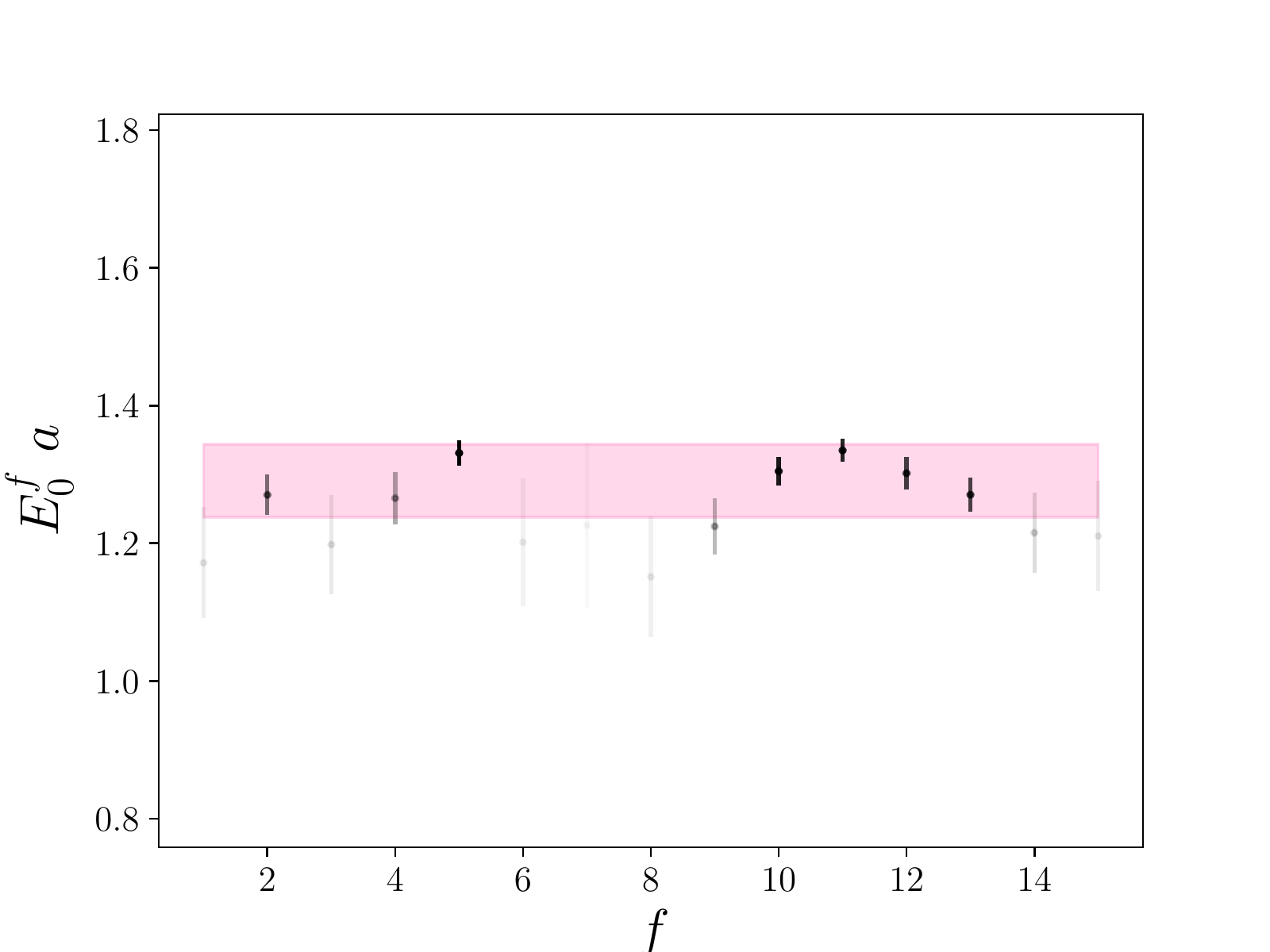} \\\vspace{-65pt}
     \begin{minipage}[c][150pt][t]{70pt} ${}^3\text{He} \newline L/a=48$ \end{minipage}
     \includegraphics[width=.40\textwidth]{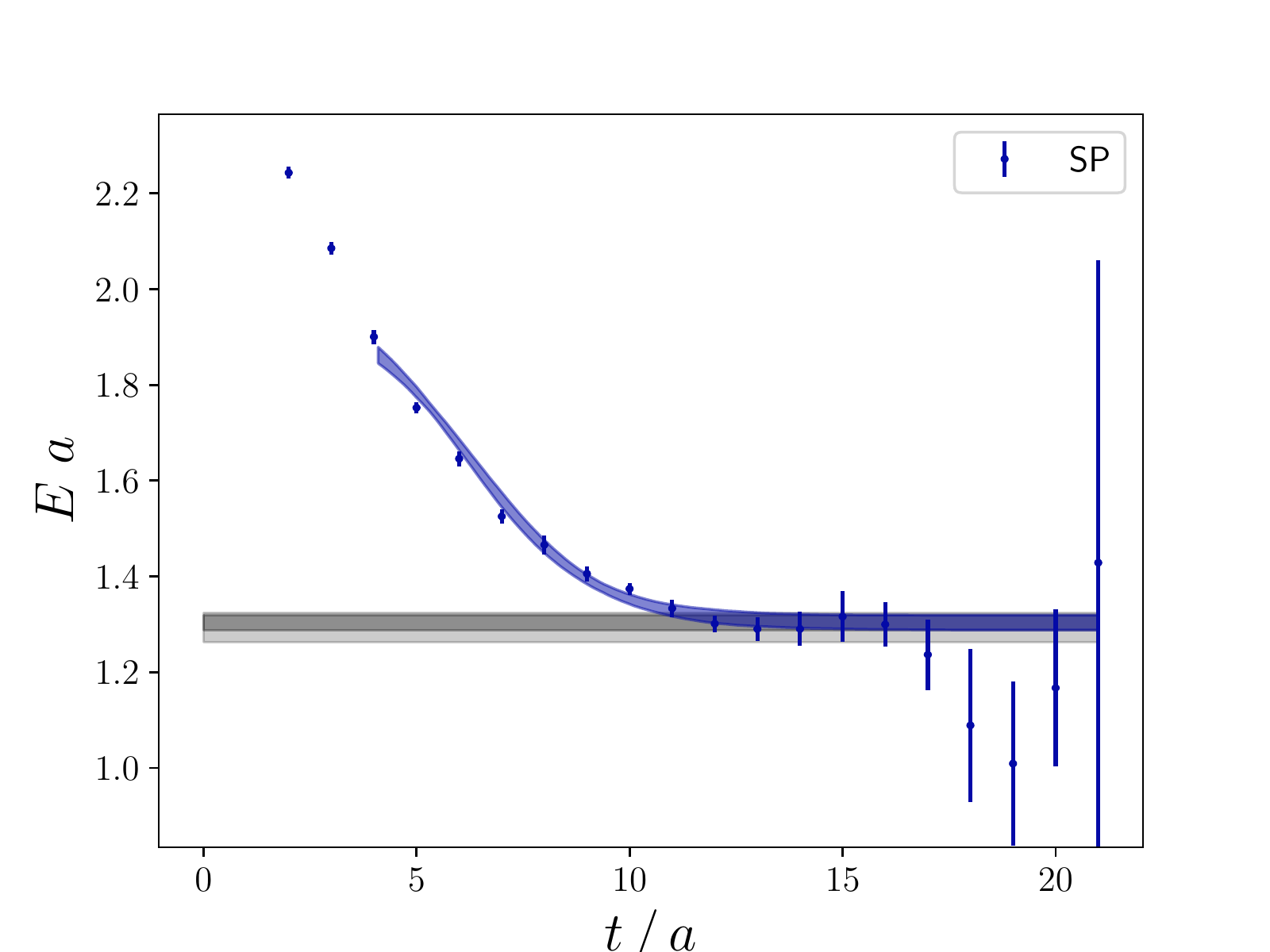} \hspace{10pt}
     \includegraphics[width=.40\textwidth]{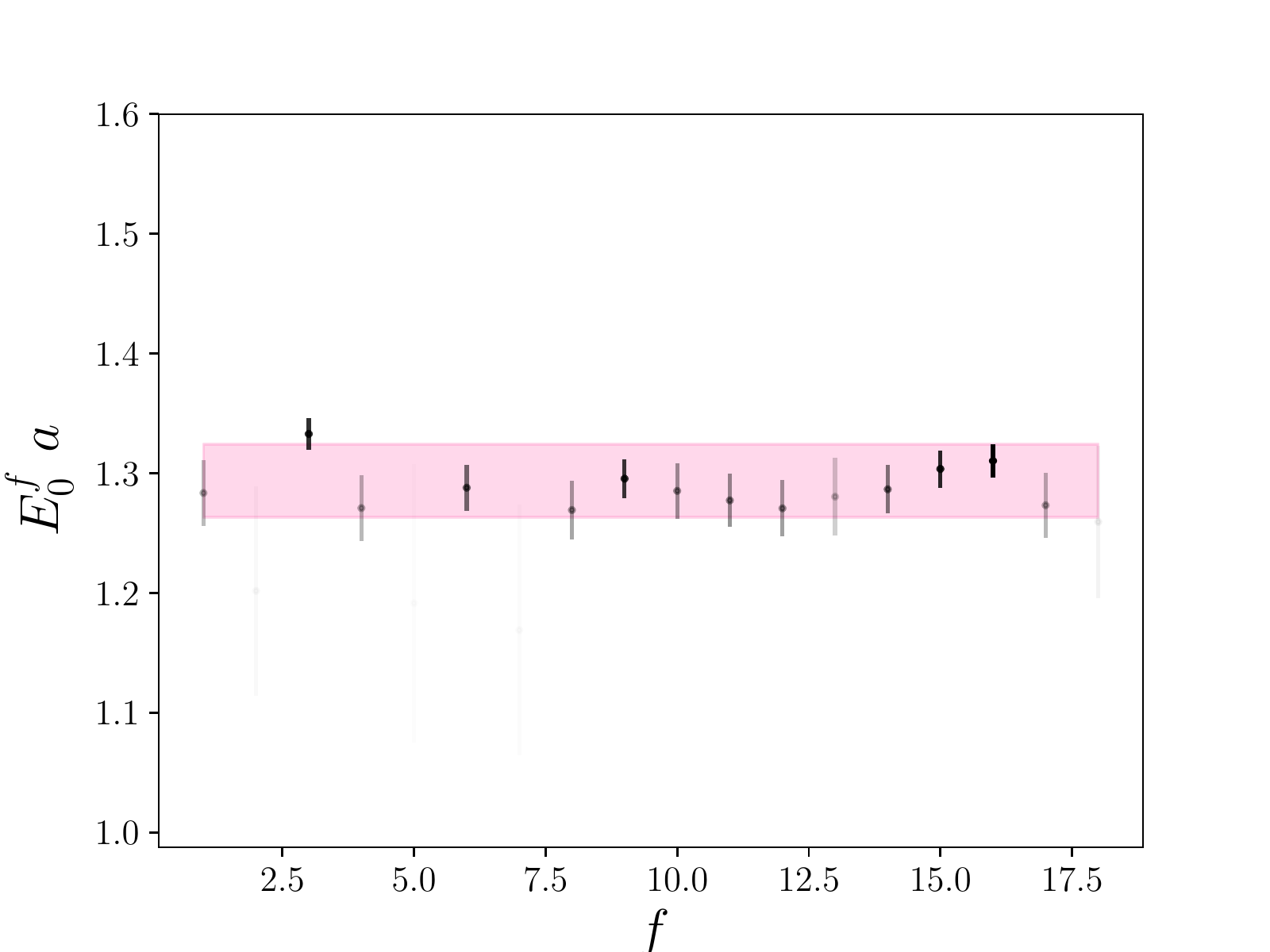} \\\vspace{-65pt}
   \caption{Fit results for three-nucleon systems ${}^3\text{H}$ and ${}^3\text{He}$. The figures are analogous to Fig.~\ref{fig:pion48a}, see Appendix~\ref{app:fits} for a definition of the fitting procedure employed.
    \label{fig:baryon3}}
\end{figure}

Figs.~\ref{fig:pion32a}-\ref{fig:pion32c} show fit results for $n \pi^+$ systems with $L/a = 32$.
Results with $L/a = 48$ are shown in Figs.~\ref{fig:pion48a}-\ref{fig:pion48c}.
Figs.~\ref{fig:kaon48a}-\ref{fig:kaon32c} show analogous fit results for $n \overline{K}^0$ systems with $L/a\in\{32,48\}$.
Single-nucleon fit results for $p$ and $n$ are shown in Fig.~\ref{fig:baryon1}.
Two-nucleon fit results are shown for $pp$ and $nn$ in Fig.~\ref{fig:baryon2a} and for $np$ systems in Fig.~\ref{fig:baryon2b}.
Three-nucleon fit results are shown in Fig.~\ref{fig:baryon3}.

\bibliography{QEDrefs}

\end{document}